\documentclass{jnmp01b}

\usepackage{amsmath}

\numberwithin{equation}{section}

\allowdisplaybreaks

\newtheorem{proposition}{Proposition}
\newtheorem*{theorem}{Theorem}

\makeatletter
\DeclareRobustCommand{\primfrac}[1]{%
  \PackageWarning{amsmath}{%
Foreign command \@backslashchar#1; %
\protect\frac\space or \protect\genfrac\space should be used instead%
  }
  \global\@xp\let\csname#1\@xp\endcsname\csname @@#1\endcsname
  \csname#1\endcsname
}
\makeatother

\begin{document}

\renewcommand{\evenhead}{M.\ Bruschi and F.\ Calogero}
\renewcommand{\oddhead}{Solvable and/or Integrable and/or Linearizable 
$N$-Body Problems}


\thispagestyle{empty}

\begin{flushleft}
\footnotesize \sf
Journal of Nonlinear Mathematical Physics \qquad 2000, V.7, N~3,
\pageref{firstpage}--\pageref{lastpage}.
\hfill {\sc Article}
\end{flushleft}

\vspace{-5mm}

\copyrightnote{2000}{M.\ Bruschi and F.\ Calogero}

\Name{Solvable and/or Integrable and/or Linearizable 
$N$-Body Problems in Ordinary (Three-Dimensional) 
Space. I}

\label{firstpage}

\Author{M. BRUSCHI and F. CALOGERO}

\Adress{Dipartimento di Fisica, Universita' di Roma ``La Sapienza,'' 
00185 Roma, Italy\\
Istituto Nazionale di Fisica Nucleare, Sezione di Roma}

\Date{Received February 9, 2000; 
Accepted April 1, 2000}

\begin{abstract}
\noindent
Several \textit{N}-body problems in ordinary (3-dimensional) space are 
introduced which are characterized by Newtonian equations of motion 
(``acceleration equal force;'' in most cases, the forces are 
velocity-dependent) and are amenable to exact treatment (``solvable'' and/or 
``integrable'' and/or ``linearizable''). These equations of motion are 
always rotation-invariant, and sometimes translation-invariant as well. In 
many cases they are Hamiltonian, but the discussion of this aspect is 
postponed to a subsequent paper. We consider ``few-body problems'' (with, 
say, \textit{N}=1,2,3,4,6,8,12,16,...) as well as ``many-body problems'' 
(\textit{N} an arbitrary positive integer). The main focus of this paper is 
on various techniques to uncover such \textit{N}-body problems. We do not 
discuss the detailed behavior of the solutions of all these problems, but we 
do identify several models whose motions are completely periodic or multiply 
periodic, and we exhibit in rather explicit form the solutions in some 
cases.
\end{abstract}



\section{Introduction} \label{I}


In this paper we exhibit several solvable and/or integrable and/or 
linearizable dynamical systems which are naturally interpretable as 
\textit{N}-body problems in ordinary (three-dimensional) space, inasmuch as 
they are characterized by equations of motion of Newtonian type 
(``acceleration equal force''),
\begin{equation*}
\ddot {\vec {r}}_{n} = \vec {f}_{n} \left( {\vec {r}_{m} ,\dot {\vec 
{r}}_{m} ;m = 1,...,N} \right) \  .
\tag{1.1}
\end{equation*}

Here and throughout, superimposed arrows denote 3-dimensional vectors, and 
superimposed dots denote differentiation with respect to the independent 
variable $t$ (``time''). Of course the 3-vector $\vec {r}_{n} \left( {t} 
\right)$ identifies the position of the \textit{n}-th particle at time $t$. 
Unless otherwise indicated, the index $n$, as well as most other indices, 
runs from 1 to $N$. In some cases $N$ is a small integer, say $N = 
1,2,3,4,6,8,12,16$ (``few-body problems''); in others, it is an arbitrary 
positive integer (``many-body problems'').

Our main purpose in this paper (the first of a series) is to introduce 
techniques whereby many such models can be manufactured. Only some of these 
models are actually exhibited herein; and only in some cases do we discuss 
in any detail the solutions of their equations of motion, namely the actual 
motions entailed by the \textit{N}-body problem under consideration. Yet we 
trust our treatment demonstrates the interest of these models. It should be 
emphasized in this respect that, until recently, essentially \textit{all} 
known solvable and/or integrable and/or linearizable \textit{N}-body models 
were embedded in one-dimensional space (except for the important but 
essentially trivial case of a collection of harmonic oscillators, 
characterized by linear equations of motion). Only recently some solvable 
and/or integrable many-body problems in two dimensional space were 
introduced [1], as well as some solvable few-body problems in 2- and 
3-dimensional space [2], by techniques completely different from those 
introduced in this paper. The dynamical behaviors exhibited by these models 
are much richer than those characterizing models in one-dimensional space, 
where scattering processes are, to say the least, not very spectacular, and 
particles cannot get past each other without colliding (which is instead not 
the case already in two-dimensional space). Indeed, a number of interesting 
phenomena are exhibited by solvable models in the plane: nontrivial 
scattering, nontrivial periodic motions, limit cycles, etc. [1] . 
Three-dimensional space allows for an even richer gamut of motions; 
moreover, we happen to live in three-dimensional space, and we are therefore 
specially interested in models that evolve in the same environment. Indeed, 
we suggest that some of the models treated herein are suitable to test the 
reliability of numerical codes for the integration of many-body motions in 
three-dimensional space, such as those used in ``realistic'' molecular 
dynamics calculations. Some of these models are also likely to have 
applicative relevance, but we do not elaborate on this aspect in the present 
paper.

These remarks underscore the relevance of the results reported herein, and 
justify our exclusive interest, in this paper, on the 3-dimensional case 
(although it will be clear in the following that certain results are easily 
extendable, for instance, to 2- and 4-dimensional spaces). But let us also 
emphasize that the results reported herein consist mainly of appropriate 
reinterpretations of more or less known findings concerning 
\textit{treatable models for the time-evolution of matrices} as well as 
\textit{appropriate parameterizations of matrices in terms of 3-vectors}, the 
combination of these two techniques to yield $N$-body problems in 
3-dimensional space being, as explained below, the main new idea underlying 
the approach introduced in this paper.

We deem that evolution equations of type (1.1) qualify as Newtonian 
equations of motion for an \textit{N}-body problem in 3-dimensional space 
\textit{iff} they are \textit{rotation-invariant}; it is otherwise trivially 
easy to manufacture solvable and/or integrable \textit{N}-body problems in 
\textit{d}-dimensional space, with arbitrary $d$, by an appropriate 
reinterpretation of known solvable and/or integrable \textit{M}-body 
problems in \textit{one-dimensional} space, with $M = Nd$.

Indeed, the equations of motion of type (1.1) we consider below are all 
\textit{rotation-invariant}, namely the ``forces'' $\vec {f}_{n} $ can 
generally be written as follows:
\begin{gather*}
\vec{f}_n  = \sum_{m=1}^{N}  [\vec{r}_{m}\varphi_{nm} + 
\dot{\vec{r}}_m \tilde{\varphi}_{nm}]
+ \sum_{m_{1},m_{2}=1}^{N}~ 
[\vec{r}_{m_{1}}\wedge \vec{r}_{m_{2}} \varphi_{nm_{1}m_{2}} 
\\*
\qquad{}+ \vec{r}_{m_{1}}\wedge~
\dot{\vec{r}}_{m_{2}}¥~ 
\tilde{\varphi}_{nm_{1}m_{2}¥}~+~\dot{\vec{r}}_{m_{1}}~\wedge~
\dot{\vec{r}}_{m_{2}}¥ ~\tilde{\varphi}_{nm_{1}m_{2}¥}]\ ,
\tag{1.2}
\end{gather*}
with the \textit{scalar} functions $\varphi ,\tilde {\varphi} $ and $\tilde 
{\tilde {\varphi} }$ depending on the $N\left( {2N + 1} \right)$ (scalar) 
products $\vec {r}_{j} \cdot \vec {r}_{k} ,\,\,\vec {r}_{j} \cdot \dot {\vec 
{r}}_{k} ,\,\,\,\dot {\vec {r}}_{j} \cdot \dot {\vec {r}}_{k} $ and on the 
$N\left( {N - 1} \right)\left( {3N - 2} \right)/2$ (pseudoscalar) triple 
products $\vec {r}_{j} \cdot \vec {r}_{k} \wedge \vec {r}_{m} ,\,\,\vec 
{r}_{j} \cdot \vec {r}_{k} \wedge \dot {\vec {r}}_{m} ,\,\,\vec {r}_{j} 
\cdot \dot {\vec {r}}_{k} \wedge \dot {\vec {r}}_{m} ,\,\,\dot {\vec 
{r}}_{j} \cdot \dot {\vec {r}}_{k} \wedge \dot {\vec {r}}_{m} $. This is the 
most general rotation-invariant force one can write in 3-dimensional space. 
In fact generally (see below) only a few of the scalar functions $\varphi 
,\tilde {\varphi} $ or $\tilde {\tilde {\varphi} }$ are actually present 
(most vanish identically), and moreover those few which are present only 
depend on a few of the scalar and pseudoscalar products listed above (and 
generally these scalar functions only depend on few of their subscripts, see 
below).

Some of the models given below are moreover \textit{translation-invariant}, 
namely they are characterized by forces $\vec {f}_{n} $ which remain 
invariant under the \textit{translation}
\begin{equation*}
\vec {r}_{n} \to \vec {r}_{n} + \vec {r}_{0} \  ,
\tag{1.3}
\end{equation*}
with $\vec {r}_{0} $ an arbitrary \textit{constant} (time-independent) 
3-vector.

Some of the models treated below involve scalar quantities evolving in time, 
in addition to three-vectors.

Several of the models considered below are Hamiltonian, but we defer (with a 
few exceptions) the discussion of this aspect to a subsequent paper of this 
series.

Most of the models considered below feature ``velocity-dependent'' forces, 
namely the forces $\vec {f}_{n} $ in the right-hand-side of (1.1) do depend 
on (at least some of) the velocities $\dot {\vec {r}}_{m} $ ; this 
dependence is generally linear or quadratic. The forces depend moreover on 
(at least some of) the ``particle coordinates'' $\vec {r}_{m} $ ; this 
dependence is generally nonlinear, in most cases strongly so (rational). 

Some of the models considered below feature forces that are 
\textit{homogeneous functions of degree one} of their arguments, so that 
(1.1) is invariant under the ``scale transformation'' 
\begin{equation*}
\vec {r}_{n} \to c\,\, \vec {r}_{n} \  ,
\tag{1.4}
\end{equation*}
with $c$ an arbitrary (of course scalar) \textit{constant}.

Several of the models considered below feature only \textit{few-body 
}forces. Particularly interesting are the cases with \textit{two-body 
}forces only,
\begin{equation*}
\ddot {\vec {r}}_{n} = \sum\limits_{m = 1}^{N} {\vec {f}_{nm} \left( {\vec 
{r}_{n} ,\dot {\vec {r}}_{n} ;\vec {r}_{m} ,\dot {\vec {r}}_{m}}  \right)} 
\  .
\tag{1.5}
\end{equation*}
We never consider ``trivial'' cases with ``triangular'' forces $\vec 
{f}_{nm} $ which vanish for $m \le n$ or for $m \ge n$ . Indeed in some 
cases the two-body forces $\vec {f}_{nm} $ in the right-hand-side of 
(1.5) 
are independent of the indices $n$ and $m$, except for the property of 
acting only among ``nearest neighbours'' (which of course need not be 
physically close to each other as they move), say
\begin{gather*}
\vec {f}_{nm} \left( {\vec {r}_{n} ,\dot {\vec {r}}_{n} ;\vec {r}_{m} ,\dot 
{\vec {r}}_{m}}  \right) = 
\delta _{m,n} \vec {f}\left( {\vec {r}_{n} ;\dot 
{\vec {r}}_{n}}  \right) + \delta _{m,n + 1} \vec {f}^{\left( { +}  
\right)}\left( {\vec {r}_{n} ,\dot {\vec {r}}_{n} ;\vec {r}_{n + 1} ,\dot 
{\vec {r}}_{n + 1}}  \right) 
\\* \qquad{}
+ \delta _{m,n - 1} \vec {f}^{\left( { -}  
\right)}\left( {\vec {r}_{n} ,\dot {\vec {r}}_{n} ;\vec {r}_{n - 1} ,\dot 
{\vec {r}}_{n - 1}}  \right) \  .
\tag{1.6}
\end{gather*}
(In these cases with ``nearest-neighbour'' forces one should also add 
``boundary conditions'' specifying the behavior of the ``first'' and 
``last'' particles; we will generally omit to do so in this paper, since we 
stop short of discussing in detail the behavior of the various models we 
introduce). And we also have cases with two-body forces altogether 
independent of the particle indices,
\begin{equation*}
\vec {f}_{nm} \left( {\vec {r}_{n} ,\dot {\vec {r}}_{n} ;\vec {r}_{m} ,\dot 
{\vec {r}}_{m}}  \right) = \vec {f}\left( {\vec {r}_{n} ,\dot {\vec {r}}_{n} 
;\vec {r}_{m} ,\dot {\vec {r}}_{m}}  \right) \  ,
\tag{1.7}
\end{equation*}
including cases where $\vec {f}$ has a very simple form (see below).

In any case, for the convenience of the hasty browser we provide, in the 
following Section \ref{II}, a representative list of the solvable and/or 
integrable and/or linearizable \textit{N}-body problems treated in this 
paper. We consider convenient to make the heuristic (useful if not quite 
precise) distinction among ``solvable,'' ``integrable'' and ``linearizable'' 
models, as follows. \textit{Solvable} models are characterized by the 
availability of a\textit{} technique of solution which requires 
\textit{purely algebraic operations} (such as inverting and diagonalizing 
finite matrices), possibly solving \textit{known linear, possibly 
nonautonomous, ODEs} in terms of \textit{known special functions} (say, of 
hypergeometric type), and perhaps the \textit{inversion of known functions 
}(as in the standard \textit{solution by quadratures}). \textit{Integrable 
}models are those for which a ``Lax pair'' approach is available, yielding 
an adequate supply of \textit{constants of the motion}. As a rule these 
latter models are also solvable, but generally this requires more labor; the 
solvable models are generally also integrable. In the Hamiltonian cases, all 
these models are generally integrable in the Liouville sense. We moreover 
introduce the notion of \textit{linearizable}  problems whose solution 
entails, in addition to the \textit{algebraic} operations required to treat 
solvable or integrable models, the solution of \textit{linear, generally 
nonautonomous, ODEs}, which, in spite of their being generally rather 
simple, might indeed give rise to quite complicated (chaotic?) behaviors. In 
the Hamiltonian cases these latter many-body problems need not be integrable 
in the Liouville sense, although the linearity of the equations to be 
finally solved entails the possibility to introduce \textit{constants of the 
motion} via the \textit{superposition principle,} which guarantees that the 
\textit{general solution} can be represented as a\textit{} linear 
combination with \textit{constant} coefficients of an appropriate set of 
specific solutions. In any case, a linearizable many-body problem is 
certainly much easier to treat than the generic many-body problem, inasmuch 
as its solution can be reduced to solving a \textit{linear first-order 
matrix ODE} (indeed, in many cases, a \textit{single linear second-order 
scalar ODE} -- albeit a \textit{nonautonomous} one -- see below).

Clearly these three categories of problems -- \textit{solvable}, 
\textit{integrable}, \textit{linearizable} - are ordered in terms of 
increasing difficulty (to solve them), so that, as indeed our language 
(including the title of this paper) indicates, problems belonging to a lower 
category generally also belong to the following ones.

As much as we tried above to delineate clearly the distinction between 
\textit{solvable}, \textit{integrable} and \textit{linearizable} models, we 
are well aware that this categorization is imprecise; for one thing, we left 
vague what ``solving'' a problem means: finding the general solution? 
solving the initial-value problem? for which class of initial data? and what 
about boundary conditions (which in some cases, see above and below, are 
essential to completely define the problem)? Moreover, as indicated above, 
we are well aware that the boundaries among the three categories we have 
introduced lack complete cogency. Yet this distinction provides an useful 
means to convey synthetically significant information on the status of the 
different models we introduce, a status whose more precise definition will 
then require a detailed case-by-case analysis of the problem under 
consideration, which clearly exceeds the scope of this presentation (lest 
its size approximate a book rather than a paper).

Our starting point to obtain solvable and/or integrable and/or linearizable 
\textit{N}-body problems \textit{in ordinary (3-dimensional) space} are 
solvable and/or integrable and/or linearizable evolution equations for 
\textit{matrices}. A survey of such models is provided in Section \ref{III} 
(which 
might appeal to some readers in its own right), and a number of matrix 
representations in terms of 3-vectors suitable to generate 
\textit{rotation-invariant N}-body problems in 3-dimensional space are 
reviewed in Section \ref{IV}. 
The material of these two sections is then used in 
Section \ref{V} to 
derive several solvable and/or integrable and/or linearizable 
\textit{N}-body problems in 3-dimensional space, including, in some cases, 
their explicit solutions and/or some information about their behaviors (for 
instance, the identification of cases with completely periodic solutions). 
As mentioned above, a representative list of these problems is also reported 
in the following Section \ref{II}: we deem 
the advantage thereby gained in terms 
of clarity of presentation, especially for the hasty browser, to 
overcompensate for the inevitable repetitiveness this entails. A final 
Section \ref{VI} outlines future researches suggested by the findings reported in 
this paper. To avoid the flow of the discourse being too cluttered by 
detailed computations some of these are relegated to a number of appendices, 
as well as some findings that do not quite fit in the mainstream of our 
presentation.


\section{List of solvable and/or integrable and/or linearizable few-body 
and many-body problems} \label{II}


In this section we display, for the convenience of the hasty browser, hence 
with minimal commentary, a representative (namely, far from complete) list 
of the \textit{N}-body problems treated in this paper. To maximize the 
\textit{user-friendly} character of this presentation, we separate 
\textit{few-body} from \textit{many-body} systems, and we indicate in each 
case whether the equations of motion presented are \textit{solvable}, 
\textit{integrable} or \textit{linearizable} (for the explanation of this 
distinction see the end of the introductory Section \ref{I}). Some of the 
\textit{many-body} problems exhibited in Section \ref{II.B} are direct 
generalizations of corresponding \textit{few-body} problems listed in 
Section \ref{II.A}; but there is no one-to-one correspondence between the two 
lists, and this justifies our separate presentation of a list of 
\textit{few-body} problems (of course any \textit{many-body} problem 
includes \textit{few-body} problems as special subcases).

Of course the number of arbitrary ``coupling constants'' featured by these 
models can generally be marginally increased or decreased by rescaling the 
dependent or independent variables.

We trust the notation below to be self-explanatory; otherwise the reader 
should go to Section \ref{V}, and to some relevant appendices, where these 
findings are more fully explained and demonstrated; to facilitate such a 
follow-up, relevant hints are provided below.


\subsection{Few-body problems} \label{II.A}


A \textit{linearizable one-body problem} (featuring an arbitrary function, 
see (5.22)):
\begin{equation*}
\ddot {\vec {r}}\left( {t} \right) = \varphi \left( {r} \right)\,\vec 
{r}\left( {t} \right) \wedge \dot {\vec {r}}\left( {t} \right) \  .
\tag{2.1}
\end{equation*}
%
%
\textit{Special cases}\\
%
%
i). The choice $\varphi \left( {r} \right) = k$ yields 
\begin{equation*}
\ddot {\vec {r}}\left( {t} \right) = \,k\,\,\vec {r}\left( {t} \right) 
\wedge \dot {\vec {r}}\left( {t} \right) \  ,
\tag{2.2}
\end{equation*}
which describes the motion in ordinary space of a particle acted upon by a 
force proportional to its angular momentum (of course the arbitrary coupling 
constant $k$ could be rescaled away). This is a \textit{solvable} equation: 
its explicit solution in terms of \textit{parabolic cylinder} functions is 
given in Appendix A (see formulas (A.19), (A.28) and (A.36)). See (5.23).


ii). The choice $\varphi \left( {r} \right) = k/r^{2}$ yields
\begin{equation*}
\ddot {\vec {r}}\left( {t} \right) = \,\,k\,\left[ {\,\vec {r}\left( {t} 
\right) \wedge \dot {\vec {r}}\left( {t} \right)} \right]/r^{2} \  ,
\tag{2.3}
\end{equation*}
which is \textit{solvable} in terms of \textit{hypergeometric} functions 
(see Appendix A, formulas (A.19), (A.28) and (A.38)). See (5.25).


iii). The choice $\varphi \left( {r} \right) = k/r^{3}$ yields 
\begin{equation*}
\ddot {\vec {r}}\left( {t} \right) = \,\,k\,\left[ {\,\vec {r}\left( {t} 
\right) \wedge \dot {\vec {r}}\left( {t} \right)} \right]/r^{3}\,,
\tag{2.4}
\end{equation*}
which is the ``Newton / Lorentz'' equation of an electrical charge in the 
magnetic (radial) field of a magnetic monopole (or of a magnetic monopole in 
the electrical field of an electrical charge). This equation is partially 
investigated in Appendix A, a deeper investigation is postponed to a 
subsequent paper.


A \textit{linearizable one-body problem} (featuring an arbitrary 
\textit{even} function, and 3 arbitrary constants) :
\begin{equation*}
\ddot {\vec {r}} = 2\,a\,\dot {\vec {r}} + b\,\vec {r} + c\,\left[ {2\,\dot 
{\vec {r}}\,\left( {\dot {\vec {r}} \cdot \vec {r}} \right) - \vec 
{r}\,\left( {\dot {\vec {r}} \cdot \dot {\vec {r}}} \right)} \right]/r^{2} + 
\vec {r} \wedge \dot {\vec {r}}\,\,\,f\left( {r} \right) \  .
\tag{2.5}
\end{equation*}
See (5.17). Subcases of this model (with $a\, = \,b\, = \,c\, = 0$) were 
considered above; other subcases are \textit{solvable}, including the case 
$f\left( {r} \right)\, = \,0$, which yields the \textit{solvable one-body 
problem} (with 3 arbitrary coupling constants):
\begin{equation*}
\ddot {\vec {r}} = 2\,a\,\dot {\vec {r}} + b\,\vec {r} + c\,\left[ {2\,\dot 
{\vec {r}}\,\left( {\dot {\vec {r}} \cdot \vec {r}} \right) - \vec 
{r}\,\left( {\dot {\vec {r}} \cdot \dot {\vec {r}}} \right)} \right]/r^{2} 
\  .
\tag{2.6}
\end{equation*}
The generic\textit{} solution of this equation of motion is 
\textit{completely periodic}, with period $T = 2\,\pi /\omega $, if $a = 0$ 
and $b\,\left( {c - 1} \right) \equiv \omega ^{2} > 0$. See (5.1), (5.2).


Another interesting case obtains setting $c = 0\,,\,\,f\left( {r} \right) = 
C$ in (2.5), getting thereby
\begin{equation*}
\ddot {\vec {r}} = 2\,a\,\dot {\vec {r}} + b\,\vec {r} + C\,\vec {r} 
\wedge \dot {\vec {r}} \  .
\tag{2.7}
\end{equation*}
See (5.18). This one-body problem is \textit{solvable} if $b\, = \,\left( 
{8/9} \right)\,a^{2}$ or $b\, = \,0$; see Appendix B.


An \textit{integrable and} \textit{linearizable} (clearly Hamiltonian) 
\textit{one-body problem} (with 2 arbitrary constants):
\begin{equation*}
\ddot {\vec {r}} = a\,\vec {r} - b\,r^{2}\vec {r}\ .
\tag{2.8}
\end{equation*}
See (5.16).


An \textit{integrable} (scalar \& 3-vector) \textit{one-body problem 
}(featuring 2 arbitrary constants, and a constant three-vector):
\begin{gather*}
\ddot {\rho} \, = 2\,c^{2}\,\left[ {\rho \,\left( {\rho ^{2} - 3\,r^{2}} 
\right) + \gamma \,\rho - \left( {\vec {C} \cdot \vec {r}} \right)} \right] 
\  ,
\tag{2.9a}
\\
\ddot {\vec {r}} = 2\,c^{2}\,\left[ {\vec {r}\,\left( { - r^{2} + 3\,\rho 
^{2}} \right) + \gamma \,\vec {r} + \rho \,\vec {C}} \right] \  .
\tag{2.9b}
\end{gather*}
Of course, only for $\vec {C} = 0$ is \textit{rotation-invariance 
}preserved. See (5.15).


A \textit{solvable} (scalar \& 3-vector) \textit{one-body problem} (with 4 
arbitrary coupling constants):
\begin{gather*}
\ddot {\rho}  = \alpha + \beta \,\rho + \gamma \,\left[ {\dot {\rho}  + 
c\,\left( {\rho ^{2} - r^{2}} \right)} \right] - c\,\left[ {3\,\rho \,\dot 
{\rho}  - 3\,\left( {\vec {r} \cdot \dot {\vec {r}}} \right) + c\,\rho 
\,\left( {\rho ^{2} - 3r^{2}} \right)} \right] \  ,
\tag{2.10a}
\\
\ddot {\vec {r}} = \beta \,\vec {r} + \gamma \,\left[ {\dot {\vec {r}} + 
2\,c\,\rho \,\vec {r}} \right] - c\,\left[ {3\,\dot {\rho} \,\vec {r} + 
3\,\rho \,\dot {\vec {r}} - \vec {r} \wedge \dot {\vec {r}} + c\,\vec 
{r}\,\left( {3\,\rho ^{2} - r^{2}} \right)} \right] \  .
\tag{2.10b}
\end{gather*}
See (5.12).


A \textit{solvable} (translation-invariant)\textit{ two-body problem} (with 
two arbitrary constants):
\begin{equation*}
\ddot {\vec {r}}^{\left( { \pm}  \right)} = \left\{ {\alpha \,\left( {\dot 
{\vec {r}}^{\left( { +}  \right)} + \dot {\vec {r}}^{\left( { -}  \right)}} 
\right) \pm C\,\left[ {\left( {\vec {r}^{\left( { +}  \right)} - \vec 
{r}^{\left( { -}  \right)}} \right) \wedge \left( {\dot {\vec {r}}^{\left( { 
+}  \right)} - \dot {\vec {r}}^{\left( { -}  \right)}} \right)} \right]} 
\right\}/2\ .
\tag{2.11}
\end{equation*}
See (5.24).


A \textit{solvable} (translation-invariant)\textit{ 2-body problem} (with 4 
arbitrary coupling constants):
\begin{gather*}
\ddot {\vec {r}}^{\left( { \pm}  \right)} = \left[ {\left( {\alpha /2} 
\right) \pm a} \right]\dot {\vec {r}}^{\left( { +}  \right)} + \left[ 
{\left( {\alpha /2} \right) \mp a} \right]\dot {\vec {r}}^{\left( { -}  
\right)} \pm \left\{ {b\vec {r} + c\left[ {2\dot {\vec {r}}\left( {\dot 
{\vec {r}} \cdot \vec {r}} \right) - \vec {r}\left( {\dot {\vec {r}} \cdot 
\dot {\vec {r}}} \right)} \right]/r^{2}} \right\}/2 \  ,
\tag{2.12a}
\\
\vec {r}\left( {t} \right) \equiv \vec {r}^{\left( { +}  \right)}\left( {t} 
\right) - \vec {r}^{\left( { -}  \right)}\left( {t} \right) \  .
\tag{2.12b}
\end{gather*}
See (5.5).

A \textit{linearizable two-body problem} (with 6 arbitrary constants):
\begin{gather*}
\ddot {\vec {r}}^{\left( {1} \right)} = 2\,\left( {\alpha \,\dot {\vec 
{r}}^{\left( {1} \right)} - \tilde {\alpha} \,\dot {\vec {r}}^{\left( {2} 
\right)}} \right) + \beta \,\vec {r}^{\left( {1} \right)} - \tilde {\beta 
}\,\vec {r}^{\left( {2} \right)}
\\
\qquad{}+ c\,\left( {\vec {r}^{\left( {1} \right)} 
\wedge \dot {\vec {r}}^{\left( {1} \right)} - \vec {r}^{\left( {2} \right)} 
\wedge \dot {\vec {r}}^{\left( {2} \right)}} \right) - \tilde {c}\,\left( 
{\vec {r}^{\left( {1} \right)} \wedge \dot {\vec {r}}^{\left( {2} \right)} + 
\vec {r}^{\left( {2} \right)} \wedge \dot {\vec {r}}^{\left( {1} \right)}} 
\right),
\tag{2.13a}
\\
\ddot {\vec {r}}^{\left( {2} \right)} = 2\,\left( {\alpha \,\dot {\vec 
{r}}^{\left( {2} \right)} + \tilde {\alpha} \,\dot {\vec {r}}^{\left( {1} 
\right)}} \right) + \beta \,\vec {r}^{\left( {2} \right)} + \tilde {\beta 
}\,\vec {r}^{\left( {1} \right)}
\\
\qquad{}+ c\,\left( {\vec {r}^{\left( {1} \right)} 
\wedge \dot {\vec {r}}^{\left( {2} \right)} + \vec {r}^{\left( {2} \right)} 
\wedge \dot {\vec {r}}^{\left( {1} \right)}} \right) + \tilde {c}\,\left( 
{\vec {r}^{\left( {1} \right)} \wedge \dot {\vec {r}}^{\left( {1} \right)} - 
\vec {r}^{\left( {2} \right)} \wedge \dot {\vec {r}}^{\left( {2} \right)}} 
\right).
\tag{2.13b}
\end{gather*}

If $\alpha = \tilde {\beta}  = 0,\,\,\tilde {\alpha}  = 3\omega 
/2,\,\,\,\beta = 2\omega ^{2}$ or $\alpha = \beta = \tilde {\beta}  = 
0,\,\,\tilde {\alpha}  = \omega /2,$ with $\omega $ an arbitrary (real, 
nonvanishing) constant, this model is \textit{solvable} and its generic 
solution is \textit{completely periodic} with period $T = 2\pi /\left| 
{\omega}  \right|$. See (5.20) and Appendix B.


A \textit{solvable 3-body problem} (with 4 arbitrary coupling constants):
\begin{gather*}
\ddot {\vec {r}}^{\left( {j} \right)} = \left( {2a + c\lambda}  \right)\dot 
{\vec {r}}^{\left( {j} \right)} + b\vec {r}^{\left( {j} \right)} 
\\
\qquad{}+ 
c\sum\limits_{k = 1,2,3,{\rm mod}\left( {3} \right)}{\left\{ {\left( {\dot {\vec 
{r}}^{\left( {k} \right)} + \lambda \vec {r}^{\left( {k} \right)}}
\right)
\left[ {\dot {\vec {r}}^{\left( {j} \right)} \cdot \vec {r}^{\left( 
{k + 1} \right)} \wedge \vec {r}^{\left( {k + 2} \right)}} \right]} 
\right\}
/\Delta \,\,\,\,,\,\,j = 1,2,3} ,
\tag{2.14a}
\\
\Delta \equiv \vec {r}^{\left( {1} \right)} \cdot \vec {r}^{\left( {2} 
\right)} \wedge \vec {r}^{\left( {3} \right)} \  .
\tag{2.14b}
\end{gather*}
See (5.9).


A \textit{linearizable} \textit{3-body problem} (with 12 arbitrary coupling 
constants; note that all indices are defined mod(3)):
\begin{gather*}
 \ddot {\vec {r}}_{n} \, = \,\frac{{2}}{{\mu _{n}} }
\{{\mu _{n + 1} \mu _{n + 2} \dot{\vec {r}}_{n + 2} \wedge \dot{\vec 
{r}}_{n + 1}} +  \sum_{m = 1}^{3} [- \frac{{1}}{{2}}a_{n,m} \dot {\vec 
{r}}_{m}
\\
\qquad{}+\left( a_{n + 1,m} \mu _{n + 2} \dot {\vec {r}}_{n + 2} - a_{n + 
2,m} \mu _{n + 1} \dot{\vec {r}}_{n + 1} + \sum\limits_{k = 1}^{3} a_{n + 
1,m} a_{n + 2,k} \vec{r}_{k}  \right) \wedge \vec{r}_{m}  
]\}\ .\tag{2.15}
\end{gather*}
See (5.27).


A \textit{solvable 4-body problem} (with 3 arbitrary coupling constants; 
translation-invariant if $b = 0$):
\begin{gather*}
\ddot {\vec {r}}^{\left( {j} \right)} = 2a\dot {\vec {r}}^{\left( {j} 
\right)} + b\vec {r}^{\left( {j} \right)} + c\sum\limits_{k = 
1,2,3,4,mod\left( {4} \right)} {\left({ -} \right)^{k}}
\\
\qquad\qquad\left\{\dot {\vec 
{r}}^{(k)}\left[\dot {\vec {r}}^{(j)} \cdot 
\left (\vec {r}^{(k+1)} - \vec {r}^{(k + 2)} \right)
\wedge \left(\vec {r}^{(k+2)} - \vec 
{r}^{(k+3)} \right) \right] \right\}/ \Delta \ ,
\tag{2.16a}
\\
\Delta \equiv \left( {\vec {r}^{\left( {2} \right)} - \vec {r}^{\left( {1} 
\right)}} \right) \cdot \left( {\vec {r}^{\left( {3} \right)} - \vec 
{r}^{\left( {1} \right)}} \right) \wedge \left( {\vec {r}^{\left( {4} 
\right)} - \vec {r}^{\left( {1} \right)}} \right) \  .
\tag{2.16b}
\end{gather*}
See (5.10).


A \textit{linearizable four-body problem} (with 8 arbitrary coupling 
constants):
\begin{gather*}
\ddot {\vec {r}}^{\left( {1, \pm}  \right)} = \left\{ {\gamma \,\dot {\vec 
{s}}^{\left( {1} \right)} - \tilde {\gamma} \,\dot {\vec {s}}^{\left( {2} 
\right)} \pm \left[ {} \right.} \right.2\,\left( {\alpha \,\dot {\vec 
{r}}^{\left( {1} \right)} - \tilde {\alpha} \,\dot {\vec {r}}^{\left( {2} 
\right)}} \right) + \beta \,\vec {r}^{\left( {1} \right)} - \tilde {\beta 
}\,\vec {r}^{\left( {2} \right)}
\\
\qquad {}+ c\,\left( {\vec {r}^{\left( {1} \right)} \wedge 
\dot {\vec {r}}^{\left( 
{1} \right)} - \vec {r}^{\left( {2} \right)} \wedge \dot {\vec {r}}^{\left( 
{2} \right)}} \right) - \tilde {c}\,\left( {\vec {r}^{\left( {1} \right)} 
\wedge \dot {\vec {r}}^{\left( {2} \right)} + \vec {r}^{\left( {2} \right)} 
\wedge \dot {\vec {r}}^{\left( {1} \right)}} \right)\left. {} \right]\left. 
{} \right\}/2 \  ,
\tag{2.17a}
\\
\ddot {\vec {r}}^{\left( {2, \pm}  \right)} = \left\{ {} \right.\gamma 
\,\dot {\vec {s}}^{\left( {2} \right)} + \tilde {\gamma} \,\dot {\vec 
{s}}^{\left( {1} \right)} \pm \left[ {} \right.2\left( {\alpha \,\dot {\vec 
{r}}^{\left( {2} \right)} + \tilde {\alpha} \,\dot {\vec {r}}^{\left( {1} 
\right)}} \right) + \beta \,\vec {r}^{\left( {2} \right)} + \tilde {\beta 
}\,\vec {r}^{\left( {1} \right)}
\\
\qquad{} + c\left( {\vec {r}^{\left( {1} \right)} 
\wedge \dot {\vec {r}}^{\left( {2} 
\right)} + \vec {r}^{\left( {2} \right)} \wedge \dot {\vec {r}}^{\left( {1} 
\right)}} \right) + \tilde {c}\left( {\vec {r}^{\left( {1} \right)} \wedge 
\dot {\vec {r}}^{\left( {1} \right)} - \vec {r}^{\left( {2} \right)} \wedge 
\dot {\vec {r}}^{\left( {2} \right)}} \right)\left. {} \right]\left. {} 
\right\}/2 \  ,
\tag{2.17b}
\\
\vec {r}^{\left( {1} \right)} \equiv \vec {r}^{\left( {1, +}  \right)} - 
\vec {r}^{\left( {1, -}  \right)},\,\,\,\vec {r}^{\left( {2} \right)} \equiv 
\vec {r}^{\left( {2, +}  \right)} - \vec {r}^{\left( {2, -}  
\right)}\ ,
\\
\vec {s}^{\left( {1} \right)} \equiv \vec {r}^{\left( {1, 
+}  \right)} + \vec {r}^{\left( {1, -}  \right)},\,\,\,\vec {s}^{\left( {2} 
\right)} \equiv \vec {r}^{\left( {2, +}  \right)} + \vec {r}^{\left( {2, -}  
\right)} \  .
\tag{2.18}
\end{gather*}
The generic solution of this model is \textit{completely periodic} with 
period $T = 2\pi /\left| {\omega}  \right|$ if there holds either one of the 
two sets of restrictions on the 4 coupling constants $\alpha ,\,\tilde 
{\alpha} ,\,\beta ,\,\tilde {\beta} $ reported above (after (2.13b), and in 
addition there hold the two constraints $\gamma = 0,\,\,\,\tilde {\gamma}  = 
m\,\omega $, with $m$ an arbitrary (nonvanishing) integer ($m \ne 0$; for $m 
= 0$, namely $\gamma = \tilde {\gamma}  = 0$, (2.18) reduces to (2.13)). See 
(5.21).


\subsection{Many-body problems} \label{II.B}


A \textit{linearizable N-body problem} (with $3N$\textit{} arbitrary 
coupling constants):
\begin{equation*}
\ddot {\vec {r}}_{n} = \sum\limits_{n_{1} = 1}^{N} {\left( {2\,a_{n - n_{1} 
} \,\dot {\vec {r}}_{n_{1}}  + b_{n - n_{1}}  \,\vec {r}_{n_{1}} }  \right)} 
+ \sum\limits_{n_{1} ,n_{2} = 1}^{N} {\left( {c_{n - n_{1} - n_{2}}  \,\vec 
{r}_{n_{1}}  \wedge \dot {\vec {r}}_{n_{2}} }  \right)} \  .
\tag{2.19}
\end{equation*}
Here, and in analogous equations below, all indices are defined ${\rm mod}\left( 
{N} \right)$. If $a_{n} = b_{n} = 0$ the model is actually 
\textit{solvable}. See (5.30).


A \textit{linearizable} (translation-invariant) \textit{(2N)-body problem 
}(with $4N$\textit{} arbitrary coupling constants):
\begin{gather*}
\ddot {\vec {r}}_{n}^{\left( { \pm}  \right)} = \left\{ {} \right.\alpha 
_{n} \,\dot {\vec {s}}_{n} \pm \sum\limits_{n_{1} = 1}^{N} {\left( {2\,a_{n 
- n_{1}}  \,\dot {\vec {r}}_{n_{1}}  + b_{n - n_{1}}  \,\vec {r}_{n_{1}} }  
\right)}
\\
\qquad{}+ \sum\limits_{n_{1} ,n_{2} = 1}^{N} {\left( {c_{n - n_{1} - n_{2} 
} \,\vec {r}_{n_{1}}  \wedge \dot {\vec {r}}_{n_{2}} }  \right)} \left. {} 
\right\}/2 \  ,
\tag{2.20a}
\\
\vec {s}_{n} \equiv \vec {r}_{n}^{\left( { +}  \right)} + \vec 
{r}_{n}^{\left( { -}  \right)} ,\,\,\,\vec {r}_{n} = \vec {r}_{n}^{\left( { 
+}  \right)} - \vec {r}_{n}^{\left( { -}  \right)} \  ,
\quad
\vec {r}_{n}^{\left( { \pm}  \right)} \equiv \left( {\vec {s}_{n} \pm \vec 
{r}_{n}}  \right)/2 \  .
\tag{2.20b}
\end{gather*}
Again, if $a_{n} = b_{n} = 0$, the model is actually \textit{solvable}. See 
(5.31).


A \textit{solvable Hamiltonian many-body problem} (with $6\,N$\textit{ 
}arbitrary coupling constants):
\begin{gather*}
\dot {\vec {q}}_{n} = \sum\limits_{n_{1} = 1}^{N} {\left( {\alpha _{n - 
n_{1}}  \,\vec {p}_{n_{1}}  + \gamma _{n - n_{1}}  \,\vec {q}_{n_{1}} }  
\right)} + \sum\limits_{n_{1} ,n_{2} ,n_{3} = 1}^{N} {\left[ {c_{n - n_{1} - 
n_{2} - n_{3}}  \,\left( {\vec {p}_{n_{1}}  \wedge \vec {q}_{n_{2}} }  
\right) \wedge \vec {q}_{n_{3}} }  \right]} \  ,
\tag{2.21a}
\\
\dot {\vec {p}}_{n} = - \sum\limits_{n_{1} = 1}^{N} {\left( {\beta _{n - 
n_{1}}  \,\vec {q}_{n_{1}}  + \gamma _{n - n_{1}}  \,\vec {p}_{n_{1}} }  
\right)} + \sum\limits_{n_{1} ,n_{2} ,n_{3} = 1}^{N} {\left[ {c_{n - n_{1} - 
n_{2} - n_{3}}  \,\left( {\vec {p}_{n_{1}}  \wedge \vec {q}_{n_{2}} }  
\right) \wedge \vec {p}_{n_{3}} }  \right]}\ .\tag{2.21b}
\end{gather*}
See (5.32).


A scalar/vector \textit{solvable N-body problem} (with $8\,N$\textit{ 
}arbitrary coupling constants):
\begin{gather*}
\ddot {\rho} _{n} = \alpha _{n} + \sum\limits_{n_{1} = 1}^{N} {\left\{ 
{\beta _{n - n_{1}}  \,\rho _{n_{1}}  + \gamma _{n - n_{1}}  \,\dot {\rho 
}_{n_{1}} }  \right\}} 
\\
\qquad{}- 3\sum\limits_{n_{1} ,n_{2} = 1}^{N} {\left\{ {c_{n 
- n_{1} - n_{2}}  \,\left[ {\rho _{n_{1}}  \,\dot {\rho} _{n_{2}}  - \left( 
{\vec {r}_{n_{1}}  \cdot \dot {\vec {r}}_{n_{2}} }  \right)} \right]} 
\right\}} 
\\
\qquad{} + \sum\limits_{n_{1} ,n_{2} ,n_{3} = 1}^{N} {\left\{ {c_{n - n_{1} - n_{2} 
- n_{3}}  \,\gamma _{n_{1}}  \,\left[ {\rho _{n_{2}}  \,\rho _{n_{3}}  - 
\left( {\vec {r}_{n_{2}}  \cdot \vec {r}_{n_{3}} }  \right)} \right]} 
\right\}}
\\
\qquad{}- \sum\limits_{n_{1} ,n_{2} ,n_{3} ,n_{4} = 1}^{N}{\left\{ {c_{n 
- n_{1} - n_{2} - n_{3} - n_{4}}  \,c_{n_{1}}  \,\rho _{n_{2}}  \,\left[ 
{\rho _{n_{3}}  \,\rho _{n_{4}}  - 3\,\left( {\vec {r}_{n_{3}}  \cdot \vec 
{r}_{n_{4}} }  \right)} \right]} \right\}} \ ,
\tag{2.22a}
\\
\ddot {\vec {r}}_{n} = \sum\limits_{n_{1} = 1}^{N} {\left\{ {\beta _{n - 
n_{1}}  \,\vec {r}_{n_{1}}  + \gamma _{n - n_{1}}  \,\dot {\vec {r}}_{n_{1} 
}}  \right\}}
\\*
\qquad{}- \sum\limits_{n_{1} ,n_{2} = 1}^{N} {\left\{ {c_{n - n_{1} - 
n_{2}}  \,\left[ {3\,\dot {\rho} _{n_{1}}  \,\vec {r}_{n_{2}}  + 3\,\rho 
_{n_{1}}  \,\dot {\vec {r}}_{n_{2}}  - \left( {\vec {r}_{n_{1}}  \wedge \dot 
{\vec {r}}_{n_{2}} }  \right)} \right]} \right\}} 
\\
\qquad{}+ 2\sum\limits_{n_{1} ,n_{2} ,n_{3} = 1}^{N} {\left\{ {c_{n - n_{1} - n_{2} 
- n_{3}}  \,\gamma _{n_{1}}  \,\rho _{n_{2}}  \,\vec {r}_{n_{3}} }  
\right\}}
\\
\qquad{} - \sum\limits_{n_{1} ,n_{2} ,n_{3} ,n_{4} = 1}^{N}{\left\{ {c_{n 
- n_{1} - n_{2} - n_{3} - n_{4}}  \,c_{n_{1}}  \,\vec {r}_{n_{2}}  \,\left[ 
{3\,\rho _{n_{3}}  \,\rho _{n_{4}}  - \left( {\vec {r}_{n_{3}}  \cdot \vec 
{r}_{n_{4}} }  \right)} \right]} \right\}} \ .
\tag{2.22b}
\end{gather*}
See (5.34).


A \textit{solvable N-body problem} (with $N^{2} + 3N$ arbitrary coupling 
constants):
\begin{gather*}
\ddot {\vec {r}}_{n} = a_{n} \,\dot {\vec {r}}_{n} - \alpha _{n} \,\gamma 
_{n} \,\vec {r}_{n} \wedge \dot {\vec {r}}_{n} + \{  - \gamma 
_{n} \,\dot {\vec {r}}_{n} \wedge \left[ {\gamma _{n} ^{2}\,\left( {\vec 
{r}_{n} \cdot \dot {\vec {r}}_{n}}  \right)\,\vec {r}_{n} + a_{n} \,\gamma 
_{n} \,\vec {r}_{n} \wedge \dot {\vec {r}}_{n}}  \right] 
\\
\qquad{}- \sum\limits_{m = 1}^{N} \{b_{nm} \,\{{\dot {\vec {r}}}_{m} + 
\gamma _{n} \,\dot {\vec {r}}_{n} \wedge  [\gamma _{n} ^{2}\, ( 
\vec {r}_{n} \cdot{\vec {r}_{m}}) \,\vec {r}_{n} + a_{n}^{2}\,\,\vec {r}_{m} 
+ a_{n} \,\gamma _{n} \,\vec {r}_{n} \wedge \vec 
{r}_{m}]\}\}\}
\\
\qquad{} / [ a_{n} \,( a_{n} ^{2} + \gamma _{n} ^{2}\,\,r_{n} ^{2} ) ]
\  .\tag{2.23}
\end{gather*}
See (5.35).


A scalar/vector \textit{solvable N-body problem} (with $4N$\textit{ 
}arbitrary coupling constants and ``nearest-neighbour'' interactions):
\begin{gather*}
\ddot {\rho} _{n} = \left( {a_{n} - a_{n + 1}}  \right)\,\tilde {c}_{n} + 
\left( {a_{n} - a_{n + 1}}  \right)\,\left( {\tilde {a}_{n} - a_{n}}  
\right)\,\rho _{n} + \tilde {c}_{n} \,\left( {b_{n} \,\rho _{n} - b_{n + 1} 
\,\rho _{n + 1}}  \right)
\\
\qquad{}- \left( {\tilde {a}_{n} - a_{n}}  \right)\,b_{n + 1} \,\left( {\rho _{n} 
\,\rho _{n + 1} - \vec {r}_{n} \cdot \vec {r}_{n + 1}}  \right) - 3\,b_{n} 
\,\left( {\dot {\rho} _{n} \,\rho _{n} - \dot {\vec {r}}_{n} \cdot \vec 
{r}_{n}}  \right)
\\
\qquad{}+ \left( {\tilde {a}_{n} - 2\,a_{n} + a_{n + 1}}  
\right)\,\left[ {\rho _{n} + b_{n} \,\left( {\rho _{n} ^{2} - r_{n} ^{2}} 
\right)} \right]
\\
\qquad{}+ b_{n + 1} \,\left( {\dot {\rho} _{n} \,\rho _{n + 1} - \dot {\vec 
{r}}_{n} \cdot \vec {r}_{n + 1}}  \right) + b_{n} \,b_{n + 1} \left[ {\rho 
_{n + 1} \,\left( {\rho _{n} ^{2} - r_{n} ^{2}} \right) - 2\,\rho _{n} 
\,\vec {r}_{n} \cdot \vec {r}_{n + 1}}  \right] 
\\
\qquad{}- b_{n} ^{2}\,\rho _{n} 
\,\left( {\rho _{n} ^{2} - 3r_{n} ^{2}} \right),\tag{2.24a}
\\
\ddot {\vec {r}}_{n} = \left( {\,a_{n} - a_{n + 1}}  \right)\,\left( {\tilde 
{a}_{n} - a_{n}}  \right)\vec {r}_{n} + \tilde {c}_{n} \,\left( {b_{n} 
\,\vec {r}_{n} - b_{n + 1} \,\vec {r}_{n + 1}}  \right)\,
\\
\qquad{}- \left( {\tilde 
{a}_{n} - a_{n}}  \right) b_{n + 1} \,\left( {\rho _{n} \,\vec {r}_{n + 1} 
+ \rho _{n + 1} \,\vec {r}_{n} + \vec {r}_{n} \wedge \vec {r}_{n + 1}}  
\right)
\\
\qquad{} - b_{n} \,\left( {3\,\dot {\rho} _{n} \,\vec {r}_{n} + 3\,\rho _{n} \,\dot 
{\vec {r}}_{n} - \dot {\vec {r}}_{n} \wedge \vec {r}_{n}}  \right) + \left( 
{\tilde {a}_{n} - 2\,a_{n} + a_{n - 1}}  \right)\,\left( {\dot {\vec 
{r}}_{n} + 2\,b_{n} \,\rho _{n} \,\vec {r}_{n}}  \right) 
\\
\qquad{}+ b_{n + 1} 
\,\left( {\rho _{n + 1} \,\dot {\vec {r}}_{n} + \dot {\rho} _{n} \,\vec 
{r}_{n} + \dot {\vec {r}}_{n} \wedge \vec {r}_{n + 1}}  \right)
\\
\qquad+ b_{n} \,b_{n + 1} \,\left[ {2\,\rho _{n + 1} \,\rho _{n} \,\vec {r}_{n} + 
\left( {\rho _{n} ^{2} - r_{n} ^{2}} \right)\,\vec {r}_{n + 1} + 2\,\rho 
_{n} \,\vec {r}_{n} \wedge \vec {r}_{n + 1}}  \right] - b_{n} ^{2}\,\left( 
{3\,\rho _{n} ^{2} - r_{n} ^{2}} \right)\,\vec {r}_{n} \ 
.\tag{2.24b}
\end{gather*}
See (5.36).


A \textit{linearizable N-body problem} (featuring $2\,N^{2} + N$ arbitrarily 
assigned functions of time):
\begin{gather*}
\ddot {\vec {r}}_{n} = \sum\limits_{n_{1} = 1}^{N} \{2\,a_{nn_{1}}  
\left( {t} \right)\,\,\dot {\vec {r}}_{n_{1}}  + b_{nn_{1}}  \left( {t} 
\right)\,\,\vec {r}_{n_{1}}  + C_{n_{1}}  \left( {t} \right)\,\,\left[ 
{2\,\left( {\vec {r}_{n_{1}}  \wedge \dot {\vec {r}}_{n}}  \right) - \left( 
{\vec {r}_{n} \wedge \dot {\vec {r}}_{n_{1}} }  \right)} \right]
\\
\qquad{}- \dot 
{C}_{n_{1}}  \left( {t} \right)\,\,\left( {\vec {r}_{n} \wedge \vec 
{r}_{n_{1}}}  \right)\} 
\\
\qquad{} + \sum\limits_{n_{1} ,n_{2} = 1}^{N} {\left\{ {2\,a_{nn_{1}}  \left( {t} 
\right)\,\,C_{n_{2}}  \left( {t} \right)\,\,\left( {\vec {r}_{n_{1}}  \wedge 
\vec {r}_{n_{2}} }  \right) - C_{n_{1}}  \,\left( {t} \right)\,C_{n_{2}}  
\left( {t} \right)\,\,\left[ {\left( {\vec {r}_{n} \wedge \vec {r}_{n_{1}}  
} \right) \wedge \vec {r}_{n_{2}} }  \right]} \right\}} \  
.\tag{2.25}
\end{gather*}
See (5.38).


An \textit{integrable N-body problem} (with ``nearest-neighbour'' 
interactions):
\begin{gather*}
\ddot {\vec {r}}_{n} = \left[ {2\,\dot {\vec {r}}_{n} \,\left( {\dot {\vec 
{r}}_{n} \cdot \vec {r}_{n}}  \right) - \vec {r}_{n} \,\left( {\dot {\vec 
{r}}_{n} \cdot \dot {\vec {r}}_{n}}  \right)} \right]/r_{n} ^{2}
\\*
\qquad{}+ \gamma\,\left\{ {\vec {r}_{n + 1} - \left[ {2\,\vec {r}_{n} \,\left( {\vec {r}_{n} 
\cdot \vec {r}_{n - 1}}  \right) - \vec {r}_{n - 1} \,\left( {\vec {r}_{n} 
\cdot \vec {r}_{n}}  \right)} \right]/r_{n - 1} ^{2}} \right\} \  .
\tag{2.26}
\end{gather*}
See (5.39).


A scalar/vector \textit{integrable N-body problem} (translation-invariant, 
with ``nearest-neigh\-bour'' interactions):
\begin{gather*}
\ddot {\rho} _{n} = c\,\left[ {\dot {\rho} _{n} \,\left( {\rho _{n + 1} - 
2\,\rho _{n} + \rho _{n - 1}}  \right) - \dot {\vec {r}}_{n} \cdot \left( 
{\vec {r}_{n + 1} - 2\,\vec {r}_{n} + \vec {r}_{n - 1}}  \right)} \right] 
\  ,\tag{2.27a}
\\
\ddot {\vec {r}}_{n} = c\,\left[ {\dot {\vec {r}}_{n} \,\left( {\rho _{n + 
1} - 2\,\rho _{n} + \rho _{n - 1}}  \right) + \dot {\rho} _{n} \,\left( 
{\vec {r}_{n + 1} - 2\,\vec {r}_{n} + \vec {r}_{n - 1}}  \right) + \dot 
{\vec {r}}_{n} \wedge \left( {\vec {r}_{n + 1} - \vec {r}_{n - 1}}  \right)} 
\right] \  .
\tag{2.27b}
\end{gather*}
See (5.40).



A \textit{linearizable scalar-vector N-body problem} (with 2 arbitrary 
coupling constants, and ``nearest-neighbour'' interactions):
\begin{gather*}
\ddot {\rho} _{n} = \,a\dot {\rho} _{n} + \tilde {\rho} _{n} \dot {\rho 
}_{n} - \vec {\tilde {r}}_{n} \cdot \dot {\vec {r}}_{n} + c\,\left\{ {\tilde 
{\rho} _{n - 1} \dot {\rho} _{n} - \vec {\tilde {r}}_{n - 1} \cdot \dot 
{\vec {r}}_{n} - \rho _{n} \tilde {\tilde {\rho} }_{n + 1} + \vec {r}_{n} 
\cdot \vec {\tilde {\tilde {r}}}_{n + 1}}  \right\} \  ,\tag{2.28a}
\\
\ddot {\vec {r}}_{n} = a\,\dot {\vec {r}}_{n} + \tilde {\rho} _{n} \dot 
{\vec {r}}_{n} + \dot {\rho} _{n} \vec {\tilde {r}}_{n} - \vec {\tilde 
{r}}_{n} \wedge \dot {\vec {r}}_{n}
\\
\qquad{}+ c\,\left\{ {\tilde {\rho} _{n - 1} 
\dot {\vec {r}}_{n} + \dot {\rho} _{n} \vec {\tilde {r}}_{n - 1} - \vec 
{\tilde {r}}_{n - 1} \wedge \dot {\vec {r}}_{n} - \rho _{n} \vec {\tilde 
{\tilde {r}}}_{n + 1} - \tilde {\tilde {\rho} }_{n + 1} \vec {r}_{n} - \vec 
{r}_{n} \wedge \vec {\tilde {\tilde {r}}}_{n + 1}}  
\right\}\ ,\tag{2.28b}
\end{gather*}
where
\begin{gather*}
\tilde {\rho} _{n} \, \equiv \,\,\left( {\rho _{n} \dot {\rho} _{n} + \vec 
{r}_{n} \cdot \dot {\vec {r}}_{n}}  \right)/\left( {\rho _{n} ^{2} + r_{n} 
^{2}} \right) \  ,\tag{2.28c}
\\
\vec {\tilde {r}}_{n} \equiv \left( {\rho _{n} \dot {\vec {r}}_{n} - \dot 
{\rho} _{n} \vec {r}_{n} - \vec {r}_{n} \wedge \dot {\vec {r}}_{n}}  
\right)/\left( {\rho _{n} ^{2} + r_{n} ^{2}} \right) \  
,\tag{2.28d}
\end{gather*}
and
\begin{gather*}
\tilde {\tilde {\rho} }_{n + 1} \equiv \tilde {\rho} _{n} \tilde {\rho} _{n 
+ 1} - \vec {\tilde {r}}_{n} \cdot \vec {\tilde {r}}_{n + 1} \  ,
\tag{2.28e}
\\
\vec {\tilde {\tilde {r}}}_{n + 1} \equiv \tilde {\rho} _{n} \vec {\tilde 
{r}}_{n + 1} + \tilde {\rho} _{n + 1} \vec {\tilde {r}}_{n} - \vec {\tilde 
{r}}_{n} \wedge \vec {\tilde {r}}_{n + 1} \  .
\tag{2.28f}
\end{gather*}
See (5.43).


A \textit{linearizable N-body problem} (with $3N$\textit{} arbitrary 
coupling constants, and ``nearest-neighbour'' interactions):
\begin{gather*}
\ddot {\rho} _{n} = \,\left( {a_{n + 1} - a_{n} - b_{n}}  \right)\dot {\rho 
}_{n} + \tilde {\rho} _{n} \dot {\rho} _{n} - \vec {\tilde {r}}_{n} \cdot 
\dot {\vec {r}}_{n} + b_{n + 1} \left( {\tilde {\rho} _{n + 1} \dot {\rho 
}_{n} - \vec {\tilde {r}}_{n + 1} \cdot \dot {\vec {r}}_{n}}  \right)
\\*
\qquad{}+ c_{n + 1} \rho _{n} - c_{n} \left( {\rho _{n} \tilde {\tilde {\rho} }_{n 
- 1} - \vec {r}_{n} \cdot \vec {\tilde {\tilde {r}}}_{n - 1}}  \right)\ ,
\tag{2.29a}
\\
\ddot {\vec {r}}_{n} = \left( {a_{n + 1} - a_{n} - b_{n}}  \right)\,\,\dot 
{\vec {r}}_{n} + \tilde {\rho} _{n} \dot {\vec {r}}_{n} + \dot {\rho} _{n} 
\vec {\tilde {r}}_{n} - \vec {\tilde {r}}_{n} \wedge \dot {\vec 
{r}}_{n}
\\
\qquad{}+ 
b_{n + 1} \left( {\tilde {\rho} _{n + 1} \dot {\vec {r}}_{n} + \dot {\rho 
}_{n} \vec {\tilde {r}}_{n + 1} - \vec {\tilde {r}}_{n + 1} \wedge \dot 
{\vec {r}}_{n}}  \right)
\\
\qquad{}+ c_{n + 1} \vec {r}_{n} - c_{n} \left( {\rho _{n} \vec {\tilde {\tilde 
{r}}}_{n - 1} + \tilde {\tilde {\rho} }_{n - 1} \vec {r}_{n} + \vec {r}_{n} 
\wedge \vec {\tilde {\tilde {r}}}_{n - 1}}  \right)\ ,
\tag{2.29b}
\end{gather*}
where $\tilde {\rho} _{n} ,\vec {\tilde {r}}_{n} $ are given again by 
(2.28c), (2.28d) and
\begin{gather*}
\tilde {\tilde {\rho} }_{n - 1} \equiv \tilde {\rho} _{n} 
\mathord{\buildrel{\lower3pt\hbox{$\scriptscriptstyle\frown$}}\over {\rho} } 
_{n - 1} - \vec {\tilde {r}}_{n} \cdot \vec 
{\mathord{\buildrel{\lower3pt\hbox{$\scriptscriptstyle\frown$}}\over {r}} 
}_{n - 1} \  ,
\tag{2.29c}
\\
\vec {\tilde {\tilde {r}}}_{n - 1} \equiv \tilde {\rho} _{n} \vec 
{\mathord{\buildrel{\lower3pt\hbox{$\scriptscriptstyle\frown$}}\over {r}} 
}_{n - 1} + 
\mathord{\buildrel{\lower3pt\hbox{$\scriptscriptstyle\frown$}}\over {\rho} } 
_{n - 1} \vec {\tilde {r}}_{n} - \vec {\tilde {r}}_{n} \wedge \vec 
{\mathord{\buildrel{\lower3pt\hbox{$\scriptscriptstyle\frown$}}\over {r}} 
}_{n - 1} \  ,
\tag{2.29d}
\\
\mathord{\buildrel{\lower3pt\hbox{$\scriptscriptstyle\frown$}}\over {\rho} } 
_{n} \, \equiv \,\,\left( {\rho _{n} \dot {\rho} _{n} + \vec {r}_{n} \cdot 
\dot {\vec {r}}_{n}}  \right)\,/\,\left( {\dot {\rho} _{n} ^{2} + \,\,\dot 
{\vec {r}}_{n} \cdot \dot {\vec {r}}_{n}}  \right) \  ,
\tag{2.29e}
\\
\vec {\mathord{\buildrel{\lower3pt\hbox{$\scriptscriptstyle\frown$}}\over 
{r}}} _{n} \equiv \left( {\dot {\rho} _{n} \vec {r}_{n} \, - \rho _{n} \dot 
{\vec {r}}_{n} + \vec {r}_{n} \wedge \dot {\vec {r}}_{n}}  
\right)\,/\,\left( {\dot {\rho} _{n} ^{2} + \,\,\dot {\vec {r}}_{n} \cdot 
\dot {\vec {r}}_{n}}  \right) \  ,
\tag{2.29f}
\end{gather*}
See (5.44).


For \textit{linearizable} equations of motion involving the $N + 5$ 
three-vectors $\vec {r}_{n} \left( {t} \right)$, $\vec {f}\left( {t} 
\right)$, $\vec {g}\left( {t} \right)$, $\vec {h}\left( {t} \right)$, $\vec 
{v}\left( {t} \right)$, $\vec {y}\left( {t} \right)$ and the $2$ scalars $\eta 
\left( {t} \right)$, $\theta \left( {t} \right)$, and featuring $2\,N^{2} + 
20\,N + 42$ arbitrary ``coupling constants,'' we refer the reader to (5.45).


A scalar/vector \textit{solvable} $N^{2}$\textit{-body problem} (with 
$4N^{2}$ arbitrary coupling constants):
\begin{gather*}
\ddot {\rho} _{nm} = a_{nm} + \sum\limits_{m_{1} = 1}^{N} [b_{m_{1} 
m} \,\rho _{nm_{1}}  + c_{m_{1} m} \,\dot {\rho}_{nm_{1}}] 
\\
\qquad{}- 
\sum\limits_{m_{1} ,m_{2} = 1}^{N} d_{m_{1} m_{2}} [\dot {\rho }_{nm_{1}}  \,\rho_{m_{2} n} - (\dot{\vec {r}}_{nm_{1}}  \cdot 
\vec {r}_{m_{2} n}) + 2\,\rho _{nm_{1}}  \,\dot{\rho}_{m_{2} n} - 
2\,(\vec{r}_{nm_{1}}  \cdot \dot{\vec {r}}_{m_{2} n})]
\\
\qquad{} + \sum\limits_{m_{1} ,m_{2} ,m_{3} = 1}^{N} {d_{m_{1} m_{2}}  \,c_{m_{3} m} 
\,\left[ {\rho _{nm_{1}}  \,\rho _{m_{2} m_{3}}  - \left( {\vec {r}_{nm_{1} 
} \cdot \vec {r}_{m_{2} m_{3}} }  \right)} \right]} 
\\
\qquad{}- \sum\limits_{m_{1} ,m_{2} ,m_{3} ,m_{4} = 1}^{N} {d_{m_{1} m_{2}}  
\,d_{m_{3} m_{4}}  \,\left[ {} \right.} \rho _{nm_{1}}  
\,\rho _{m_{2} m_{3}}  \,\rho _{m_{4} m} 
\\ \qquad{}
 - \rho _{nm_{1}}  \left( {\vec {r}_{m_{2} m_{3}}  \cdot \vec {r}_{m_{4} m} 
} \right) - \rho _{m_{2} m_{3}}  \left( {\vec {r}_{nm_{1}}  \cdot \vec 
{r}_{m_{4} m}}  \right) - \rho _{m_{4} m} \,\left( {\vec {r}_{nm_{1}}  \cdot 
\vec {r}_{m_{2} m_{3}} }  \right)
\\ \qquad{}
 + \left( {\vec {r}_{nm_{1}}  \wedge \vec 
{r}_{m_{2} m_{3}} }  \right) \cdot \vec {r}_{m_{4} m} \left. {} 
\right]\ ,\tag{2.30a}
\\ 
\ddot {\vec {r}}_{nm} = \sum\limits_{m_{1} = 1}^{N} {\left[ {b_{m_{1} m} 
\,\vec {r}_{nm_{1}}  + c_{m_{1} m} \,\dot {\vec {r}}_{nm_{1}} }  \right]} + 
\sum\limits_{m_{1} ,m_{2} = 1}^{N} {d_{m_{1} m_{2}}  \,\left[ {\dot {\vec 
{r}}_{nm_{1}}  \wedge \vec {r}_{m_{2} n} + 2\,\vec {r}_{nm_{1}}  \wedge \dot 
{\vec {r}}_{m_{2} n}}  \right]}
\\ \qquad{}
 - \sum\limits_{m_{1} ,m_{2} ,m_{3} = 1}^{N} d_{m_{1} m_{2}}  \,c_{m_{3} m} 
\,\vec {r}_{nm_{1}}  \wedge \vec{r}_{m_{2} m_{3}}
\\ \qquad{}
- \sum\limits_{m_{1} ,m_{2} ,m_{3} ,m_{4} = 1}^{N} {} d_{m_{3} m_{4}}  
d_{m_{1} m_{2}}\left[ 
{} \right.\vec {r}_{nm_{1}}  \,\left\{ \rho _{m_{2} m_{3}}  \,\rho_{m_{4} 
m} - \left( \vec {r}_{m_{2} m_{3}}  \cdot \vec{r}_{m_{4} m}  \right)
\right\}
\\ \qquad{}
 + \vec {r}_{m_{2} m_{3}}  \,\left\{ {\rho _{nm_{1}}  \,\rho _{m_{4} m} - 
\left( {\vec {r}_{nm_{1}}  \cdot \vec {r}_{m_{4} m}}  \right)} \right\} + 
\vec {r}_{m_{4} m} \,\left\{ {\rho _{nm_{1}}  \,\rho _{m_{2} m_{3}}  - 
\left( {\vec {r}_{nm_{1}}  \cdot \vec {r}_{m_{2} m_{3}} }  \right)} 
\right\}\left. {} \right].\tag{2.30b}
\end{gather*}
See (5.46).


A \textit{solvable} (Hamiltonian) \textit{N-body problem} (with $2\,N^{2} + 
N$ arbitrary coupling constants):
\begin{gather*}
\dot {\vec {q}}_{n} = \sum\limits_{m = 1}^{N} {\left( {a_{nm} \,\vec {q}_{m} 
+ b_{nm} \,\vec {p}_{m} + 4\,\lambda \,\left[ {\vec {q}_{n} \wedge \left[ 
{\vec {q}_{m} \wedge \vec {p}_{m}}  \right]} \right]} \right)}\  
,
\\
\dot {\vec {p}}_{n} = \sum\limits_{m = 1}^{N} {\left( {c_{nm} 
\,\vec {q}_{m} - a_{mn} \,\vec {p}_{m} + 4\,\lambda \,\left[ {\vec {p}_{n} 
\wedge \left[ {\vec {q}_{m} \wedge \vec {p}_{m}}  \right]} \right]} 
\right)} \ 
,\tag{2.31}
\end{gather*}
where $b_{nm} \, = \,b_{mn} $ , $c_{nm} \, = \,c_{mn} $ .


For the corresponding Hamiltonian, see (F.8b).

\section{Survey of solvable and/or integrable and/or linearizable matrix 
models} \label{III}


In this section we survey solvable and/or integrable and/or linearizable 
matrix evolution equations. We focus throughout on square matrices, whose 
symbols are underlined. Most of the results reported in this section are 
probably well known.


\subsection{An explicitly solvable matrix evolution equation}
\label{III.A}


In this subsection we detail the explicit solution of the following matrix 
evolution equation:
\begin{equation*}
\underline {\ddot {M}} = 2a\underline {\dot {M}} + b\underline {M} + 
c\underline {\dot {M}} \,\underline {M} ^{ - 1}\underline {\dot 
{M}} \ .\tag{3.1}
\end{equation*}

In writing (3.1) we implicitly assume the matrix $\underline {M} $ to be 
invertible; it is clear from the results reported below that this property 
is generally maintained by the time-evolution (3.1), unless the solution 
develops a singularity (this is possible in some cases, as can be easily 
seen from the solutions reported below). The 3 quantities $a,b$ and $c$ are 
arbitrary (scalar, possibly complex) constants (the factor of 2 in front of 
$a$ is introduced for notational convenience). In fact, cases in which $a,b$ 
and $c$ depend on the time $t$ can also be treated, but for simplicity we 
refrain from doing so.

The matrix evolution equation (3.1) can be solved in explicit form for 
arbitrary initial conditions. For $c \ne 1$ the solution reads
\begin{gather*}
\underline {M} \left( {t} \right) = {\rm exp}\left( a\,\gamma \,t \right) \{ 
\,{\rm cosh}\left( \Delta \,t \right) 
\\* \qquad{}
+ \Delta^{- 1} {\rm sinh} \left( \Delta \,t 
\right) \left[{\underline{\dot {M}}} \left( 0 \right)\,\left[ \gamma 
\,\underline {M} \left( 0 \right) \right]^{ - 1} - a \right]\,\}^{\,\gamma}
\,\underline {M} \left( 0 \right) \  
,\tag{3.2a}
\\
\gamma = 1/\left( {1 - c} \right) \  ,\tag{3.2b}
\\
\Delta = \left[ {a^{2} + b\,\left( {1 - c} \right)} \right]^{1/2}\ . \tag{3.2c}
\end{gather*}

Hence $\underline {M} \left( {t} \right)$ is completely periodic with period 
$T = 2\pi /\omega $ , for arbitrary initial conditions, if
\begin{gather*}
a\,\gamma = i\,m\omega \  ,\tag{3.3a}
\\
\Delta = i\,n\,\omega \  ,\tag{3.3b}
\end{gather*}
with $\omega $ an arbitrary (nonvanishing) \textit{real} constant, and $m,n$ 
two arbitrary integers (positive or negative, at least one of them 
nonvanishing). These two conditions entail, if $m \ne 0$, a single 
restriction on the 3 parameters $a,b$ and $c$:
\begin{equation*}
b = a^{2}\,\left( {1 - \left\{ {\,n/\left[ {m\,\left( {c - 1} \right)} 
\right]} \right\}^{2}} \right)/\left( {c - 1} \right)\ . \tag{3.3c}
\end{equation*}
However, in this case with $m \ne 0$, the condition (3.3a) with (3.2b) 
entails that $a$ and $c$ cannot both be \textit{real}. Perhaps more 
interesting is the case with $m = 0$, entailing (see (3.3a)
\begin{equation*}
a = 0 \  . \tag{3.4a}
\end{equation*}

Then \textit{all} solutions of (3.1) are periodic, with period 
\begin{equation*}
T = 2\,\pi \,\left[ {b\,\left( {c - 1} \right)} \right]^{ - 1/2} \  ,
\tag{3.4b}
\end{equation*}
if this quantity is \textit{real}. It is moreover clear from (3.2a) that in 
this \textit{real} case, with $a = 0,\,\,$ and $\,\,b,c,T$ \textit{real}, a 
necessary and sufficient condition to exclude that the matrix $\underline 
{M} \left( {t} \right)$ become singular (or, equivalently, noninvertible) at 
any time is that {\it none} of the eigenvalues of the (generally 
\textit{nonsymmetrical}) matrix $\underline {\dot {M}} \left( {0} 
\right)\,\left[ {\underline {M} \left( {0} \right)} \right]^{ - 1}$ be 
\textit{real}.

The behavior of the solution (3.2) in more general cases is sufficiently 
clear from the explicit formula (3.2) not to require any additional 
elaboration here.

The solution of (3.1) in the special case 
\begin{equation*}
c = 1
\tag{3.5a}
\end{equation*}
could be obtained by a limiting procedure from (3.2), but it deserves to be 
separately displayed. If
\begin{equation*}
a \ne 0 \tag{3.5b}
\end{equation*}
it reads
\begin{gather*}
\underline {M} \left( {t} \right) = {\rm exp}\left\{ {\,\left[ {b/\left( {2\,a} 
\right)} \right]\,\,\left[ { - t + {\rm exp}\left( {a\,t} \right)\,\,a^{ - 
1}\,{\rm sinh}\,\left( {a\,t} \right)} \right]\,} \right\}
\\ \qquad{}
{\rm exp}\left\{ 
{\,{\rm exp}\left( {a\,t} \right)\,\,a^{ - 1}{\rm sinh}\,\left( {a\,t} \right)\,\left( 
{\underline {\dot {M}} \left( {0} \right)\,\,\left[ {\underline {M} \left( 
{0} \right)} \right]^{ - 1}} \right)} \right\}\,\,\underline {M} \left( {0} 
\right)\ .
\tag{3.5c}
\end{gather*}
This solution is periodic with (real) period $T = 2\pi /\omega $ , for 
arbitrary initial conditions, iff
\begin{gather*}
a = i\,m\omega \  ,
\tag{3.6a}
\\
b = 2\,n\,m\omega ^{2} \  ,
\tag{3.6b}
\end{gather*}
with $\omega $ real (nonvanishing) and $m,n$ two arbitrary integers, except 
for the restriction $m \ne 0$, which entails that in this case $a$ must be 
imaginary, see (3.6a).

Again, we do not elaborate on the behavior of the solution (3.5c) in more 
general cases than the periodic one.

Finally, in the case
\begin{equation*}
c = 1\,,\,\,\,\,a\, = 0 \  ,\tag{3.8a}
\end{equation*}
the solution of (3.1) reads simply
\begin{equation*}
\underline {M} \left( {t} \right) = {\rm exp}\left( {b\,t^{2}/2} \right) 
{\rm exp}\left\{ 
{t\,\underline {\dot {M}} \left( {0} \right)\,\left[ {\underline {M} \left( 
{0} \right)} \right]^{ - 1}} \right\}\,\underline {M} \left( {0} \right) 
\  .\tag{3.8b}
\end{equation*}


\subsection{A class of linearizable matrix evolution equations}
\label{III.B}


In this subsection we present a technique to manufacture \textit{new 
}(generally nontrivial) \textit{linearizable} matrix evolution equations 
from \textit{known} (possibly trivially solvable) \textit{linearizable 
}matrix evolution equations. Some aspects of this technique might be new; we 
indicate below, and accordingly give due credit for, those aspects we know 
not to be new.

In Appendix A we treat in some detail a specific, very simple, example, as 
indicated at the end of this subsection. To make that appendix 
self-contained, we explain there again the essence of the results presented 
in this subsection, since in that very simple context this can be done quite 
tersely. Hence the reader might find it profitable to scan Appendix A (at 
least its opening part), before delving in the treatment given below.

Our treatment is based on the following \textit{Lemma}.

Assume that the $N + 5$ square matrices $\underline {u} _{n} ,\,\,\underline 
{f} ,\,\,\underline {g} ,\,\,\underline {h} ,\,\,\underline {v} 
,\,\,\underline {y} $ satisfy the $N + 5$ matrix ODEs
\begin{gather*}
\underline{\tilde {U}} _{n} ( \underline {u} _{m}^{(0 )} ,\underline{u}_{m}^{(1)} ,\underline {u} 
_{m}^{(2)} ,...,m = 1,...,N;\underline{f} ^{(0)},\underline {f}^{(1)},
\underline {f}^{(2)},...;
\\ \qquad
\underline{g}^{(0)},\underline{g} ^{(1)},\underline{g}^{(2)},...;
\underline{h}^{(0)},\underline{h}^{(1)},\underline{h} ^{(2 )},...;t) = 0\ ,
\tag{3.9a}
\\[1ex]
\underline{\tilde{F}} (\underline{u}_{m}^{(0)} ,\underline{u}_{m}^{(1)},
\underline{u}_{m}^{(2)},...,m = 1,...,N;
\underline{f}^{(0)},\underline{f}^{(1)},\underline{f}^{(2)},...;
\\ \qquad
\underline{g}^{(0)},\underline{g}^{(1)},\underline{g}^{(2)},...;\underline{h} 
^{(0)},\underline{h}^{(1)},\underline {h} 
^{(2)},...;t) = 0 \  ,
\tag{3.9b}
\\[1ex]
\underline{\tilde{G}} (\underline{u} _{m}^{(0)} 
,\underline{u}_{m}^{(1)} ,\underline{u} _{m}^{(2)} ,...,m = 1,...,N;
\underline{f}^{(0)},\underline{f}^{(1)},\underline{f}^{(2)},...;
\\ \qquad
\underline{g}^{(0)},\underline{g}^{(1)},\underline{g}^{(2)},...;
\underline{h}^{(0)},\underline{h}^{(1)},\underline{h} 
^{(2)},...;t) = 0 \  ,\tag{3.9c}
\\[1ex]
\underline{\tilde{H}} (\underline{u}_{m}^{(0)} 
,\underline{u}_{m}^{(1)} ,\underline{u}_{m}^{(2)},...,m = 1,...,N;
\underline{f}^{(0)},\underline 
{f}^{(1)},\underline{f}^{(2}),...;
\\ \qquad
\underline{g}^{(0)},\underline{g}^{(1)},\underline{g}^{(2)},...;\underline{h} 
^{(0)},\underline{h}^{(1)},\underline{h} 
^{(2)},...;t) = 0 \  ,
\tag{3.9d}
\\[1ex]
\sum\limits_{j = 0}^{J_{1}}  \underline{\tilde {V}}_{j} 
(\underline {u}_{m}^{(0)} ,\underline{u}_{m}^{(1)} ,
\underline{u}_{m}^{(2)} ,...,m = 
1,...,N;\underline{f}^{(0)},\underline{f}^{(1)},\underline{f}^{(2)},...;
\\ \qquad
\underline{g}^{(0)},\underline{g}^{(1)},\underline{g}^{(2)},...;
\underline{h}^{(0)},\underline{h}^{(1)},\underline{h}^{(2)},...;t )\,
\underline{v}^{(j)}=0\ ,\tag{3.9e}
\\[1ex]
\sum\limits_{j = 0}^{J_{2}}  \underline{y}^{(j)}\, \underline{\tilde{Y}}_{j} 
(\underline{u} _{m}^{(0)} ,\underline{u}_{m}^{(1)} ,\underline{u}_{m}^{(2)}
,...,m = 1,...,N;\underline {f} ^{(0)},\underline{f}^{(1)},\underline{f}^{(2)}
,...;
\\ \qquad
\underline{g}^{(0)},\underline{g}^{(1)},\underline{g}^{(2)},...;\underline{h} 
^{(0)},\underline{h}^{(1)},\underline{h} 
^{(2)},...;t)=0\  .\tag{3.9f}
\end{gather*}
Here $\underline {u} _{n}^{\left( {j} \right)} \equiv d^{j}\underline {u} 
_{n} /dt^{j},\,\,j = 0,1,2,...$, with analogous formulas for $\underline{f} 
^{(j)},\,\,\underline{g}^{(j)},$ $\underline{h}^{(j)},\,\,\underline{v} 
^{(j)},\,\,\underline{y}^{(j)}$, and the 
functions $\underline {\tilde {U}} _{n} \,,\,\,\underline {\tilde {F}} 
\,,\,\,\underline {\tilde {G}} \,,\,\,\underline {\tilde {H}} 
\,,\,\,\underline {\tilde {V}} _{j} \,,\,\,\underline {\tilde {Y}} _{j} $, 
are ``scalar/matrix functions of matrices,'' namely functions built only out 
of their matrix arguments (whose ordering is of course important) and of 
scalars (including, possibly, the independent variable $t$), so that they 
satisfy the identity
\begin{gather*}
\underline{W} ^{-1}\,\underline{\tilde {Z}} (\underline{u} 
_{m}^{(0)} ,\underline{u}_{m}^{(1)} 
,\underline{u}_{m}^{(2)} ,...,m = 1,...,N;\underline{f} 
^{(0)},\underline{f}^{(1)},\underline{f} 
^{(2)},...;
\\ \qquad
\underline {g} ^{(0)},\underline 
{g}^{(1)},\underline{g} ^{(2)},...;\underline{h}^{(0)},\underline{h}^{(1)},
\underline{h}^{(2)},...;t )\,\underline{W} \,=
\\[1ex]
\underline{\tilde{Z}} (\underline{W}^{-1} \underline{u}_{m}^{(0)} 
\underline{W} ,\underline{W}^{ - 1}\underline 
{u} _{m}^{(1)} \underline{W} ,\underline{W} ^{ - 
1}\underline{u}_{m}^{(2)} \underline {W} ,...,m = 
1,...,N;
\\* \qquad
\underline {W}^{ - 1} \underline {f} ^{(0)}\underline 
{W} ,\underline {W} ^{ - 1}\underline {f} ^{(1)}\underline 
{W} ,\underline {W} ^{ - 1}\underline {f} ^{(2)}\underline {W} ,...;
\\[1ex]
\underline{W}^{-1}\underline{g}^{(0)}\underline{W} 
,\underline {W} ^{ - 1}\underline {g} ^{( 1)}\underline {W} 
,\underline {W}^{ - 1}\underline {g} ^{(2)}\underline {W} 
,...;\underline {W} ^{ - 1}\underline {h} ^{(0)}\underline 
{W} ,\underline {W} ^{ - 1}\underline {h} ^{(1)}\underline 
{W} ,\underline {W} ^{ - 1}\underline {h} ^{(2)}\underline 
{W} ,...;t)\ ,\tag{3.10}
\end{gather*}
where $\underline {\tilde {Z}} $ denotes any one of these functions. The 
notation employed to write these equations is meant to suggest (in a sense 
which will become clear below) that the equation $\underline {\tilde {Z}} = 
0$ plays the primary role in determining the matrix $\underline {z} $ , with 
the obvious correspondence (if $\underline {\tilde {Z}} = \underline {\tilde 
{U}} _{n} $, then $\underline {z} = \underline {u} _{n} $; if $\underline 
{\tilde {Z}} = \underline {\tilde {F}} $, then $\underline {z} = \underline 
{f} $; and so on). Note moreover the qualitative difference among the 
equations satisfied by the $N + 3$ matrices $\underline {u} _{n} 
\,,\,\,\underline {f} \,,\,\,\underline {g} \,,\,\,\underline {h} $, 
(3.9a,b,c,d), and those satisfied by $\underline {v} $ and $\underline {y} $ 
, (3.9e,f) (in this latter equations the parameters $J_{1} $ and $J_{2} $ 
are of course two nonnegative integers). The different role of the $N$ 
matrices $\underline {u} _{n} $, and of the $3$ matrices $\underline {f} 
\,,\,\,\underline {g} \,,\,\,\underline {h} $, will become clear below.

Consider now $N + 5$ square matrices $\underline {U} _{n} \,,\,\,\underline 
{F} \,,\,\,\underline {G} \,,\,\,\underline {H} \,,\,\,\underline {V} 
\,,\,\,\underline {Y} $, related to the $N + 5$ square matrices $\underline 
{u} _{n} \,,\,\,\underline {f} \,,\,\,\underline {g} \,,\,\,\underline {h} 
\,,\,\,\underline {v} \,,\,\,\underline {y} $ as follows:
\begin{gather*}
\underline {u} _{n} = \underline {W} \,\underline {U} _{n} \,\underline {W} 
^{ - 1},\,\,\,\,\,\underline {f} = \underline {W} \,\underline {F} 
\,\underline {W} ^{ - 1},\,\,\,\,\,\underline {g} = \underline {W} 
\,\underline {G} \,\underline {W} ^{ - 1},\,\,\,\,\,\underline {h} = 
\underline {W} \,\underline {H} \,\underline {W} ^{ - 1},
\tag{3.11a}
\\
\underline {U} _{n} = \underline {W} ^{ - 1}\,\underline {u} _{n} 
\,\underline {W} ,\,\,\,\,\,\underline {F} = \underline {W} ^{ - 
1}\,\underline {f} \,\underline {W} ,\,\,\,\,\,\underline {G} = \underline 
{W} ^{ - 1}\,\underline {g} \,\underline {W} ,\,\,\,\,\,\underline {H} = 
\underline {W} ^{ - 1}\,\underline {h} \,\underline {W} ,
\tag{3.11b}
\\
\underline {v} = \underline {W} \,\underline {V} ,\,\,\,\,\,\underline {y} = 
\underline {Y} \,\underline {W} ^{ - 1} \  ,
\tag{3.12a}
\\
\underline {V} = \underline {W} ^{ - 1}\,\underline {v} 
,\,\,\,\,\,\underline {Y} = \underline {y} \,\underline {W} \  ,
\tag{3.12b}
\end{gather*}
with
\begin{gather*}
\underline {\dot {W}} = \underline {g} \,\underline {v} + \underline {f} 
\,\underline {W} + \underline {W} \,\underline {y} \,\underline {h} 
\,\underline {W} \  ,
\tag{3.13a}
\\
\underline {\dot {W}} = \underline {W} \,\underline {K} \  ,
\tag{3.13b}
\end{gather*}
where
\begin{equation*}
\underline {K} = \underline {F} + \underline {G} \,\underline {V} + 
\underline {Y} \,\underline {H} \  .
\tag{3.14}
\end{equation*}
Note that (3.13b) with (3.14) follows from (3.13a) via (3.11a) and 
(3.12a).

To complete the definition of the (invertible!) matrix $\underline {W} 
\equiv \underline {W} \left( {t} \right)$ one must supplement the matrix ODE 
(3.13) with an ``initial condition,'' say $\underline {W} \left( {0} \right) 
= \underline {W} _{0} $. It will remain our privilege to make an appropriate 
choice for the (invertible!) matrix $\underline {W} _{0} $; generally it 
will be convenient to set simply $\underline {W} _{0} = \underline {1} $ .

We can now formulate the first of the two \textit{Propositions} contained in 
our \textit{Lemma}.
\begin{proposition}
The $N + 5$ square matrices $\underline {U} _{n} \,,\,\,\underline {F} 
\,,\,\,\underline {G} \,,\,\,\underline {H} \,,\,\,\underline {V} 
\,,\,\,\underline {Y} $ can be obtained from the $N + 5$ square matrices 
$\underline {u} _{n} \,,\,\,\underline {f} \,,\,\,\underline {g} 
\,,\,\,\underline {h} \,,\,\,\underline {v} \,,\,\,\underline {y} $ by 
algebraic operations (inversion and multiplication of matrices) and by 
solving a \emph{linear second-order matrix ODE} (which in some cases 
reduces to a \emph{first-order matrix ODE}, see below). If the matrices 
$\,\underline {f} \,,\,\,\underline {g} \,,\,\,\underline {h} 
\,,\,\,\underline {v} \,,\,\,\underline {y} $ are time-dependent, this 
\emph{linear ODE} is generally \emph{nonautonomous}.
\end{proposition}

Let us then define matrices $\underline {U} _{n}^{\left[ {j} \right]} 
\,,\,\,\underline {F} ^{\left[ {j} \right]}\,,\,\,\underline {G} ^{\left[ 
{j} \right]}\,,\,\,\underline {H} ^{\left[ {j} \right]}$ and $\underline {V} 
^{\left\{ {j} \right\}},\,{}^{\left\{ {j} \right\}}\underline {Y} $ , via 
the following recursive formulas:
\begin{gather*}
\underline {Z} ^{\left[ {0} \right]} = \underline {Z} 
,\,\,\,\,\,\,\underline {Z} ^{\left[ {j + 1} \right]} = \underline {\dot 
{Z}} ^{\left[ {j} \right]}\, - \,\left[ {\underline {Z} ^{\left[ {j} 
\right]},\underline {K}}  \right] \  ,
\tag{3.15a}
\\
\underline {Z} ^{\left\{ {0} \right\}} = \underline {Z} 
,\,\,\,\,\,\underline {Z} ^{\left\{ {j + 1} \right\}} = \underline {\dot 
{Z}} ^{\left\{ {j} \right\}}\, + \,\underline {K} \,\underline {Z} ^{\left\{ 
{j} \right\}} \  ,
\tag{3.15b}
\\
{}^{\left\{ {0} \right\}}\underline {Z} = \underline {Z} 
,\,\,\,\,\,{}^{\left\{ {j + 1} \right\}}\underline {Z} = {}^{\left\{ {j} 
\right\}}\underline {\dot {Z}} \, - \,{}^{\left\{ {j} \right\}}\underline 
{Z} \,\underline {K} \  ,
\tag{3.15c}
\end{gather*}
so that
\begin{gather*}
\underline {Z} ^{\left[ {1} \right]} = \underline {\dot {Z}} - \left[ 
{\underline {Z} ,\underline {K}}  \right],\,\,\,\,\,\underline {Z} ^{\left[ 
{2} \right]} = \underline {\ddot {Z}} - 2\left[ {\underline {\dot {Z}} 
,\underline {K}}  \right] - \left[ {\underline {Z} ,\underline {\dot {K}}}  
\right] + \left[ {} \right.\left[ {\underline {Z} ,\underline {K}}  
\right],\,\underline {K} \left. {} \right] \  ,...,
\tag{3.16a}
\\
\underline {Z} ^{\left\{ {1} \right\}} = \underline {\dot {Z}} + \underline 
{K} \,\underline {Z} ,\,\,\,\,\,\underline {Z} ^{\left\{ {2} \right\}} = 
\underline {\ddot {Z}} + 2\underline {K} \,\underline {\dot {Z}} + 
\underline {K} \,\underline {Z} + \underline {K} ^{2}\,\underline {Z} \  
,...,
\tag{3.16b}
\\
{}^{\left\{ {1} \right\}}\underline {Z} = \underline {\dot {Z}} - \underline 
{Z} \,\underline {K} ,\,\,\,\,\,{}^{\left\{ {2} \right\}}\underline {Z} = 
\underline {\ddot {Z}} - 2\underline {\dot {Z}} \,\underline {K} - 
\underline {Z} \,\underline {K} + \underline {Z} \,\underline {K} ^{2} \  
,... \  .
\tag{3.16c}
\end{gather*}
Here and throughout $\left[ {\underline {A} ,\underline {B}}  \right]$ 
denotes the \textit{commutator} of the two matrices $\underline {A} $ and 
$\underline {B} ,\,\,\left[ {\underline {A} ,\underline {B}}  \right] \equiv 
\underline {A} \,\underline {B} - \underline {B} \,\underline {A} $ .

We can then formulate the second \textit{Proposition} of our \textit{Lemma}.
\begin{proposition}
If the $N + 5$ square matrices $\underline {u} _{n} \,,\,\,\underline {f} 
\,,\,\,\underline {g} \,,\,\,\underline {h} \,,\,\,\underline {v} 
\,,\,\,\underline {y} $ satisfy the $N + 5$ matrix ODEs (3.9), then the $N + 
5$ square matrices $\underline {U} _{n} \,,\,\,\underline {F} 
\,,\,\,\underline {G} \,,\,\,\underline {H} \,,\,\,\underline {V} 
\,,\,\,\underline {Y} $ satisfy the $N + 5$ matrix ODEs
\begin{gather*}
\underline{\tilde {U}}_{n} (\underline{U} _{m}^{[0]} ,\underline{U} _{m}^{[1]} ,\underline {U} 
_{m}^{[2]} ,...,m = 1,...,N;\underline {F}^{[0]},\underline {F} ^{[ 1]},
\underline{F} ^{[2]},...;
\\ \qquad
\underline{G} ^{[0]},\underline{G} ^{[1]},\underline {G}^{[2]},...;
\underline {H}^{[0]},\underline {H} ^{[ 1]},\underline {H} 
^{[2]},...;t) = 0,\tag{3.17a}
\\[1ex]
\underline {\tilde {F}} ( \underline {U} _{m}^{\left[ {0} \right]} 
,\underline {U} _{m}^{\left[ {1} \right]} ,\underline {U} _{m}^{\left[ {2} 
\right]} ,...,m = 1,...,N;\underline {F} ^{\left[ {0} \right]},\underline 
{F} ^{\left[ {1} \right]},\underline {F} ^{\left[ {2} 
\right]},...;
\\ \qquad
\underline {G} ^{\left[ {0} \right]},\underline {G} ^{\left[ 
{1} \right]},\underline {G} ^{\left[ {2} \right]},...;\underline {H} 
^{\left[ {0} \right]},\underline {H} ^{\left[ {1} \right]},\underline {H} 
^{\left[ {2} \right]},...;t) = 0,
\tag{3.17b}
\\[1ex]
\underline {\tilde {G}} ( \underline {U} _{m}^{\left[ {0} \right]} 
,\underline {U} _{m}^{\left[ {1} \right]} ,\underline {U} _{m}^{\left[ {2} 
\right]} ,...,m = 1,...,N;\underline {F} ^{\left[ {0} \right]},\underline 
{F} ^{\left[ {1} \right]},\underline {F} ^{\left[ {2} 
\right]},...;
\\ \qquad
\underline {G} ^{\left[ {0} \right]},\underline {G} ^{\left[ 
{1} \right]},\underline {G}^{\left[ {2} \right]},...;\underline {H} 
^{\left[{0} \right]},\underline {H}^{\left[ {1} \right]},\underline {H} 
^{\left[ {2} \right]},...;t) = 0,\tag{3.17c}
\\[1ex]
\underline {\tilde {H}} ( \underline {U} _{m}^{\left[ {0} \right]} 
,\underline {U} _{m}^{\left[ {1} \right]} ,\underline {U} _{m}^{\left[ {2} 
\right]} ,...,m = 1,...,N;\underline {F} ^{\left[ {0} \right]},\underline 
{F} ^{\left[ {1} \right]},\underline {F} ^{\left[ {2} 
\right]},...;
\\ \qquad
\underline {G} ^{\left[ {0} \right]},\underline {G} ^{\left[ 
{1} \right]},\underline {G} ^{\left[ {2} \right]},...;\underline {H} 
^{\left[ {0} \right]},\underline {H} ^{\left[ {1} \right]},\underline {H} 
^{\left[ {2} \right]},...;t) = 0,\tag{3.17d}
\\[1ex]
\sum\limits_{j = 0}^{J_{1}}  \underline{\tilde {V}}_{j} ( 
\underline{U} _{m}^{[0]} ,\underline{U}_{m}^{\left[{1} 
\right]} ,\underline{U}_{m}^{\left[ {2} \right]} ,...,m = 
1,...,N;\underline {F}^{\left[ {0} \right]},\underline {F}^{\left[ {1} 
\right]},\underline {F} ^{\left[ {2} \right]},...;
\\ \qquad
\underline {G}^{\left[ 
{0} \right]},\underline{G}^{\left[ {1} \right]},\underline{G}^{\left[ 
{2} \right]},...;\underline {H}^{\left[ {0} \right]},\underline {H} 
^{\left[{1} \right]},\underline {H}^{\left[{2}\right]},...;t)\,
\underline{V}^{\left\{ {j} \right\}}=0,\tag{3.17e}
\\[1ex]
\sum\limits_{j = 0}^{J_{2}}  {{}^{\{j}\}} \underline {Y} 
\,\underline {\tilde{Y}}_{j} ( \underline{U} _{m}^{\left[{0} 
\right]} ,\underline{U} _{m}^{\left[{1} \right]} ,\underline {U} 
_{m}^{\left[ {2} \right]} ,...,m = 1,...,N;\underline {F}^{\left[ {0} 
\right]},\underline {F} ^{\left[ {1} \right]},\underline {F} ^{\left[ {2} 
\right]},...;
\\ \qquad
\underline{G}^{\left[ {0} \right]},\underline {G} ^{\left[ 
{1} \right]},\underline {G} ^{\left[ {2} \right]},...;\underline {H} 
^{\left[ {0} \right]},\underline {H} ^{\left[ {1} \right]},\underline {H} 
^{\left[ {2} \right]},...;t)=0, 
\tag{3.17f}
\end{gather*}
obtained from (3.9) by replacing, wherever they appear, the matrices 
$\underline {u} _{n}^{\left( {j} \right)}$, $\underline {f} ^{\left( {j} 
\right)},$ $\underline {g} ^{\left( {j} \right)}$, $\underline {h} ^{\left( 
{j} \right)}$, $\underline {v} ^{\left( {j} \right)}$, $\underline {y} 
^{\left( {j} \right)}$ respectively with the matrices $\underline {U} 
_{n}^{\left[ {j} \right]} ,\,\underline {F}^{\left[ {j} 
\right]},\,\underline {G}^{\left[ {j} \right]},\,\underline {H}^{\left[ 
{j} \right]},$ $\underline {V} ^{\left\{ {j} \right\}},$ $\,{}^{\left\{ {j} 
\right\}}\underline {Y} $.
\end{proposition}
\begin{proof}[Proof of the Lemma] 
It follows easily from the relations (3.11) and 
(3.12), taking into account (3.13) and (3.14). Indeed the matrix Riccati 
equation (3.13a) is \textit{linearizable} via the position
\begin{equation*}
\underline {W} = - \left( {\underline {y} \,\underline {h}}  \right)^{ - 
1}\,\underline {\dot {M}} \,\underline {M} ^{ - 1}\ ,
\tag{3.18}
\end{equation*}
which yields for $\underline {M} $ the \textit{linear ODE}
\begin{equation*}
\underline {\ddot {M}} = \left( {\underline {\dot {y}} \,\underline {h} + 
\underline {y} \,\underline {\dot {h}} + \underline {y} \,\underline {h} 
\,\underline {f}}  \right)\left( {\underline {y} \,\underline {h}}  
\right)^{ - 1}\,\underline {\dot {M}} - \underline {y} \,\underline {h} 
\,\underline {g} \,\underline {v} \,\underline {M} \  .
\tag{3.19}
\end{equation*}

Hence, if the 5 matrices $\,\underline {f} \,,\,\,\underline {g} 
\,,\,\,\underline {h} \,,\,\,\underline {v} \,,\,\,\underline {y} $ are 
known, the matrix $\underline {W} $ can be evaluated, via (3.18), by solving 
the \textit{linear} (generally \textit{nonautonomous}) matrix ODE 
(3.19). 
Then the $N + 5$ matrices $\underline {U} _{n} \,,\,\,\underline {F} 
\,,\,\,\underline {G} \,,\,\,\underline {H} \,,\,\,\underline {V} 
\,,\,\,\underline {Y} $ can be obtained from the, assumedly known, $N + 5$ 
matrices $\underline {u} _{n} \,,\,\,\underline {f} \,,\,\,\underline {g} 
\,,\,\,\underline {h} \,,\,\,\underline {v} \,,\,\,\underline {y} $ via 
(3.11b) and (3.12b).

The first part of the \textit{Lemma} is thereby proven.

The second part of the \textit{Lemma} is implied by the formulas
\begin{equation*}
\underline {z} ^{\left( {j} \right)} = \underline {W} \,\underline {Z} 
^{\left[ {j} \right]}\,\underline {W} ^{ - 1} \  ,
\tag{3.20a}
\end{equation*}
where $\underline {z} $ stands for $\underline {u} _{n} ,\,\underline {f} 
,\,\underline {g} $ or $\underline {h} $ , and correspondingly $\underline 
{Z} $ stands for $\underline {U} _{n} ,\,\underline {F} ,\,\underline {G} $ 
or $\underline {H} $, and
\begin{gather*}
\underline {v} ^{\left( {j} \right)} = \underline {W} \,\,\underline {V} 
^{\left\{ {j} \right\}} \  ,
\tag{3.20b}
\\
\underline {y} ^{\left( {j} \right)} = {}^{\left\{ {j} \right\}}\underline 
{Y} \,\,\underline {W} ^{ - 1}\ ,
\tag{3.20c}
\end{gather*}
which, via (3.10), (3.11) and (3.12), clearly entail the equivalence of 
(3.9) and (3.17).

As for the validity of these formulas, (3.20), they are an immediate 
consequence of the definitions (3.15) and of the formulas (3.11) and (3.12), 
as can be easily verified by time-differentiating them and using (3.13) and 
(3.14).

The \textit{Lemma} is thereby proven. 
\end{proof}

Note that we are implicitly assuming that the 2 matrices $\underline {W} $ 
and $\underline {y} \,\underline {h} $ are invertible, see (3.11), (3.12) 
and (3.18). A breakdown of one of these two conditions at some time $t_{c} $ 
will generally show up as a singularity of the corresponding solutions of 
(3.17). But note that the second condition (invertibility of $\underline {y} 
\,\underline {h} $) is invalid but irrelevant in the special subcase in 
which $\underline {y} \,\underline {h} $ vanishes identically, see the 
\textit{Remark} (iii) below. In this special case the (first-order) matrix 
evolution equation (3.13a) is already \textit{linear}, hence there is no 
need to introduce via (3.18) the matrix $\underline {M} $.

Now we are ready to formulate and prove the main result of this subsection.

\begin{theorem}
If the matrix ODEs (3.9) are \emph{linearizable}, the matrix ODEs (3.17) 
are also \emph{linearizable}. 
\end{theorem}
\begin{proof}
The second \textit{Proposition} of the \textit{Lemma 
}entails that the solutions of (3.17) can be obtained by first solving the 
ODEs (3.9) to determine the $N + 5$ matrices $\underline {u} _{n} 
,\,\underline {f} ,\,\underline {g} ,\,\underline {h} ,\,\underline {v} 
,\,\underline {y} $ and by then obtaining from these, according to the first 
\textit{Proposition} of the \textit{Lemma}, the $N + 5$ matrices $\underline 
{U} _{n} ,\,\underline {F} ,\,\underline {G} ,\,\underline {H} ,\,\underline 
{V} ,\,\underline {Y} $. 
\end{proof}

\subsubsection*{Remarks}

(i) Note that, even if the equations (3.9) satisfied by the $N + 5$ matrices 
$\underline {u} _{n} ,\,\underline {f} ,\,\underline {g} ,\,\underline {h} 
,\,\underline {v} ,\,\underline {y} $ are \textit{linear} (and possibly 
quite trivial, see examples below), the equations (3.17) satisfied by the $N 
+ 5$ matrices $\underline {U} _{n} ,\,\underline {F} ,\,\underline {G} 
,\,\underline {H} ,\,\underline {V} ,\,\underline {Y} $ are generally 
\textit{nonlinear}. Also note that, in the context of the initial-value 
problem, the convenient choice $\underline {W} _{0} = \underline {1} $, see 
above (paragraph after (3.14)), entails that the initial conditions for the 
$N + 5$ matrices $\underline {u} _{n} ,\,\underline {f} ,\,\underline {g} 
,\,\underline {h} ,\,\underline {v} ,\,\underline {y} $ can be explicitly 
obtained from the initial conditions for the $N + 5$ matrices $\underline 
{U} _{n} ,\,\underline {F} ,\,\underline {G} ,\,\underline {H} ,\,\underline 
{V} ,\,\underline {Y} $ via (3.20) with (3.15,16).

(ii) Not all the equations (3.9a,b,c,d,e,f) need be differential, for 
instance (3.9b) might read
\begin{equation*}
\underline {f} = \underline {\tilde {f}} \left( {\underline {u} _{m}^{\left( 
{j} \right)} ,m = 1,...,N;\underline {g} ^{\left( {j} \right)};\underline 
{h} ^{\left( {j} \right)};j = 0,1,2,...} \right) \  ,
\tag{3.21}
\end{equation*}
with $\underline {\tilde {f}} $ a given ``scalar/matrix function'' of its 
matrix arguments, yielding thereby an explicit definition of the matrix 
$\underline {f} $ in terms of the matrices $\underline {u} _{n} 
,\,\underline {g} ,\,\underline {h} $ and their time-derivatives, and 
likewise an explicit definition of the matrix $\underline {F} $ in term of 
the matrices $\underline {U} _{n} ,\,\underline {G} ,\,\underline {H} $ and 
their time-derivatives, see (3.7b) and (3.20a); and so on (see examples 
below). But this applies only to the equations (3.9a,b,c,d), and likewise to 
their counterparts (3.17a,b,c,d), not to (3.9e,f) and (3.17e,f) (except in 
the trivial cases discussed in the following \textit{Remark} (iii) ).

(iii) The linear homogeneous equations satisfied by $\underline {v} $ and 
$\underline {y} $, see (3.9e,f), and likewise the (generally nonlinear) 
equations satisfied by $\underline {V} $ and $\underline {Y} $ , see 
(3.17e,f), clearly admit the trivial solutions $\underline {v} = \underline 
{V} = 0$ or $\underline {y} = \underline {Y} = 0$ (the vanishing of 
$\underline {v} $ entails the vanishing of $\underline {V} $ , and 
viceversa, and likewise for $\underline {y} $ and $\underline {Y} $ , see 
(3.12)). This corresponds to reduced versions of the above \textit{Lemma}, 
which are obtained by setting to zero one of the pairs of functions 
$\underline {v} ,\,\underline {V} $ and $\underline {y} ,\,\underline {Y} $ 
(or possibly both pairs), as indeed entailed by the special cases of (3.9e) 
and correspondingly of (3.17e), respectively of (3.9f) and correspondingly 
of (3.17f), with $\underline {\tilde {V}} _{0} = 1,\,\underline {\tilde {V}} 
_{j} = 0$ for $j > 0$ , respectively $\underline {\tilde {Y}} _{0} = 
1,\,\underline {\tilde {Y}} _{j} = 0$ for $j > 0$. In the first case, 
$\underline {v} = \underline {V} = 0$, the matrices $\underline {g} $ and 
$\underline {G} $ can also be ignored, see (3.14), and one can then ignore 
the equations (3.9e) and (3.17e) as well as (3.9c) and (3.17c). In the 
second case, $\underline {y} = \underline {Y} = 0$, $\underline {h} $ and 
$\underline {H} $ can likewise be ignored, see (3.14), as well as the 
equations (3.9f), (3.9b) and (3.9d), (3.17d). If both conditions hold, 
$\underline {v} = \underline {V} = 0$ and $\underline {y} = \underline {Y} = 
0$, then the only matrices that play a significant role are $\underline {u} 
_{n} $ and $\underline {f} $ , and correspondingly $\underline {U} _{n} $ 
and $\underline {F} $, and the only relevant equations are (3.9a,b) and 
(3.17a,b); in this special case $\underline {K} = \underline {F} $, see 
(3.14).

\subsubsection*{Examples}

We now exhibit two classes of \textit{linearizable 
second-order matrix ODEs} satisfied by the $N + 5$ matrices $\underline {U} 
_{n} ,\,\,\underline {F} ,\,\,\underline {G} ,\,\,\underline {H} 
,\,\,\underline {V} ,\,\,\underline {Y} $, obtained from this 
\textit{Theorem} by making specific choices for the functions $\underline 
{\tilde {U}} _{n} ,\,\,\underline {\tilde {F}} ,\,\,\underline {\tilde {G}} 
,\,\,\underline {\tilde {H}} ,\,\,\underline {\tilde {V}} ,$ $\underline 
{\tilde {Y}} $, see (3.9) and (3.17). Once and for all, let us emphasize 
that these are representative examples, selected to display the type of 
\textit{linearizable matrix ODEs} encompassed by our approach, and chosen 
with an eye to the manufacture of \textit{linearizable many-body problems}, 
see the following Sections \ref{IV} and \ref{V}; 
it will of course be trivially easy for 
the diligent reader to manufacture additional examples.

The \textit{first class of examples} we consider is characterized by 
\textit{linearizable} (in fact, \textit{linear})\textit{} matrix ODEs of 
type (3.9) which read as follows:
\begin{gather*}
\mu _{n}^{\left( {u} \right)} \underline {\ddot {u}} _{n} = \sum\limits_{m = 
1}^{N} \left[ a_{nm}^{\left( {uu} \right)} \,\underline {\dot {u}} _{m} + 
b_{nm}^{\left( {uu} \right)} \,\underline {u}_{m}   \right] + a_{n}^{\left( 
{uf} \right)} \,\underline {\dot {f}} + b_{n}^{\left( {uf} \right)} 
\,\underline {f}
\\ \qquad{}
+ a_{n}^{\left( {ug} \right)} \underline {\dot {g}} + 
b_{n}^{\left( {ug} \right)} \,\underline {g} + a_{n}^{\left( {uh} \right)} 
\,\underline {\dot {h}} + b_{n}^{\left( {uh} \right)} \,\underline 
{h} \tag{3.22a}
\\
\mu ^{\left( {f} \right)}\underline {\ddot {f}} = \sum\limits_{m = 1}^{N} 
{\left[ {a_{m}^{\left( {fu} \right)} \,\underline {\dot {u}} _{m} + 
b_{m}^{\left( {fu} \right)} \,\underline {u}_{m} }  \right] +}  a^{\left( 
{ff} \right)}\,\underline {\dot {f}} + b^{\left( {ff} \right)}\,\underline 
{f}
\\ \qquad{}
+ a^{\left( {fg} \right)}\,\underline {\dot {g}} + b^{\left( {fg} 
\right)}\,\underline {g} + a^{\left( {fh} \right)}\,\underline {\dot {h}} + 
b^{\left( {fh} \right)}\,\underline {h} \  ,
\tag{3.22b}
\\
\mu ^{\left( {g} \right)}\underline {\ddot {g}} = \sum\limits_{m = 1}^{N} 
{\left[ {a_{m}^{\left( {gu} \right)} \,\underline {\dot {u}} _{m} + 
b_{m}^{\left( {gu} \right)} \,\underline {u}_{m} }  \right] +}  a^{\left( 
{gf} \right)}\,\underline {\dot {f}} + b^{\left( {gf} \right)}\,\underline 
{f}
\\ \qquad{}
+ a^{\left( {gg} \right)}\,\underline {\dot {g}} + b^{\left( {gg} 
\right)}\,\underline {g} + a^{\left( {gh} \right)}\,\underline {\dot {h}} + 
b^{\left( {gh} \right)}\,\underline {h} \  ,
\tag{3.22c}
\\
\mu ^{\left( {h} \right)}\underline {\ddot {h}} = \sum\limits_{m = 1}^{N} 
{\left[ {a_{m}^{\left( {hu} \right)} \,\underline {\dot {u}} _{m} + 
b_{m}^{\left( {hu} \right)} \,\underline {u}_{m} }  \right] +}  a^{\left( 
{hf} \right)}\,\underline {\dot {f}} + b^{\left( {hf} \right)}\,\underline 
{f}
\\ \qquad{}
+ a^{\left( {hg} \right)}\,\underline {\dot {g}} + b^{\left( {hg} 
\right)}\,\underline {g} + a^{\left( {hh} \right)}\,\underline {\dot {h}} + 
b^{\left( {hh} \right)}\,\underline {h} \  ,
\tag{3.22d}
\\
\mu^{\left({v}\right)} \underline{\ddot {v}} = \{{\sum\limits_{m = 
1}^{N} [\tilde{a}_{m}^{({vu})} \,\underline{\dot {u}}
_{m}^{2} + \tilde {b}_{m}^{\left( {vu} \right)} \,\underline{u} _{m}^{2}}  
] + \tilde {a}^{\left( {vf} \right)}\,\underline{\dot {f}}^{2} + 
\tilde{b}^{\left( {vf} \right)}\,\underline {f}^{2} + \tilde{a}^{\left( 
{vg} \right)}\,\underline{\dot{g}}^{2}
\\ \qquad{}
+ \tilde {b}^{\left( {vg} 
\right)}\,\underline {g} ^{2} + \tilde {a}^{\left( {vh} \right)}\,\underline 
{\dot {h}}^{2} + \tilde {b}^{\left( {vh} \right)}\,\underline {h} ^{2}  
\}\,\underline {\dot {v}}
\\ \qquad{}
 + \{\sum\limits_{m = 1}^{N} \left[ a_{m}^{\left( {vu} \right)} 
\,\underline {\dot {u}} _{m}^{2} + b_{m}^{\left( {vu} \right)} \,\underline 
{u} _{m}^{2}  \right] + a^{\left( {vf} \right)}\,\underline {\dot {f}} ^{2} 
+ b^{\left( {vf} \right)}\,\underline {f}^{2} + a^{\left( {vg} 
\right)}\,\underline {\dot{g}}^{2} + b^{\left( {vg} \right)}\,\underline 
{g}^{2}
\\ \qquad{}
+ a^{\left( {vh} \right)}\,\underline{\dot {h}}^{2} + b^{\left( 
{vh} \right)}\, \underline{h}^{2}\}\,\underline {v} \ ,
\tag{3.22e}
\\
\mu^{\left( {y} \right)}\underline{\ddot {y}} =\underline {\dot {y}} \quad 
\{ \sum\limits_{m = 1}^{N} [\tilde{a}_{m}^{\left( {yu} 
\right)} \,\underline {\dot {u}} _{m}^{2} + \tilde{b}_{m}^{\left( {yu} 
\right)} \,\underline{u} _{m}^{2} ] + \tilde {a}^{\left( {yf} 
\right)}\,\underline {\dot {f}} ^{2} + \tilde {b}^{\left( {yf} 
\right)}\,\underline {f} ^{2} + \tilde {a}^{\left( {yg} \right)}\,\underline 
{\dot {g}}^{2}
\\ \qquad{}
+ \tilde{b}^{\left( {yg} \right)}\,\underline{g} ^{2} + 
\tilde{a}^{\left( {yh} \right)}\,\underline {\dot {h}} ^{2} + \tilde 
{b}^{\left( {yh} \right)}\,\underline {h} ^{2} \}
\\ \qquad{}
+ \underline {y} \{ \sum\limits_{m = 1}^{N} \left[ a_{m}^{\left( 
{yu} \right)} \,\underline {\dot {u}}_{m}^{2} + b_{m}^{\left( {yu} \right)} 
\,\underline {u} _{m}^{2}  \right] + a^{\left( {yf} \right)}\,\underline 
{\dot {f}} ^{2} + b^{\left( {yf} \right)}\,\underline {f} ^{2} + a^{\left( 
{yg} \right)}\,\underline {\dot {g}} ^{2}
\\ \qquad{}
+ b^{\left( {yg} \right)}\,\underline {g} ^{2} + 
a^{\left({yh}\right)}\,
\underline{\dot {h}}^{2} + b^{\left( {yh} \right)}\,\underline {h} ^{2}\}\ .
\tag{3.22f}
\end{gather*}
These equations contain $N + 5$ quantities of type $\mu $, and $\left( 
{2N^{2} + 20N + 42} \right)$ quantities of type $a,\,b$ (variously decorated 
with lower indices and upper symbols of identification); our treatment would 
apply even if all these quantities were (arbitrarily!) given functions of 
time, but for simplicity we assume hereafter that they are (arbitrarily!) 
given constants. Then the $N + 3$ matrices $\underline {u} _{n} 
,\,\,\underline {f} ,\,\,\underline {g} ,\,\,\underline {h} $ can be 
obtained from (3.22a,b,c,d) via purely algebraic operations, and the $2$ 
matrices $\underline {v} $ and $\underline {y} $ can subsequently be 
obtained by solving the \textit{linear nonautonomous} matrix ODEs (3.22e,f).

The corresponding (generally nonlinear!) \textit{linearizable} matrix ODEs 
satisfied by the $N + 5$ matrices $\underline {U} _{n} ,\,\,\underline {F} 
,\,\,\underline {G} ,\,\,\underline {H} ,\,\,\underline {V} ,\,\,\underline 
{Y} $ read as follows:
\begin{gather*}
\mu _{n}^{\left( {u} \right)} \,\underline {\ddot {U}} _{n} = 2\,\left[ 
{\underline {\dot {U}} _{n} ,\underline {K}}  \right] + \left[ {\underline 
{U} _{n} ,\underline {\dot {K}}}  \right] - \left[ {\,\left[ {\underline {U} 
_{n} ,\underline {K}}  \right],\,\underline {K}}  \right]
\\ \qquad{}
+ \sum\limits_{m = 
1}^{N} {\left[ {a_{nm}^{\left( {uu} \right)} \,\left\{ {\underline {\dot 
{U}} _{m} - \left[ {\underline {U} _{m} ,\underline {K}}  \right]} \right\} 
+ b_{nm}^{\left( {uu} \right)} \,\underline {U} _{m}}  \right]}
\\ \qquad{}
 + a_{n}^{\left( {uf} \right)} \,\left\{ {\underline {\dot {F}} - \left[ 
{\underline {F} ,\underline {K}}  \right]} \right\} + b_{n}^{\left( {uf} 
\right)} \,\underline {F} + a_{n}^{\left( {ug} \right)} \,\left\{ 
{\underline {\dot {G}} - \left[ {\underline {G} ,\underline {K}}  \right]} 
\right\}
\\ \qquad{}
+ b_{n}^{\left( {ug} \right)} \,\underline {G} + a_{n}^{\left( {uh} 
\right)} \,\left\{ {\underline {\dot {H}} - \left[ {\underline {H} 
,\underline {K}}  \right]} \right\} + b_{n}^{\left( {uh} \right)} 
\,\underline {H} \  ,
\tag{3.23a}
\\
\mu ^{\left( {f} \right)}\,\underline {\ddot {F}} = 2\,\left[ {\underline 
{\dot {F}} ,\underline {K}}  \right] + \left[ {\underline {F} ,\underline 
{\dot {K}}}  \right] - \left[ {\,\left[ {\underline {F} ,\underline {K}}  
\right],\,\underline {K}}  \right] 
\\ \qquad{}
+ \sum\limits_{m = 1}^{N} {\left[ 
{a_{m}^{\left( {fu} \right)} \,\left\{ {\underline {\dot {U}} _{m} - \left[ 
{\underline {U} _{m} ,\underline {K}}  \right]} \right\} + b_{m}^{\left( 
{fu} \right)} \,\underline {U}_{m} }  \right] }
 + a^{\left( {ff} \right)}\,\left\{ {\underline {\dot {F}} - \left[ 
{\underline {F} ,\underline {K}}  \right]} \right\} 
+ b^{\left( {ff} 
\right)}\,\underline {F} 
\\ \qquad{}
+ a^{\left( {fg} \right)}\,\left\{ {\underline 
{\dot {G}} - \left[ {\underline {G} ,\underline {K}}  \right]} \right\} + 
b^{\left( {fg} \right)}\,\underline {G}
 + a^{\left( {fh} \right)}\left\{ 
{\underline {\dot {H}} - \left[ {\underline {H} ,\underline {K}}  \right]} 
\right\} + b^{\left( {fh} \right)}\,\underline {H} \ ,
\tag{3.23b}
\\
\mu ^{\left( {g} \right)}\,\underline {\ddot {G}} = 2\,\left[ {\underline 
{\dot {G}} ,\underline {K}}  \right] + \left[ {\underline {G} ,\underline 
{\dot {K}}}  \right] - \left[ {\,\left[ {\underline {G} ,\underline {K}}  
\right],\,\underline {K}}  \right] 
\\ \qquad{}
+ \sum\limits_{m = 1}^{N} {\left[ 
{a_{m}^{\left( {gu} \right)} \,\left\{ {\underline {\dot {U}} _{m} - \left[ 
{\underline {U} _{m} ,\underline {K}}  \right]} \right\} + b_{m}^{\left( 
{gu} \right)} \,\underline {U} _{m}}  \right]} 
 + a^{\left( {gf} \right)}\,\left\{ {\underline {\dot {F}} - \left[ 
{\underline {F} ,\underline {K}}  \right]} \right\} + b^{\left( {gf} 
\right)}\,\underline {F} 
\\ \qquad{}
+ a^{\left( {gg} \right)}\,\left\{ {\underline 
{\dot {G}} - \left[ {\underline {G} ,\underline {K}}  \right]} \right\} + 
b^{\left( {gg} \right)}\,\underline {G}
 + a^{\left( {gh} \right)}\left\{ 
{\underline {\dot {H}} - \left[ {\underline {H} ,\underline {K}}  \right]} 
\right\} + b^{\left( {gh} \right)}\,\underline {H} \  ,
\tag{3.23c}
\\
\mu ^{\left( {h} \right)}\,\underline {\ddot {H}} = 2\,\left[ {\underline 
{\dot {H}} ,\underline {K}}  \right] + \left[ {\underline {H} ,\underline 
{\dot {K}}}  \right] - \left[ {\,\left[ {\underline {H} ,\underline {K}}  
\right],\,\underline {K}}  \right] 
\\ \qquad{}
+ \sum\limits_{m = 1}^{N} {\left[ 
{a_{m}^{\left( {hu} \right)} \,\left\{ {\underline {\dot {U}} _{m} - \left[ 
{\underline {U} _{m} ,\underline {K}}  \right]} \right\} + b_{m}^{\left( 
{hu} \right)} \,\underline {U} _{m}}  \right]} 
 + a^{\left( {hf} \right)}\,\left\{ {\underline {\dot {F}} - \left[ 
{\underline {F} ,\underline {K}}  \right]} \right\} + b^{\left( {hf} 
\right)}\,\underline {F} 
\\ \qquad{}
+ a^{\left( {hg} \right)}\,\left\{ {\underline 
{\dot {G}} - \left[ {\underline {G} ,\underline {K}}  \right]} \right\} + 
b^{\left( {hg} \right)}\,\underline {G}
+ a^{\left( {hh} \right)} \left\{ 
{\underline {\dot {H}} - \left[ {\underline {H} ,\underline {K}}  \right]} 
\right\} + b^{\left( {hh} \right)}\,\underline {H} \  ,
\tag{3.23d}
\\
\mu ^{\left( {v} \right)}\,\underline {\ddot {V}} = - 2\,\underline {K} 
\,\underline {\dot {V}} - \underline {\dot {K}} \,\underline {V} - 
\underline {K} ^{2}\,\underline {V} + \left\{ {} \right.\sum\limits_{m = 
1}^{N} {\left[ {\tilde {a}_{m}^{\left( {vu} \right)} \,\left\{ {\underline 
{\dot {U}} _{m} - \left[ {\underline {U} _{m} ,\underline {K}}  \right]} 
\right\}^{2} + \tilde {b}_{m}^{\left( {vu} \right)} \,\underline {U} 
_{m}^{2}}  \right]} 
\\ \qquad{}
 + \tilde {a}^{\left( {vf} \right)}\,\left\{ {\underline {\dot {F}} - \left[ 
{\underline {F} ,\underline {K}}  \right]} \right\}^{2} + \tilde {b}^{\left( 
{vf} \right)}\,\underline {F} ^{2} + \tilde {a}^{\left( {vg} 
\right)}\,\left\{ {\underline {\dot {G}} - \left[ {\underline {G} 
,\underline {K}}  \right]} \right\}^{2} + \tilde {b}^{\left( {vg} 
\right)}\,\underline {G} ^{2} 
\\ \qquad{}
+ \tilde {a}^{\left( {vh} \right)}\,\left\{ 
{\underline {\dot {H}} - \left[ {\underline {H} ,\underline {K}}  \right]} 
\right\}^{2}
+ \tilde {b}^{\left( {vh} \right)}\,\underline {H} ^{2}\left. 
{} \right\}\,\left\{ {\underline {\dot {V}} + \underline {K} \,\underline 
{V}}  \right\}
\\ \qquad{}
 + \left\{ {} \right.\sum\limits_{m = 1}^{N} {\left[ {a_{m}^{\left( {vu} 
\right)} \,\left\{ {\underline {\dot {U}} _{m} - \left[ {\underline {U} _{m} 
,\underline {K}}  \right]} \right\}^{2} + b_{m}^{\left( {vu} \right)} 
\,\underline {U} _{m}^{2}}  \right]} 
\\ \qquad{}
 + a^{\left( {vf} \right)}\,\left\{ {\underline {\dot {F}} - \left[ 
{\underline {F} ,\underline {K}}  \right]} \right\}^{2} + b^{\left( {vf} 
\right)}\,\underline {F} ^{2} + a^{\left( {vg} \right)}\,\left\{ {\underline 
{\dot {G}} - \left[ {\underline {G} ,\underline {K}}  \right]} \right\}^{2} 
+ b^{\left( {vg} \right)}\,\underline {G} ^{2} 
\\ \qquad{}
+ a^{\left( {vh} 
\right)}\,\left\{ {\underline {\dot {H}} - \left[ {\underline {H} 
,\underline {K}}  \right]} \right\}^{2}
+ b^{\left( {vh} 
\right)}\,\underline {H} ^{2}\left. {} \right\}\,\underline {V} \ ,
\tag{3.23e}
\\
\mu ^{\left( {y} \right)}\,\underline {\ddot {Y}} =2\,\underline {\dot {Y}} 
\,\underline{K} + \underline {Y} \,\underline {\dot {K}} - \underline {Y} 
\,\underline {K} ^{2}
\\ \qquad{}
+ \left\{ {\underline {\dot {Y}}} - \underline {Y} 
\,\underline {K}  \right\}\{ {} \sum\limits_{m = 1}^{N}\left[ {a_{m}^{\left( {yu} \right)} \,\{ \underline {\dot {U}} _{m} - 
\left[ \underline {U} _{m} ,\underline {K}  \right]} \}^{2}
+ b_{m}^{\left( {yu} \right)} \,\underline {U} _{m}^{2}  \right] 
\\ \qquad{}
 + \tilde {a}^{\left( {yf} \right)}\,\left\{ {\underline {\dot {F}} - \left[ 
{\underline {F} ,\underline {K}}  \right]} \right\}^{2} + \tilde {b}^{\left( 
{yf} \right)}\,\underline {F} ^{2} + \tilde {a}^{\left( {yg} 
\right)}\,\left\{ {\underline {\dot {G}} - \left[ {\underline {G} 
,\underline {K}}  \right]} \right\}^{2} + \tilde {b}^{\left( {yg} 
\right)}\,\underline {G} ^{2}
\\ \qquad{}
 + \tilde {a}^{\left( {yh} \right)}\left\{ 
{\underline {\dot {H}} - \left[ {\underline {H} ,\underline {K}}  \right]} 
\right\}^{2} + \tilde {b}^{\left( {yh} \right)}\,\underline {H} ^{2}\left. 
{} \right\}
\\ \qquad{}
 + \underline {Y} \,\left\{ {} \right.\quad \sum\limits_{m = 1}^{N} {\left[ 
{a_{m}^{\left( {yu} \right)} \,\left\{ {\underline {\dot {U}} _{m} - \left[ 
{\underline {U} _{m} ,\underline {K}}  \right]} \right\}^{2} + b_{m}^{\left( 
{yu} \right)} \,\underline {U} _{m}^{2}}  \right]} 
\\ \qquad{}
 + a^{\left( {yf} \right)}\,\left\{ {\underline {\dot {F}} - \left[ 
{\underline {F} ,\underline {K}}  \right]} \right\}^{2} + b^{\left( {yf} 
\right)}\,\underline {F} ^{2} + a^{\left( {yg} \right)}\,\left\{ {\underline 
{\dot {G}} - \left[ {\underline {G} ,\underline {K}}  \right]} \right\}^{2} 
+ b^{\left( {yg} \right)}\,\underline {G} ^{2}
\\ \qquad{}
 + a^{\left( {yh} 
\right)}\left\{ {\underline {\dot {H}} - \left[ {\underline {H} 
,\underline {K}}  \right]} \right\}^{2} + b^{\left( {yh} 
\right)}\,\underline {H} ^{2}\left. {} \right\}\ .
\tag{3.23f}
\end{gather*}
In these formulas $\underline {K} $ is of course defined by (3.14).

\bigskip

\noindent
The \textit{second class of examples} we consider is characterized by (much 
simpler!) \textit{linearizable} (or \textit{linear}, see below) matrix ODEs 
of type (3.9) which read as follows:
\begin{gather*}
\underline {\ddot {u}} _{n} = \underline {\tilde {u}} \left( {\underline {u} 
_{m} ,\,\underline {\dot {u}} _{m} ,\,\,m = 1,...,N;t} \right) \  ,
\tag{3.24a}
\\
\underline {f} = \underline {\tilde {f}} \left( {\underline {u} _{m} 
,\,\underline {\dot {u}} _{m} ,\,\,m = 1,...,N;t} \right) \  .
\tag{3.24b}
\end{gather*}

Here $\underline {\tilde {u}} $ is a given scalar/matrix function, such that 
(3.24a) is a \textit{solvable} or \textit{linearizable} system of ODEs (see 
for instance the preceding Section \ref{III.A}, or below), 
and $\underline {\tilde 
{f}} $ is an (arbitrarily) given scalar/matrix function. The corresponding 
ODEs of type (3.17) then read 
\begin{gather*}
\underline {\ddot {U}} _{n} = 2\,\left[ {\underline {\dot {U}} _{n} 
,\,\underline {F}}  \right] + \left[ {\underline {U} _{n} ,\,\underline 
{\dot {F}}}  \right] - \left[ {\,\left[ {\underline {U} _{n} ,\,\underline 
{F}}  \right],\,\,\underline {F}}  \right] + \underline {\tilde 
{u}}
\\ \qquad
\left( {\underline {U} _{m} ,\,\underline {\dot {U}} _{m} - \left[ 
{\underline {U} _{m} ,\,\,\underline {F}}  \right],\,\,m = 1,...,N;t} 
\right) \  ,
\tag{3.25a}
\\
\underline {F} = \underline {\tilde {f}} \left( {\underline {U} _{m} 
,\,\underline {\dot {U}} _{m} - \left[ {\underline {U} _{m} ,\,\,\underline 
{F}}  \right],\,\,m = 1,...,N;t} \right) \  .
\tag{3.25b}
\end{gather*}
Note that we are here in the special case mentioned in the \textit{Remark 
}(iii), see above, corresponding to the choice $\underline {v} = \underline 
{V} = 0$ and $\underline {y} = \underline {Y} = 0$.

Also note that, if $\underline {f} $ depends on the matrices $\underline 
{\dot {u}} _{m} $ (besides the matrices $\underline {u} _{m} $), see 
(3.24b), hence $\underline {F} $ depends on the matrices $\underline {\dot 
{U}} _{m} $ (besides the matrices $\underline {U} _{m} $ ; see (3.20a) and 
(3.16a), then the ODE (3.25a) contains the second time-derivative of the 
matrices $\underline {U} _{m} $ in the right-hand-side as well (from the 
second term).

For the special choice 
\begin{equation*}
\underline {\tilde {u}} _{n} = \sum\limits_{m = 1}^{N} {\left[ {2\,a_{nm} 
\left( {t} \right)\,\,\underline {\dot {u}} _{m} + b_{nm} \left( {t} 
\right)\,\,\underline {u_{m}} }  \right]} \  ,
\tag{3.26}
\end{equation*}
which entails of course that the ODEs (3.24a) are \textit{linear}, the ODEs 
(3.25a) read
\begin{gather*}
\underline {\ddot {U}} _{n} = 2\,\left[ {\underline {\dot {U}} _{n} 
,\,\underline {F}}  \right] + \left[ {\underline {U} _{n} ,\,\underline 
{\dot {F}}}  \right] - \left[ {\,\left[ {\underline {U} _{n} ,\,\underline 
{F}}  \right],\,\,\underline {F}}  \right] 
\\ \qquad{}
+\sum\limits_{m = 1}^{N} {\left[ 
{2\,a_{nm} \left( {t} \right)\,\,\left\{ {\underline {\dot {U}} _{m} - 
\left[ {\underline {U} _{n} ,\,\underline {F}}  \right]} \right\} + b_{nm} 
\left( {t} \right)\,\,\underline {U} _{m}}  \right]} \  .
\tag{3.27}
\end{gather*}
The factor $2$ in the right-hand-sides of the last two equations is 
introduced for notational convenience, see below. In the last equation, 
(3.27), the matrix $\underline {F} $ is of course always given by 
(3.25b), 
with an arbitrarily chosen $\underline {\tilde {f}} $. For instance, if we 
make for this function the simple choice 
\begin{equation*}
\underline {\tilde {f}} \left( {\underline {u} _{m} ,\,\underline {\dot {u}} 
_{m} ,\,\,m = 1,...,N;t} \right) = \sum\limits_{m = 1}^{N} {\left[ {c_{m} 
\left( {t} \right)\,\,\underline {u} _{m}}  \right]} \  ,
\tag{3.28a}
\end{equation*}
which entails
\begin{equation*}
\underline {F} = \sum\limits_{m = 1}^{N} {\left[ {c_{m} \left( {t} 
\right)\,\,\underline {U} _{m}}  \right]} \  ,
\tag{3.28b}
\end{equation*}
then the \textit{linearizable} matrix ODEs satisfied by the $N$ matrices 
$\underline {U} _{n} $ read as follows:
\begin{gather*}
\underline{\ddot {U}} _{n} = \sum\limits_{m = 1}^{N} \{ 
\,2\,\,a_{nm} \left({t} \right)\,\,\,\underline {\dot {U}} _{m} + b_{nm} 
\left( {t} \right)\,\,\underline {U} _{m}
\\ \quad{}
+ 2\,\,c_{m} \left( {t} 
\right)\,\,\left[ \underline {\dot {U}} _{n} ,\,\,\underline {U} _{m}  
\right] + c_{m} \left( {t} \right)\,\left[ \underline {U}_{n} 
,\,\,\underline {\dot {U}} _{m}  \right] + \dot {c}_{m} \left( {t} 
\right)\left[\underline{U}_{n} ,\,\underline{U}_{m}  \right]\} 
\\ \quad{}
 - \sum\limits_{m_{1} ,m_{2} = 1}^{N} {\left\{ {2\,a_{nm_{1}}  \left( {t} 
\right)\,c_{m_{2}}  \left( {t} \right)\,\left[ {\underline {U} _{m_{1}}  
,\,\underline {U} _{m_{2}} }  \right] + c_{m_{1}}  \left( {t} 
\right)\,c_{m_{2}}  \left( {t} \right)\,\left[ {\,\left[ {\underline {U} 
_{n} ,\,\,\underline {U} _{m_{1}} }  \right],\,\,\underline {U} _{m_{2}} }  
\right]} \right\}}.
\tag{3.29}
\end{gather*}

These equations deserve further elaboration, which for simplicity is 
hereafter restricted to the case with time-independent constants $a_{nm} 
,\,b_{nm} $ and $c_{m} $. Then the above ODEs becomes
\begin{gather*}
\underline {\ddot {U}} _{n} = \sum\limits_{m = 1}^{N} {\left\{ {2\,a_{nm} 
\,\,\underline {\dot {U}} _{m} + b_{nm} \,\,\underline {U_{m}}  + 2\,c_{m} 
\,\left[ {\underline {\dot {U}} _{n} ,\,\,\underline {U} _{m}}  \right] + 
c_{m} \,\left[ {\underline {U} _{n} ,\,\,\underline {\dot {U}} _{m}}  
\right]} \right\}} 
\\ \qquad{}
 - \sum\limits_{m_{1} ,m_{2} = 1}^{N} {\left\{ {2\,a_{nm_{1}}  \,c_{m_{2}}  
\,\left[ {\underline {U} _{m_{1}}  ,\,\underline {U} _{m_{2}} }  \right] + 
c_{m_{1}}  \,c_{m_{2}}  \,\left[ {\,\left[ {\underline {U} _{n} 
,\,\,\underline {U} _{m_{1}} }  \right],\,\,\underline {U} _{m_{2}} }  
\right]} \right\}} \ ,
\tag{3.30}
\end{gather*}
and the corresponding ODEs satisfied by the matrices $\underline {u} _{n} $ 
read
\begin{equation*}
\underline {\ddot {u}} _{n} = \sum\limits_{m = 1}^{N} {\left[ {2\,\,a_{nm} 
\,\,\underline {\dot {u}} _{m} + b_{nm} \,\,\underline {u_{m}} }  \right]} 
\  ,
\tag{3.31}
\end{equation*}
and can therefore be solved by purely algebraic operations. Particularly 
simple is the ``diagonal case'' characterized by the restrictions $a_{nm} = 
\delta _{nm} \,a_{n} \,\,,\,\,b_{nm} = \delta _{nm} \,b_{n} $, which entail 
that the ODEs (3.31) uncouple and their solution reads
\begin{gather*}
\underline {u} _{n} \left( {t} \right) = \,\Delta _{n} ^{ - 1} {\rm exp}\left( 
{a_{n} \,t} \right)\,\,\left[ {\Delta _{n} \,\underline {u} _{n} \left( {0} 
\right)\,\,{\rm cosh}\left( {\Delta _{n} \,t} \right) + \underline {\dot {u}} _{n} 
\left( {0} \right)\,\,\,{\rm sinh}\left( {\Delta _{n} \,t} \right)} \right] \  
,\tag{3.32a}
\\
\Delta _{n} = \left( {a_{n} ^{2} + b_{n}}  \right)^{1/2}.
\tag{3.32b}
\end{gather*}

Even simpler is the case with $a_{n} = a,\,\,\,b_{n} = b$, hence $\Delta 
_{n} = \Delta = \left( {a^{2} + b} \right)^{1/2}$. Then the evolution of the 
matrix $\underline {W} \left( {t} \right)$, see (3.13a), (3.24b) and 
(3.28a), (3.32), reads simply
\begin{equation*}
\underline {\dot {W}} \left( {t} \right) = {\rm exp}\left( {a\,t} 
\right)\,\,\left[ {\underline {A} \,\,\,{\rm cosh}\left( {\Delta \,t} \right) + 
\underline {B} \,\,\,\Delta ^{ - 1}\,{\rm sinh}\left( {\Delta \,t} \right)} 
\right]\,\,\,\,\underline {W} \left( {t} \right) \  ,
\tag{3.33}
\end{equation*}
with $\underline {A} = \sum\limits_{m = 1}^{N} {\left[ {c_{m} \,\,\underline 
{u} _{m} \left( {0} \right)} \right]} ,\,\,\,\,\underline {B} = 
\sum\limits_{m = 1}^{N} {\left[ {c_{m} \,\,\underline {\dot {u}} _{m} \left( 
{0} \right)} \right]} $ two \textit{constant} matrices. But let us emphasize 
that the simplicity of this case, see (3.33), has a rather trivial origin: 
indeed the nonlinearity of the matrix ODEs (3.30) in this special case 
$a_{nm} = \delta _{nm} \,a, \quad b_{nm} = \delta _{nm} \,b$ is in a way 
marginal, since in this case the ODEs (3.30), which can of course be 
rewritten as follows,
\begin{gather*}
\underline {\ddot {U}} _{n} = 2\,a\,\,\underline {\dot {U}} _{n} + 
b\,\,\underline {U_{n}}  + 2\,\left[ {\underline {\dot {U}} _{n} 
,\,\,\underline {F}}  \right] + \left[ {\underline {U} _{n} ,\,\,\underline 
{\dot {F}}}  \right] - 2\,a\,\left[ {\underline {U} _{n} ,\,\underline {F}}  
\right]
\\ \qquad{}
- \left[ {\,\left[ {\underline {U} _{n} ,\,\,\underline {F}}  
\right],\,\,\underline {F}}  \right]\ ,
\tag{3.34a}
\\
\underline {F} = \sum\limits_{m = 1}^{N} {\left[ {c_{m} \,\underline {U} 
_{m}}  \right]} \  ,
\tag{3.34b}
\end{gather*}
are easily seen (by multiplying (3.34a) by $c_{n} $ and summing over $n$ 
using (3.34b) to imply a single (uncoupled) ODE for the quantity 
$\underline {F} $,
\begin{equation*}
\underline {\ddot {F}} = 2\,a\,\underline {\dot {F}} + b\,\underline {F} + 
\left[ {\underline {\dot {F}} ,\,\,\underline {F}}  \right]\ .
\tag{3.34c}
\end{equation*}
Note that, once this equation has been solved for $\underline {F} $, the 
ODEs (3.34a) for the matrices $\underline {U} _{n} $ become \textit{linear 
}(albeit \textit{nonautonomous}).

The matrix ODE (3.34) coincides (up to a trivial rescaling) with the special 
case of the system (3.30) corresponding to $N = 1$. Let us conclude this 
subsection by focusing on this interesting case. There is then a single 
matrix $\underline {U} _{1} \equiv \underline {U} \left( {t} \right)$ which 
satisfies the matrix evolution equation
\begin{equation*}
\underline {\ddot {U}} = 2\,a\,\underline {\dot {U}} + b\,\underline {U} + 
c\,\left[ {\underline {\dot {U}} ,\,\underline {U}}  \right] \  ,
\tag{3.35}
\end{equation*}
or, more generally,
\begin{equation*}
\underline {\ddot {U}} = 2\,\,a\,\,\underline {\dot {U}} + b\,\,\underline 
{U} + \left[ {\,\underline {\dot {U}} ,\,\,\,\underline {\tilde {f}} \left( 
{\underline {U}}  \right)} \right] \  ,
\tag{3.36}
\end{equation*}
if we retain the freedom to make an arbitrary choice (rather than the 
special choice (3.28)) for the scalar/matrix function $\underline {\tilde 
{f}} $ , see (3.24b) and (3.25b). Note that, independently of this choice, 
the corresponding evolution equation for $\underline {u} \left( {t} \right)$ 
reads simply
\begin{equation*}
\underline {\ddot {u}} = 2\,\,a\,\,\underline {\dot {u}} + b\,\,\underline 
{u} \  ,
\tag{3.37a}
\end{equation*}
hence its solution reads (see (3.32))
\begin{gather*}
\underline {u} \left( {t} \right) = \,\Delta ^{ - 1}{\rm exp}\left( {a\,t} 
\right)\,\,\left[ {\,\Delta \,\underline {u} \left( {0} 
\right)\,\,{\rm cosh}\left( {\Delta \,t} \right) + \underline {\dot {u}} \left( 
{0} \right)\,\,\,{\rm sinh}\left( {\Delta \,t} \right)} \right] \  ,
\tag{3.37b}
\\
\Delta = \left( {a^{2} + b} \right)^{1/2},
\tag{3.37c}
\end{gather*}
and the corresponding evolution equation for $\underline {W} $ coincides of 
course with (3.33), now with $\underline {A} = c\,\underline {u} \left( {0} 
\right),\,\,\,\underline {B} = c\,\underline {\dot {u}} \left( {0} \right)$ 
(two constant matrices).

In Appendix A we discuss in some detail the matrix evolution equation 
(3.36) 
with $a = b = 0$ and for $2 \times 2$ matrices. In this special case, and 
for special choices of the function $\tilde {f}$ (including $\tilde 
{f}\left( {u} \right) = c\,u$ which yields (3.35), the corresponding 
evolution equation for $\underline {W} $ can be explicitly solved in terms 
of known special functions (see Appendix A).

In Appendix B an additional analysis of the matrix evolution equation 
(3.35) 
with $a,\,b$ and $c$ three arbitrary (possibly complex) \textit{constants 
}is given, and in particular it is shown that, if $a = \left( {3/2} 
\right)\,i\omega ,\,\,\,b = 2\,\omega ^{2}$ , or if $a = i\omega /2,\,\,b = 
0$, with $\omega $ an arbitrary (nonvanishing) \textit{real} constant, this 
matrix ODE is \textit{solvable} and \textit{all} its solutions are 
\textit{completely periodic}, with period $T = 2\,\pi /\omega $.

The matrix ODE (3.35) is not new: its \textit{linearizability} (essentially 
by the same approach as explained above) was already demonstrated by I. Z. 
Golubchik and V. V. Sokolov [3] (see also the reference in their paper to a 
private communication by E. V. Ferapontov relevant to the special case 
treated in Appendix A).

Let us end this subsection by pointing out that the examples given above are 
focussed on \textit{second-order ODEs} for the $N + 5$ matrices $\underline 
{U} _{n} ,\,\,\underline {F} ,\,\,\underline {G} ,\,\,\underline {H} 
,\,\,\underline {V} ,\,\,\underline {Y} $ (or for a subset of them), 
obtained starting from \textit{second-order ODEs} for the $N + 5$ matrices 
$\underline {u} _{n} ,\,\,\underline {f} ,\,\,\underline {g} ,\,\,\underline 
{h} ,\,\,\underline {v} ,\,\,\underline {y} $ (or for a subset of them). But 
it is also of interest to consider \textit{first-order matrix ODEs}, both 
because of their possible applicative relevance, and because such equations 
may also be connected to our main goal in this paper, namely to construct 
models of many-body problems in ordinary (three-dimensional) space. Indeed 
there are two ways in which such a goal may be realized by starting from 
\textit{first-order matrix ODEs} of the kind yielded by the technique 
introduced in this subsection: such equations may be appropriately 
interpreted as Hamiltonian equations of motion (rather than Newtonian 
equations) for a many-body problem, or they may be used to obtain (by 
time-differentiation and appropriate substitutions) new \textit{second-order 
matrix ODEs}, which can then be appropriately interpreted as Newtonian 
equations of motion for a many-body problem. Each of these two possibilities 
is illustrated by an example, whose presentation is relegated to the two 
Appendices F and G.


\subsection{Some integrable matrix evolution equations related to the 
nonabelian Toda model}
\label{III.C}


In this subsection we consider various coupled systems of matrix evolution 
equations, which coincide, or are closely related, with the 
\textit{integrable} evolution equations of the so-called nonabelian Toda 
lattice [4]
\begin{equation*}
\underline {\ddot {G}} _{n} = \underline {\dot {G}} _{n} \,\left[ 
{\underline {G} _{n}}  \right]^{ - 1}\,\underline {\dot {G}} _{n} + \gamma 
\,\left\{ {\,\underline {G} _{n + 1} - \underline {G} _{n} \,\left[ 
{\underline {G} _{n - 1}}  \right]^{ - 1}\,\underline {G} _{n}}  \right\} 
\  .
\tag{3.38}
\end{equation*}
Here $\underline {G} _{n} \equiv \underline {G} _{n} \left( {t} \right)$ is 
a time-dependent square matrix, labeled by the index $n$, and $\gamma $is a 
(``coupling'') constant (possibly complex, see below), which could be 
eliminated via the scale transformation $\underline {g} _{n} \to \gamma 
^{n}\,\underline {G} _{n} $, or, if it is positive, via the time rescaling 
$t \to \gamma ^{ - 1/2}\,t$.

In this paper we do not discuss the actual solution of this matrix evolution 
equation: it suffices at this stage for us to know that this is an 
\textit{integrable} equation [4]. But let us emphasize that this evolution 
equation should be completed by prescriptions ``at the $n$-boundaries,'' 
such as, say, $\underline {G} _{0} \left( {t} \right) = \underline {G} _{N + 
1} \left( {t} \right) = 0$ (``free ends'') or $\underline{G}_{0} \left( 
{t} \right) = \underline{G}_{N} \left( {t} \right),\,\,\underline {G}_{1} 
\left( {t} \right) = \underline{G} _{N + 1} \left( {t} \right)$ 
(``periodic''), if (3.38) is to hold for $n = 1,...,N$; of course these 
prescriptions are relevant to determine the solution.

Another, perhaps more interesting, \textit{integrable} matrix model reads
\begin{equation*}
\underline {\ddot {Q}} _{n} = \underline {\dot {Q}} _{n} \left[ {\underline 
{Q} _{n + 1} - \underline {Q} _{n}}  \right] - \left[ {\underline {Q} _{n} - 
\underline {Q} _{n - 1}}  \right]\,\underline {\dot {Q}} _{n} \  .
\tag{3.39a}
\end{equation*}
Here of course $\underline {Q} _{n} \equiv \underline {Q} _{n} \left( {t} 
\right)$ is again a square matrix.

The integrability of this matrix evolution equation is entailed by its 
relation to the integrable equation (3.38). Indeed it is clear that, via the 
positions [4]
\begin{gather*}
\underline {A} _{n} \left( {t} \right) = \left[ {\underline {G} _{n} \left( 
{t} \right)} \right]^{ - 1}\,\underline {G} _{n + 1} \left( {t} \right) 
\  ,
\tag{3.40a}
\\
\underline {B} _{n} \left( {t} \right) = \left[ {\underline {G} _{n} \left( 
{t} \right)} \right]^{ - 1}\,\underline {\dot {G}} _{n} \left( {t} \right) 
\  ,
\tag{3.40b}
\end{gather*}
(3.38) can be rewritten as [4]
\begin{gather*}
\underline {\dot {A}} _{n} = \underline {A} _{n} \,\underline {B} _{n + 1} - 
\underline {B} _{n} \,\underline {A} _{n} \  ,
\tag{3.41a}
\\
\underline {\dot {B}} _{n} = \gamma \,\left[ {\underline {A} _{n} - 
\underline {A} _{n - 1}}  \right] \  .
\tag{3.41b}
\end{gather*}
We now set
\begin{equation*}
\underline {Q} _{n} = \sum\limits_{m = m_{0}} ^{n} {\underline {B} _{m}}  
\  ,
\tag{3.42a}
\end{equation*}
entailing
\begin{equation*}
\underline {B} _{n} = \underline {Q} _{n} - \underline {Q} _{n - 1} \  ,
\tag{3.42b}
\end{equation*}
hence, via (3.41b),
\begin{equation*}
\gamma \,\underline {A} _{n} = \underline {\dot {Q}} _{n} \  .
\tag{3.42c}
\end{equation*}
Insertion of (3.42c,b) in (3.42c,b) yields precisely (3.39a). 

Of course also this equation must eventually be completed with ``end-point'' 
prescriptions, namely appropriate definitions for $\underline {Q} _{0} 
\left( {t} \right)$ and $\underline {Q} _{N + 1} \left( {t} \right)$ 
(assuming again that the index $n$ in (3.34a) runs from $1$ to $N$); 
consistently with this ambiguity we left undefined the lower limit $m_{0} 
$ of the sum in the right-hand-side of (3.42a).

Note that (3.39a) is invariant under the \textit{translation} $\underline 
{Q} _{n} \left( {t} \right) \to \underline {Q} _{n} \left( {t} \right) + 
\underline {Q} _{0} ,\,\,\,\underline {\,\,\dot {Q}} _{0} = 0$.

An arbitrary (``coupling'') constant $c$ can be reintroduced in (3.39a) via 
the rescaling $\underline {Q} \to c\underline {\,Q} $ , so that it read
\begin{equation*}
\underline {\ddot {Q}} _{n} = c\,\left\{ {\underline {\dot {Q}} _{n} 
\,\left[ {\underline {Q} _{n + 1} - \underline {Q} _{n}}  \right] - \left[ 
{\underline {Q} _{n} - \underline {Q} _{n - 1}}  \right]\,\underline {\dot 
{Q}} _{n}}  \right\} \  .\tag{3.39b}
\end{equation*}

Note that the \textit{ansatz} $\underline {Q} _{n} \left( {t} \right) = 
n\,\,\underline {U} \left( {t} \right)$\textit{} is compatible with this 
evolution equation, and it yields for $\underline {U} \left( {t} \right)$ 
the simplest one of the evolution equations discussed in the preceding 
subsection, see (3.35) (with $a = b = 0$) and Appendix A.

Finally, we report another \textit{integrable} first-order matrix evolution 
equation, which is the simplest instance of a class of integrable nonlinear 
matrix evolution equations [5]. It reads (see (5.1) of [5])
\begin{equation*}
\underline {\dot {A}} _{n} = c\,\left\{ {\underline {A} _{n - 1} 
\,\underline {A} _{n} - \underline {A} _{n} \,\underline {A} _{n + 1}}  
\right\}\ ,\tag{3.43a} \label{eq11}
\end{equation*}
and it yields, via the position
\begin{equation*}
\underline {A} _{n} = \underline {\dot {U}} _{n} \,\left[ {\underline {U} 
_{n}}  \right]^{ - 1},\,\,\,\,\,\underline {\dot {U}} _{n} = \underline {A} 
_{n} \,\,\underline {U} _{n} \  ,\tag{3.43b} \label{eq12}
\end{equation*}
the second-order \textit{linearizable} matrix evolution equation
\begin{equation*}
\underline {\ddot {U}} _{n} = \underline {\dot {U}} _{n} \,\left[ 
{\underline {U} _{n}}  \right]^{ - 1}\,\underline {\dot {U}} _{n} + 
c\,\left\{ {\underline {\dot {U}} _{n - 1} \,\left[ {\underline {U} _{n - 1} 
} \right]^{ - 1}\,\underline {\dot {U}} _{n} - \underline {\dot {U}} _{n} 
\,\left[ {\underline {U} _{n}}  \right]^{ - 1}\,\underline {\dot {U}} _{n + 
1} \,\left[ {\underline {U} _{n + 1}}  \right]^{ - 1}\,\underline {U} _{n}}  
\right\}.
\tag{3.43c}
\end{equation*}
Here of course $\underline {U} _{n} \equiv \underline {U} _{n} \left( {t} 
\right)$. This equation is \textit{linearizable} because, to solve it, one 
must solve firstly the \textit{integrable} equation (\ref{eq11}) and then a 
\textit{linear nonautonomous matrix ODE} (see the second of the (\ref{eq12})).

The structure of this evolution equations entails that, if $\underline {U} 
_{n} \left( {t} \right)$ satisfies it, then $\underline {\tilde {U}} _{n} 
\left( {t} \right)\, = \,\underline {U} _{n} \left( {\tau}  \right)$ with 
$\tau = \left[ {{\rm exp}\left( {a\,t} \right) - 1} \right]/a$ satisfies the more 
general evolution equation
\begin{gather*}
\underline {\ddot {\tilde {U}}} _{n} = a\,\,\underline {\dot {\tilde {U}}} 
_{n} + \underline {\dot {\tilde {U}}} _{n} \,\left[ {\underline {\tilde {U}} 
_{n}}  \right]^{ - 1}\,\underline {\dot {\tilde {U}}} _{n} 
\\ \qquad{}
+ c\,\left\{ {\,\underline {\dot {\tilde {U}}} _{n - 1} \,\left[ 
{\underline {\tilde {U}} _{n - 1}}  \right]^{ - 1}\,\underline {\dot {\tilde 
{U}}} _{n} - \underline {\dot {\tilde {U}}} _{n} \,\left[ {\underline 
{\tilde {U}} _{n}}  \right]^{ - 1}\,\underline {\dot {\tilde {U}}} _{n + 1} 
\,\left[ {\underline {\tilde {U}} _{n + 1}}  \right]^{ - 1}\,\underline 
{\tilde {U}} _{n}}  \right\}.\tag{3.43d}
\end{gather*}
This suggests that, if $a = \pm i\,\omega ,\,\,\,\omega > 0$ , the generic 
solution of this latter equation is completely periodic with period $T = 
2\pi /\omega $ .


\subsection{Some other solvable and/or integrable and/or linearizable 
matrix evolution equations}
\label{III.D}


In this subsection we present some other \textit{solvable and/or integrable 
and/or linearizable} matrix evolution equations.


A \textit{solvable} equation reads as follows:
\begin{equation*}
\underline {\ddot {U}} = \alpha \,\underline {1} + \beta \,\underline {U} + 
\gamma \,\left( {\underline {\dot {U}} + c\,\underline {U} ^{2}} \right) - 
c\,\left( {\underline {\dot {U}} \,\underline {U} + 2\underline {U} 
\,\underline {\dot {U}} + c\,\underline {U} ^{3}} \right) \  ,
\tag{3.44} \label{eq1}
\end{equation*}
with $\alpha ,\,\beta ,\,\gamma ,\,c$ arbitrary constants. Indeed by setting
\begin{equation*}
c\,\underline {U} = \underline {V} ^{ - 1}\,\underline {\dot {V}} \  ,
\tag{3.45}
\end{equation*}
one easily sees that (\ref{eq1}) becomes the \textit{linear},\textit{} explicitly 
\textit{solvable},\textit{} matrix ODE
\begin{equation*}
\underline {{V}} = c\,\alpha \,\underline {V} + c\,\beta \,\underline {\dot 
{V}} + \gamma \,\underline {\ddot {V}} \  .
\tag{3.46}
\end{equation*}
It is easily seen that, if the 4 constants $\alpha ,\,\beta ,\,\gamma ,\,c$ 
are real, a necessary and sufficient condition to guarantee that all 
solutions of (3.46), hence as well of (3.44), be completely periodic, is 
validity of the equalities $\alpha = \gamma = 0$, together with the 
inequality $c\,\beta < 0$.


A \textit{linearizable} system of coupled nonlinear matrix ODEs reads
\begin{equation*}
\underline {\ddot {U}} _{n} = \underline {\dot {U}} _{n} \,\left[ 
{\underline {U} _{n}}  \right]^{ - 1}\,\underline {\dot {U}} _{n} + 
\sum\limits_{m} {a_{nm} \,\underline {\dot {U}} _{m} \,\left[ {\underline 
{U} _{m}}  \right]^{ - 1}\,\underline {U} _{n}}  \  ,
\tag{3.47}
\end{equation*}
where the time-dependent square matrices $\underline {U} _{n} \equiv 
\underline {U} _{n} \left( {t} \right)$ are labeled by the index $n$ and the 
(scalar) quantities $a_{nm} $ are arbitrary constants (there are $N^{2}$ of 
them, if the indices $n,m$ range from $1$ to $N$). Indeed by setting
\begin{gather*}
\underline {V} _{n} = \underline {\dot {U}} _{n} \,\left[ {\underline {U} 
_{n}}  \right]^{ - 1} \  ,
\tag{3.48a}
\\
\underline {\dot {U}} _{n} = \underline {V} _{n} \,\underline {U} _{n} \  
,
\tag{3.48b}
\end{gather*}
one gets for $\underline {V} _{n} \left( {t} \right)$ the \textit{linear}, 
explicitly \textit{solvable}, matrix ODE
\begin{equation*}
\underline {\dot {V}} _{n} = \sum\limits_{m} {a_{nm} \,\underline {V} _{m}}  
\  .
\tag{3.49}
\end{equation*}

Hence to solve (3.47) one first solves, explicitly, this \textit{linear 
}equation with constant coefficients, and then the \textit{linear 
nonautonomous ODE} (3.48b).


The following simple matrix evolution equation is \textit{integrable}
\begin{equation*}
\underline {\ddot {U}} = c^{2}\,\left( {2\,\underline {U} ^{3} + \underline 
{C} \,\underline {U} + \underline {U} \,\underline {C}}  \right) \  ,
\tag{3.50}
\end{equation*}
where $c$ is an arbitrary scalar quantity (the factor $2$ could of course be 
rescaled away), and the arbitrary matrix $\underline {C} $ is also constant 
($\underline {\dot {C}} = 0$; one could of course set $\underline {C} = 0$, 
or $\underline {C} = C\,\,\underline {1} $). Indeed it can be obtained from 
a special case of the \textit{integrable} equation (3.39b), as follows. 
Consider the following special (``periodic'') solution of (3.39b):
\begin{equation*}
\underline {Q} _{2n} \left( {\,t} \right) = \underline {A} \left( {t} 
\right)\,,\,\,\,\,\,\,\,\underline {Q} _{2n + 1} \left( {\,t} \right) = 
\underline {B} \left( {t} \right) \  ,\tag{3.51}
\end{equation*}
so that the 2 matrices $\underline {A} \left( {t} \right),\,\,\underline {B} 
\left( {t} \right)$ satisfy the equations
\begin{equation*}
\underline {\ddot {A}} = c\,\,\left\{ {\,\underline {\dot {A}} 
\,\,\underline {B} + \underline {B} \,\,\underline {\dot {A}} - \underline 
{\dot {A}} \,\,\underline {A} - \underline {A} \,\,\underline {\dot {A}} \,} 
\right\}\,,\,\,\,\,\,\underline {\,\ddot {B}} = c\,\,\left\{ {\underline 
{\dot {B}} \,\,\underline {A} + \underline {A} \,\,\underline {\dot {B}} - 
\underline {\dot {B}} \,\,\underline {B} - \underline {B} \,\,\underline 
{\dot {B}} \,} \right\} \  .\tag{3.52}
\end{equation*}
Now set
\begin{gather*}
\underline {S} = \underline {A} + \underline {B} \,,\,\,\,\,\,\,\underline 
{D} = \underline {A} - \underline {B} \  ,
\tag{3.53a}
\\
\underline {A} = \left( {\underline {S} + \underline {D}}  
\right)/2\,,\,\,\,\,\,\,\underline {B} = \left( {\underline {S} - \underline 
{D}}  \right)/2 \  ,
\tag{3.53b}
\end{gather*}
so that
\begin{gather*}
\underline {\ddot {S}} = c\,\left\{ {\underline {\dot {A}} \,\underline {B} 
+ \underline {A} \,\underline {\dot {B}} + \underline {\dot {B}} 
\,\underline {A} + \underline {B} \,\underline {\dot {A}} - \underline {\dot 
{A}} \,\underline {A} - \underline {A} \,\underline {\dot {A}} - \underline 
{\dot {B}} \,\underline {B} - \underline {B} \,\underline {\dot {B}}}  
\right\} \  ,
\tag{3.54a}
\\
\underline {\ddot {D}} = c\,\left\{ {\underline {\dot {A}} \,\underline {B} 
- \underline {A} \,\underline {\dot {B}} - \underline {\dot {B}} 
\,\underline {A} + \underline {B} \,\underline {\dot {A}} - \underline {\dot 
{A}} \,\underline {A} - \underline {A} \,\underline {\dot {A}} + \underline 
{\dot {B}} \,\underline {B} + \underline {B} \,\underline {\dot {B}}}  
\right\} \  .
\tag{3.54b}
\end{gather*}
It is now clear that (3.54a) can be integrated once to yield
\begin{equation*}
\underline {\dot {S}} = - c\,\left\{ {\underline {D} ^{2} + \underline {C}}  
\right\}\ ,
\tag{3.55a}
\end{equation*}
while (3.54b) can be rewritten as follows:
\begin{equation*}
\underline {\ddot {D}} = - c\,\left\{ {\underline {\dot {S}} \,\underline 
{D} + \underline {D} \,\underline {\dot {S}}}  \right\} \  
.\tag{3.55b}
\end{equation*}
But, via (3.55a) and the identification 
\begin{equation*}
\underline {D} \left( {t} \right) = \underline {U} \left( {t} \right) \  
,\tag{3.55c}
\end{equation*}
this last equation yields precisely (3.50). Note that the initial data for 
(3.50), $\underline {U} _{0} $ and $\underline {\dot {U}} _{0} $, determine, 
via (3.55c), $\underline {D} _{0} $ and $\underline {\dot {D}} _{0} $; 
moreover (3.55a), with $\underline {C} $ assigned (\underline 
{arbitrarily!)} and $\underline {U} _{0} $ (hence $\underline {D} _{0} $, 
see (3.55c) given, determines $\underline {\dot {S}} _{0} $, while 
$\underline {S} _{0} $ can be assigned arbitrarily, consistently with the 
translation-invariance of (3.55). From the initial data for $\underline {D} 
\left( {t} \right)$and $S\left( {t} \right)$ one obtains, via (3.55b), the 
initial data for $\underline {A} \left( {t} \right)$ and $B\left( {t} 
\right)$, namely $\underline {A} \left( {0} \right)$, $\underline {\dot {A}} 
\left( {0} \right)$, $B\left( {0} \right)$, $\underline {\dot {B}} \left( 
{0} \right)$; one then solves (3.52), or equivalently, via (3.51), (3.39b), 
and in this manner one finally gets $\underline {A} \left( {t} \right)$, 
$B\left( {t} \right)$, hence, via (3.53a) and (3.55c), the solution 
$\underline {U} \left( {t} \right)$ of (3.50) ( with $c$ an arbitrarily 
assigned scalar constant and $\underline {C} $ an arbitrarily assigned 
constant matrix).


A \textit{solvable} matrix evolution equation reads as follows:
\begin{gather*}
\underline {\ddot {U}} = 2\,a\,\left( {\underline {U} \,\underline {\dot 
{U}} + \underline {\dot {U}} \,\underline {U}}  \right) - 
2\,a^{2}\,\underline {U} ^{3} - 4\,b\,\underline {U} ^{2} + 3\,b\,\underline 
{\dot {U}} - 2\,b^{2}\,\underline {U} 
\\ \qquad{}
- 2\,a\,\left( {a\,\underline {U} ^{2} 
+ b\,\underline {U} - \underline {\dot {U}}}  \right)^{1/2}\,\underline {U} 
\,\left( {a\,\underline {U} ^{2} + b\,\underline {U} - \underline {\dot {U}} 
} \right)^{1/2} \  ,
\tag{3.56}
\end{gather*}
where $a,\,b$\textbf{} are 2 arbitrary scalar constants.

To demonstrate the solvability of this equation, we start from the matrix 
evolution equation
\begin{equation*}
\underline {\dot {M}} = a\,\underline {M} ^{2} + b\,\underline {M} \  ,
\tag{3.57}
\end{equation*}
whose solution reads
\begin{equation*}
\underline {M} \left( {t} \right) = \left\{ {\left[ {\underline {M} \left( 
{0} \right)} \right]^{ - 1}\,{\rm exp}\left( { - b\,t} \right) + \left( {b/a} 
\right)\,\left[ {{\rm exp}\left( { - b\,t} \right) - 1} \right]} \right\}^{ - 1} 
\  .
\tag{3.58}
\end{equation*}
Now set
\begin{equation*}
\underline {M} \left( {t} \right) = \left( {{\begin{array}{*{20}c}
 {\underline {A} \left( {t} \right)} \hfill & {\underline {B} \left( {t} 
\right)} \hfill \\
 {\underline {B} \left( {t} \right)} \hfill & {\underline {A} \left( {t} 
\right)} \hfill \\
\end{array}} } \right) \  ,
\tag{3.59}
\end{equation*}
a position which is clearly compatible with (3.57) and (3.58), and which 
yields for the 2 matrices $\underline {A} \left( {t} \right)$ and 
$\underline {B} \left( {t} \right)$ the equations
\begin{equation*}
\underline {\dot {A}} = a\,\left( {\underline {A} ^{2} + \underline {B} 
^{2}} \right) + b\,\underline {A} \,,\,\,\,\,\,\,\,\underline {\dot {B}} = 
a\,\left( {\underline {A} \,\underline {B} + \underline {B} \,\underline {A} 
} \right) + b\,\underline {B} \  ,
\tag{3.60}
\end{equation*}
the first of which can be solved for $\underline {B} $ , yielding
\begin{equation*}
\underline {B} = \left( {\underline {\dot {A}} /a - \underline {A} ^{2} - 
b\,\underline {A} /a} \right)^{1/2} \  .
\tag{3.61}
\end{equation*}

The matrix ODE (3.56) is then easily obtained, via the identification 
$\underline {U} \left( {t} \right) = \underline {A} \left( {t} \right)$, by 
time-differentiating the first, and using the second, of the (3.60), as well 
as (3.61).


A somewhat analogous \textit{integrable} matrix ODE reads as follows:
\begin{equation*}
\underline {\ddot {U}} = 2\,c\,\left\{ {c\,\underline {U} ^{3} + \left[ 
{c\,\underline {U} ^{} + \underline {\dot {U}}}  \right]^{1/2}\,\underline 
{U} \,\left[ {c\,\underline {U} ^{} + \underline {\dot {U}}}  \right]^{1/2}} 
\right\} \  .
\tag{3.62}
\end{equation*}
The starting point to get it is the \textit{integrable equation} (3.43a), in 
the special (``periodic'') case with $\underline {A} _{n \pm 3} = \underline 
{A} _{n} $. Indeed it obtains via the following steps: (i) consider the 
special case of this integrable equation characterized by the additional 
restriction $\underline {A} _{0} + \underline {A} _{1} + \underline {A} _{2} 
= 0$, whose compatibility with the equations of motion is easily verified; 
(ii) use this restriction to eliminate, say, $\underline {A} _{2} $; (iii) 
set $\underline {A} _{0} + \underline {A} _{1} = \underline {S} $, 
$\underline {A} _{0} - \underline {A} _{1} = \underline {D} $; (iv) express 
$\underline {S} $ via $\underline {D} $ and $\underline {\dot {D}} $, 
thereby obtaining a second-order ODE for $\underline {D} $ that does not 
contain $\underline {S} $; (v) finally make the identification $\underline 
{D} = - 2\underline {U} $.


Next, we report the \textit{solvable} matrix evolution equations
\begin{gather*}
\underline {\ddot {U}} _{n} = \sum\limits_{m = 1}^{N} \{ \left( 
\tilde{d}_{nm} \,a_{m} \,\alpha_{m} + b_{nm}  \right)\,\underline {\dot 
{U}} _{m} + \tilde{d}_{nm} \,c_{m} \,\left[ \underline {\dot {U}}_{m} 
,\,\underline{V} _{m} \right]
\\ \qquad{}
+ \tilde{d}_{nm} \,c_{m} \,\alpha _{m} 
\,\left[ \underline{U}_{m} ,\,\underline{\dot {U}} _{m}  \right]
\} \  ,
\tag{3.63a}
\end{gather*}
where the matrices $\underline {V} _{n} $ are obtained, in terms of the 
matrices $\underline {U} _{m} $ and their time-derivatives $\underline {\dot 
{U}} _{m} $, by solving [6] the matrix equations
\begin{equation*}
a_{n} \,\underline {V} _{n} + c_{n} \,\left[ {\underline {U} _{n} 
,\,\underline {V} _{n}}  \right] = \sum\limits_{m = 1}^{N} {\left\{ {d_{nm} 
\,\left[ {\underline {\dot {U}} _{m} - \sum\limits_{{m}' = 1}^{N} {b_{m{m}'} 
\,\underline {U} _{{m}'}} }  \right]} \right\}} \  .
\tag{3.63b}
\end{equation*}
Here the $N\,\left( {3 + 2N} \right)$constants $a_{n} ,\,\alpha _{n} 
,\,c_{n} ,\,\tilde {d}_{nm} ,\,b_{nm} $ are essentially arbitrary, while the 
$N^{2}$constants $d_{nm} $ are the matrix elements of the matrix $\underline 
{D} $ which is the inverse of the matrix $\underline {\tilde {D}} $ having 
matrix elements $\tilde {d}_{nm} $. The solvability of these equations of 
motion is demonstrated in Appendix H.


Next, we report here two \textit{solvable} and one\textit{ linearizable 
}matrix evolution equations, referring for their derivation (and the display 
of more general cases) to Appendix E.


\textit{Solvable} equations:
\begin{gather*}
\underline {\ddot {U}} = c + a\,\alpha \,\underline {U} + \left( {\alpha - 
a} \right)\,\underline {\dot {U}} + b\,\alpha \,\underline {U} ^{2} - 
2\,b\,\underline {\dot {U}} \,\underline {U} \  ,
\tag{3.64}
\\
\underline {\ddot {U}} _{n} = \left[ {a_{n} - a_{n + 1}}  \right]\,\tilde 
{c} + \left[ {a_{n + 1} - a_{n}}  \right]\,\left[ {a_{n} - \tilde {a}_{n}}  
\right]\,\underline {U} _{n} + \tilde {c}_{n} \,\left[ {b_{n} \,\underline 
{U} _{n} - b_{n + 1} \,\underline {U} _{n + 1}}  \right]
\\ \qquad{}
 + \left[ {a_{n} - \tilde {a}_{n}}  \right]\,b_{n + 1} \,\underline {U} _{n 
+ 1} \,\underline {U} _{n} - 2\,b_{n} \,\underline {\dot {U}} _{n} 
\,\underline {U} _{n} 
\\ \qquad{}
+\left[ {\tilde {a}_{n} + a_{n + 1} - 2a_{n} + b_{n + 1} \,\underline {U} 
_{n + 1} - b_{n} \,\underline {U} _{n}}  \right]\,\left[ {\underline {\dot 
{U}} _{n} + b_{n} \,\underline {U} _{n} ^{2}} \right] \  .
\tag{3.65}
\end{gather*}


\textit{Linearizable} equation:
\begin{gather*}
\underline {\ddot {M}} _{n} = \left[ {a_{n + 1} - a_{n} - b_{n}}  
\right]\,\underline {\dot {M}} _{n} + \left\{ {\underline {\dot {M}} _{n} 
\,\left[ {\underline {M} _{n}}  \right]^{ - 1} + b_{n + 1} \,\underline 
{\dot {M}} _{n + 1} \,\left[ {\underline {M} _{n + 1}}  \right]^{ - 1}} 
\right\}\,\underline {\dot {M}} _{n} 
\\ \qquad{}
 + c_{n + 1} \underline {M} _{n} - c_{n} \,\underline {\dot {M}} _{n} 
\,\left[ {\underline {M} _{n}}  \right]^{ - 1}\,\underline {M} _{n - 1} 
\,\left[ {\underline {\dot {M}} _{n - 1}}  \right]^{ - 1}\,\underline {M} 
_{n} \  .\tag{3.66}
\end{gather*}


Finally we report the first-order \textit{integrable} ``Nahm 
equations'' [7]
\begin{equation*}
\underline {\dot {M}} _{n} = \,c\,\left[ {\underline {M} _{n + 1} 
,\,\,\underline {M} _{n + 2}}  \right]\,,\,\,\,\,\,\,\,n = 
1,\,2,\,3\,\,\,{\rm mod}\left( {3} \right) \  ,\tag{3.67}
\end{equation*}
where the constant $c$ could of course be rescaled away (see below). A 
simple way to obtain from these equations a set of 3 coupled 
\textit{linearizable} second-order matrix ODEs is via the positions
\begin{equation*}
c\,\underline {\dot {M}} _{n} = \mu _{n} \,\,\underline {\dot {U}} _{n} + 
\sum\limits_{m = 1}^{3} {a_{n,m} \,\underline {U} _{m}}  ,\,\,\,\,\,\,\,n = 
1,\,2,\,3\,\,\,{\rm mod}\left( {3} \right) \  ,
\tag{3.68}
\end{equation*}
which transform (3.67) into
\begin{gather*}
\mu _{n} \,\underline {\ddot {U}} _{n} = - \sum\limits_{m = 1}^{3} {a_{n,m} 
} \,\underline {\dot {U}} _{m} 
\\ \qquad{}
+ \left[ {\mu _{n + 1} \,\underline {\dot 
{U}} _{n + 1} + \sum\limits_{m = 1}^{3} {a_{n + 1,m}}  \,\underline {U} _{m} 
,\,\,\mu _{n + 2} \,\underline {\dot {U}} _{n + 2} + \sum\limits_{m = 1}^{3} 
{a_{n + 2,m}}  \,\underline {U} _{m}}  \right]
\\
n = 
1,\,2,\,3\,\,\,{\rm mod}\left( {3} \right)\ .
\tag{3.69}
\end{gather*}
Clearly we are assuming here, for simplicity, that the 3 quantities $\mu 
_{n} $, as well as the 9 quantities $a_{n,m} $, are (arbitrary) constants. 
These equations, (3.69), are categorized as \textit{linearizable}, since to 
solve them one must first solve the \textit{integrable} ODEs (3.67)\textit{ 
}for $\underline {M} _{n} \left( {t} \right)$ and then the \textit{linear 
}(generally nonautonomous) ODEs (3.68) for $\underline {U} _{n} \left( {t} 
\right)$.

\subsection{Duplications, quadruplications, multiplications}
\label{III.E}


In this subsection we describe some tricks whereby, from an evolution 
equation for one or more matrices, one can formally get (coupled) evolution 
equations involving a larger number of matrices.  Most of these tricks can 
be iterated, or used sequentially, to get equations involving more and more 
matrices.  These tricks are introduced and explained below by showing how 
they work on some specific examples; they are of course applicable much 
more generally.


\textit{Complexification}.  A standard method to double the number of 
(dependent) variables is to complexify.  For instance setting 
\begin{equation*}
\underline {U} = \underline {X} + i~\underline {Y} ,~a = \alpha + 
i~\tilde {\alpha} ,~ b = \beta + i\,\tilde{\beta} ,\,\,\, c = \gamma + 
i\,\tilde{\gamma} \  ,\tag{3.70} 
\end{equation*}
one gets from (3.35) the \textit{linearizable} system of 2 \textit{real} coupled matrix evolution equations
\begin{gather*}
 \underline {\ddot {X}} = 2\,\left( {\alpha \,\underline {\dot {X}} - 
\tilde {\alpha} \,\underline {\dot {Y}}} \right) + \left( {\beta 
\,\underline {X} - \tilde {\beta} \,\underline {Y}} \right) + \gamma 
\,\,\left\{ {\,\left[ {\underline {\dot {X}} ,\,\underline {X}} \right] - 
\left[ {\underline {\dot {Y}} ,\,\underline {Y}} \right]\,} \right\}
\\ \qquad{}
-  \tilde {\gamma } \left\{ {\,\,\left[ {\underline {\dot {X}} ,\,\underline 
{Y}} \right] + \left[ {\underline {\dot {Y}} ,\,\underline {X}} \right]\,} 
\right\} \  ,\tag{3.71a}
\\
 \underline {\ddot {Y}} = 2\,\left( {\tilde {\alpha} 
\,\underline {\dot {X}} + \alpha \,\underline {\dot {Y}}} \right) + \left( 
{\tilde {\beta }\,\underline {X +} \beta \,\underline {Y}} \right) + \gamma 
\,\,\left\{ {\,\left[ {\underline {\dot {X}} ,\,\underline {Y}} \right] + 
\left[ {\underline {\dot {Y}} ,\,\underline {X}} \right]\,} \right\}
\\ \qquad{}
+  \tilde {\gamma} \,\,\left\{ {\,\left[ {\underline {\dot {X}} ,\,\underline 
{X}} \right]
- \left[ {\underline {\dot {Y}} ,\,\underline {Y}} \right]\,} 
\right\} \  .  \tag{3.71b}
\end{gather*}
Note that there may well be other motivations to complexify than merely the 
purpose to double the number of dependent (matrix) variables under 
consideration; for instance we have seen above that there are cases with 
\textit{complex} coupling constants when \textit{all} the solutions of a 
matrix evolution equation are completely periodic (see above, second 
paragraph after (3.37c), and Appendix B).


\textit{Association} with other \textit{(solvable and/or integrable and/or 
linearizable}) equations\textit{.} For instance consider the 
\textit{solvable} equation (3.1), which we rewrite here with a trivial 
notational change:
\begin{equation*}
 \underline {\ddot {D}} = 2\,a\,\underline {\dot {D}} + b\,\underline {D} 
+ c\,\underline {\dot {D}} \,\underline {D} ^{ - 1}\,\underline {\dot {D}} 
\  .  \tag{3.72}
\end{equation*}
Now \textit{associate} to this the (trivially \textit{solvable}) matrix 
evolution equation
\begin{equation*}
 \underline {\ddot {S}} = \alpha \,\underline {\dot {S}} \  , 
\tag{3.73a}
\end{equation*}
whose general solution reads
\begin{equation*}
\underline {S} \left( {t} \right) = \underline {S} \left( {0} \right) + 
\underline {\dot {S}} \left( {0} \right)\,\left[ {{\rm exp}\left( {\alpha \,t} 
\right) - 1} \right]/\alpha \  .  \tag{3.73b}
\end{equation*}
Then set
\begin{gather*}
 \underline {U} ^{\left( { +} \right)}\left( 
{t} \right) = \left[ {\underline {S} \left( {t} \right) + \underline {D} 
\left( {t} \right)} \right]/2,\,\,\,\,\,\underline {U} ^{\left( { -} 
\right)}\left( {t} \right) = \left[ {\underline {S} \left( {t} \right) - 
\underline {D} \left( {t} \right)} \right]/2 \  , \tag{3.74a}
\\
 \underline {S} \left( {t} \right) = \underline 
{U} ^{\left( { +} \right)}\left( {t} \right) + \underline {U} ^{\left( { -} 
\right)}\left( {t} \right),\,\,\,\,\,\underline {D} \left( {t} \right) = 
\underline {U} ^{\left( { +} \right)}\left( {t} \right) - \underline {U} 
^{\left( { -} \right)}\left( {t} \right) \  , \tag{3.74b}
\end{gather*}
and thereby obtain for $\underline {U} ^{\left( { +} 
\right)}\left( {t} \right)$ and $\underline {U} ^{\left( { -} 
\right)}\left( {t} \right)$ the system of (clearly \textit{solvable}) 
coupled matrix equations
\begin{equation*}
 \underline {\ddot {U}} ^{\left( { \pm} 
\right)} = \left[ {\left( {\alpha /2} \right) \pm a} \right]\,\underline 
{\dot {U}} ^{\left( { +} \right)} + \left[ {\left( {\alpha /2} \right) \mp 
a} \right]\,\underline {\dot {U}} ^{\left( { -} \right)} \pm \left\{ 
{b\,\underline {D} + c\,\underline {\dot {D}} \,\underline {D} ^{ - 
1}\,\underline {\dot {D}}} \right\}/2 \ , \tag{3.75}
\end{equation*}
of course with $D\left( {t} \right)$ defined in terms of 
$\underline {U} ^{\left( { +} \right)}\left( {t} \right)$ and $\underline 
{U} ^{\left( { - } \right)}\left( {t} \right)$ by the second of the 
(3.74b).
\noindent
An important advantage of this kind of trick is to yield 
\textit{translation-invariant} models; indeed (3.75) is invariant 
under the translation $\underline {U} ^{\left( { \pm} \right)}\left( {t} 
\right) \to \underline {U} ^{\left( { \pm} \right)}\left( {t} \right) + 
\underline {U} _{0} $ , with $\underline {U} _{0} $ an arbitrary constant 
matrix.  


\textit{Quadruplication}.  Given an evolution equation 
satisfied by the matrix $\underline {M} \left( {t} \right)$, it is easy to 
get equations for 4 matrices, say $\underline {A} \left( {t} 
\right),\,\,\underline {B} \left( {t} \right),\,\,\underline {C} \left( {t} 
\right),\,\,\underline {D} \left( {t} \right)$, via the position
\begin{equation*}
\underline {M} \left( {t} \right) = \left( 
{{\begin{array}{*{20}c} {\underline {A} \left( {t} \right)} \hfill & 
{\underline {B} \left( {t} \right)} \hfill \\ 
{\underline {C} \left( {t} 
\right)} \hfill & {\underline {D} \left( {t} \right)} \hfill \\ 
\end{array}} } \right) \  .  \tag{3.76} 
\end{equation*}

The following related formulas are worth recording, for future use:
\begin{gather*}
\underline {M} ^{ - 1} = \left( 
{{\begin{array}{*{20}c} {\left( {\underline {A} - \underline {B} \underline 
{D} ^{ - 1}\,\underline {C}} \right)^{ - 1}} \hfill & { - \underline {A} ^{ 
- 1}\,\underline {B} \,\left( {\underline {D} - \underline {C} \,\underline 
{A} ^{ - 1}\,\underline {B}} \right)^{ - 1}} \hfill \\ { - \underline {D} 
^{ - 1}\,\underline {C} \,\left( {\underline {A} - \underline {B} 
\,\underline {D} ^{ - 1}\,\underline {C}} \right)^{ - 1}} \hfill & {\left( 
{\underline {D} - \underline {C} \,\underline {A} ^{ - 1}\,\underline {B}} 
\right)^{ - 1}} \hfill \\ \end{array}} } \right) \  , \tag{3.77} 
\\
 \underline {M} \,\underline {\tilde {M}} = 
\left( {{\begin{array}{*{20}c} {\underline {A} \,\underline {\tilde {A}} + 
\underline {B} \,\underline {\tilde {C}}} \hfill & {\underline {A} 
\,\underline {\tilde {B}} + \underline {B} \,\underline {\tilde {D}}} 
\hfill \\ {\underline {C} \,\underline {\tilde {A}} + \underline {D} 
\,\underline {\tilde {C}}} \hfill & {\underline {C} \,\underline {\tilde 
{B}} + \underline {D} \,\underline {\tilde {D}}} \hfill \\ \end{array}} } 
\right) \  , \tag{3.78}
\\
 \underline {\tilde {M}} \,\underline {M} ^{ - 
1}\,\underline {\tilde {M}} \equiv \left( {{\begin{array}{*{20}c} 
{\underline {\hat {A}}} \hfill & {\underline {\hat {B}}} \hfill \\ 
{\underline {\hat {C}}} \hfill & {\underline {\hat {D}}} \hfill \\ 
\end{array}} } \right) \  , \tag{3.79a}
\\
 \underline {\hat {A}} = \left( {\underline 
{\tilde {A}} - \,\underline {\tilde {B}} \underline {D} ^{ - 1}\,\underline 
{C}} \right)\,\left( {\underline {A} - \underline {B} \,\underline {D} ^{ - 
1}\,\underline {C}} \right)^{ - 1}\,\underline {\tilde {A}} 
\\ \qquad{}
+ \left( 
{\underline {\tilde {B}} - \underline {\tilde {A}} \,\underline {A} ^{ - 
1}\,\underline {B}} \right)\,\left( {\underline {D} - \underline {C} 
\,\underline {A} ^{ - 1}\,\underline {B}} \right)^{ - 1}\,\underline 
{\tilde {C}} \  , \tag{3.79b}
\\
 \underline {\hat {B}} = \left( {\underline 
{\tilde {A}} - \underline {\tilde {B}} \underline {D} ^{ - 1}\,\underline 
{C}} \right)\,\left( {\underline {A} - \underline {B} \,\underline {D} ^{ - 
1}\,\underline {C}} \right)^{ - 1}\,\underline {\tilde {B}}
\\ \qquad{}
+ \left( 
{\underline {\tilde {B}} - \underline {\tilde {A}} \,\underline {A} ^{ - 
1}\,\underline {B}} \right)\,\left( {\underline {D} - \underline {C} 
\,\underline {A} ^{ - 1}\,\underline {B}} \right)^{ - 1}\,\underline 
{\tilde {D}} \  , \tag{3.79c}
\\
\underline {\hat {C}} = \left( {\underline 
{\tilde {C}} - \underline {\tilde {D}} \,\underline {D} ^{ - 1}\,\underline 
{C}} \right)\,\left( {\underline {A} - \underline {B} \,\underline {D} ^{ - 
1}\,\underline {C}} \right)^{ - 1}\,\underline {\tilde {A}}
\\ \qquad{}
+ \left( 
{\underline {\tilde {D}} - \underline {\tilde {C}} \,\underline {A} ^{ - 
1}\,\underline {B}} \right)\,\left( {\underline {D} - \underline {C} 
\,\underline {A} ^{ - 1}\,\underline {B}} \right)^{ - 1}\,\underline 
{\tilde {C}} \  , \tag{3.79d} \label{eq16} 
\\
\underline {\hat {D}} = \left( {\underline 
{\tilde {C}} - \,\underline {\tilde {D}} \underline {D} ^{ - 1}\,\underline 
{C}} \right)\,\left( {\underline {A} - \underline {B} \,\underline {D} ^{ - 
1}\,\underline {C}} \right)^{ - 1}\,\underline {\tilde {B}}
\\ \qquad{}
+ \left( 
{\underline {\tilde {D}} - \underline {\tilde {C}} \,\underline {A} ^{ - 
1}\,\underline {B}} \right)\,\left( {\underline {D} - \underline {C} 
\,\underline {A} ^{ - 1}\,\underline {B}} \right)^{ - 1}\,\underline 
{\tilde {D}} \  .  \tag{3.79e}  \label{eq17} 
\end{gather*}

The special case of these formulas with $\underline {B} = \underline {C} $ 
and $\underline {A} = \underline {D} $ is also worth recording 
(``\textit{duplication}''):
\begin{gather*}
 \underline {M} = \left( 
{{\begin{array}{*{20}c} {\underline {A}} \hfill & {\underline {B}} \hfill 
\\ {\underline {B}} \hfill & {\underline {A}} \hfill \\ \end{array}} } 
\right) \  , \tag{3.80}
\\
 \underline {M} ^{ - 1} = \left( 
{{\begin{array}{*{20}c} {\left( {\underline {A} - \underline {B} 
\,\underline {A} ^{ - 1}\,\underline {B}} \right)^{ - 1}} \hfill & { - 
\underline {A} ^{ - 1}\,\underline {B} \,\left( {\underline {A} - 
\underline {B} \,\underline {A} ^{ - 1}\,\underline {B}} \right)^{ - 1}} 
\hfill \\ { - \underline {A} ^{ - 1}\,\underline {B} \,\left( {\underline 
{A} - \underline {B} \,\underline {A} ^{ - 1}\,\underline {B}} \right)^{ - 
1}} \hfill & {\left( {\underline {A} - \underline {B} \,\underline {A} ^{ - 
1}\,\underline {B}} \right)^{ - 1}} \hfill \\ \end{array}} } \right) 
\tag{3.81}
\\ \qquad
 = \left( {{\begin{array}{*{20}c} {1} \hfill & 
{ - \underline {A} ^{ - 1}\,\underline {B}} \hfill \\ { - \underline {A} ^{ 
- 1}\,\underline {B}} \hfill & {1} \hfill \\ \end{array}} } \right)\,\left( 
{{\begin{array}{*{20}c} {\left( {\underline {A} - \underline {B} 
\,\underline {A} ^{ - 1}\,\underline {B}} \right)^{ - 1}} \hfill & {0} 
\hfill \\ {0} \hfill & {\left( {\underline {A} - \underline {B} 
\,\underline {A} ^{ - 1}\,\underline {B}} \right)^{ - 1}} \hfill \\ 
\end{array}} } \right),\tag{3.82}
\\
\underline {\tilde {M}} \,\underline {M} ^{ - 
1}\underline {\tilde {M}} \equiv \left( {{\begin{array}{*{20}c} {\underline 
{\hat {A}}} \hfill & {\underline {\hat {B}}} \hfill \\ {\underline {\hat 
{B}}} \hfill & {\underline {\hat {A}}} \hfill \\ \end{array}} } 
\right)~~,\tag{3.83a}
\\
\underline {\hat {A}} = \left( {\underline {\tilde {A}} - \,\underline 
{\tilde {B}} \underline {A} ^{ - 1}\,\underline {B}}  \right)\,\left( 
{\underline {A} - \underline {B} \,\underline {A} ^{ - 1}\,\underline {B}}  
\right)^{ - 1}\,\underline {\tilde {A}}
\\ \qquad{}
+ \left( {\underline {\tilde {B}} - 
\underline {\tilde {A}} \,\underline {A} ^{ - 1}\,\underline {B}}  
\right)\,\left( {\underline {A} - \underline {B} \,\underline {A} ^{ - 
1}\,\underline {B}}  \right)^{ - 1}\,\underline {\tilde 
{B}}\tag{3.83b}
\\
\underline {\hat {B}} = \left( {\underline {\tilde {A}} - \underline {\tilde 
{B}} \,\underline {A} ^{ - 1}\,\underline {B}}  \right)\,\left( {\underline 
{A} - \underline {B} \,\underline {A} ^{ - 1}\,\underline {B}}  \right)^{ - 
1}\,\underline {\tilde {B}}
\\ \qquad{}
+ \left( {\underline {\tilde {B}} - \underline 
{\tilde {A}} \,\underline {A} ^{ - 1}\,\underline {B}}  \right)\,\left( 
{\underline {A} - \underline {B} \,\underline {A} ^{ - 1}\,\underline {B}}  
\right)^{ - 1}\,\underline {\tilde {A}}\ .\tag{3.83c}
\end{gather*}
For instance insertion of the \textit{duplication} formula 
\begin{equation*}
\underline {U} \left( {t} \right) = \left( {{\begin{array}{*{20}c}
{\underline {U} ^{\left( {1} \right)}\left( {t} \right)} \hfill & 
{\underline {U} ^{\left( {2} \right)}\left( {t} \right)} \hfill \\
{\underline {U} ^{\left( {2} \right)}\left( {t} \right)} \hfill & 
{\underline {U} ^{\left( {1} \right)}\left( {t} \right)} \hfill \\
\end{array}} } \right)\tag{3.84}
\end{equation*}
into (3.1), yields the following system of 2 coupled matrix equations:
\begin{gather*}
\underline {\ddot {U}} ^{\left( {1} \right)} = 2\,a\,\underline {\dot {U}} 
^{\left( {1} \right)} + b\,\underline {U} ^{\left( {1} \right)}
\\ \qquad{}
+ c\,\left\{ 
{\underline {\dot {U}} ^{\left( {1} \right)}\,\underline {F} \,\underline 
{\dot {U}} ^{\left( {1} \right)} + \underline {\dot {U}} ^{\left( {2} 
\right)}\,\underline {F} \,\underline {\dot {U}} ^{\left( {2} \right)} - 
\left[ {\underline {\dot {U}} ^{\left( {1} \right)}\,\underline {G} 
\,\underline {\dot {U}} ^{\left( {2} \right)} + \underline {\dot {U}} 
^{\left( {2} \right)}\,\underline {G} \,\underline {\dot {U}} ^{\left( {1} 
\right)}} \right]} \right\} \  ,
\tag{3.85a}
\\
\underline {\ddot {U}} ^{\left( {2} \right)} = 2\,a\,\underline {\dot {U}} 
^{\left( {2} \right)} + b\,\underline {U} ^{\left( {2} \right)}
\\ \qquad{}
+ c\,\left\{ 
{\underline {\dot {U}} ^{\left( {1} \right)}\,\underline {F} \,\underline 
{\dot {U}} ^{\left( {2} \right)} + \underline {\dot {U}} ^{\left( {2} 
\right)}\,\underline {F} \,\underline {\dot {U}} ^{\left( {1} \right)} - 
\left[ {\underline {\dot {U}} ^{\left( {1} \right)}\,\underline {G} 
\,\underline {\dot {U}} ^{\left( {1} \right)} + \underline {\dot {U}} 
^{\left( {2} \right)}\,\underline {G} \,\underline {\dot {U}} ^{\left( {2} 
\right)}} \right]} \right\} \  ,\tag{3.85b}
\\
\underline {F} \equiv \left\{ {\underline {U} ^{\left( {1} \right)} - 
\underline {U} ^{\left( {2} \right)}\left[ {\underline {U} ^{\left( {1} 
\right)}} \right]^{ - 1}\underline {U} ^{\left( {2} \right)}} \right\}^{ - 
1} \ ,\tag{3.85c}
\\
\underline {G} \equiv \left[ {\underline {U} ^{\left( {1} \right)}} 
\right]^{ - 1}\underline {U} ^{\left( {2} \right)}\underline {F} \  .
\tag{3.85d}
\end{gather*}

\textit{Multiplication.} Finally, let us indicate two tricks whereby, from 
an equation involving one matrix, one can obtain equations involving many 
matrices. Again, we illustrate each of these techniques via an example, but 
it is clear that they can be used quite generally (also applied taking as 
starting point equations involving more than a single matrix).

\textit{First trick.} Let us again consider the \textit{linearizable} matrix 
equation (3.35) satisfied by the matrix $\underline {U} \left( {t} \right)$, 
and let us set
\begin{equation*}
\underline {U} \left( {t} \right) = \sum\limits_{j = 1}^{J} {\eta _{j} 
\,\underline {U} _{j} \left( {t} \right)} \  ,\tag{3.86a}
\end{equation*}
as well as
\begin{equation*}
a = \sum\limits_{j = 1}^{J} {\eta _{j} \,a_{j} ,\,\,\,\,\,b = \sum\limits_{j 
= 1}^{J} {\eta _{j} \,b_{j} ,\,\,\,\,\,c = \sum\limits_{j = 1}^{J} {\eta 
_{j} \,c_{j}} } }  \  ,\tag{3.86b}
\end{equation*}
where the $J$ quantities $\eta _{j} $ are the elements on an Abelian algebra 
satisfying the multiplication law
\begin{equation*}
\eta _{j} \,\eta _{k} = \eta _{k} \,\eta _{j} = \eta _{j + k} ;\,\,\,j,k = 
1,...,J,\,{\rm mod}\left( {J} \right)\ .\tag{3.87}
\end{equation*}
Here $J$ is an arbitrary positive integer. A standard representation of this 
algebra is
\begin{equation*}
\eta _{j} = {\rm exp}\left( {2\,\pi \,i\,j/J} \right) \ .\tag{3.88}
\end{equation*}
It is then clear that insertion of the \textit{ansatz} (3.86) into (3.35) 
yields the following system of $J$coupled equations for the $J$ matrices 
$\underline {U} _{j} \left( {t} \right)$:
\begin{equation*}
\underline {\ddot {U}} _{j} = \sum\limits_{k = 1}^{J} {\left\{ {2a_{j - k} 
\,\underline {\dot {U}} _{k} + b_{j - k} \,\underline {U} _{k}}  \right\}} + 
\sum\limits_{k_{1} ,k_{2} = 1}^{J} {\left\{ {c_{j - k_{1} - k_{2}}  \,\left[ 
{\underline {\dot {U}} _{k_{1}}  ,\,\underline {U} _{k_{2}} }  \right]} 
\right\}} \ ,\tag{3.89}
\end{equation*}
of course with all indices defined mod($J$). 

Now assume that $\underline {U} \left( {a,\,b,\,c;\,\underline {U} \left( 
{0} \right),\,\underline {\dot {U}} \left( {0} \right);\,t} \right)$ is the 
solution of (3.35) corresponding to the given constants $a,b,c$ and to the 
initial conditions $\underline {U} \left( {0} \right),\,\,\underline {\dot 
{U}} \left( {0} \right)$; then (see Appendix C for a proof) the solution of 
(3.89) with initial condition $\underline {U} _{j} \left( {0} 
\right),\,\,\underline {\dot {U}} _{j} \left( {0} \right),\,\,\,\,j = 
1,...,J$ is given by the (rather explicit!) formula
\begin{gather*}
\underline {U} _{j} \left( {t} \right) = J^{ - 1}\,\sum\limits_{k = 1}^{J} 
{\left\{ {} \right.{\rm exp}\left( { - 2\,\pi \,i\,j\,k/J} \right)} 
\underline {U} \,\left[ {} \right.a = \sum\limits_{k_{1} = 1}^{J} {a_{k_{1} 
} \,{\rm exp}\left( {2\,\pi \,i\,k_{1} \,k/J} \right),\,\,}
\\ \qquad{}
b = \sum\limits_{k_{2} = 1}^{J} {b_{k_{2}}  \,{\rm exp}\left( {2\,\pi \,i\,k_{2} 
\,k/J} \right)\,,\,\,\,c = \sum\limits_{k_{3} = 1}^{J} {c_{k_{3}}  \,} 
{\rm exp}\left( {2\,\pi \,i\,k_{3} \,k/J} \right);}
\\ \qquad{}
\underline {U} \left( {0} \right) = \sum\limits_{k_{4} = 1}^{J} {\underline 
{U} _{k_{4}}  \left( {0} \right)\,{\rm exp}\left( {2\,\pi \,i\,k_{4} \,k/J} 
\right)\,} ,
\\ \qquad{}
\underline {\dot {U}} \left( {0} \right) = 
\sum\limits_{k_{5} = 1}^{J} {\underline {U} _{k_{5}}  \left( {0} 
\right)\,{\rm exp}\left( {2\,\pi \,i\,k_{5} \,k/J} \right)} \left. {} 
\right];t\left. {} \right\} \  .\tag{3.90}
\end{gather*}

Hence (3.89) is a \textit{linearizable} equation, no less than (3.35) 
(indeed \textit{solvable} for $2 \times 2$ matrices if all the constants 
$a_{j} $ and $b_{j} $ vanish, $a_{j} = b_{j} = 0$; see Appendix A).

Let us however emphasize that this trick, by its very nature, generally 
yields coupled equations that can be uncoupled by a linear transformation. 
Indeed it is easily seen (using (C.8)) that the coupled ODEs (3.89) are 
transformed into the uncoupled ODEs
\begin{equation*}
\underline {\ddot {\tilde {U}}} _{n} = 2\,\tilde {a}_{n} \,\underline {\dot 
{\tilde {U}}} _{n} + \tilde {b}_{n} \,\underline {\tilde {U}} _{n} + \tilde 
{c}_{n} \,\left[ {\underline {\dot {\tilde {U}}} _{n} ,\,\underline {\tilde 
{U}} _{n}}  \right] \  ,\tag{3.91}
\end{equation*}
via the following transformation among tilded and untilded quantities:
\begin{equation*}
z_{n} = J^{ - 1}\,\sum\limits_{j = 1}^{J} {\tilde {z}_{j} \,{\rm exp}\left( 
{2\,\pi \,i\,n\,j/J} \right)\,,\,\,\,\,\,\,\,\tilde {z}_{n} = \sum\limits_{j 
= 1}^{J} {z_{j} \,{\rm exp}\left( { - 2\,\pi \,i\,n\,j/J} \right)}}\ .\tag{3.92}
\end{equation*}


\textit{Second trick}. This is a natural extension of the 
\textit{quadruplication} technique mentioned above. To illustrate it, we 
start from the following \textit{solvable} matrix ODE:
\begin{equation*}
\underline {\ddot {M}} = \underline {A} + \underline {M} \,\underline {B} + 
\underline {\dot {M}} \,\underline {C} + \underline {M} \,\underline {D} 
\,\underline {M} \,\underline {C} - \underline {\dot {M}} \,\underline {D} 
\,\underline {M} - 2\underline {M} \,\underline {D} \,\underline {\dot {M}} 
- \underline {M} \,\underline {D} \,\underline {M} \,\underline {D} 
\,\underline {M} \ .\tag{3.93}
\end{equation*}

Here $\underline {A} ,\,\underline {B} ,\,\underline {C} ,\,\underline {D} $ 
are 4 constant matrices (which can be chosen arbitrarily, except for the 
condition that $\underline {D} $ be invertible, see below), and the 
solvability of (3.93) is demonstrated by introducing the matrix $\underline 
{V} \left( {t} \right)$ via the position
\begin{equation*}
\underline {M} = \underline {D} ^{ - 1}\underline {V} ^{ - 1}\underline 
{\dot {V}} \ ,\tag{3.94}
\end{equation*}
and then noticing that $\underline {V} \left( {t} \right)$ satisfies the 
following third-order \textit{linear} matrix ODE (with constant 
coefficients, hence solvable):
\begin{equation*}
\underline {{V}} = \underline {V} \,\underline {D} \,\underline {A} + 
\underline {\dot {V}} \,\underline {B} + \underline {\ddot {V}} \,\underline 
{C} \  .\tag{3.95}
\end{equation*}

Note that, up to trivial notational changes, (3.93) reduces to (3.44) in the 
special case $\underline {A} = \alpha \,\underline {1} ,\,\,\underline {B} = 
\beta \,\underline {1} ,\,\,\underline {C} = \gamma \,\underline {1} ,\,\,\underline {D} = c\underline {1} $.

We now assume the matrix $\underline {M} $ to be itself made of matrices, 
and we denote, with self-explanatory notation, its matrix elements as 
$\underline {U} _{nm} $; likewise we denote as $a_{nm} \,\underline {1} 
$, $b_{nm} \,\underline {1}$, $c_{nm} \,\underline {1}$, $d_{nm} \,\underline 
{1}$ the matrix elements of the matrices $\underline {A} ,\,\underline {B} 
,\,\underline {C} ,\,\underline {D} $ (the choice of multiples of the unit 
matrix is merely for simplicity). In this manner we obtain from (393) the 
following \textit{solvable} system of $N^{2}$ coupled matrix ODEs for the 
$N^{2}$ matrices $\underline {U} _{nm}$ (we assume here the indices $n,m$ to 
range from $1$ to $N$): 
\begin{gather*}
\underline {\ddot {U}} _{nm} = a_{nm} \,\underline {1} + \sum\limits_{m_{1} 
= 1}^{N} {\left[ {b_{m_{1} m} \,\underline {U} _{nm_{1}}  + c_{m_{1} m} 
\,\underline {\dot {U}} _{nm_{1}} }  \right]}
\\ \qquad{}
+ \sum\limits_{m_{1} ,m_{2} 
,m_{3} = 1}^{N} {d_{m_{1} m_{2}}  \,c_{m_{3} m} \,\underline {U} _{nm_{1}}  
\,\underline {U} _{m_{2} m_{3}} }
\\ \qquad{}
- \sum\limits_{m_{1} ,m_{2} = 1}^{N} {d_{m_{1} m_{2}}  \,\left[ {\underline 
{\dot {U}} _{nm_{1}}  \,\underline {U} _{m_{2} m} + 2\,\underline {U} 
_{nm_{1}}  \,\underline {\dot {U}} _{m_{2} m}}  \right]}
\\ \qquad{}
+  \sum\limits_{m_{1} ,m_{2} ,m_{3} ,m_{4} = 1}^{N} {d_{m_{1} m_{2}}  
\,d_{m_{3} m_{4}}  \,\underline {U} _{nm_{1}}  \,\underline {U} _{m_{2} 
m_{3}}  \,\underline {U} _{m_{4} m}} \ .\tag{3.96}
\end{gather*}

\section{Reduction of matrices to 3-vectors}
\label{IV}


In this section we review various convenient parameterizations of matrices 
in terms of 3-vectors (and, if need be, scalars).  Obviously some of these 
results could be trivially extended to vectors of higher, or lower, 
dimensionality than 3; but, as already mentioned, in this paper we prefer 
to focus exclusively on 3-dimensional (ordinary!) space.  We always denote 
(square) matrices by underlining their symbols, and 3-vectors by 
superimposed arrows.  Because of the structure of the matrix evolution 
equations of Section \ref{III}, we are particularly interested in 
parameterizations of matrices in terms of one or more 3-vectors (and 
possibly some scalars), which belong to one (or more) of the following 
three categories. 


We term parameterizations of {\it type (i)}, for {\it invertible }matrices, those which are preserved under the 
operation $\underline {\tilde {M}} \,\,\underline {M} ^{ - 1}\,\underline 
{\tilde {M}} $, namely are such that, if both $\underline {M} $ and 
$\underline {\tilde {M}} $ are so parametrized in terms of one or more 
3-vectors, the combination $\underline {\tilde {M}} \,\,\underline {M} ^{ - 
1}\,\underline {\tilde {M}} $ admits the \textit{same }parameterization in 
terms of (appropriately defined) 3-vectors.


We term parameterizations of \textit{type (ii)} those which are preserved 
for commutators, namely are such that, if both $\underline {M} $ and 
$\underline {\tilde {M}} $ are so parametrized in terms of one or more 
3-vectors, the commutator $\left[ {\underline {M} ,\,\underline {\tilde 
{M}} } \right] \equiv \underline {M} \,\underline {\tilde {M}} - \underline 
{\tilde {M}} \,\underline {M} $ admits the \textit{same }parameterization in 
terms of (appropriately defined) 3-vectors.


Finally we term parameterizations of \textit{type (iii) }those which are 
preserved under the product operation, namely are such that, if both 
$\underline {M} $ and $\underline {\tilde {M}} $ are so parametrized in 
terms of one or more 3-vectors, the product $\underline {M} \underline 
{\tilde {M}} $ admits the \textit{same }parameterization in terms of 
(appropriately defined) 3-vectors.  Obviously parameterizations of 
\textit{type (iii) }are also of \textit{type (ii)}, and, for 
\textit{invertible} matrices, of \textit{type (i) }as well.


We indicate with the symbol $\dot { = }\,$ the one-to-one correspondence 
that the parameterization under consideration institutes among matrices and 
3-vectors.  For instance the more common parameterization we use, for ($2 
\times 2$)-matrices, reads
\begin{equation*}
\underline {M} = \rho \,\underline {1} + i\,\vec {r} \cdot \underline 
{\vec {\sigma }} \  , \tag{4.1a}
\end{equation*}
where $\rho $ is a scalar and the 3 matrices $\underline {\sigma 
} _{x} ,\,\underline {\sigma } _{y} ,\,\underline {\sigma } _{z} $ are the 
standard Pauli matrices,
\begin{equation*}
 \underline {\sigma } _{x} = \left( {{\begin{array}{*{20}c} {0} \hfill & 
{1} \hfill \\ {1} \hfill & {0} \hfill \\ \end{array} }} 
\right)\,,\,\,\,\underline {\sigma } _{y} = \left( {{\begin{array}{*{20}c} 
{0} \hfill & {i} \hfill \\ { - i} \hfill & {0} \hfill \\ \end{array} }} 
\right)\,,\,\,\,\,\underline {\sigma } _{z} = \left( 
{{\begin{array}{*{20}c} {1} \hfill & {0} \hfill \\ {0} \hfill & { - 1} 
\hfill \\ \end{array} }} \right)\,\,\,.  \  \tag{4.2}
\end{equation*}
(Note that in the following the unit matrix $\underline {1} = \left( 
{{\begin{array}{*{20}c} {1} \hfill & {0} \hfill \\ {0} \hfill & {1} \hfill 
\\ \end{array} }} \right)$is sometimes omitted).  So in this case, in 
correspondence to (4.1a), we write
\begin{equation*}
 \underline {M} \dot { = }\left( {\rho ,\,\vec {r}} \right)\ , 
\tag{4.1b}
\end{equation*}
and, via standard calculations (see Appendix D, where we report 
for convenience a number of standard formulas involving $\underline {\sigma 
} $-matrices), we also have
\begin{gather*}
 \underline {M} ^{ - 1}\dot { = }\left( {\rho , - \vec {r}} 
\right)/\left( {\rho ^{2} + r^{2}} \right) \  , \tag{4.1c}
\\
\underline {M} \,\underline {\tilde {M}} \dot { = }\left( {\rho \,\tilde 
{\rho } - \vec {r} \cdot \vec {\tilde {r}},\,\rho \,\vec {\tilde {r}} + 
\tilde {\rho }\,\vec {r} - \vec {r} \wedge \vec {\tilde {r}}} \right) \  
, \tag{4.1d}
\\
 \underline {\tilde {M}} \,\underline {M} ^{ - 1}\,\underline {\tilde 
{\tilde {M}}} \dot { = } (\tilde {\rho }\,\rho \,\tilde {\tilde {\rho }} + 
\tilde {\rho }\,\left( 
{\mathord{\buildrel{\lower3pt\hbox{$\scriptscriptstyle\rightharpoonup$}}\over 
{r}} \cdot \vec {\tilde {\tilde {r}}}} \right) - \rho \,\left( {\vec 
{\tilde {r}} \cdot \vec {\tilde {\tilde {r}}}} \right) + \tilde {\tilde 
{\rho }}\,\left( {\vec {r} \cdot 
\mathord{\buildrel{\lower3pt\hbox{$\scriptscriptstyle\rightharpoonup$}}\over 
{\tilde {r}}} } \right) 
\\* \qquad{}
+ \vec {r} \cdot \left( {\tilde {\vec {r}} 
\wedge \tilde {\tilde {\vec {r}}}} \right),\,\,\vec {\tilde {r}}\,\left[ 
{\rho \,\tilde {\tilde {\rho }} + \left( {\vec {r} \cdot \vec {\tilde 
{\tilde {r}}}} \right)} \right] - \vec {r}\,\left[ {\tilde {\rho }\,\tilde 
{\tilde {\rho }} + \left( {\vec {\tilde {r}} \cdot \vec {\tilde {\tilde 
{r}}}} \right)} \right]
\\* \qquad{}
 + \vec {\tilde {\tilde {r}}}\left[ {\tilde 
{\rho }\,\rho + \left( {\vec {\tilde {r}} \cdot \vec {r}} \right)} \right] 
+ \left[ {\tilde {\rho }\,\vec {r} \wedge \vec {\tilde {\tilde {r}}} - \rho 
\,\vec {\tilde {r}} \wedge \vec {\tilde {\tilde {r}}} + \tilde {\tilde 
{\rho }}\,\vec {\tilde {r}} \wedge \vec {r}} \right])/\left( {\rho ^{2} + 
r^{2}} \right) \  .  \tag{4.1e}
\end{gather*}
The formula (4.1d) shows that this parameterization belongs to \textit{type 
(iii)}.


Let us then restrict consideration to the class of 
\textit{traceless }($2 \times 2$)-matrices, that admit the parameterization
\begin{equation*}
 \underline {M} = i\,\vec {r} \cdot \underline {\vec {\sigma }} 
\tag{4.3a}
\end{equation*}
(namely the special case of (4.1) with $\rho = 0$).  The formulas written 
above entail that this parameterization belongs both to \textit{type (i) 
}and to \textit{type (ii)}, but not to \textit{type (iii)}.  The relevant 
formulas read:
\begin{gather*}
 \underline {M} \dot { = }\vec {r} \  , \tag{4.3b}
\\
 \underline {M} ^{ - 1}\dot { = } - \vec {r}/r^{2} \  , \tag{4.3c}
\\
 \underline {\tilde {M}} \,\underline {M} ^{ - 1}\,\underline {\tilde 
{M}} \dot { = }\left[ {2\,\vec {\tilde {r}}\,\left( {\vec {\tilde {r}} 
\cdot \vec {r}} \right) - \vec {r}\,\left( {\vec {\tilde {r}} \cdot \vec 
{\tilde {r}}} \right)} \right]/r^{2} \  , \tag{4.3d}
\\
 \left[ {\underline {M} ,\,\underline {\tilde {M}} } \right]\dot { = } - 
2\vec {r} \wedge \vec {\tilde {r}} \  .  \tag{4.3c}
\end{gather*}


The next parameterization we consider is, in terms of 2 three-vectors, for 
(invertible) antisymmetrical ($4 \times 4$)-matrices.  It reads
\begin{equation*}
 \underline {M} = \left( {{\begin{array}{*{20}c} {0} \hfill & {x^{\left( 
{1} \right)}} \hfill & {y^{\left( {1} \right)}} \hfill & {z^{\left( {1} 
\right)}} \hfill \\ { - x^{\left( {1} \right)}} \hfill & {0} \hfill & 
{z^{\left( {2} \right)}} \hfill & { - y^{\left( {2} \right)}} \hfill \\ { - 
y^{\left( {1} \right)}} \hfill & { - z^{\left( {2} \right)}} \hfill & {0} 
\hfill & {x^{\left( {2} \right)}} \hfill \\ { - z^{\left( {1} \right)}} 
\hfill & {y^{\left( {2} \right)}} \hfill & { - x^{\left( {2} \right)}} 
\hfill & {0} \hfill \\ \end{array} }} \right) \  , \tag{4.4a}
\end{equation*}
which we write
\begin{equation*}
 \underline {M} \dot { = }\left( {\vec {r}^{\left( {1} \right)},\,\vec 
{r}^{\left( {2} \right)}} \right) \  .  \tag{4.4b}
\end{equation*}
It is then easy to verify that
\begin{gather*}
 \underline {M} ^{ - 1}\dot { = } - \,\left( {\vec {r}^{\left( {2} 
\right)},\,\vec {r}^{\left( {1} \right)}} \right)/\left( {\vec {r}^{\left( 
{1} \right)} \cdot \vec {r}^{\left( {2} \right)}} \right) \  , 
\tag{4.4c}
\\
 \underline {\tilde {M}} \,\underline {M} ^{ - 1}\,\underline {\tilde 
{M}} \dot { = }\left( {} \right.\vec {\tilde {r}}^{\left( {1} 
\right)}\left[ {\left( {\vec {\tilde {r}}^{\left( {1} \right)} \cdot \vec 
{r}^{\left( {2} \right)}} \right) + \left( {\vec {\tilde {r}}^{\left( {2} 
\right)} \cdot \vec {r}^{\left( {1} \right)}} \right)} \right] - \vec 
{r}^{\left( {1} \right)}\,\left( {\vec {\tilde {r}}^{\left( {1} \right)} 
\cdot \vec {\tilde {r}}^{\left( {2} \right)}} \right),
\\ \qquad
 \vec {\tilde {r}}^{(2)}\,\left[ {\left( {\vec {\tilde {r}}^{\left( 
{1} \right)} \cdot \vec {r}^{\left( {2} \right)}} \right) + \left( {\vec 
{\tilde {r}}^{\left( {2} \right)} \cdot \vec {r}^{\left( {1} \right)}} 
\right)} \right] - \vec {r}^{\left( {2} \right)}\,\left( {\vec {\tilde 
{r}}^{\left( {1} \right)} \cdot \vec {\tilde {r}}^{\left( {2} \right)}} 
\right)\left.  {} \right)/\left( {\vec {r}^{\left( {1} \right)} \cdot \vec 
{r}^{\left( {2} \right)}} \right) \  .  \tag{4.4d}
\end{gather*}
The last formula shows that this is a parameterization of \textit{type (i)} 
(provided the 2 three-vectors $\vec {r}^{\left( {1} \right)},\,\,\vec 
{r}^{\left( {2} \right)}$ are not orthogonal).


The special case of this parameterization with $\vec {r}^{\left( {1} 
\right)} = \vec {r},\,\,\,\vec {r}^{\left( {2} \right)} = \lambda \,\vec 
{r}$,
\begin{equation*}
 \underline {M} = \left( {{\begin{array}{*{20}c} {0} \hfill & {x} \hfill & {y} \hfill & {z} \hfill 
\\ { - x} \hfill & {0} \hfill & {\lambda \,z} \hfill & { - \lambda \,y} 
\hfill \\ { - y} \hfill & { - \lambda \,z} \hfill & {0} \hfill & {\lambda 
\,x} \hfill \\ { - z} \hfill & {\lambda \,y} \hfill & { - \lambda \,x} 
\hfill & {0} \hfill \\ \end{array}}} \right) \  , \tag{4.5a}
\end{equation*}
with $\lambda $ an arbitrary (nonvanishing) constant, is also of 
\textit{type (i)}, yielding
\begin{gather*}
\underline {M} \dot { = }\vec {r} \  , 
\tag{4.5b}
\\
 \underline {\tilde {M}} \,\underline {M} ^{ - 
1}\,\underline {\tilde {M}} \dot { = }\left[ {2\,\vec {\tilde {r}}\,\left( 
{\vec {\tilde {r}} \cdot \vec {r}} \right) - \vec {r}\,\left( {\vec {\tilde 
{r}} \cdot \vec {\tilde {r}}} \right)} \right]/r^{2} \  ; 
\tag{4.5c}
\end{gather*}
but these two formulas merely reproduce (4.3b) and (4.3d).


A parameterization of ($3 \times 3$)-matrices in terms of 3-vectors takes 
the natural form
\begin{equation*}
\underline {M} = \left( 
{{\begin{array}{*{20}c} {x^{\left( {1} \right)}} \hfill & {y^{\left( {1} 
\right)}} \hfill & {z^{\left( {1} \right)}} \hfill \\ {x^{\left( {2} 
\right)}} \hfill & {y^{\left( {2} \right)}} \hfill & {z^{\left( {2} 
\right)}} \hfill \\ {x^{\left( {3} \right)}} \hfill & {y^{\left( {3} 
\right)}} \hfill & {z^{\left( {3} \right)}} \hfill \\ \end{array} }} 
\right) \  , \tag{4.6a}
\end{equation*}
which we write as 
\begin{equation*}
 \underline {M} = \left( 
{{\begin{array}{*{20}c} {\vec {r}^{\left( {1} \right)}} \hfill \\ {\vec 
{r}^{\left( {2} \right)}} \hfill \\ {\vec {r}^{\left( {3} \right)}} \hfill 
\\ \end{array} }} \right) \  .  \tag{4.6b}
\end{equation*}
Then clearly
\begin{equation*}
\underline {M} ^{ - 1} = \left( 
{{\begin{array}{*{20}c} {u^{\left( {1} \right)}_{x} } \hfill & {u^{\left( 
{2} \right)}_{x} } \hfill & {u^{\left( {3} \right)}_{x} } \hfill \\ 
{u^{\left( {1} \right)}_{y} } \hfill & {u^{\left( {2} \right)}_{y} } \hfill 
& {u^{\left( {3} \right)}_{y} } \hfill \\ {u^{\left( {1} \right)}_{z} } 
\hfill & {u^{\left( {2} \right)}_{z} } \hfill & {u^{\left( {3} \right)}_{z} 
} \hfill \\ \end{array} }} \right) \  , \tag{4.6c}
\end{equation*}
which we can also write
\begin{equation*}
 \underline {M} ^{ - 1} = \left( {\vec 
{u}^{\left( {1} \right)},\,\vec {u}^{\left( {2} \right)},\,\vec {u}^{\left( 
{3} \right)}} \right) \  , \tag{4.6d}
\end{equation*}
where the $3$ three-vectors $\vec {u}^{\left( {j} \right)}$ are defined, in 
terms of the $3$ three-vectors $\vec {r}^{\left( {k} \right)}$ , so that
\begin{equation*}
 \vec {u}^{\left( {j} \right)} \cdot \vec {r}^{\left( {k} \right)} = 
\delta _{jk} ;\,\,\,j,k = 1,2,3 \  , \tag{4.6e}
\end{equation*}
which also entail
\begin{gather*}
 \label{eq24} \sum\limits_{j = 1}^{3} {u_{x}^{\left( {j} 
\right)} \,x^{\left( {j} \right)} = 1} \  , \tag{4.6f}
\\
 \sum\limits_{j = 1}^{3} {u_{y}^{\left( {j} 
\right)} \,x^{\left( {j} \right)} = \sum\limits_{j = 1}^{3} {u_{z}^{\left( 
{j} \right)} \,x^{\left( {j} \right)} = 0} } \  , \tag{4.6g}
\end{gather*}
as well as the analogous 6 equalities obtained by cyclic 
permutations of the $x,\,y,\,z$ components of the 3-vectors $\vec 
{u}^{\left( {j} \right)} \equiv \left( {u_{x}^{\left( {j} \right)} 
,u_{y}^{\left( {j} \right)} ,u_{z}^{\left( {j} \right)} } \right)$ and 
$\vec {r}^{\left( {j} \right)} \equiv \left( {x^{\left( {j} 
\right)},y^{\left( {j} \right)},\,z^{\left( {j} \right)}} \right)$ .

An explicit definition of the 3-vectors $\vec {u}^{\left( {j} \right)}$ 
reads
\begin{gather*}
 \vec {u}^{\left( {j} \right)} = \vec 
{r}^{\left( {j + 1} \right)} \wedge \vec {r}^{\left( {j + 2} 
\right)}/\Delta ;\,\,\,j = 1,\,2,\,3,\,\,\,{\rm mod}\left( {3} \right) \  , 
\tag{4.6h}
\\
\Delta = \vec {r}^{\left( {1} \right)} \cdot 
\vec {r}^{\left( {2} \right)} \wedge \vec {r}^{\left( {3} \right)} \  .  
\tag{4.6i}
\end{gather*}
Note that $\Delta $ coincides (up to a factor $1/6$ , and possibly a sign) 
with the volume of the tetrahedron of vertices $\vec {0},\vec {r}^{\left( 
{1} \right)},\vec {r}^{\left( {2} \right)},\vec {r}^{\left( {3} \right)}$.

Hence for this parameterization we can write
\begin{equation*}
 \underline {M} \dot { = }\left( {\vec 
{r}^{\left( {j} \right)},\,\,j = 1,\,2,\,3} \right) \  , 
\tag{4.6l}
\end{equation*}
and
\begin{equation*}
\underline {\tilde {M}} \,\underline {M} ^{ - 
1}\,\underline {\tilde {\tilde {M}}} \dot { = }\left( {\vec {v}^{\left( {j} 
\right)},\,\,j = 1,\,2,\,3} \right) \  , \tag{4.6m}
\end{equation*}
with
\begin{equation*}
 \vec {v}^{\left( {j} \right)} = 
\sum\limits_{k = 1,2,3,mod\left( {3} \right)} {\left[ {\vec {\tilde 
{r}}^{\left( {j} \right)} \cdot \vec {r}^{\left( {k + 1} \right)} \wedge 
\vec {\tilde {\tilde {r}}}^{\left( {k + 2} \right)}} \right]\,\,} \vec 
{\tilde {\tilde {r}}}^{\left( {k} \right)}/\Delta \  , \tag{4.6n}
\end{equation*}
with $\Delta $ defined by (4.6i).

The formula (4.6m) with (4.6n) holds \textit{a fortiori }if $\underline 
{\tilde {\tilde {M}}} = \underline {\tilde {M}} $, hence it is clear that 
this parameterization is of \textit{type (i)}.

A special case of this parameterization is obtained by replacing the 3 
three-vectors $\vec {r}^{\left( {j} \right)}$ as follows:
\begin{equation*}
 \vec {r}^{\left( {j} \right)} \to \vec {\bar 
{r}} \equiv \vec {r}^{\left( {j} \right)} - \sum\limits_{k = 1}^{3} {\vec 
{r}^{\left( {k} \right)}/3} \  .  \tag{4.7a}
\end{equation*}
Then the treatment given above remains applicable, with the constraint on 
the (new) input vectors $\vec {\bar {r}}^{\left( {j} \right)}$ to have zero 
sum,
\begin{equation*}
 \sum\limits_{j = 1}^{3} {\vec {\bar 
{r}}^{\left( {j} \right)} = 0} \  .  \tag{4.7b}
\end{equation*}
It is then clear that the (new) vectors $\vec {\bar {v}}^{\left( {j} 
\right)}$, see (4.6m,n),
\begin{equation*}
\vec {\bar {v}}^{\left( {j} \right)} = 
\sum\limits_{k = 1,2,3,{\rm mod}\left( {3} \right)} {\left[ {\vec {\bar {\tilde 
{r}}}^{\left( {j} \right)} \cdot \vec {\bar {r}}^{\left( {k + 1} \right)} 
\wedge \vec {\bar {\tilde {\tilde {r}}}}^{\left( {k + 2} \right)}} 
\right]\,\,} \vec {\bar {\tilde {\tilde {r}}}}^{\left( {k} \right)}/\Delta 
\  , \tag{4.7c}
\end{equation*}
also satisfy the condition to have zero sum,
\begin{equation*}
\sum\limits_{j = 1}^{3} {\vec {\bar {v}}^{\left( {j} \right)} = 0} \  .  
\tag{4.7d}
\end{equation*}
Hence this parameterization is also of \textit{type (i)}.  It has the 
advantage to yield, in terms of the original 3-vectors $\vec {r}^{\left( 
{j} \right)}$, \textit{translation-invariant }equations.


The next parameterization we consider is, for ($4 \times 4$)-matrices, in 
terms of $4$ three-vectors.  It reads
\begin{equation*}
\underline {M} = \left( {{\begin{array}{*{20}c} {\rho } \hfill & {x^{\left( 
{1} \right)}} \hfill & {y^{\left( {1} \right)}} \hfill & {z^{\left( {1} 
\right)}} \hfill \\ {\rho } \hfill & {x^{\left( {2} \right)}} \hfill & 
{y^{\left( {2} \right)}} \hfill & {z^{\left( {2} \right)}} \hfill \\ {\rho 
} \hfill & {x^{\left( {3} \right)}} \hfill & {y^{\left( {3} \right)}} 
\hfill & {z^{\left( {3} \right)}} \hfill \\ {\rho } \hfill & {x^{\left( {4} 
\right)}} \hfill & {y^{\left( {4} \right)}} \hfill & {z^{\left( {4} 
\right)}} \hfill \\ \end{array} }} \right) \  , \tag{4.8a}
\end{equation*}
which we denote as follows:
\begin{equation*}
 \underline {M} \dot { = }\left( {\rho 
\,,\,\vec {r}^{\left( {j} \right)},\,\,j = 1,...,4} \right) \  .  
\tag{4.8b}
\end{equation*}
It is a matter of standard vector and matrix algebra to obtain the 
corresponding formula for the ($4 \times 4$)-matrix $\underline {\tilde 
{M}} \,\underline {M} ^{ - 1}\underline {\tilde {M}} $:
\begin{gather*}
\underline {\tilde {M}} \,\underline {M} ^{ - 
1}\,\underline {\tilde {M}} \,\,\dot { = }\,\left( {\tilde {\rho }^{2}\rho 
^{ - 1}\,,\,\,\vec {v}^{\left( {j} \right)},\,\,j = 1,...,4} \right) \  
, \tag{4.8c}
\\
 \vec {v}^{\left( {j} \right)} = 
\sum\limits_{k = 1,2,3,4,{\rm mod}\left( {4} \right)} \left( { - } 
\right)^{k}\,\vec {\tilde {r}} ^{\left( {k} \right)}\,\{ {\tilde 
{\rho }}\rho ^{ - 1}\,\vec {r}^{\left( {k + 1} \right)} \cdot \left( \vec 
{r}^{\left( {k + 2} \right)} \wedge \vec {r}^{\left( {k + 3} \right)}
\right)
\\ \qquad
 + \vec {\tilde {r}}^{\left( {j} \right)} \cdot \left[ \left( {\vec 
{r}}^{\left( {k + 1} \right)} - \vec {r}^{\left( {k + 2} \right)} \right) 
\wedge \left( {\vec {r}}^{\left( {k + 2} \right)} - \vec {r}^{\left( {k + 3} 
\right)} \right) \right] / \Delta \  ,\tag{4.8d}
\\
 \Delta = \left( {\vec {r}^{\left( {2} 
\right)} - \vec {r}^{\left( {1} \right)}} \right) \cdot \left[ {\left( 
{\vec {r}^{\left( {3} \right)} - \vec {r}^{\left( {1} \right)}} \right) 
\wedge \left( {\vec {r}^{\left( {4} \right)} - \vec {r}^{\left( {1} 
\right)}} \right)} \right] \  .  \tag{4.8e}
\end{gather*}
Note that the quantity $\Delta $ defined by this formula is 
\textit{translation-invariant} and coincides, up to a factor $1/6$ and 
possibly a sign, with the volume of the tetrahedron of vertices $\vec 
{r}^{\left( {j} \right)},\,j = 1,2,3,4$.

These formulas entail that this parameterization is of \textit{type (i)}.


The last parameterization we consider is applicable to \textit{antisymmetric 
(}$3 \times 3$)-matrices.  It reads
\begin{equation*}
 \underline {M} = \left( 
{{\begin{array}{*{20}c} {0} \hfill & {x} \hfill & {y} \hfill \\ { - x} 
\hfill & {0} \hfill & {z} \hfill \\ { - y} \hfill & { - z} \hfill & {0} 
\hfill \\ \end{array} }} \right) \  , \tag{4.9a}
\end{equation*}
which we denote as follows:
\begin{equation*}
 \underline {M} \dot { = }\vec {r} \  .  
\tag{4.9b}
\end{equation*}

The following formulas are then easy to verify:
\begin{gather*}
 \underline {M} \,\underline {\tilde {M}} = 
\left( {{\begin{array}{*{20}c} { - x\,\tilde {x} - y\,\tilde {y}} \hfill & 
{ - y\,\tilde {z}} \hfill & {x\,\tilde {z}} \hfill \\ { - z\,\tilde {y}} 
\hfill & { - x\,\tilde {x} - z\,\tilde {z}} \hfill & { - x\,\tilde {y}} 
\hfill \\ {z\,\tilde {x}} \hfill & { - y\,\tilde {x}} \hfill & { - 
y\,\tilde {y} - z\,\tilde {z}} \hfill \\ \end{array} }} \right) \  , 
\tag{4.9c}
\\
 \left[ {\underline {M} ,\,\underline {\tilde 
{M}} } \right]\dot { = } - \vec {r} \wedge \vec {\tilde {r}} \  , 
\tag{4.9d}
\\
 \underline {M} ^{2m + 1}\dot { = }\left( { - } \right)^{m}\,r^{2m}\,\vec 
{r},\,\,\,\,m = 0,\,1,\,2,...  \  .  \tag{4.9e}
\end{gather*}

These formulas entail that this parameterization is of \textit{type (ii) 
}(note, however, that (4.9e) entails that this parameterization is also 
preserved for any \textit{odd }power of the matrix $\underline {M} $ ).



\section{Results}
\label{V}


In this section we report, and in some cases briefly analyze, several 
\textit{solvable and/or integrable and/or linearizable few- and many-body 
problems in ordinary (3-dimensional) space}, which are easily obtained by 
combining the results described in the preceding two sections. We do not 
display the calculations that underlie these findings; they are quite 
straightforward, although sometimes tedious. Some of these results have been 
already synthetically highlighted in Section \ref{II}.

The presentation is conveniently split in two parts, dealing respectively 
with \textit{few-body problems }(Section \ref{V.A}) 
and \textit{many-body problems 
}(Section \ref{V.B}). The clarity thereby gained overcompensates for the minor 
repetitions entailed by this separation.

Clearly the results displayed herein are far from exhausting, or 
systematically presenting, \textit{all }the \textit{solvable and/or 
integrable and/or linearizable N-body problems} that can be obtained by 
straightforward applications of the results and methods described in the 
preceding two sections. The diligent reader will have no difficulty in 
manufacturing many other solvable and/or integrable and/or linearizable 
\textit{N}-body problems, after having become familiar with the methodology 
to obtain and investigate such problems, as described below (indeed, several 
hints in this direction are provided).


\subsection{Few-body problems}
\label{V.A}


In this subsection we consider several \textit{solvable and/or integrable 
and/or linearizable few-body problems}, obtained by applying some of the 
parameterizations of Section \ref{IV} to some of the \textit{solvable and/or 
integrable and/or linearizable} matrix ODEs of Section \ref{III}.


A simple \textit{solvable one-body problem }is obtained by applying the 
parameterization (4.3) to the \textit{solvable }matrix evolution equation 
(3.1). Its equation of motion reads
\begin{equation*}
\ddot {\vec {r}} = 2\,a\,\dot {\vec {r}} + b\,\vec {r} + c\,\left[ {2\,\dot 
{\vec {r}}\,\left( {\dot {\vec {r}} \cdot \vec {r}} \right) - \vec 
{r}\,\left( {\dot {\vec {r}} \cdot \dot {\vec {r}}} \right)} \right]/r^{2} 
\  .\tag{5.1}
\end{equation*}
Its general solution reads (see (3.2))
\bigskip
\begin{gather*}
\vec {r}\left( {t} \right) = \vec {r}\left( {0} \right)\,\left[ {\varphi 
^{\left( { + } \right)}\left( {t} \right) + i\,B\,\varphi ^{\left( { - } 
\right)}\left( {t} \right)} \right] - i\,\dot {\vec {r}}\left( {0} 
\right)\,\varphi ^{\left( { - } \right)}\left( {t} \right)/C \  
,\tag{5.2a}
\\
\varphi ^{\left( { \pm } \right)}\left( {t} \right) = exp\left( {a\,\gamma 
\,t} \right)\left\{ {} \right.\,\left[ {{\rm cosh}\left( {\Delta \,t} \right) + 
\left( {A + i\,C} \right)\,\Delta ^{ - 1}\,{\rm sinh}\left( {\Delta \,t} \right)} 
\right]^{\,\gamma }
\\ \qquad{}
\left. { \pm \left[ {{\rm cosh}\left( {\Delta \,t} \right) + \left( {A - i\,C} 
\right)\,\Delta ^{ - 1}\,{\rm sinh}\left( {\Delta \,t} \right)} \right]^{\,\gamma 
}\,} \right\}/2,\tag{5.2b}
\\
\gamma = 1/\left( {1 - c} \right) \  ,
\tag{5.2c}
\\
\Delta = \left[ {a^{2} + b\,\left( {1 - c} \right)} \right]^{1/2} \  ,
\tag{5.2d}
\\
A = - a + \left( {1 - c} \right)\,\left[ {\dot {\vec {r}}\left( {0} \right) 
\cdot \vec {r}\left( {0} \right)} \right]/\left[ {r\left( {0} \right)} 
\right]^{2} \  ,\tag{5.2e}
\\
B = \left[ {\dot {\vec {r}}\left( {0} \right) \cdot \vec {r}\left( {0} 
\right)} \right]/\left| {\dot {\vec {r}}\left( {0} \right) \wedge \vec 
{r}\left( {0} \right)} \right| \  ,
\tag{5.2f}
\\
C = \left| {\dot {\vec {r}}\left( {0} \right) \wedge \vec {r}\left( {0} 
\right)} \right|/\left[ {r\left( {0} \right)^{2}} \right] \  .
\tag{5.2g}
\end{gather*}

Here the symbol $\left| {\vec {v}} \right|$ indicates the modulus of the 
3-vector $\vec {v}$, so that $\left| {\vec {v}} \right|^{2} \equiv v_{x} 
^{2} + v_{y} ^{2} + v_{z} ^{2}$ (irrespective of whether the 3-vector $\vec 
{v}$ is real or complex).

The behavior of this system can be read from this explicit formula (see also 
Section \ref{III.A}). Of course for the model (5.1) to admit of a ``physical'' 
interpretation the constants $a,\,b,\,c$ must be \textit{real}, as well as 
the initial conditions $\vec {r}\left( {0} \right),\,\,\dot {\vec {r}}\left( 
{0} \right)$; then $\varphi ^{\left( { + } \right)}\left( {t} \right)$ is 
\textit{real}, $\varphi ^{\left( { - } \right)}\left( {t} \right)$ is 
\textit{imaginary }(see (5.2b)), and of course $\vec {r}\left( {t} \right)$ 
is \textit{real}. Note that all solutions of this problem are periodic, with 
the period given by (3.4b), if $a = 0$ and $b\,\left( {c - 1} \right) > 0$.


A \textit{solvable 2-body problem} obtains from (5.1) by complexification:
\begin{equation*}
\vec {r}\left( {t} \right) = \vec {r}^{\left( {1} \right)}\left( {t} \right) 
+ i\,\vec {r}^{\left( {2} \right)}\left( {t} \right),\,\,\,a = \alpha + 
i\,\tilde {\alpha },\,\,\,b = \beta + i\,\tilde {\beta },\,\,\,c = \gamma + 
i\,\tilde {\gamma } \  .\tag{5.3}
\end{equation*}
Its equations of motion read as follows:
\bigskip
\begin{gather*}
\ddot {\vec {r}}^{\left( {1} \right)} = 2\,\alpha \,\dot {\vec {r}}^{\left( 
{1} \right)} - 2\,\tilde {\alpha }\,\dot {\vec {r}}^{\left( {2} \right)} + 
\beta \,\vec {r}^{\left( {1} \right)} - \tilde {\beta }\,\vec {r}^{\left( 
{2} \right)}
\\ \qquad{}
+ \frac{{\gamma \,\vec {R}^{\left( {1} \right)} - \tilde 
{\gamma }\,\vec {R}^{\left( {2} \right)}}}{{\left( {\vec {r}^{\left( {1} 
\right)} \cdot \vec {r}^{\left( {1} \right)} - \vec {r}^{\left( {2} \right)} 
\cdot \vec {r}^{\left( {2} \right)}} \right)^{2} + 4\,\left( {\vec 
{r}^{\left( {1} \right)} \cdot \vec {r}^{\left( {2} \right)}} \right)^{2}}} 
\  ,\tag{5.4a}
\\
\ddot {\vec {r}}^{\left( {2} \right)} = 2\,\alpha \,\dot {\vec {r}}^{\left( 
{2} \right)} + 2\,\tilde {\alpha }\,\dot {\vec {r}}^{\left( {1} \right)} + 
\beta \,\vec {r}^{\left( {2} \right)} + \tilde {\beta }\,\vec {r}^{\left( 
{1} \right)}
\\ \qquad{}
+ \frac{{\,\gamma \,\vec {R}^{\left( {2} \right)} + \tilde 
{\gamma }\,\vec {R}^{\left( {1} \right)}}}{{\left( {\vec {r}^{\left( {1} 
\right)} \cdot \vec {r}^{\left( {1} \right)} - \vec {r}^{\left( {2} \right)} 
\cdot \vec {r}^{\left( {2} \right)}} \right)^{2} + 4\,\left( {\vec 
{r}^{\left( {1} \right)} \cdot \vec {r}^{\left( {2} \right)}} \right)^{2}}} 
\  ,\tag{5.4b}
\end{gather*}
where
\begin{gather*}
\vec {R}^{\left( {1} \right)} = \left( {\vec {r}^{\left( {1} \right)} \cdot 
\vec {r}^{\left( {1} \right)} - \vec {r}^{\left( {2} \right)} \cdot \vec 
{r}^{\left( {2} \right)}} \right)\,\vec {\rho }^{\left( {1} \right)} + 
2\,\left( {\vec {r}^{\left( {1} \right)} \cdot \vec {r}^{\left( {2} 
\right)}} \right)\,\vec {\rho }^{\left( {2} \right)} \  
,\tag{5.4c}
\\
\vec {R}^{\left( {2} \right)} = \left( {\vec {r}^{\left( {1} \right)} \cdot 
\vec {r}^{\left( {1} \right)} - \vec {r}^{\left( {2} \right)} \cdot \vec 
{r}^{\left( {2} \right)}} \right)\,\vec {\rho }^{\left( {2} \right)} - 
2\,\left( {\vec {r}^{\left( {1} \right)} \cdot \vec {r}^{\left( {2} 
\right)}} \right)\,\vec {\rho }^{\left( {1} \right)} \  
,\tag{5.4d}
\\
\vec {\rho }^{\left( {1} \right)} = 2\,\,\left( {\dot {\vec {r}}^{\left( {1} 
\right)} \cdot \vec {r}^{\left( {1} \right)} - \dot {\vec {r}}^{\left( {2} 
\right)} \cdot \vec {r}^{\left( {2} \right)}} \right)\,\dot {\vec 
{r}}^{\left( {1} \right)} - 2\,\,\left( {\dot {\vec {r}}^{\left( {1} 
\right)} \cdot \vec {r}^{\left( {2} \right)} - \dot {\vec {r}}^{\left( {2} 
\right)} \cdot \vec {r}^{\left( {1} \right)}} \right)\,\dot {\vec 
{r}}^{\left( {2} \right)}
\\ \qquad{}
 - \left( {\dot {\vec {r}}^{\left( {1} \right)} \cdot \dot {\vec 
{r}}^{\left( {1} \right)} - \dot {\vec {r}}^{\left( {2} \right)} \cdot \dot 
{\vec {r}}^{\left( {2} \right)}} \right)\,\vec {r}^{\left( {1} \right)} + 
2\,\left( {\dot {\vec {r}}^{\left( {1} \right)} \cdot \dot {\vec 
{r}}^{\left( {2} \right)}} \right)\,\vec {r}^{\left( {2} \right)}\ ,
\tag{5.4e}
\\
\vec {\rho }^{\left( {2} \right)} = 2\,\,\left( {\dot {\vec {r}}^{\left( {1} 
\right)} \cdot \vec {r}^{\left( {1} \right)} - \dot {\vec {r}}^{\left( {2} 
\right)} \cdot \vec {r}^{\left( {2} \right)}} \right)\,\dot {\vec 
{r}}^{\left( {2} \right)} + 2\,\,\left( {\dot {\vec {r}}^{\left( {1} 
\right)} \cdot \vec {r}^{\left( {2} \right)} - \dot {\vec {r}}^{\left( {2} 
\right)} \cdot \vec {r}^{\left( {1} \right)}} \right)\,\dot {\vec 
{r}}^{\left( {1} \right)}
\\ \qquad{}
 - \left( {\dot {\vec {r}}^{\left( {1} \right)} \cdot \dot {\vec 
{r}}^{\left( {1} \right)} - \dot {\vec {r}}^{\left( {2} \right)} \cdot \dot 
{\vec {r}}^{\left( {2} \right)}} \right)\,\vec {r}^{\left( {2} \right)} - 
2\,\left( {\dot {\vec {r}}^{\left( {1} \right)} \cdot \dot {\vec 
{r}}^{\left( {2} \right)}} \right)\,\vec {r}^{\left( {1} 
\right)}\ .\tag{5.4f}
\end{gather*}

An ``unphysical'' aspect of these equations of motion is the appearance of 
certain components of the force which are independent of the coordinate and 
velocity of the particle on which the force acts (we refer for instance to 
the terms $ - 2\,\tilde {\alpha }\,\dot {\vec {r}}^{\left( {2} \right)}$ and 
$ - \tilde {\beta }\,\vec {r}^{\left( {2} \right)}$ in the right-hand-side 
of (5.4a)). \textit{This phenomenon is characteristic of several equations 
considered in this paper, and will not be highlighted again in the 
following.}

It is obviously easy to get from (5.2) and (5.3) the explicit solution of 
these equations of motion; again we forsake any discussion of their behavior 
(see Section \ref{III.A}, and note that conditions on the ``coupling constants'' 
are identified there which are sufficient to guarantee that\textbf{\textit{ 
}}\textit{all} solutions of this 2-body problem be completely periodic).


A \textit{solvable translation-invariant 2-body problem }is obtained by 
applying the reduction (4.3) to the coupled matrix evolution equations 
(3.75) (instead of (3.1)). It reads
\begin{gather*}
\ddot {\vec {r}}^{\left( { \pm } \right)} = \left[ {\left( {\alpha /2} 
\right) \pm a} \right]\,\dot {\vec {r}}^{\left( { + } \right)} + \left[ 
{\left( {\alpha /2} \right) \mp a} \right]\,\dot {\vec {r}}^{\left( { - } 
\right)} 
\\ \qquad{}
\pm \left\{ {b\,\vec {r} + c\,\left[ {2\,\dot {\vec {r}}\,\left( 
{\dot {\vec {r}} \cdot \vec {r}} \right) - \vec {r}\,\left( {\dot {\vec {r}} 
\cdot \dot {\vec {r}}} \right)} \right]/r^{2}} \right\}/2 \  
,\tag{5.5a}
\end{gather*}
where
\begin{equation*}
\vec {r}\left( {t} \right) = \vec {r}^{\left( { + } \right)}\left( {t} 
\right) - \vec {r}^{\left( { - } \right)}\left( {t} \right) \  .
\tag{5.5b}
\end{equation*}
These equations of motion are \textit{translation-invariant. }Their solution 
is given, via (5.5b) and
\begin{equation*}
\vec {s}\left( {t} \right) = \vec {r}^{\left( { + } \right)}\left( {t} 
\right) + \vec {r}^{\left( { - } \right)}\left( {t} \right) \  ,
\tag{5.5c}
\end{equation*}
by (5.2) and (see (3.73b))
\begin{equation*}
\vec {s}\left( {t} \right) = \vec {s}\left( {0} \right) + \dot {\vec 
{s}}\left( {0} \right)\,\left[ {{\rm exp}\left( {\alpha \,t} \right) - 1} 
\right]/\alpha \  ,\tag{5.6a}
\end{equation*}
which correspond to the trivially \textit{solvable }equation of motion 
satisfied by $\vec {s}\left( {t} \right)~,$ 
\begin{equation*}
\ddot {\vec {s}} = \alpha \,\dot {\vec {s}} \  .
\tag{5.6b}
\end{equation*}

A \textit{solvable translation-invariant 4-body problem }obtains from the 
previous one by complexification (see (5.3)). It is left for the diligent 
reader to write out the corresponding equations of motion (a tedious task) 
and to analyze the behavior of their solutions (it is easy to identify 
restrictions on the ``coupling constants'' which guarantee that all 
solutions remain confined or are completely periodic).


All the models presented above can be generalized by using the 
parameterization (4.1) rather than (4.3). Every 3-vector $\vec {r}\left( {t} 
\right)$ gets then associated with a scalar $\rho \left( {t} \right)$. The 
diligent reader will have no difficulty to obtain these results.


Another \textit{solvable 2-body problem }is obtained by applying the 
parameterization (4.4) to the matrix evolution equation (3.1). Its equations 
of motion read
\begin{gather*}
\ddot {\vec {r}}^{\left( {1} \right)} = 2\,a\,\dot {\vec {r}}^{\left( {1} 
\right)} + b\,\vec {r}^{\left( {1} \right)} + c\,\left\{ {\dot {\vec 
{r}}^{\left( {1} \right)}\,\left[ {\left( {\dot {\vec {r}}^{\left( {1} 
\right)} \cdot \vec {r}^{\left( {2} \right)}} \right) + \left( {\dot {\vec 
{r}}^{\left( {2} \right)} \cdot \vec {r}^{\left( {1} \right)}} \right)} 
\right] - \vec {r}^{\left( {1} \right)}\,\left( {\dot {\vec {r}}^{\left( {1} 
\right)} \cdot \dot {\vec {r}}^{\left( {2} \right)}} \right)} 
\right\}
\\ \qquad{}
/\left( {\vec {r}^{\left( {1} \right)} \cdot \vec {r}^{\left( {2} 
\right)}} \right) \  ,\tag{5.7a}
\\
\ddot {\vec {r}}^{\left( {2} \right)} = 2\,a\,\dot {\vec {r}}^{\left( {2} 
\right)} + b\,\vec {r}^{\left( {2} \right)} + c\,\left\{ {\dot {\vec 
{r}}^{\left( {2} \right)}\,\left[ {\left( {\dot {\vec {r}}^{\left( {2} 
\right)} \cdot \vec {r}^{\left( {1} \right)}} \right) + \left( {\dot {\vec 
{r}}^{\left( {1} \right)} \cdot \vec {r}^{\left( {2} \right)}} \right)} 
\right] - \vec {r}^{\left( {2} \right)}\,\left( {\dot {\vec {r}}^{\left( {2} 
\right)} \cdot \dot {\vec {r}}^{\left( {1} \right)}} \right)} 
\right\}
\\ \qquad{}
/\left( {\vec {r}^{\left( {2} \right)} \cdot \vec {r}^{\left( {1} 
\right)}} \right) \  .\tag{5.7b}
\end{gather*}

The task of providing an explicit solution of these equations of motion is 
purely algebraic, but rather tedious. We leave it as an exercise for the 
diligent reader.

The reduction $\vec {r}^{\left( {1} \right)} = \vec {r},\,\,\,\vec 
{r}^{\left( {2} \right)} = \lambda \,\,\vec {r}$ \ is clearly compatible with 
these equations of motion, and it yields back (5.1).


A \textit{solvable 4-body problem }can be obtained from (5.7) via the 
complexification (5.3); the task of writing the corresponding equations of 
motion, and of identifying the cases in which all solutions are confined, 
multiply periodic or completely periodic, is left as an easy exercise for 
the diligent reader.


A \textit{solvable translation-invariant 4-body problem }can be obtained, as 
above, by applying the parameterization (4.4) to the coupled matrix evolution 
equations (3.75) (rather than to (3.1)), and a \textit{solvable 
translation-invariant 8-body problem }can subsequently be obtained by 
complexification. It is an easy task (left for the diligent reader) to write 
the corresponding equations of motion, and to analyze them at least to the 
extent of identifying the cases in which all solutions are confined, 
multiply periodic or completely periodic.


Next, let us consider the \textit{solvable 3-body problem }that is obtained 
by applying the parameterization (4.6) to the matrix evolution equation 
(3.1). The corresponding equations of motion read
\begin{gather*}
\ddot{\vec {r}}^{\left( {j} \right)} = 2\,a\,\dot {\vec {r}}^{\left( {j} 
\right)} + b\,\vec{r}^{\left( {j} \right)} + c\sum\limits_{k = 
1,2,3,{\rm mod}\left( {3} \right)} \{ \dot{\vec {r}}^{\left( {k} 
\right)}\,[ {\dot {\vec {r}}}^{\left( {j} \right)} \cdot \vec 
{r}^{\left( {k + 1} \right)} \wedge \vec {r}^{\left( {k + 2} \right)}] \}
\\ \qquad{}
/\Delta \  ,
\quad j = 1,2,3,\tag{5.8a}
\end{gather*}
with
\begin{equation*}
\Delta \equiv \vec {r}^{\left( {1} \right)} \cdot \vec {r}^{\left( {2} 
\right)} \wedge \vec {r}^{\left( {3} \right)} \  .\tag{5.8b}
\end{equation*}
Via the simple transformation 
\begin{equation*}
\vec {r}^{\left( {j} \right)}\left( {t} \right) \to {\rm exp}\left( {\lambda \,t} 
\right)\,\vec {r}^{\left( {j} \right)}\left( {t} \right)\tag{5.9a}
\end{equation*}
this model takes the more general form
\begin{gather*}
\ddot {\vec {r}}^{\left( {j} \right)} = \left( {2\,a + \lambda \,\left( {c - 
2} \right)} \right)\,\dot {\vec {r}}^{\left( {j} \right)} + \left( {b + 
2\,\lambda \,a + \lambda ^{2}\,\left( {c - 1} \right)} \right)\,\vec 
{r}^{\left( {j} \right)}
\\ \qquad{}
 + c\sum\limits_{k = 1,2,3,mod\left( {3} \right)} {\left\{ {\left( {\dot 
{\vec {r}}^{\left( {k} \right)} + \lambda \,\vec {r}^{\left( {k} \right)}} 
\right)\,\left[ {\dot {\vec {r}}^{\left( {j} \right)} \cdot \vec {r}^{\left( 
{k + 1} \right)} \wedge \vec {r}^{\left( {k + 2} \right)}} \right]} 
\right\}/\Delta } \  ,
\quad
j = 1,2,3,\tag{5.9b}
\end{gather*}
with $\Delta $ always defined by (5.8b), and it turns out, remarkably, to be 
closely related to (and to some extent more general than) an already known 
solvable 3-body problem [2]. It is an amusing task (left for the diligent 
reader) to compare the explicit solutions given in [2b] with those 
entailed by the treatment given herein. Also interesting are the various 
generalizations of this model that can be introduced in close analogy with 
the treatment given above (by \textit{complexification }and by 
\textit{association }with another solvable model so as to get 
\textit{translation-invariant }equations); in this manner one gets 
\textit{solvable 6- and 12-body problems}, and easily identifies the cases 
in which their solutions are confined, multiply periodic or completely 
periodic (for instance, all solutions of the 6-body problem obtained from 
(5.9b) by complexification are completely periodic, with period $T = \pi 
/\omega $, if $\lambda = b = 0,\,a = \pm i\omega ,\omega > 0$).


The equations of motion (5.9b) (as well as, \textit{a fortiori}, (5.9a)) are 
clearly consistent with the restriction that the ``center of mass'' $\vec 
{R} = \left( {1/3} \right)\,\sum\limits_{j = 1}^{3} {\vec {r}^{\left( {j} 
\right)}} $ be at rest in the origin, $\vec {R} = 0$ .The special solutions 
of (5.9b), or (5.8), that fulfill this constraint, correspond to the 
solution of the \textit{solvable translation-invariant 3-body problem }that 
is obtained by formally replacing in (5.9b) or (5.8) every 3-vector $\vec 
{r}^{\left( {j} \right)}$ with $\vec {\bar {r}}^{\left( {j} \right)} \equiv 
\vec {r}^{\left( {j} \right)} - \vec {R}$ .


Next we consider the \textit{solvable 4-body problem }that is obtained by 
applying the parameterization (4.8) with $\rho = 1$ to the matrix evolution 
equation (3.1) with $b\, = \,0\,$. Its equations of motion read
\begin{gather*}
\ddot {\vec {r}}^{\left( {j} \right)} = 2\,a\,\dot {\vec {r}}^{\left( {j} 
\right)} + c\sum\limits_{k = 1,2,3,4,{\rm mod}\left( {4} \right)} \left( { - } 
\right)^{k}
\\ \qquad{}
\left\{ {\dot {\vec {r}}^{\left( {k} \right)}\,\left[ {\,\dot 
{\vec {r}}^{\left( {j} \right)} \cdot \left( {\vec {r}^{\left( {k + 1} 
\right)} - \vec {r}^{\left( {k + 2} \right)}} \right) \wedge \left( {\vec 
{r}^{\left( {k + 2} \right)} - \vec {r}^{\left( {k + 3} \right)}} \right)} 
\right]} \right\}/\Delta \  ,
\tag{5.10a}
\\
\Delta \equiv \left( {\vec {r}^{\left( {2} \right)} - \vec {r}^{\left( {1} 
\right)}} \right) \cdot \left( {\vec {r}^{\left( {3} \right)} - \vec 
{r}^{\left( {1} \right)}} \right) \wedge \left( {\vec {r}^{\left( {4} 
\right)} - \vec {r}^{\left( {1} \right)}} \right) \  .
\tag{5.10b}
\end{gather*}
Again, this model (which, for $b = 0$, is \textit{translation-invariant}) 
turns out, remarkably, to be closely related to (and to some extent more 
general than) an already known solvable 4-body problem [2]. It is an amusing 
task (left for the diligent reader) to compare the explicit solutions given 
in [2b] with those entailed by the treatment given herein. Also 
interesting are the various generalizations of this model that can be 
introduced in close analogy with the treatment given above (by 
\textit{complexification }and by \textit{association }with another solvable 
model so as to get \textit{translation-invariant }equations); in this manner 
one gets \textit{solvable 8- and 16-body problems}, and easily identifies 
the cases in which their solutions are confined, multiply periodic or 
completely periodic (for instance, all solutions of the solvable 8-body 
problem obtained by complexification from (5.10) are completely periodic 
with period $T = \pi /\omega $, if $b = 0,a = \pm i\omega ,\omega > 0$).


Next, let us exhibit the \textit{solvable 2-body problem }that is obtained 
by applying the simple parameterization (4.3) to the \textit{solvable }system 
of 2 coupled matrix evolution equations~(3.85): 
\begin{gather*}
\ddot {\vec {r}}^{\left( {j} \right)} = 2\,a\,\dot {\vec {r}}^{\left( {j} 
\right)} + b\,\vec {r}^{\left( {j} \right)} + c\,\left\{ {} \right.\dot 
{\vec {r}}^{\left( {j} \right)}\,\left[ {q\dot {q} - 4p\dot {p}} \right] + 
2\,\dot {\vec {r}}^{\left( {j + 1} \right)}\,\left[ {q\,\dot {p} - p\,\dot 
{q}} \right]
\\* \qquad{}
 + \vec {r}^{\left( {j} \right)}\,\left[ {2\,p\,\tilde {p} - q\,\tilde 
{q}/2} \right] + \vec {r}^{\left( {j + 1} \right)}\,\left[ {p\,\tilde {q} - 
q\,\tilde {p}} \right]\left. {} \right\}/d,
\quad
j = 1,2,{\rm mod}\left( {2} \right) \  ,
\tag{5.11a}
\\
q \equiv \left( {\vec {r}^{\left( {1} \right)}} \right)^{2} + \left( {\vec 
{r}^{\left( {2} \right)}} \right)^{2},\,\,\,p \equiv \vec {r}^{\left( {1} 
\right)} \cdot \vec {r}^{\left( {2} \right)},\,\,\,\tilde {q} \equiv 
2\,\left[ {\dot {\vec {r}}^{\left( {1} \right)} \cdot \dot {\vec 
{r}}^{\left( {1} \right)} + \dot {\vec {r}}^{\left( {2} \right)} \cdot \dot 
{\vec {r}}^{\left( {2} \right)}} \right],\,\,\,
\\
\tilde {p} \equiv 2\,\dot 
{\vec {r}}^{\left( {1} \right)} \cdot \dot {\vec {r}}^{\left( {2} \right)},
\tag{5.11b}
\\
d \equiv q^{2} - 4\,p^{2} \equiv \left( {\vec {r}^{\left( {1} \right)} + 
\vec {r}^{\left( {2} \right)}} \right)^{2}\,\left( {\vec {r}^{\left( {1} 
\right)} - \vec {r}^{\left( {2} \right)}} \right)^{2} \  .
\tag{5.11c}
\end{gather*}

This model is presented as a simple example of the kind of results that can 
be obtained by using the \textit{quadruplication }(in fact, in this case, 
only \textit{duplication}) technique described in Section \ref{III.E}; clearly 
many more models (which, however, become more and more complicated) can be 
obtained by iterated uses of these techniques, which can be moreover 
combined with those used above (\textit{complexification}, and 
\textit{association }with other solvable models, in particular to 
manufacture \textit{translation-invariant }models). The explicit solution of 
this model (from (3.2), (3.84) and (4.3a)) is a straightforward, if tedious, 
task, that we leave as an exercise for the diligent reader.


Next we consider the scalar/vector \textit{solvable one-body problem }which 
is obtained by applying the parameterization (4.1) to the solvable matrix 
evolution equation (3.44). It is characterized by the following equations of 
motion:
\begin{gather*}
\ddot {\rho } = \alpha + \beta \,\rho + \gamma \,\left[ {\dot {\rho } + 
c\,\left( {\rho ^{2} - r^{2}} \right)} \right] - c\,\left[ {3\,\rho \,\dot 
{\rho } - 3\,\left( {\vec {r} \cdot \dot {\vec {r}}} \right) + c\,\rho 
\,\left( {\rho ^{2} - 3r^{2}} \right)} \right] \  ,
\tag{5.12a}
\\
\ddot {\vec {r}} = \beta \,\vec {r} + \gamma \,\left[ {\dot {\vec {r}} + 
2\,c\,\rho \,\vec {r}} \right] - c\,\left[ {3\,\dot {\rho }\,\vec {r} + 
3\,\rho \,\dot {\vec {r}} - \vec {r} \wedge \dot {\vec {r}} + c\,\vec 
{r}\,\left( {3\,\rho ^{2} - r^{2}} \right)} \right] \  .
\tag{5.12b}
\end{gather*}
Note that in this case the 3-vector equation (5.12b) is coupled to the 
scalar equation (5.12a). One can of course complexify this model by setting, 
say, 
\begin{gather*}
\vec {r} = \vec {r}^{\left( {1} \right)} + i\,\vec {r}^{\left( {2} 
\right)},\,\,\,\rho = \rho ^{\left( {1} \right)} + i\,\rho ^{\left( {2} 
\right)},
\\
\alpha = a + i\,\tilde {a},\,\,\,\beta = b + i\,\tilde 
{b},\,\,\,\gamma = c + i\,\tilde {c},\,\,\,c = C + i\,\tilde {C},
\tag{5.13}
\end{gather*}
getting thereby a \textit{solvable 2-body problem}. And one can perform an 
additional doubling via the technique of \textit{association}, see Section 
\ref{III.E}, getting thereby a \textit{solvable translation-invariant 4-body 
problem}.


Another \textit{solvable one-body problem} is obtained by applying the 
parameterization (4.1) to the \textit{solvable }matrix evolution equation 
(3.56). We do not report the corresponding equations, that the very diligent 
reader will easily get (using (D.8)). The remarks given above, after 
(5.12b), are then again applicable.


Next, we report the equations of motion of the \textit{scalar/vector} 
\textit{solvable one-body problem }that is obtained by applying the 
parameterization (4.1) to the \textit{solvable }matrix evolution equation 
(3.64):
\begin{gather*}
\ddot {\rho } = c + \left( {\alpha - a} \right)\,\dot {\rho } + a\,\alpha 
\,\rho + b\,\alpha \,\left( {\rho ^{2} - r^{2}} \right) - 2\,b\,\left( {\dot 
{\rho }\,\rho - \dot {\vec {r}} \cdot \vec {r}} \right) \  ,
\tag{5.14a}
\\
\ddot {\vec {r}} = \left( {\alpha - a} \right)\,\dot {\vec {r}} + a\,\alpha 
\,\vec {r} + 2\,b\,\alpha \,\rho \,\vec {r} - 2\,b\,\left( {\dot {\rho 
}\,\vec {r} + \rho \,\dot {\vec {r}}} \right) - 2\,b\,\vec {r} \wedge \dot 
{\vec {r}} \  .
\tag{5.14b}
\end{gather*}
Note the similarity of these equations of motion to (5.12). The remarks made 
after that equation are as well applicable now.


Application of the parameterization (4.1) to the\textit{ integrable }matrix 
evolution equation (3.50) (with the position $\underline {C} = \gamma 
\,\underline {1} + i\,\vec {C} \cdot \underline {\vec {\sigma }} $) yields 
the following \textit{scalar/vector integrable one-body problem :}
\begin{gather*}
\ddot {\rho } = 2\,c^{2}\,\left[ {\rho \,\left( {\rho ^{2} - 3\,r^{2}} 
\right) + \gamma \,\rho - \left( {\vec {C} \cdot \vec {r}} \right)} \right] 
\  ,\tag{5.15a}
\\
\ddot {\vec {r}} = 2\,c^{2}\,\left[ { - \vec {r}\,\left( {r^{2} - 3\,\rho 
^{2}} \right) + \gamma \,\vec {r} + \rho \,\vec {C}} \right] \  .
\tag{5.15b}
\end{gather*}
It is easily seen that these equations are yielded by the Hamiltonian
\begin{gather*}
H\left( {\vec {p},\pi ;\vec {r},\rho } \right) = \left( {p^{2} - \pi ^{2}} 
\right)/2 + c^{2}\,
\\ \qquad{}
\left[ { - 2\,\gamma \,\left( {r^{2}\rho ^{2}} \right) - 
4\,\rho \,\,\left( {\vec {C}.\vec {r}} \right) + r^{4} - 6\,r^{2}\,\rho ^{2} 
+ \rho ^{4}} \right]\,/\,2 \  .\tag{5.15c}
\end{gather*}
Of course only in the case with $\vec {C} = 0$ is 
\textit{rotation-invariance }preserved. Then one can moreover set $\rho = 0$ 
, so that (5.15a) is trivially satisfied, and (5.15b) reads
\begin{equation*}
\ddot {\vec {r}} = a\,\vec {r} - b\,r^{2}\,\vec {r}\ ,\tag{5.16a}
\end{equation*}
where, for notational simplicity, we set $c^{2} = b/2$ , $2\,c^{2}\gamma = 
a$ . This is of course the one-body problem yielded by the standard 
Hamiltonian
\begin{equation*}
H\left( {\vec {p},\vec {r}} \right) = p^{2}/2 + V\left( {r} 
\right)\tag{5.16b}
\end{equation*}
with
\begin{equation*}
V\left( {r} \right) = - a\,r^{2}/2 + b\,r^{4}/4 \  ,\tag{5.16c}
\end{equation*}
whose integrability is well-known (the standard one-body Hamiltonian (5.16b) 
is of course integrable for any spherically symmetrical potential energy 
$V\left( {r} \right)$). Clearly all solutions of this model are confined if 
$b > 0$ . An analysis of these motions is beyond the scope of the present 
paper, although we plan to take a look in the future at the interplay of the 
technique of solution entailed by the finding reported above, with the 
standard method to solve one-body spherically symmetrical problems in 
three-dimensional space.


We leave as an exercise for the diligent reader the derivation of the 
equations of motion entailed by the application of the parameterization (4.3) 
to the integrable matrix evolution equation (3.62).


Next we report the \textit{linearizable one-body problem }that is obtained 
by applying the parameterization (4.3) to the integrable matrix equation 
(A.8b) (using (D.10)):
\begin{equation*}
\ddot {\vec {r}} = 2\,a\,\dot {\vec {r}} + b\,\vec {r} + k\left[ {2\,\dot 
{\vec {r}}\,\left( {\dot {\vec {r}} \cdot \vec {r}} \right) - \vec 
{r}\,\left( {\dot {\vec {r}} \cdot \dot {\vec {r}}} \right)} \right]/r^{2} + 
\,\frac{{\tilde {f}\left( {i\,r} \right) - \tilde {f}\left( { - i\,r} 
\right)}}{{ir}}\vec {r} \wedge \dot {\vec {r}} \  .
\tag{5.17}
\end{equation*}
Here $\tilde {f}\left( {r} \right)$ is an arbitrary function (in fact, only 
its odd part contributes). If we restrict attention to the simplest case in 
which $\tilde {f}\left( {r} \right)$ is linear in $r$, $\tilde {f}\left( {r} 
\right) = Cr/2$, and we moreover set $k = 0$, we get:
\begin{equation*}
\ddot {\vec {r}} = 2 a\dot {\vec {r}} + b\, \vec {r} + C\, \vec {r} 
\wedge \dot {\vec {r}} \  .\tag{5.18}
\end{equation*}
Note that the above \textit{linearizable one-body problem} can be directly 
obtained by applying the parameterization (4.3) to the matrix evolution 
equation (3.35) (with the trivial notational change $c \to C$). A discussion 
of this problem and its solution can be found in Appendix B.


Complexification of equation (5.18) via the positions
\begin{equation*}
\vec {r} = \vec {r}^{\left( {1} \right)} + i\,\vec {r}^{\left( {2} 
\right)},\,\,a = \alpha + i\,\tilde {\alpha },\,\,\,b = \beta + i\,\tilde 
{\beta },\,\,\,C = c + i\,\tilde {c}\ ,\tag{5.19}
\end{equation*}
yields the \textit{integrable 2-body problem }characterized by the equations 
of motion
\begin{gather*}
\ddot {\vec {r}}^{\left( {1} \right)} = 2\,\left( {\alpha \,\dot {\vec 
{r}}^{\left( {1} \right)} - \tilde {\alpha }\,\dot {\vec {r}}^{\left( {2} 
\right)}} \right) + \beta \,\vec {r}^{\left( {1} \right)} - \tilde {\beta 
}\,\vec {r}^{\left( {2} \right)} + c\,\left( {\vec {r}^{\left( {1} \right)} 
\wedge \dot {\vec {r}}^{\left( {1} \right)} - \vec {r}^{\left( {2} \right)} 
\wedge \dot {\vec {r}}^{\left( {2} \right)}} \right)
\\ \qquad{}
- \tilde {c}\,\left( 
{\vec {r}^{\left( {1} \right)} \wedge \dot {\vec {r}}^{\left( {2} \right)} + 
\vec {r}^{\left( {2} \right)} \wedge \dot {\vec {r}}^{\left( {1} \right)}} 
\right)\ ,\tag{5.20a}
\\
\ddot {\vec {r}}^{\left( {2} \right)} = 2\,\left( {\alpha \,\dot {\vec 
{r}}^{\left( {2} \right)} + \tilde {\alpha }\,\dot {\vec {r}}^{\left( {1} 
\right)}} \right) + \beta \,\vec {r}^{\left( {2} \right)} + \tilde {\beta 
}\,\vec {r}^{\left( {1} \right)} + c\,\left( {\vec {r}^{\left( {1} \right)} 
\wedge \dot {\vec {r}}^{\left( {2} \right)} + \vec {r}^{\left( {2} \right)} 
\wedge \dot {\vec {r}}^{\left( {1} \right)}} \right)
\\ \qquad{}
+ \tilde {c}\,\left( 
{\vec {r}^{\left( {1} \right)} \wedge \dot {\vec {r}}^{\left( {1} \right)} - 
\vec {r}^{\left( {2} \right)} \wedge \dot {\vec {r}}^{\left( {2} \right)}} 
\right)\ .\tag{5.20b}
\end{gather*}
For an analysis of some aspects of the solution of this problem we refer to 
Appendix B, where in particular it is shown that, at least in the 2 cases 
characterized by the restrictions $\alpha = \tilde {\beta } = 0,\,\,\,\tilde 
{\alpha } = 3\,\omega /2,\,\,\beta = 2\,\omega ^{2}$ or $\alpha = \beta = 
\tilde {\beta } = 0,\,\,\,\tilde {\alpha } = \omega /2\,,$ with $\omega $ an 
arbitrary (real, nonvanishing) constant, this model is \textit{solvable} and 
\textit{all }its solutions are completely periodic with period $T = 2\pi 
/\omega $.


\textit{Association }of (5.20) with an appropriate, trivially solvable, 
model of type (5.6) yields the \textit{linearizable translation-invariant 
4-body problem }characterized by the equations of motion
\begin{gather*}
\ddot {\vec {r}}^{\left( {1, \pm } \right)} = \left\{ {\,\gamma \,\dot {\vec 
{s}}^{\left( {1} \right)} - \tilde {\gamma }\,\dot {\vec {s}}^{\left( {2} 
\right)} \pm \left[ {} \right.} \right.2\,\left( {\alpha \,\dot {\vec 
{r}}^{\left( {1} \right)} - \tilde {\alpha }\,\dot {\vec {r}}^{\left( {2} 
\right)}} \right) + \beta \,\vec {r}^{\left( {1} \right)} - \tilde {\beta 
}\,\vec {r}^{\left( {2} \right)}
\\ \qquad{}
+c\, \left( {\vec {r}^{\left( {1} \right)} \wedge \dot {\vec {r}}^{\left( 
{1} \right)} - \vec {r}^{\left( {2} \right)} \wedge \dot {\vec {r}}^{\left( 
{2} \right)}} \right) - \tilde {c}\,\left( {\vec {r}^{\left( {1} \right)} 
\wedge \dot {\vec {r}}^{\left( {2} \right)} + \vec {r}^{\left( {2} \right)} 
\wedge \dot {\vec {r}}^{\left( {1} \right)}} \right)\left. {} \right]\left. 
{} \right\}/2 \  ,\tag{5.21a}
\\
\ddot {\vec {r}}^{\left( {2, \pm } \right)} = \left\{ {} \right.\gamma 
\,\dot {\vec {s}}^{\left( {2} \right)} + \tilde {\gamma }\,\dot {\vec 
{s}}^{\left( {1} \right)} \pm \left[ {} \right.2\,\left( {\alpha \,\dot 
{\vec {r}}^{\left( {2} \right)} + \tilde {\alpha }\,\dot {\vec {r}}^{\left( 
{1} \right)}} \right) + \beta \,\vec {r}^{\left( {2} \right)} + \tilde 
{\beta }\,\vec {r}^{\left( {1} \right)}
\\ \qquad{}
 + c\,\left( {\vec {r}^{\left( {1} \right)} \wedge \dot {\vec {r}}^{\left( 
{2} \right)} + \vec {r}^{\left( {2} \right)} \wedge \dot {\vec {r}}^{\left( 
{1} \right)}} \right) + \tilde {c}\,\left( {\vec {r}^{\left( {1} \right)} 
\wedge \dot {\vec {r}}^{\left( {1} \right)} - \vec {r}^{\left( {2} \right)} 
\wedge \dot {\vec {r}}^{\left( {2} \right)}} \right)\left. {} \right]\left. 
{} \right\}/2 \  ,\tag{5.21b}
\\
\vec {r}^{\left( {1} \right)} \equiv \vec {r}^{\left( {1, + } \right)} - 
\vec {r}^{\left( {1, - } \right)},\,\,\,\vec {r}^{\left( {2} \right)} \equiv 
\vec {r}^{\left( {2, + } \right)} - \vec {r}^{\left( {2, - } 
\right)}\ ,
\\
\vec {s}^{\left( {1} \right)} \equiv \vec {r}^{\left( {1, 
+ } \right)} + \vec {r}^{\left( {1, - } \right)},\,\,\,\vec {s}^{\left( {2} 
\right)} \equiv \vec {r}^{\left( {2, + } \right)} + \vec {r}^{\left( {2, - } 
\right)} \  .\tag{5.21c}
\end{gather*}
This model is \textit{solvable} and \textit{all }its\textit{ }solutions are 
completely periodic with period $T = 2\pi /\omega $ if there holds either 
one of the 2 sets of restrictions on the 4 coupling constants $\alpha 
,\tilde {\alpha },\beta ,\tilde {\beta }$ reported above (after (5.20b)), 
and in addition there hold the 2 constraints $\gamma = 0,\tilde {\gamma } = 
m\,\omega $, with $m$ an arbitrary integer ($m \ne 0$; for $m = 0$, namely 
$\gamma = \tilde {\gamma } = 0$, (5.21) reduces to (5.20)). 


Last but not least, let us display a class of \textit{one-body problems,}
\begin{equation*}
\ddot {\vec {r}}\left( {t} \right) = \varphi \left( {r} \right)\,\vec 
{r}\left( {t} \right) \wedge \dot {\vec {r}}\left( {t} \right) \  
,\tag{5.22}
\end{equation*}
where $\varphi \left( {r} \right)$ is an arbitrary function. The above 
equation can be obtained via the parameterization (4.3) from the integrable 
matrix equation (3.36) (with $a = b = 0$, see Appendix A for details). The 
simplest choice $\varphi \left( {r} \right) = k$ yields
\begin{equation*}
\ddot {\vec {r}}\left( {t} \right) = \,k\,\,\vec {r}\left( {t} \right) 
\wedge \dot {\vec {r}}\left( {t} \right)
\tag{5.23}
\end{equation*}
which is a subcase of (5.18) and has a remarkably neat physical 
interpretation (in spite of its parity-violating character): it describes 
the motion in ordinary space of a particle acted upon by a force 
proportional to its angular momentum (of course the arbitrary coupling 
constant $k$ could be rescaled away). This is a \textit{solvable} equation: 
its explicit solution in terms of \textit{parabolic cylinder} functions is 
given in Appendix A (see (A.19),(A.28) and (A.36)).


A \textit{solvable translation-invariant 2-body problem }can be obtained 
from (5.23) via the positions (5.5b,c), with $\vec {s}\left( {t} 
\right)$ satisfying the \textit{solvable }equation of motion (5.6b). Its 
equations of motion read ($k = C$):
\begin{equation*}
\ddot {\vec {r}}^{\left( { \pm } \right)} = \left\{ {\alpha \,\left( {\dot 
{\vec {r}}^{\left( { + } \right)} + \dot {\vec {r}}^{\left( { - } \right)}} 
\right) \pm C\,\left[ {\left( {\vec {r}^{\left( { + } \right)} - \vec 
{r}^{\left( { - } \right)}} \right) \wedge \left( {\dot {\vec {r}}^{\left( { 
+ } \right)} - \dot {\vec {r}}^{\left( { - } \right)}} \right)} \right]} 
\right\}/2.\tag{5.24}
\end{equation*}


Next we consider the choice (in (5.11))
\begin{equation*}
\varphi \left( {r} \right) = k/r^{2} \  ,\tag{5.25a}
\end{equation*}
yielding
\begin{equation*}
\ddot {\vec {r}}\left( {t} \right) = \frac{{\,k\,\,\vec {r}\left( {t} 
\right) \wedge \dot {\vec {r}}\left( {t} \right)}}{{r^{2}}} \  .
\tag{5.25b}
\end{equation*}
Also in this case we have an explicit solution in terms of 
\textit{hypergeometric} functions (see Appendix A, formulas (A.19), (A.28) 
and (A.38)).


A ``physical'' case obtains from (5.22) via the choice:
\begin{equation*}
\varphi \left( {r} \right) = k/r^{3} \  ,\tag{5.26a}
\end{equation*}
yielding
\begin{equation*}
\ddot {\vec {r}}\left( {t} \right) = \frac{{\,k\,\,\vec {r}\left( {t} 
\right) \wedge \dot {\vec {r}}\left( {t} \right)}}{{r^{3}}} \  ,
\tag{5.26b}
\end{equation*}
namely the ``Newton / Lorentz'' equation of motion of an electrical charge 
in the magnetic (radial) field of a magnetic monopole (or of a magnetic 
monopole in the electrical field of an electrical charge). This equation is 
partially investigated in Appendix A, a more complete investigation is 
postponed to a subsequent paper.


We end this subsection with two \textit{linearizable three-body problems} 
that are obtained from the ``Nahm equations'' (3.67). The first one is 
obtained by applying the parameterization (4.3) to the linearizable matrix 
equation (3.69) and it reads:
\begin{gather*}
\ddot {\vec{r}}_{n} \, = \,\frac{{2}}{{\mu _{n} }} \{ \mu_{n + 1} \mu 
_{n + 2} \dot {\vec {r}}_{n + 2} \wedge \dot {\vec {r}}_{n + 1}
\\ \qquad{}
 +  \sum\limits_{m = 1}^{3}[ - \frac{{1}}{{2}}a_{n,m} \dot {\vec 
{r}}_{m}
\\ \qquad{}
+ \left( a_{n + 1,m} \mu _{n + 2} \dot {\vec {r}}_{n + 2} - a_{n + 
2,m} \mu _{n + 1} \dot {\vec {r}}_{n + 1} + \sum\limits_{k = 1}^{3} a_{n + 
1,m} a_{n + 2,k} \vec {r}_{k}  \right) \wedge \vec {r}_{m} ]\}\ .
\tag{5.27}
\end{gather*}
In the above equations as well as in the following ones up to the end of 
this Subsection \ref{V.A}, all indices are defined mod(3).


The second \textit{linearizable three-body problem} is obtained from (3.67) 
via \textit{duplication }(see (3.80)) and by again applying the 
parameterization (4.3). The relevant formulas read:
\begin{gather*}
\underline {M} _{n} \, = \left( {\,{\begin{array}{*{20}c}
 {\underline {A} _{n} } \hfill & {\underline {B} _{n} } \hfill \\
 {\underline {B} _{n} } \hfill & {\underline {A} _{n} } \hfill \\
\end{array} }} \right) \  ,
\tag{5.28a}
\\
\underline {A} _{n} \, = \,\frac{{i}}{{c}}\vec {u}_{n} \cdot \vec {\sigma 
}_{n} \  ,\tag{5.28b}
\\
\underline {B} _{n} \, = \,i\,\vec {r}_{n} \cdot \vec {\sigma }_{n} \  ,
\tag{5.28c}
\end{gather*}

Their insertion in (3.67) yields, after some labor, the following Newtonian 
equations of motion:
\begin{gather*}
\ddot {\vec {r}}_{n} \, = \,\left[ {a_{n} + c^{2}\left( {r_{n + 1}^{2} + 
r_{n + 2}^{2} } \right)} \right]\vec {r}_{n} - \left[ {b_{n,n + 1} + 
c^{2}\left( {\vec {r}_{n} \cdot \vec {r}_{n + 1} } \right)} \right]\vec 
{r}_{n + 1}
\\ \qquad{}
- \left[ {b_{n,n + 2} + c^{2}\left( {\vec {r}_{n} \cdot \vec 
{r}_{n + 2} } \right)} \right]\vec {r}_{n + 2} \  ,\tag{5.29a}
\end{gather*}
where:
\begin{gather*}
a_{n} \, = \,\left\{ {} \right.\,u_{n + 1}^{2} + u_{n + 2}^{2} + \left( 
{\gamma _{n,n} - \gamma _{n + 1,n + 1} - \gamma _{n + 2,n + 2} } 
\right)\left[ {\left( {\vec {r}_{n + 1} \cdot \vec {u}_{n + 1} } \right) + 
\left( {\vec {r}_{n + 2} \cdot \vec {u}_{n + 2} } \right)} \right]
\\ \qquad{}
 - \gamma _{n,n + 1} \left[ {\left( {\vec {r}_{n} \cdot \vec {u}_{n + 1} } 
\right) + \left( {\vec {r}_{n + 1} \cdot \vec {u}_{n} } \right)} \right] - 
\gamma _{n,n + 2} \left[ {\left( {\vec {r}_{n} \cdot \vec {u}_{n + 2} } 
\right) + \left( {\vec {r}_{n + 2} \cdot \vec {u}_{n} } \right)} 
\right]\left. {} \right\}/\Delta ^{2} \  , \tag{5.29b}
\\
b_{n.m} \, = \,\left\{ {} \right.\,\vec {u}_{n} \cdot \vec {u}_{m} + 
\frac{{1}}{{2}}\left( {\gamma _{m,m} - \gamma _{m + 1,m + 1} - \gamma _{m + 
2,m + 2} } \right)\left[ {\left( {\vec {r}_{n} \cdot \vec {u}_{m} } \right) 
+ \left( {\vec {r}_{m} \cdot \vec {u}_{n} } \right)} \right]
\\ \qquad{}
 - \gamma _{m,n} \left[ {\left( {\vec {r}_{n + 1} \cdot \vec {u}_{n + 1} } 
\right) + \left( {\vec {r}_{n + 2} \cdot \vec {u}_{n + 2} } \right)} \right] 
\\ \qquad{}
+ \gamma _{m,2n - m} \left[ {\left( {\vec {r}_{n} \cdot \vec {u}_{2n - m} } 
\right) + \left( {\vec {r}_{2n - m} \cdot \vec {u}_{n} } \right)} 
\right]\left. {} \right\}/\Delta ^{2} \ ,
\tag{5.29c}
\\
\gamma _{n,m} \, = \,\,\dot {\vec {r}}_{n} \cdot \vec {r}_{m} \  ,
\tag{5.29d}
\\
\vec {u}_{n} \, = \,\frac{{1}}{{3}}\left( {\gamma _{n,n} - \gamma _{n + 1,n 
+ 1} - \gamma _{n + 2,n + 2} } \right)\vec {r}_{n} + \gamma _{n + 1,n} \vec 
{r}_{n + 1} + \gamma _{n + 2,n} \vec {r}_{n + 2} \ ,
\tag{5.29e}
\\
\Delta \, = \,\vec {r}_{1} \cdot \vec {r}_{2} \wedge \vec {r}_{3} \  .
\tag{5.29f}
\end{gather*}
Note that an apparent paradox arises, namely, setting $c\, = \,0$ in the 
above formulas, the ``free motion'' equation $\ddot {\vec {r}}\, = \,0$ is 
not recovered (in contrast to what seems suggested by (3.67)): this is due 
to the definition (5.28b), where the coupling constant \textit{c} appears in 
the denominator.


\subsection{Many-body problems}
\label{V.B}


In this subsection we consider several \textit{solvable and/or integrable 
and/or linearizable many-body problems}, obtained by applying the (first) 
\textit{multiplication }technique of Section \ref{III.E} and Appendix C to 
some of 
the models of the preceding Section \ref{V.A}, or some of the parameterizations of 
Section \ref{IV} to some of the matrix ODEs of Section \ref{III}.


The first model we report is obtained by applying the first 
\textit{multiplication }technique of Section \ref{III.E}, or 
Appendix C, to the 
\textit{solvable }equation of motion (5.18).

The relevant equations of motion then read
\begin{equation*}
\ddot {\vec {r}}_{n} = \sum\limits_{n_{1} = 1}^{N} {\left( {2\,a_{n - n_{1} 
} \,\dot {\vec {r}}_{n_{1} } + b_{n - n_{1} } \,\vec {r}_{n_{1} } } \right)} 
+ \sum\limits_{n_{1} ,n_{2} = 1}^{N} {\left( {c_{n - n_{1} - n_{2} } \,\vec 
{r}_{n_{1} } \wedge \dot {\vec {r}}_{n_{2} } } \right)} \  .
\tag{5.30}
\end{equation*}
These equations of motion are obtained from (5.18) via the positions $\vec 
{r} = \sum\limits_{n = 1}^{N} {\eta _{n} \,\vec {r}_{n} } $, $a = 
\sum\limits_{n = 1}^{N} {\eta _{n} \,a_{n} } $and so on 
(see Section \ref{III.E}, 
or Appendix C), which entail that the indices $n,n_{1} ,n_{2} $, see (5.30), 
run from $1$ to $N$ and are defined mod$\left( {N} \right)$. For instance for 
$N = 3$ these equations of motion, when displayed in longhand, read
\begin{gather*}
\ddot {\vec {r}}_{1} = 2\,\left( {a_{1} \,\dot {\vec {r}}_{3} + a_{2} \,\dot 
{\vec {r}}_{2} + a_{3} \,\dot {\vec {r}}_{1} } \right) + b_{1} \,\vec 
{r}_{3} + b_{2} \,\vec {r}_{2} + b_{3} \,\vec {r}_{1} \,
\\ \qquad{}
+ 2 \left[ {c_{1} 
\,\left( {\vec {r}_{3} \wedge \dot {\vec {r}}_{3} } \right) + c_{2} \,\left( 
{\vec {r}_{1} \wedge \dot {\vec {r}}_{1} } \right) + c_{3} \,\left( {\vec 
{r}_{2} \wedge \dot {\vec {r}}_{2} } \right)} \right]
\\ \qquad{}
 + c_{1} \,\left( {\vec {r}_{1} \wedge \dot {\vec {r}}_{2} + \vec {r}_{2} 
\wedge \dot {\vec {r}}_{1} } \right) + c_{2} \,\left( {\vec {r}_{2} \wedge 
\dot {\vec {r}}_{3} + \vec {r}_{2} \wedge \dot {\vec {r}}_{3} } \right) + 
c_{3} \,\left( {\vec {r}_{1} \wedge \dot {\vec {r}}_{3} + \vec {r}_{3} 
\wedge \dot {\vec {r}}_{1} } \right) \  ,
\tag{5.30a}
\\
\ddot {\vec {r}}_{2} = 2\,\left( {a_{1} \,\dot {\vec {r}}_{1} + a_{2} \,\dot 
{\vec {r}}_{3} + a_{3} \,\dot {\vec {r}}_{2} } \right) + b_{1} \,\vec 
{r}_{1} + b_{2} \,\vec {r}_{3} + b_{3} \,\vec {r}_{2} 
\\ \qquad{}
+ 2 \left[ {c_{1} 
\,\left( {\vec {r}_{2} \wedge \dot {\vec {r}}_{2} } \right) + c_{2} \,\left( 
{\vec {r}_{3} \wedge \dot {\vec {r}}_{3} } \right) + c_{3} \,\left( {\vec 
{r}_{1} \wedge \dot {\vec {r}}_{1} } \right)} \right]
\\ \qquad
 + c_{1} \,\left( {\vec {r}_{1} \wedge \dot {\vec {r}}_{3} + \vec {r}_{3} 
\wedge \dot {\vec {r}}_{1} } \right) + c_{2} \,\left( {\vec {r}_{1} \wedge 
\dot {\vec {r}}_{2} + \vec {r}_{1} \wedge \dot {\vec {r}}_{2} } \right) + 
c_{3} \,\left( {\vec {r}_{2} \wedge \dot {\vec {r}}_{3} + \vec {r}_{3} 
\wedge \dot {\vec {r}}_{2} } \right) \  ,
\tag{5.30b}
\\
\ddot {\vec {r}}_{3} = 2\,\left( {a_{1} \,\dot {\vec {r}}_{2} + a_{2} \,\dot 
{\vec {r}}_{1} + a_{3} \,\dot {\vec {r}}_{3} } \right) + b_{1} \,\vec 
{r}_{2} + b_{2} \,\vec {r}_{1} + b_{3} \,\vec {r}_{3} 
\\ \qquad{}
+ 2 \left[ {c_{1} 
\,\left( {\vec {r}_{1} \wedge \dot {\vec {r}}_{1} } \right) + c_{2} \,\left( 
{\vec {r}_{2} \wedge \dot {\vec {r}}_{2} } \right) + c_{3} \,\left( {\vec 
{r}_{3} \wedge \dot {\vec {r}}_{3} } \right)} \right]
\\ \qquad{}
 + c_{1} \,\left( {\vec {r}_{2} \wedge \dot {\vec {r}}_{3} + \vec {r}_{2} 
\wedge \dot {\vec {r}}_{3} } \right) + c_{2} \,\left( {\vec {r}_{1} \wedge 
\dot {\vec {r}}_{3} + \vec {r}_{3} \wedge \dot {\vec {r}}_{1} } \right) + 
c_{3} \,\left( {\vec {r}_{1} \wedge \dot {\vec {r}}_{2} + \vec {r}_{2} 
\wedge \dot {\vec {r}}_{1} } \right) \  .
\tag{5.30c}
\end{gather*}
Let us however recall that the equations obtained via this technique can 
generally be uncoupled via a linear transformation (see (3.92)).


\textit{Association }of the model (5.30) with the trivially solvable model 
characterized by the equations of motion (see (5.6))
\begin{equation*}
\ddot {\vec {s}}_{n} = \alpha _{n} \,\dot {\vec {s}}_{n} \  ,
\tag{5.31a}
\end{equation*}
yields, via the positions,
\begin{equation*}
\vec {s}_{n} = \vec {r}_{n}^{\left( { + } \right)} + \vec {r}_{n}^{\left( { 
- } \right)} ,\,\,\,\vec {r}_{n} = \vec {r}_{n}^{\left( { + } \right)} - 
\vec {r}_{n}^{\left( { - } \right)} \  ,
\quad
\vec {r}_{n} ^{\left( { \pm } \right)} = \left( {\vec {s}_{n} \pm \vec 
{r}_{n} } \right)/2 \  ,
\tag{5.31b}
\end{equation*}
the following equations of motion characterizing a \textit{linearizable 
translation-invariant many-body problem}:
\begin{equation*}
\ddot {\vec {r}}_{n} ^{\left( { \pm } \right)} = \left\{ {} \right.\alpha 
_{n} \,\dot {\vec {s}}_{n} \pm \sum\limits_{n_{1} = 1}^{N} {\left( {2\,a_{n 
- n_{1} } \,\dot {\vec {r}}_{n_{1} } + b_{n - n_{1} } \,\vec {r}_{n_{1} } } 
\right)} + \sum\limits_{n_{1} ,n_{2} = 1}^{N} {\left( {c_{n - n_{1} - n_{2} 
} \,\vec {r}_{n_{1} } \wedge \dot {\vec {r}}_{n_{2} } } \right)} \left. {} 
\right\}/2 \  .
\tag{5.31c}
\end{equation*}

In these equations of motion for the $2\,N$ three-vectors $\vec 
{r}_{n}^{\left( { \pm } \right)} $ , the three-vectors $\vec {s}_{n} $ and 
$\vec {r}_{n} $ are defined by (5.31b). Of course all indices are again 
defined mod$\left( {N} \right)$.


Likewise, let us mention the \textit{solvable Hamiltonian many-body problem 
}characterized by the equations of motion
\begin{gather*}
\dot {\vec {q}}_{n} = \sum\limits_{n_{1} = 1}^{N} {\left( {\alpha _{n - 
n_{1} } \,\vec {p}_{n_{1} } + \gamma _{n - n_{1} } \,\vec {q}_{n_{1} } } 
\right)} + \sum\limits_{n_{1} ,n_{2} ,n_{3} = 1}^{N} {\left[ {c_{n - n_{1} - 
n_{2} - n_{3} } \,\left( {\vec {p}_{n_{1} } \wedge \vec {q}_{n_{2} } } 
\right) \wedge \vec {q}_{n_{3} } } \right]} \  ,\tag{5.32a}
\\
\dot {\vec {p}}_{n} = - \sum\limits_{n_{1} = 1}^{N} {\left( {\beta _{n - 
n_{1} } \,\vec {q}_{n_{1} } + \gamma _{n - n_{1} } \,\vec {p}_{n_{1} } } 
\right)} + \sum\limits_{n_{1} ,n_{2} ,n_{3} = 1}^{N} {\left[ {c_{n - n_{1} - 
n_{2} - n_{3} } \,\left( {\vec {p}_{n_{1} } \wedge \vec {q}_{n_{2} } } 
\right) \wedge \vec {p}_{n_{3} } } \right]}~,\tag{5.32b}
\end{gather*}
whose derivation is relegated to the end of Appendix F. But we exhibit here 
in longhand the form these equations of motion take for $N = 2$:
\begin{gather*}
\dot {\vec {q}}_{1} = \alpha _{1} \,\vec {p}_{2} + \alpha _{2} \,\vec 
{p}_{1} + \gamma _{1} \,\vec {q}_{2} + \gamma _{2} \,\vec {q}_{1} 
\,
\\ \qquad{}
+ c_{1} \left[ {\left( {\vec {p}_{1} \wedge \vec {q}_{1} } \right) \wedge \vec 
{q}_{2} + \left( {\vec {p}_{1} \wedge \vec {q}_{2} } \right) \wedge \vec 
{q}_{1} + \left( {\vec {p}_{2} \wedge \vec {q}_{1} } \right) \wedge \vec 
{q}_{1} + \left( {\vec {p}_{2} \wedge \vec {q}_{2} } \right) \wedge \vec 
{q}_{2} } \right]
\\ \qquad{}
 + c_{2} \,\left[ {\left( {\vec {p}_{1} \wedge \vec {q}_{1} } \right) \wedge 
\vec {q}_{1} + \left( {\vec {p}_{1} \wedge \vec {q}_{2} } \right) \wedge 
\vec {q}_{2} + \left( {\vec {p}_{2} \wedge \vec {q}_{1} } \right) \wedge 
\vec {q}_{2} + \left( {\vec {p}_{2} \wedge \vec {q}_{2} } \right) \wedge 
\vec {q}_{1} } \right] \ ,\tag{5.33a}
\\ \qquad{}
\dot {\vec {q}}_{2} = \alpha _{1} \,\vec {p}_{1} + \alpha _{2} \,\vec 
{p}_{2} + \gamma _{1} \,\vec {q}_{1} + \gamma _{2} \,\vec {q}_{2}  
\,
\\ \qquad{}
+ c_{1} \left[ {\left( {\vec {p}_{1} \wedge \vec {q}_{1} } \right) \wedge \vec 
{q}_{1} + \left( {\vec {p}_{1} \wedge \vec {q}_{2} } \right) \wedge \vec 
{q}_{2} + \left( {\vec {p}_{2} \wedge \vec {q}_{1} } \right) \wedge \vec 
{q}_{2} + \left( {\vec {p}_{2} \wedge \vec {q}_{2} } \right) \wedge \vec 
{q}_{1} } \right]
\\ \qquad{}
 + c_{2} \,\left[ {\left( {\vec {p}_{1} \wedge \vec {q}_{1} } \right) \wedge 
\vec {q}_{2} + \left( {\vec {p}_{1} \wedge \vec {q}_{2} } \right) \wedge 
\vec {q}_{1} + \left( {\vec {p}_{2} \wedge \vec {q}_{1} } \right) \wedge 
\vec {q}_{1} + \left( {\vec {p}_{2} \wedge \vec {q}_{2} } \right) \wedge 
\vec {q}_{2} } \right] \ .\tag{5.33b}
\end{gather*}


Next, we report the scalar/vector \textit{solvable many-body problem} 
characterized by the equations of motion
\begin{gather*}
\ddot {\rho }_{n} = \alpha _{n} + \sum\limits_{n_{1} = 1}^{N} {\left\{ 
{\beta _{n - n_{1} } \,\rho _{n_{1} } + \gamma _{n - n_{1} } \,\dot {\rho 
}_{n_{1} } } \right\}} 
\\* \qquad{}
- 3\sum\limits_{n_{1} ,n_{2} = 1}^{N} {\left\{ {c_{n 
- n_{1} - n_{2} } \,\left[ {\rho _{n_{1} } \,\dot {\rho }_{n_{2} } - \left( 
{\vec {r}_{n_{1} } \cdot \dot {\vec {r}}_{n_{2} } } \right)} \right]} 
\right\}}
\\ \qquad{}
 + \sum\limits_{n_{1} ,n_{2} ,n_{3} = 1}^{N} {\left\{ {c_{n - n_{1} - n_{2} 
- n_{3} } \,\gamma _{n_{1} } \,\left[ {\rho _{n_{2} } \,\rho _{n_{3} } - 
\left( {\vec {r}_{n_{2} } \cdot \vec {r}_{n_{3} } } \right)} \right]} 
\right\}} 
\\ \qquad{}
- \sum\limits_{n_{1} ,n_{2} ,n_{3} ,n_{4} = 1}^{N} {\left\{ {c_{n 
- n_{1} - n_{2} - n_{3} - n_{4} } \,c_{n_{1} } \,\rho _{n_{2} } \,\left[ 
{\rho _{n_{3} } \,\rho _{n_{4} } - 3\,\left( {\vec {r}_{n_{3} } \cdot \vec 
{r}_{n_{4} } } \right)} \right]} \right\}} \ ,\tag{5.34a}
\\ 
\ddot {\vec {r}}_{n} = \sum\limits_{n_{1} = 1}^{N} {\left\{ {\beta _{n - 
n_{1} } \,\vec {r}_{n_{1} } + \gamma _{n - n_{1} } \,\dot {\vec {r}}_{n_{1} 
} } \right\}} 
\\* \qquad{}
- \sum\limits_{n_{1} ,n_{2} = 1}^{N} {\left\{ {c_{n - n_{1} - 
n_{2} } \,\left[ {3\,\dot {\rho }_{n_{1} } \,\vec {r}_{n_{2} } + 3\,\rho 
_{n_{1} } \,\dot {\vec {r}}_{n_{2} } - \left( {\vec {r}_{n_{1} } \wedge \dot 
{\vec {r}}_{n_{2} } } \right)} \right]} \right\}}
\\ \qquad{}
 + 2\sum\limits_{n_{1} ,n_{2} ,n_{3} = 1}^{N} {\left\{ {c_{n - n_{1} - n_{2} 
- n_{3} } \,\gamma _{n_{1} } \,\rho _{n_{2} } \,\vec {r}_{n_{3} } } 
\right\}}
\\ \qquad{}
 - \sum\limits_{n_{1} ,n_{2} ,n_{3} ,n_{4} = 1}^{N} {\left\{ {c_{n 
- n_{1} - n_{2} - n_{3} - n_{4} } \,c_{n_{1} } \,\vec {r}_{n_{2} } \,\left[ 
{3\,\rho _{n_{3} } \,\rho _{n_{4} } - \left( {\vec {r}_{n_{3} } \cdot \vec 
{r}_{n_{4} } } \right)} \right]} \right\}} \ ,\tag{5.34b}
\end{gather*}
which are obtained by applying to (5.12) the (first) \textit{multiplication 
}technique described at the end of Section \ref{III.E} and in Appendix C. Hence 
these equations of motion can be uncoupled by using a linear transformation 
of type (3.92).


Next, we report the equations of motion of the \textit{solvable many-body 
problem }which is obtained by applying the parameterization (4.3) to the 
special case of the \textit{solvable }matrix evolution equation (3.63) with 
$\underline {D} = \underline {\tilde {D}} = \underline {1} $ (see Appendix 
H):
\begin{gather*}
 \ddot {\vec {r}}_{n} = a_{n} \,\dot {\vec {r}}_{n} - \alpha _{n} \,\gamma 
_{n} \,\vec {r}_{n} \wedge \dot {\vec {r}}_{n} + \left\{ {} \right. - \gamma 
_{n} \,\dot {\vec {r}}_{n} \wedge \left[ {\gamma _{n} ^{2}\,\left( {\vec 
{r}_{n} \cdot \dot {\vec {r}}_{n} } \right)\,\vec {r}_{n} + a_{n} \,\gamma 
_{n} \,\vec {r}_{n} \wedge \dot {\vec {r}}_{n} } \right]
\\ \qquad{}
- \sum\limits_{m = 1}^{N} \{ b_{nm} \, \{ \dot{\vec {r}}_{m} + 
\gamma_{n} \,\dot {\vec{r}}_{n} \wedge [ \gamma_{n}^{2}\,\left( 
\vec{r}_{n} \cdot \vec{r}_{m}  \right)\, \vec{r}_{n} + a_{n} ^{2}\,\,
\vec{r}_{m} + a_{n} \,\gamma_{n} \,\vec{r}_{n} \wedge \vec 
{r}_{m}  ] \} \} \}
\\ \qquad{}
/\left[ a_{n} 
\,\left( a_{n}^{2} + \gamma_{n}^{2}\,\,r_{n}^{2} \right) \right]\ .
\tag{5.35}
\end{gather*}

Here the $N\,\left( {3 + N} \right)$ constants $a_{n} ,\,\alpha _{n} 
,\,\gamma _{n} ,\,b_{nm} $ are arbitrary. An analysis of the motions 
entailed by this model is straightforward (see Appendix H), but it exceeds 
the scope of this paper. Note that even the one-body case ($N = 1$; or, 
equivalently, $b_{nm} = \delta _{nm} \,b_{n} $) is nontrivial (namely, its 
equation of motion is nontrivial; of course in some sense all 
\textit{solvable }models are trivial!).


Next, we report the equations of motion characterizing the scalar/vector 
\textit{solvable many-body problem }which is obtained by applying the 
parameterization (4.1) to the \textit{solvable }matrix evolution equation 
(3.65):
\begin{gather*}
\ddot {\rho }_{n} = \left( {a_{n} - a_{n + 1} } \right)\,\tilde {c}_{n} + 
\left( {a_{n} - a_{n + 1} } \right)\,\left( {\tilde {a}_{n} - a_{n} } 
\right)\,\rho _{n} + \tilde {c}_{n} \,\left( {b_{n} \,\rho _{n} - b_{n + 1} 
\,\rho _{n + 1} } \right)
\\ \qquad{}
- \left( {\tilde {a}_{n} - a_{n} } \right)\,b_{n + 1} \,\left( {\rho _{n} 
\,\rho _{n + 1} - \vec {r}_{n} \cdot \vec {r}_{n + 1} } \right) - 3\,b_{n} 
\,\left( {\dot {\rho }_{n} \,\rho _{n} - \dot {\vec {r}}_{n} \cdot \vec 
{r}_{n} } \right) \,
\\ \qquad{}
+ \left( {\tilde {a}_{n} - 2\,a_{n} + a_{n + 1} } 
\right) \left[ {\dot {\rho }_{n} + b_{n} \,\left( {\rho _{n} ^{2} - r_{n} 
^{2}} \right)} \right]
\\ \qquad{}
+ b_{n + 1} \,\left( {\dot {\rho }_{n} \,\rho _{n + 1} - \dot {\vec 
{r}}_{n} \cdot \vec {r}_{n + 1} } \right) + b_{n} \,b_{n + 1} \left[ {\rho 
_{n + 1} \,\left( {\rho _{n} ^{2} - r_{n} ^{2}} \right) - 2\,\rho _{n} 
\,\vec {r}_{n} \cdot \vec {r}_{n + 1} } \right]
\\ \qquad{}
- b_{n} ^{2}\,\rho _{n} 
\,\left( {\rho _{n} ^{2} - 3r_{n} ^{2}} \right)\ ,\tag{5.36a}
\\
\ddot {\vec {r}}_{n} = \left( {\,a_{n} - a_{n + 1} } \right)\,\left( {\tilde 
{a}_{n} - a_{n} } \right)\vec {r}_{n} + \tilde {c}_{n} \,\left( {b_{n} 
\,\vec {r}_{n} - b_{n + 1} \,\vec {r}_{n + 1} } \right)
\\ \qquad{}
 - \,b_{n + 
1} \left( 
{\tilde {a}_{n} - a_{n} } \right)\,\left( {\rho _{n} \,\vec {r}_{n + 1} + 
\rho _{n + 1} \,\vec {r}_{n} + \vec {r}_{n} \wedge \vec {r}_{n + 1} } 
\right)
\\ \qquad{}
 - b_{n} \,\left( {3\,\dot {\rho }_{n} \,\vec {r}_{n} + 3\,\rho _{n} \,\dot 
{\vec {r}}_{n} - \dot {\vec {r}}_{n} \wedge \vec {r}_{n} } \right) + \left( 
{\tilde {a}_{n} - 2\,a_{n} + a_{n + 1} } \right)\,\left( {\dot {\vec 
{r}}_{n} + 2\,b_{n} \,\rho _{n} \,\vec {r}_{n} } \right) 
\\ \qquad{}
+ b_{n + 1} \left( {\rho _{n + 1} \,\dot {\vec {r}}_{n} + \dot {\rho }_{n} \,\vec 
{r}_{n} + \dot {\vec {r}}_{n} \wedge \vec {r}_{n + 1} } \right)
\\ \qquad{}
 + b_{n} \,b_{n + 1} \,\left[ {2\,\rho _{n + 1} \,\rho _{n} \,\vec {r}_{n} + 
\left( {\rho _{n} ^{2} - r_{n} ^{2}} \right)\,\vec {r}_{n + 1} + 2\,\rho 
_{n} \,\vec {r}_{n} \wedge \vec {r}_{n + 1} } \right] - b_{n} ^{2}\,\left( 
{3\,\rho _{n} ^{2} - r_{n} ^{2}} \right)\,\vec {r}_{n} \ 
.\tag{5.36b}
\end{gather*}


Next, we report the equations of motion that are obtained by applying the 
first technique of \textit{multiplication }of Section \ref{III.E} and Appendix C 
to the \textit{integrable }one-body problem (5.15) (with $\vec {C} = 0$, to 
restore \textit{rotation-invariance}, and via the positions $2\,c^{2} = 
\sum\limits_{n = 1}^{N} {\eta _{n} \,b_{n} } $, $2\,c^{2}\,\gamma = 
\sum\limits_{n = 1}^{N} {\eta _{n} \,a_{n} } $ as well of course as $\rho = 
\sum\limits_{n = 1}^{N} {\eta _{n} \,\rho _{n} ,\vec {r} = \sum\limits_{n = 
1}^{N} {\eta _{n} \,\vec {r}_{n} } } $). They read:
\begin{gather*}
\ddot {\rho }_{n} = \sum\limits_{n_{1} = 1}^{N} {\left( {a_{n - n_{1} } 
\,\rho _{n_{1} } } \right) + \sum\limits_{n_{1} ,n_{2} ,n_{3} = 1}^{N} 
{\left\{ {b_{n - n_{1} - n_{2} - n_{3} } \,\rho _{n_{1} } \,\left[ {\rho 
_{n_{2} } \,\rho _{n_{3} } - 3\,\left( {\vec {r}_{n_{2} } \cdot \vec 
{r}_{n_{3} } } \right)} \right]} \right\}} } \ ,\tag{5.37a}
\\
\vec {r}_{n} = \sum\limits_{n_{1} = 1}^{N} {\left( {a_{n - n_{1} } \,\vec 
{r}_{n_{1} } } \right) + \sum\limits_{n_{1} ,n_{2} ,n_{3} = 1}^{N} {\left\{ 
{b_{n - n_{1} - n_{2} - n_{3} } \,\vec {r}_{n_{1} } \,\left[ {3\,\rho 
_{n_{2} } \,\rho _{n_{3} } - \left( {\vec {r}_{n_{2} } \cdot \vec {r}_{n_{3} 
} } \right)} \right]} \right\}} } \ .\tag{5.37b}
\end{gather*}

In these equations all indices are defined mod$\left( {N} \right)$. This 
scalar/vector \textit{many-body problem }is then of course an 
\textit{integrable }one, but susceptible to being uncoupled via a linear 
transformation of type (3.92).


We plan to discuss in a separate paper the \textit{integrable }class of 
(rotation-invariant!) 3-dimensional many-body problems with quadratic and 
quartic interactions obtainable by applying to the integrable matrix 
equation (3.50) other techniques of association and multiplication (see 
Section \ref{III.E}) as well as appropriate parameterizations of matrices in terms 
of 3-vectors (see Section \ref{IV}).


Next we report the \textit{linearizable many-body problem }whose equations 
of motion are obtained by applying the parameterization (4.3) to the 
\textit{linearizable }matrix evolution equations (3.29), and therefore read 
as follows:
\begin{gather*}
\ddot{\vec {r}}_{n} = \sum\limits_{n_{1} = 1}^{N} \{ 2\,a_{nn_{1} } 
\left( {t} \right)\,\,\dot {\vec{r}}_{n_{1}} + b_{nn_{1}} \left( {t} 
\right)\,\,\vec{r}_{n_{1} } + C_{n_{1}} \left( {t} \right)\,\,\left[ 
2\,\left( \vec{r}_{n_{1} } \wedge \dot {\vec {r}}_{n}  \right) - \left( 
\vec{r}_{n} \wedge \dot {\vec {r}}_{n_{1}} \right) \right]
\\ \qquad{}
- \dot {C}_{n_{1}} \left( {t} \right)\,\,\left( \vec{r}_{n} \wedge \vec 
{r}_{n_{1}}  \right) \}
\\ \qquad{}
 + \sum\limits_{n_{1} ,n_{2} = 1}^{N} {\left\{ {2\,a_{nn_{1} } \,\left( {t} 
\right)\,C_{n_{2} } \left( {t} \right)\,\,\left( {\vec {r}_{n_{1} } \wedge 
\hat {r}_{n_{2} } } \right) - C_{n_{1} } \left( {t} \right)\,\,C_{n_{2} } 
\left( {t} \right)\,\,\left[ {\left( {\vec {r}_{n} \wedge \vec {r}_{n_{1} } 
} \right) \wedge \vec {r}_{n_{2} } } \right]} \right\}} \ 
.\tag{5.38}
\end{gather*}
The connection with the notation of (3.29) is: $C_{n} = 2\,c_{n} $. Note 
that, as in (3.29), we are allowing in this case all the $N\,\left( {2N + 1} 
\right)$ quantities $a_{nm} \left( {t} \right),\,\,b_{nm} \left( {t} 
\right),\,\,C_{n} \left( {t} \right)$ to be \textit{time-dependent}.


We leave as an exercise for the diligent reader to derive the more general 
(and of course also \textit{linearizable}) equations of motion that are 
obtained by applying to (3.29) the parameterization (4.1) (rather than 
(4.3)).


Next we report the \textit{integrable many-body problem }whose equations of 
motion are obtained by applying the parameterization (4.3) to (3.38), leaving 
again as an exercise for the diligent reader to derive the more general (and 
of course also \textit{integrable}) equations of motion that are obtained by 
applying to (3.38) the parameterization (4.1) (rather than (4.3)). They read:
\begin{gather*}
\ddot {\vec {r}}_{n} = \left[ {2\,\dot {\vec {r}}_{n} \,\left( {\dot {\vec 
{r}}_{n} \cdot \vec {r}_{n} } \right) - \vec {r}_{n} \,\left( {\dot {\vec 
{r}}_{n} \cdot \dot {\vec {r}}_{n} } \right)} \right]/r_{n} ^{2}
\\ \qquad
+ \gamma 
\,\left\{ {\vec {r}_{n + 1} - \left[ {2\,\vec {r}_{n} \,\left( {\vec {r}_{n} 
\cdot \vec {r}_{n - 1} } \right) - \vec {r}_{n - 1} \,\left( {\vec {r}_{n} 
\cdot \vec {r}_{n} } \right)} \right]/r_{n - 1} ^{2}} \right\} \  
.\tag{5.39}
\end{gather*}
Note the ``nearest-neighbour'' character of these equations of motion (we do 
not repeat here, nor below in analogous cases, the comments on the need that 
these equations of motion be supplemented by ``boundary conditions at the 
$n$-ends:'' see Sections \ref{I} and \ref{III.C}).


Another scalar/vector \textit{integrable many-body problem }is obtained by 
applying the parameterization (4.1) to (3.39b) (note that in this case the 
parameterization (4.3) is not applicable; on the other hand the simple, 
quadratic, character of the nonlinearity, see below, makes the following 
equations of motion particularly suitable for the \textit{multiplication} 
techniques of Section \ref{III.E}; 
but we leave this additional extension in the 
\textit{many-body }direction as an exercise for the diligent reader). The 
equations of motion read 
\begin{gather*}
\ddot {\rho }_{n} = c\,\left[ {\dot {\rho }_{n} \,\left( {\rho _{n + 1} - 
2\,\rho _{n} + \rho _{n - 1} } \right) - \dot {\vec {r}}_{n} \cdot \left( 
{\vec {r}_{n + 1} - 2\,\vec {r}_{n} + \vec {r}_{n - 1} } \right)} \right] 
\  ,\tag{5.40a}
\\
\ddot{\vec{r}}_{n} = c\,[ \dot{\vec {r}}_{n} \,\left( \rho_{n + 
1} - 2\,\rho_{n} + \rho_{n - 1}  ) + \dot {\rho }_{n} \,( 
\vec{r}_{n + 1} - 2\,\vec{r}_{n} + \vec{r}_{n - 1} ) \right.
\\ \qquad{}
+ \dot 
{\vec{r}}_{n} \wedge ( \vec{r}_{n + 1} - \vec{r}_{n - 1} )] \ .\tag{5.40b}
\end{gather*}

Note the \textit{translation-invariant }character of these equations of 
motion (i. e., invariant under $\rho _{n} \to \rho _{n} + \rho _{0} $ , 
$\vec {r}_{n} \to \vec {r}_{n} + \vec {r}_{0} $ with $\rho _{0} ,\vec 
{r}_{0} $ time-independent but otherwise arbitrary).


Let us also report the generalization of these equations that are obtained 
by first replacing formally in (5.40) $\rho _{n} \left( {t} \right),\,\vec 
{r}_{n} \left( {t} \right)$ with, say, $\tilde {\rho }_{n} \left( {\tau } 
\right),\,\vec {\tilde {r}}_{n} \left( {\tau } \right)$ (also replacing, of 
course, derivatives with respect to $t$ with derivatives with respect to 
$\tau $), then setting $\tilde {\rho }_{n} \left( {\tau } \right) = 
{\rm exp}\left( { - a\,t} \right)\,\rho _{n} \left( {t} \right)$, $\vec {\tilde 
{r}}_{n} \left( {\tau } \right) = {\rm exp}\left( { - a\,t} \right)\,\vec {r}_{n} 
\left( {t} \right)$, $\tau = \left[ {{\rm exp}\left( {a\,t} \right) - 1} 
\right]/a$ and then also performing a standard complexification ($\rho _{n} = 
\rho _{n}^{\left( {1} \right)} + i\,\rho _{n}^{\left( {2} \right)} $, $\vec 
{r}_{n} = \vec {r}_{n}^{\left( {1} \right)} + i\,\vec {r}_{n}^{\left( {2} 
\right)} $, $c = \gamma + i\,\tilde {\gamma }$ , $a = \lambda + i\,\omega 
$). We thus obtain the more general scalar/vector \textit{integrable 
many-body problem }characterized by the following equations of motion:
\begin{gather*}
\ddot {\rho }_{n}^{\left( {1} \right)} = - \lambda \,\dot {\rho 
}_{n}^{\left( {1} \right)} + \omega \,\dot {\rho }_{n}^{\left( {2} \right)} 
+ \gamma \,A_{n} - \tilde {\gamma }\,\tilde {A}_{n} + \left( {\lambda 
\,\gamma - \omega \,\tilde {\gamma }} \right)\,B_{n} - \left( {\lambda 
\,\tilde {\gamma } + \omega \,\gamma } \right)\,\tilde {B}_{n} \ ,
\tag{5.41a}
\\
\ddot {\rho }_{n}^{\left( {2} \right)} = - \lambda \,\dot {\rho 
}_{n}^{\left( {2} \right)} - \omega \,\dot {\rho }_{n}^{\left( {1} \right)} 
+ \tilde {\gamma }\,A_{n} + \gamma \,\tilde {A}_{n} + \left( {\lambda 
\,\tilde {\gamma } + \omega \,\gamma } \right)\,B_{n} + \left( {\lambda 
\,\gamma - \omega \,\tilde {\gamma }} \right)\,\tilde {B}_{n} \  ,
\tag{5.41b}
\\
\ddot {\vec {r}}_{n}^{\left( {1} \right)} = - \lambda \,\dot {\vec 
{r}}_{n}^{\left( {1} \right)} + \omega \,\dot {\vec {r}}_{n}^{\left( {2} 
\right)} + \gamma \,\vec {C}_{n} - \tilde {\gamma }\,\vec {\tilde {C}}_{n} + 
\left( {\lambda \,\gamma - \omega \,\tilde {\gamma }} \right)\,\vec {D}_{n} 
- \left( {\lambda \,\tilde {\gamma } + \omega \,\gamma } \right)\,\vec 
{\tilde {D}}_{n} \  ,\tag{5.41c}
\\
\ddot {\vec {r}}_{n}^{\left( {2} \right)} = - \lambda \,\dot {\vec 
{r}}_{n}^{\left( {2} \right)} - \omega \,\dot {\vec {r}}_{n}^{\left( {1} 
\right)} + \tilde {\gamma }\,\vec {C}_{n} + \gamma \,\vec {\tilde {C}}_{n} + 
\left( {\lambda \,\tilde {\gamma } + \omega \,\gamma } \right)\,\vec {D}_{n} 
+ \left( {\lambda \,\gamma - \omega \,\tilde {\gamma }} \right)\,\vec 
{\tilde {D}}_{n} \  ,\tag{5.41d}
\\
A_{n} \equiv A_{n}^{\left( {1} \right)} - A_{n}^{\left( {2} \right)} 
,\,\,\,\tilde {A}_{n} \equiv \tilde {A}_{n}^{\left( {1} \right)} + \tilde 
{A}_{n}^{\left( {2} \right)} ;\,\,\,B_{n} \equiv B_{n}^{\left( {1} \right)} 
- B_{n}^{\left( {2} \right)} ,\,\,\,\tilde {B}_{n} \equiv \tilde 
{B}_{n}^{\left( {1} \right)} + \tilde {B}_{n}^{\left( {2} \right)}  ,
\tag{5.41e}
\\
A_{n}^{\left( {j} \right)} \equiv \dot {\rho }_{n}^{\left( {j} \right)} 
\,\left( {\rho _{n + 1}^{\left( {j} \right)} - 2\,\rho _{n}^{\left( {j} 
\right)} + \rho _{n - 1}^{\left( {j} \right)} } \right) + \dot {\vec 
{r}}_{n}^{\left( {j} \right)} \cdot \left( {\vec {r}_{n + 1}^{\left( {j} 
\right)} - 2\,\vec {r}_{n}^{\left( {j} \right)} + \vec {r}_{n - 1}^{\left( 
{j} \right)} } \right),\,\,\,j = 1,\,2 \ ,\tag{5.41f}
\\
\tilde {A}_{n}^{\left( {j} \right)} \equiv \dot {\rho }_{n}^{\left( {j + 1} 
\right)} \,\left( {\rho _{n + 1}^{\left( {j} \right)} - 2\,\rho _{n}^{\left( 
{j} \right)} + \rho _{n - 1}^{\left( {j} \right)} } \right) + \dot {\vec 
{r}}_{n}^{\left( {j + 1} \right)} \cdot \left( {\vec {r}_{n + 1}^{\left( {j} 
\right)} - 2\,\vec {r}_{n}^{\left( {j} \right)} + \vec {r}_{n - 1}^{\left( 
{j} \right)} } \right)\ ,
\\ \qquad{}
j = 1,\,2,\,\,{\rm mod}\left( {2} \right) \  ,
\tag{5.41g}
\\
B_{n}^{\left( {j} \right)} \equiv \rho _{n}^{\left( {j} \right)} \,\left( 
{\rho _{n + 1}^{\left( {j} \right)} - 2\,\rho _{n}^{\left( {j} \right)} + 
\rho _{n - 1}^{\left( {j} \right)} } \right) + \vec {r}_{n}^{\left( {j} 
\right)} \cdot \left( {\vec {r}_{n + 1}^{\left( {j} \right)} - 2\,\vec 
{r}_{n}^{\left( {j} \right)} + \vec {r}_{n - 1}^{\left( {j} \right)} } 
\right),\,\,\,\,j = 1,\,2 \  ,\tag{5.41h}
\\
\tilde {B}_{n}^{\left( {j} \right)} \equiv \rho _{n}^{\left( {j + 1} 
\right)} \,\left( {\rho ^{\left( {j} \right)}_{n + 1} - 2\,\rho _{n}^{\left( 
{j} \right)} + \rho _{n - 1}^{\left( {j} \right)} } \right) + \vec 
{r}_{n}^{\left( {j + 1} \right)} \cdot \left( {\vec {r}_{n + 1}^{\left( {j} 
\right)} - 2\,\vec {r}_{n}^{\left( {j} \right)} + \vec {r}_{n - 1}^{\left( 
{j} \right)} } \right)\ ,
\\ \qquad{}
j = 1,\,2,\,\,{\rm mod}\left( {2} \right) 
\  ,\tag{5.41i}
\\
\vec {C}_{n} \equiv \vec {C}_{n}^{\left( {1} \right)} - \vec {C}_{n}^{\left( 
{2} \right)} ,\,\,\,\vec {\tilde {C}}_{n} \equiv \vec {\tilde 
{C}}_{n}^{\left( {1} \right)} + \vec {\tilde {C}}_{n}^{\left( {2} \right)} 
;\,\,\,\vec {D}_{n} \equiv \vec {D}_{n}^{\left( {1} \right)} - \vec 
{D}_{n}^{\left( {2} \right)} ,\,\,\,\vec {\tilde {D}}_{n} \equiv \vec 
{\tilde {D}}_{n}^{\left( {1} \right)} + \vec {\tilde {D}}_{n}^{\left( {2} 
\right)} \  ,\tag{5.41j}
\\
\vec {C}_{n}^{\left( {j} \right)} \equiv \dot {\vec {r}}_{n}^{\left( {j} 
\right)} \,\left( {\rho _{n + 1}^{\left( {j} \right)} - 2\,\rho _{n}^{\left( 
{j} \right)} + \rho _{n - 1}^{\left( {j} \right)} } \right) + \dot {\rho 
}_{n}^{\left( {j} \right)} \left( {\vec {r}_{n + 1}^{\left( {j} \right)} - 
2\,\vec {r}_{n}^{\left( {j} \right)} + \vec {r}_{n - 1}^{\left( {j} \right)} 
} \right)
\\ \qquad{}
 + i\,\dot {\vec {r}}_{n}^{\left( {j} \right)} \wedge \left( {\vec {r}_{n + 
1}^{\left( {j} \right)} - \vec {r}_{n - 1}^{\left( {j} \right)} } 
\right)\ ,\,\,\,\,j = 1,\,2 \  ,\tag{5.41k}
\\
\vec {\tilde {C}}_{n}^{\left( {j} \right)} \equiv \dot {\vec 
{r}}_{n}^{\left( {j + 1} \right)} \,\left( {\rho _{n + 1}^{\left( {j} 
\right)} - 2\,\rho _{n}^{\left( {j} \right)} + \rho _{n - 1}^{\left( {j} 
\right)} } \right) + \dot {\rho }_{n}^{\left( {j + 1} \right)} \,\left( 
{\vec {r}_{n + 1}^{\left( {j} \right)} - 2\,\vec {r}_{n}^{\left( {j} 
\right)} + \vec {r}_{n - 1}^{\left( {j} \right)} } \right)
 + i\,\dot {\vec {r}}_{n}^{\left( {j + 1} \right)}
\\ \qquad{}
\wedge \left( {\vec 
{r}_{n + 1}^{\left( {j} \right)} - \vec {r}_{n - 1}^{\left( {j} \right)} } 
\right) \  ,
\,\,\,\,j = 1,\,2,\,\,{\rm mod} \left( {2} \right)\ ,\tag{5.41l}
\\
\vec {D}_{n}^{\left( {j} \right)} \equiv \vec {r}_{n}^{\left( {j} \right)} 
\,\left( {\rho _{n + 1}^{\left( {j} \right)} - 2\,\rho _{n}^{\left( {j} 
\right)} + \rho _{n - 1}^{\left( {j} \right)} } \right) + \rho _{n}^{\left( 
{j} \right)} \,\left( {\vec {r}_{n + 1}^{\left( {j} \right)} - 2\,\vec 
{r}_{n}^{\left( {j} \right)} + \vec {r}_{n - 1}^{\left( {j} \right)} } 
\right)
\\ \qquad{}
+ i\,\vec {r}_{n}^{\left( {j} \right)} \wedge \left( {\vec {r}_{n + 
1}^{\left( {j} \right)} - \vec {r}_{n - 1}^{\left( {j} \right)} } 
\right)\ ,\,\,\,\,j = 1,\,2 \  ,\tag{5.41m}
\\
\vec {\tilde {D}}_{n}^{\left( {j} \right)} \equiv \vec {r}_{n}^{\left( {j + 
1} \right)} \,\left( {\rho _{n + 1}^{\left( {j} \right)} - 2\,\rho 
_{n}^{\left( {j} \right)} + \rho _{n - 1}^{\left( {j} \right)} } \right) + 
\rho _{n}^{\left( {j + 1} \right)} \,\left( {\vec {r}_{n + 1}^{\left( {j} 
\right)} - 2\,\vec {r}_{n}^{\left( {j} \right)} + \vec {r}_{n - 1}^{\left( 
{j} \right)} } \right)
\\ \qquad{}
+ i\,\vec {r}_{n}^{\left( {j + 1} \right)} \wedge \left( {\vec {r}_{n + 
1}^{\left( {j} \right)} - \vec {r}_{n - 1}^{\left( {j} \right)} } \right)\ ,
\,\,\,\,j = 1,\,2,\,\,{\rm mod}\left( {2} \right)\ .
\tag{5.41n}
\end{gather*}
Note that, as implied by the derivation of these equations of motion, we can 
assert that \textit{all }their solutions are \textit{completely periodic}, 
with period $T = 2\pi /\omega $ , if $\lambda = 0,\,\,\omega \ne 0$.

Next, we report the scalar/vector\textit{ linearizable many-body problem 
}characterized by the equations of motion
\begin{gather*}
\ddot {\rho }_{n} = \left[ {\dot {\rho }_{n} ^{2}\,\rho _{n} + 2\,\dot {\rho 
}_{n} \,\left( {\dot 
{\mathord{\buildrel{\lower3pt\hbox{$\scriptscriptstyle\rightharpoonup$}}\over 
{r}} }_{n} \cdot \vec {r}_{n} } \right) - \rho _{n} \,\left( {\dot {\vec 
{r}}_{n} \cdot \dot {\vec {r}}_{n} } \right)} \right]/\left( {\rho _{n} ^{2} 
+ r_{n} ^{2}} \right)
\\ \qquad{}
 + \sum\limits_{m} \{ a_{nm} \,\left[ \dot{\rho }_{m} 
\,\rho_{m} \,\rho_{n} + \dot{\rho }_{m} \,\left( 
\vec{r} _{m} \cdot \vec{r}_{n}  \right) - \rho_{m} \,\left( \dot{\vec 
{r}}_{m} \cdot \vec{r}_{n}  \right)\right.
\\ \qquad{}
+ \rho_{n} \, ( \dot{\vec 
{r}}_{m} \cdot \vec{r}_{m} ) ] / ( \rho_{m}^{2} + 
r_{m}^{2}) \} \  ,\tag{5.42a}
\\
\ddot {\vec {r}}_{n} = \left\{ {} \right.\dot {\vec {r}}_{n} \,\left[ {\dot 
{\rho }_{n} \,\rho _{n} + \left( {\dot {\vec {r}}_{n} \cdot \vec {r}_{n} } 
\right)} \right] - \vec {r}_{n} \,\left[ {\dot {\rho }_{n} ^{2} + \left( 
{\dot {\vec {r}}_{n} \cdot \dot {\vec {r}}_{n} } \right)} \right]
\\ \qquad{}
+ \dot 
{\vec {r}}_{n} \,\left[ {\dot {\rho }_{n} \,\rho _{n} + \left( {\dot {\vec 
{r}}_{n} \cdot \vec {r}_{n} } \right)} \right]\left. {} \right\}/\left( 
{\rho _{n} ^{2} + r_{n} ^{2}} \right)
\\ \qquad{}
+ \sum\limits_{m} {\left[ {} 
\right.a_{nm} \,\left\{ {} \right.\dot {\vec {r}}_{m} \,\left[ {\rho _{m} 
\,\rho _{n} + \left( {\vec {r}_{m} \cdot \vec {r}_{n} } \right)}
\right]}
\\ \qquad{}
 - \vec {r}_{m} \,\left[ {\dot {\rho }_{m} \,\rho _{n} + \left( {\dot {\vec 
{r}}_{m} \cdot \vec {r}_{n} } \right)} \right] + \vec {r}_{n} \,\left[ {\dot 
{\rho }_{m} \,\rho _{m} + \left( {\dot {\vec {r}}_{m} \cdot \vec {r}_{m} } 
\right)} \right] 
\\ \qquad{}
+ \left[ {\dot {\rho }_{m} \,\vec {r}_{m} \wedge \vec 
{r}_{n} - \rho _{m} \,\dot {\vec {r}}_{m} \wedge \vec {r}_{n} + \rho _{n} 
\,\dot {\vec {r}}_{m} \wedge \vec {r}_{m} } \right]\left. {} \right\}/\left( 
{\rho _{m} ^{2} + r_{m} ^{2}} \right)\left. {} \right]\ ,\tag{5.42b}
\end{gather*}
which are obtained by applying the parameterization (4.1) to the 
\textit{linearizable }matrix ODE (3.47).


Next, we report the \textit{linearizable scalar-vector many-body problem 
}that is obtained by applying the parameterization (4.1) to the 
\textit{linearizable }matrix ODE (3.43d). Its equations of motion read
\begin{gather*}
\ddot {\rho }_{n} = \,a\dot {\rho }_{n} + \tilde {\rho }_{n} \dot {\rho 
}_{n} - \vec {\tilde {r}}_{n} \cdot \dot {\vec {r}}_{n} + c\,\left\{ {\tilde 
{\rho }_{n - 1} \dot {\rho }_{n} - \vec {\tilde {r}}_{n - 1} \cdot \dot 
{\vec {r}}_{n} - \rho _{n} \tilde {\tilde {\rho }}_{n + 1} + \vec {r}_{n} 
\cdot \vec {\tilde {\tilde {r}}}_{n + 1} } \right\}
\ ,\tag{5.43a}
\\
\ddot {\vec {r}}_{n} = a\,\dot {\vec {r}}_{n} + \tilde {\rho }_{n} \dot 
{\vec {r}}_{n} + \dot {\rho }_{n} \vec {\tilde {r}}_{n} - \vec {\tilde 
{r}}_{n} \wedge \dot {\vec {r}}_{n}
\\ \qquad{}
+ c \left\{ {\tilde {\rho }_{n - 1} 
\dot {\vec {r}}_{n} + \dot {\rho }_{n} \vec {\tilde {r}}_{n - 1} - \vec 
{\tilde {r}}_{n - 1} \wedge \dot {\vec {r}}_{n} - \rho _{n} \vec {\tilde 
{\tilde {r}}}_{n + 1} - \tilde {\tilde {\rho }}_{n + 1} \vec {r}_{n} - \vec 
{r}_{n} \wedge \vec {\tilde {\tilde {r}}}_{n + 1} } 
\right\}\ ,\tag{5.43b}
\end{gather*}
with
\begin{gather*}
\tilde {\rho }_{n} \, \equiv \,\,\left( {\rho _{n} \dot {\rho }_{n} + \vec 
{r}_{n} \cdot \dot {\vec {r}}_{n} } \right)/\left( {\rho _{n} ^{2} + r_{n} 
^{2}} \right)\ ,\tag{5.43c}
\\
\vec {\tilde {r}}_{n} \equiv \left( {\rho _{n} \dot {\vec {r}}_{n} - \dot 
{\rho }_{n} \vec {r}_{n} - \vec {r}_{n} \wedge \dot {\vec {r}}_{n} } 
\right)/\left( {\rho _{n} ^{2} + r_{n} ^{2}} \right) \  
,\tag{5.43d}
\end{gather*}
and
\begin{gather*}
\tilde {\tilde {\rho }}_{n + 1} \equiv \tilde {\rho }_{n} \tilde {\rho }_{n 
+ 1} - \vec {\tilde {r}}_{n} \cdot \vec {\tilde {r}}_{n + 
1}~~,\tag{5.43e}
\\
\vec {\tilde {\tilde {r}}}_{n + 1} \equiv \tilde {\rho }_{n} \vec {\tilde 
{r}}_{n + 1} + \tilde {\rho }_{n + 1} \vec {\tilde {r}}_{n} - \vec {\tilde 
{r}}_{n} \wedge \vec {\tilde {r}}_{n + 1} \  .\tag{5.43f}
\end{gather*}
We do not write in explicit form the equations obtainable from this one by 
complexification; their interest is related to the fact that, if in (5.17a) 
$a = \pm i\,\omega ,\,\omega > 0$, its generic solution is presumably 
completely periodic, with period $T = 2\pi /\omega $.


The diligent reader may try and derive the more general models that are 
obtained by employing the parameterization (4.6), which is applicable to 
(3.43d) (see (4.6m)) and yields equations of motion only involving 
3-vectors, but in a more complicated manner.


Next, we report the equations of motion of the \textit{linearizable 
scalar-vector many-body problem }that is obtained by applying the 
parameterization (4.1) to the \textit{linearizable }matrix ODE (3.66):
\begin{gather*}
\ddot {\rho }_{n} = \,\left( {a_{n + 1} - a_{n} - b_{n} } \right)\dot {\rho 
}_{n} + \tilde {\rho }_{n} \dot {\rho }_{n} - \vec {\tilde {r}}_{n} \cdot 
\dot {\vec {r}}_{n} + b_{n + 1} \left( {\tilde {\rho }_{n + 1} \dot {\rho 
}_{n} - \vec {\tilde {r}}_{n + 1} \cdot \dot {\vec {r}}_{n} } 
\right)
\\ \qquad{}
 + c_{n + 1} \rho _{n} - c_{n} \left( {\rho _{n} \tilde {\tilde {\rho }}_{n 
- 1} - \vec {r}_{n} \cdot \vec {\tilde {\tilde {r}}}_{n - 1} } 
\right)\ ,\tag{5.44a}
\\
\ddot {\vec {r}}_{n} = \left( {a_{n + 1} - a_{n} - b_{n} } \right)\,\dot 
{\vec {r}}_{n} + \tilde {\rho }_{n} \dot {\vec {r}}_{n} + \dot {\rho }_{n} 
\vec {\tilde {r}}_{n} - \vec {\tilde {r}}_{n} \wedge \dot {\vec 
{r}}_{n}
\\ \qquad{}
+ b_{n + 1} \left( {\tilde {\rho }_{n + 1} \dot {\vec {r}}_{n} + \dot {\rho 
}_{n} \vec {\tilde {r}}_{n + 1} - \vec {\tilde {r}}_{n + 1} \wedge \dot 
{\vec {r}}_{n} } \right)
\\ \qquad{}
 + c_{n + 1} \vec {r}_{n} - c_{n} \left( {\rho _{n} \vec {\tilde {\tilde 
{r}}}_{n - 1} + \tilde {\tilde {\rho }}_{n - 1} \vec {r}_{n} + \vec {r}_{n} 
\wedge \vec {\tilde {\tilde {r}}}_{n - 1} } \right)\ ,\tag{5.44b}
\end{gather*}
where $\tilde {\rho }_{n} ,\vec {\tilde {r}}_{n} $ are given again by 
(5.43c), (5.43d) and
\begin{gather*}
\tilde {\tilde {\rho }}_{n - 1} \equiv \tilde {\rho }_{n} 
\mathord{\buildrel{\lower3pt\hbox{$\scriptscriptstyle\frown$}}\over {\rho }} 
_{n - 1} - \vec {\tilde {r}}_{n} \cdot \vec 
{\mathord{\buildrel{\lower3pt\hbox{$\scriptscriptstyle\frown$}}\over {r}} 
}_{n - 1} \  ,\tag{5.44c}
\\
\vec {\tilde {\tilde {r}}}_{n - 1} \equiv \tilde {\rho }_{n} \vec 
{\mathord{\buildrel{\lower3pt\hbox{$\scriptscriptstyle\frown$}}\over {r}} 
}_{n - 1} + 
\mathord{\buildrel{\lower3pt\hbox{$\scriptscriptstyle\frown$}}\over {\rho }} 
_{n - 1} \vec {\tilde {r}}_{n} - \vec {\tilde {r}}_{n} \wedge \vec 
{\mathord{\buildrel{\lower3pt\hbox{$\scriptscriptstyle\frown$}}\over {r}} 
}_{n - 1} \  ,\tag{5.44d}
\\
\mathord{\buildrel{\lower3pt\hbox{$\scriptscriptstyle\frown$}}\over {\rho }} 
_{n} \, \equiv \,\,\left( {\rho _{n} \dot {\rho }_{n} + \vec {r}_{n} \cdot 
\dot {\vec {r}}_{n} } \right)\,/\,\left( {\dot {\rho }_{n} ^{2} + \,\,\dot 
{\vec {r}}_{n} \cdot \dot {\vec {r}}_{n} } \right) \  
,\tag{5.44e}
\\
\vec {\mathord{\buildrel{\lower3pt\hbox{$\scriptscriptstyle\frown$}}\over 
{r}} }_{n} \equiv \left( {\dot {\rho }_{n} \vec {r}_{n} \, - \rho _{n} \dot 
{\vec {r}}_{n} + \vec {r}_{n} \wedge \dot {\vec {r}}_{n} } 
\right)\,/\,\left( {\dot {\rho }_{n} ^{2} + \,\,\dot {\vec {r}}_{n} \cdot 
\dot {\vec {r}}_{n} } \right) \  .\tag{5.44f}
\end{gather*}

  The highly nonlinear character of these equations of motion should be noted.We leave as exercise for the diligent reader to write the (only 3-vector but 
more complicated) equations of motion that obtain by using the 
parameterization (4.6), instead of (4.1).

Next, we display the scalar/vector \textit{linearizable many-body problem 
}that is obtained by applying the parameterization (4.3) and (4.1) to the 
\textit{linearizable }matrix ODEs (3.23), via the positions $\underline {U} 
_{n} = i\,\vec {r}_{n} \cdot \underline {\vec {\sigma }} $ , $\underline {F} 
= i\,\vec {f} \cdot \vec {\sigma }$, $\underline {G} = i\,\vec {g} \cdot 
\vec {\sigma }$, $\underline {H} = i\,\vec {h} \cdot \vec {\sigma }$, 
$\underline {V} = \eta + i\,\vec {v} \cdot \underline {\vec {\sigma }} $ , 
$\underline {Y} = \theta + i\,\vec {y} \cdot \underline {\vec {\sigma}}$:
\begin{gather*}
\mu^{(u)}_{n} \ddot{\vec {r}}_{n} = 4 \vec{k} \wedge \dot {\vec 
{r}}_{n} + 2 \dot {\vec {k}} \wedge \vec{r}_{n} - 4 \vec{k} \wedge 
(\vec{k} \wedge \vec{r}_{n})
\\ \quad{}
+ \sum\limits_{m = 1}^{N} \left[ a_{m}^{\left({uu} \right)} \left\{ {\dot {\vec {r}}_{m} - 2\,\vec {k} 
\wedge \vec {r}_{m} } \right\} + b_{m}^{\left({uu} \right)} \,\vec{r}_{m} \right]
\\ \quad{}
 + a_{n}^{\left( {uf} \right)} \,\left\{ {\dot {\vec {f}}} - 2\,\vec {k} 
\wedge \vec {f} \right\} + b_{n}^{\left( {uf} \right)} \,\vec {f} + 
a_{n}^{\left( {ug} \right)} \,\left\{ {\dot {\vec {g}}} - 2\,\vec {k} \wedge 
\vec {g} \right\}
\\ \quad{}
+ b_{n}^{\left(ug \right)} \,\vec {g} + a_{n}^{\left( 
{uh} \right)} \,\left\{ {\dot {\vec {h}}} - 2\,\vec {k} \wedge \vec {h} 
\right\} + b_{n}^{\left( {uh} \right)} \,\vec {h} \  
,\tag{5.45a}
\\
\mu ^{\left( {f} \right)}\,\ddot {\vec {f}} = 4\,\vec {k} \wedge \dot {\vec 
{f}} + 2\,\dot {\vec {k}} \wedge \vec {f} - 4\,\vec {k} \wedge \left( {\vec 
{k} \wedge \vec {f}} \right) + \sum\limits_{m = 1}^{N} {\left[ 
{a_{m}^{\left( {fu} \right)} \,\left\{ {\dot {\vec {r}}_{m} - 2\,\vec {k} 
\wedge \vec {r}_{m} } \right\} + b_{m}^{\left( {fu} \right)} \,\vec {r}_{m} 
} \right]}
\\ \quad{}
 + a_{}^{\left( {ff} \right)} \,\left\{ {\dot {\vec {f}} - 2\,\vec {k} 
\wedge \vec {f}} \right\} + b_{}^{\left( {ff} \right)} \,\vec {f} + 
a_{}^{\left( {fg} \right)} \,\left\{ {\dot {\vec {g}} - 2\,\vec {k} \wedge 
\vec {g}} \right\}
\\ \quad{}
+ b_{}^{\left( {fg} \right)} \,\vec {g} + a_{}^{\left( 
{fh} \right)} \,\left\{ {\dot {\vec {h}} - 2\,\vec {k} \wedge \vec {h}} 
\right\} + b_{}^{\left( {fh} \right)} \,\vec {h}\ ,\tag{5.45b}
\\
\mu ^{\left( {g} \right)}\,\ddot {\vec {g}} = 4\,\vec {k} \wedge \dot {\vec 
{g}} + 2\,\dot {\vec {k}} \wedge \vec {g} - 4\vec {k} \wedge \left( {\vec 
{k} \wedge \vec {g}} \right) + \sum\limits_{m = 1}^{N} {\left[ 
{a_{m}^{\left( {gu} \right)} \,\left\{ {\dot {\vec {r}}_{m} - 2\,\vec {k} 
\wedge \vec {r}_{m} } \right\} + b_{m}^{\left( {gu} \right)} \vec {r}_{m} } 
\right]}
\\ \quad{}
 + a_{}^{\left( {gf} \right)} \,\left\{ {\dot {\vec {f}} - 2\,\vec {k} 
\wedge \vec {f}} \right\} + b_{}^{\left( {gf} \right)} \,\vec {f} + 
a_{}^{\left( {gg} \right)} \,\left\{ {\dot {\vec {g}} - 2\,\vec {k} \wedge 
\vec {g}} \right\}
\\ \quad{}
+ b_{}^{\left( {gg} \right)} \,\vec {g} + a_{}^{\left( 
{gh} \right)} \,\left\{ {\dot {\vec {h}} - 2\,\vec {k} \wedge \vec {h}} 
\right\} + b_{}^{\left( {gh} \right)} \,\vec {h}\ ,\tag{5.45c}
\\
\mu ^{\left( {h} \right)}\,\ddot {\vec {h}} = 4\,\vec {k} \wedge \dot {\vec 
{h}} + 2\,\dot {\vec {k}} \wedge \vec {h} - 4\vec {k} \wedge \left( {\vec 
{k} \wedge \vec {h}} \right) + \sum\limits_{m = 1}^{N} {\left[ 
{a_{m}^{\left( {hu} \right)} \,\left\{ {\dot {\vec {r}}_{m} - 2\,\vec {k} 
\wedge \vec {r}_{m} } \right\} + b_{m}^{\left( {hu} \right)} \vec {r}_{m} } 
\right]}
\\ \quad{}
 + a_{}^{\left( {hf} \right)} \,\left\{ {\dot {\vec {f}} - 2\,\vec {k} 
\wedge \vec {f}} \right\} + b_{}^{\left( {hf} \right)} \,\vec {f} + 
a_{}^{\left( {hg} \right)} \,\left\{ {\dot {\vec {g}} - 2\,\vec {k} \wedge 
\vec {g}} \right\}
\\ \quad{}
+ b_{}^{\left( {hg} \right)} \,\vec {g} + a_{}^{\left( 
{hh} \right)} \,\left\{ {\dot {\vec {h}} - 2\,\vec {k} \wedge \vec {h}} 
\right\} + b_{}^{\left( {hh} \right)} \,\vec {h}\ ,\tag{5.45d}
\\
\mu ^{\left( {v} \right)}\,\ddot {\vec {v}} = 2\,\vec {k} \wedge \dot {\vec 
{v}} + \dot {\vec {k}} \wedge \vec {v} + k^{2}\,\vec {v} 
\\ \quad{}
- \left\{ {\dot 
{\vec {v}} - \vec {k} \wedge \vec {v}} \right\}\,\left\{ {} \right.\,\sum\limits_{m = 1}^{N} {\left[ {\tilde {a}_{m}^{\left( {vu} \right)} 
\,\left\{ {\dot {\vec {r}}_{m} - 2\,\vec {k} \wedge \vec {r}_{m} } 
\right\}^{2} + \tilde {b}_{m}^{\left( {vu} \right)} \,r_{m} ^{2}} \right]} 
\\ \quad{}
 + \tilde {a}^{\left( {vf} \right)}\,\left\{ {\dot {\vec {f}} - 2\,\vec {k} 
\wedge \vec {f}} \right\}^{2} + \tilde {b}^{\left( {vf} \right)}\,f^{2} + 
\tilde {a}^{\left( {vg} \right)}\,\left\{ {\dot {\vec {g}} - 2\,\vec {k} 
\wedge \vec {g}} \right\}^{2} + \tilde {b}^{\left( {vg} 
\right)}\,g^{2}
\\ \quad{}
 + \tilde {a}^{\left( {vh} \right)}\,\left\{ {\dot {\vec {h}} - 2\,\vec {k} 
\wedge \vec {h}} \right\}^{2} + \tilde {b}^{\left( {vh} 
\right)}\,h^{2}\left. {} \right\} 
- \vec {v}\,\left\{ {} 
\right.\sum\limits_{m = 1}^{N} {\left[ {a_{m}^{\left( {vu} \right)} 
\,\left\{ {\dot {\vec {r}}_{m} - 2\,\vec {k} \wedge \vec {r}_{m} } 
\right\}^{2} + b_{m}^{\left( {vu} \right)} \,r_{m} ^{2}} \right]}
\\ \quad{}
 + a^{\left( {vf} \right)}\,\left\{ {\dot {\vec {f}} - 2\,\vec {k} \wedge 
\vec {f}} \right\}^{2} + b^{\left( {vf} \right)}\,f^{2} + a^{\left( {vg} 
\right)}\,\left\{ {\dot {\vec {g}} - 2\,\vec {k} \wedge \vec {g}} 
\right\}^{2} + b^{\left( {vg} \right)}\,g^{2}
\\ \quad{}
 + a^{\left( {vh} \right)}\,\left\{ {\dot {\vec {h}} - 2\,\vec {k} \wedge 
\vec {h}} \right\}^{2} + b^{\left( {vh} \right)}\,h^{2}\left. {} \right\}\ ,
\tag{5.45e}
\\
\mu ^{\left( {v} \right)}\,\ddot {\vec {y}} = - 2\,\vec {k} \wedge \dot 
{\vec {y}} - \dot {\vec {k}} \wedge \vec {y} + k^{2}\,\vec {y}
\\ \quad{}
- \left\{ 
{\dot {\vec {y}} + \vec {k} \wedge \vec {y}} \right\}\,\left\{ {} 
\right.\sum\limits_{m = 1}^{N} {\left[ {\tilde {a}_{m}^{\left( {vy} \right)} 
\,\left\{ {\dot {\vec {r}}_{m} - 2\,\vec {k} \wedge \vec {r}_{m} } 
\right\}^{2} + \tilde {b}_{m}^{\left( {yu} \right)} \,r_{m} ^{2}} \right]} 
\\ \quad{}
 + \tilde {a}^{\left( {yf} \right)}\,\left\{ {\dot {\vec {f}} - 2\,\vec {k} 
\wedge \vec {f}} \right\}^{2} + \tilde {b}^{\left( {yf} \right)}\,f^{2} + 
\tilde {a}^{\left( {yg} \right)}\,\left\{ {\dot {\vec {g}} - 2\,\vec {k} 
\wedge \vec {g}} \right\}^{2} + \tilde {b}^{\left( {yg} \right)}\,g^{2}
\\ \quad{}
 + \tilde {a}^{\left( {yh} \right)}\,\left\{ {\dot {\vec {h}} - 2\,\vec {k} 
\wedge \vec {h}} \right\}^{2} + \tilde {b}^{\left( {yh} 
\right)}\,h^{2}\left. {} \right\} - \vec {y}\,\left\{ {} 
\right.\sum\limits_{m = 1}^{N} {\left[ {a_{m}^{\left( {yu} \right)} 
\,\left\{ {\dot {\vec {r}}_{m} - 2\,\vec {k} \wedge \vec {r}_{m} } 
\right\}^{2} + b_{m}^{\left( {yu} \right)} \,r_{m} ^{2}} \right]} 
\\ \quad{}
 + a^{\left( {yf} \right)}\,\left\{ {\dot {\vec {f}} - 2\,\vec {k} \wedge 
\vec {f}} \right\}^{2} + b^{\left( {yf} \right)}\,f^{2} + a^{\left( {yg} 
\right)}\,\left\{ {\dot {\vec {g}} - 2\,\vec {k} \wedge \vec {g}} 
\right\}^{2} 
\\ \quad{}
+ b^{\left( {yg} \right)}\,g^{2} + a^{\left( {yh} 
\right)}\,\left\{ {\dot {\vec {h}} - 2\,\vec {k} \wedge \vec {h}} 
\right\}^{2} + b^{\left( {yh} \right)}\,h^{2}\left. {} 
\right\}\ ,\tag{5.45f}
\\
\mu ^{\left( {v} \right)}\,\ddot {\eta } = - 2\,\left[ {\gamma \,\dot {\eta 
} - \left( {\vec {k} \cdot \dot {\vec {v}}} \right)} \right] - \left[ {\dot 
{\gamma }\,\eta - \left( {\dot {\vec {k}} \cdot \vec {v}} \right)} \right] + 
\left( {k^{2} - \gamma ^{2}} \right)\,\eta
\\ \quad{}
 - \dot {\eta } \left\{ {} 
\right.\sum\limits_{m = 1}^{N} {\left[ {\tilde {a}_{m}^{\left( {vu} \right)} 
\,\left\{ {\dot {\vec {r}}_{m} - 2\,\vec {k} \wedge \vec {r}_{m} } 
\right\}^{2} + \tilde {b}_{m}^{\left( {vu} \right)} \,r_{m} ^{2}} \right]} 
\\ \quad{}
 + \tilde {a}^{\left( {vf} \right)}\,\left\{ {\dot {\vec {f}} - 2\,\vec {k} 
\wedge \vec {f}} \right\}^{2} + \tilde {b}^{\left( {vf} \right)}\,f^{2} + 
\tilde {a}^{\left( {vg} \right)}\,\left\{ {\dot {\vec {g}} - 2\,\vec {k} 
\wedge \vec {g}} \right\}^{2} + \tilde {b}^{\left( {vg} 
\right)}\,g^{2}
\\ \quad{}
 + \tilde {a}^{\left( {vh} \right)}\,\left\{ {\dot {\vec {h}} - 2\,\vec {k} 
\wedge \vec {h}} \right\}^{2} + \tilde {b}^{\left( {vh} 
\right)}\,h^{2}\left. {} \right\} - \eta \,\left\{ {} \right.\sum\limits_{m 
= 1}^{N} {\left[ {a_{m}^{\left( {vu} \right)} \,\left\{ {\dot {\vec {r}}_{m} 
- 2\,\vec {k} \wedge \vec {r}_{m} } \right\}^{2} + b_{m}^{\left( {vu} 
\right)} \,r_{m} ^{2}} \right]}
\\ \quad{}
 + a^{\left( {vf} \right)}\,\left\{ {\dot {\vec {f}} - 2\,\vec {k} \wedge 
\vec {f}} \right\}^{2} + b^{\left( {vf} \right)}\,f^{2} + a^{\left( {vg} 
\right)}\,\left\{ {\dot {\vec {g}} - 2\,\vec {k} \wedge \vec {g}} 
\right\}^{2} + b^{\left( {vg} \right)}\,g^{2},
\\ \quad{}
 + a^{\left( {vh} \right)}\,\left\{ {\dot {\vec {h}} - 2\,\vec {k} \wedge 
\vec {h}} \right\}^{2} + b^{\left( {vh} \right)}\,h^{2}\left. {} \right\}\ ,
\tag{5.45g}
\\
\mu ^{\left( {v} \right)}\,\ddot {\theta } = 2\,\left[ {\gamma \,\dot 
{\theta } - \left( {\vec {k} \cdot \dot {\vec {y}}} \right)} \right] + 
\left[ {\dot {\gamma }\,\theta - \left( {\dot {\vec {k}} \cdot \vec {y}} 
\right)} \right] + \left( {k^{2} - \gamma ^{2}} \right)\,\theta 
\\ \quad{}
- \dot 
{\theta }\,\left\{ {} \right.\sum\limits_{m = 1}^{N} {\left[ {\tilde 
{a}_{m}^{\left( {vy} \right)} \,\left\{ {\dot {\vec {r}}_{m} - 2\,\vec {k} 
\wedge \vec {r}_{m} } \right\}^{2} + \tilde {b}_{m}^{\left( {yu} \right)} 
\,r_{m} ^{2}} \right]} 
\\ \quad{}
 + \tilde {a}^{\left( {yf} \right)}\,\left\{ {\dot {\vec {f}} - 2\,\vec {k} 
\wedge \vec {f}} \right\}^{2} + \tilde {b}^{\left( {yf} \right)}\,f^{2} + 
\tilde {a}^{\left( {yg} \right)}\,\left\{ {\dot {\vec {g}} - 2\,\vec {k} 
\wedge \vec {g}} \right\}^{2} + \tilde {b}^{\left( {yg} \right)}\,g^{2}
\\ \quad{}
 + \tilde {a}^{\left( {yh} \right)}\,\left\{ {\dot {\vec {h}} - 2\,\vec {k} 
\wedge \vec {h}} \right\}^{2} + \tilde {b}^{\left( {yh} 
\right)}\,h^{2}\left. {} \right\} - \theta \,\left\{ {} 
\right.\sum\limits_{m = 1}^{N} {\left[ {a_{m}^{\left( {yu} \right)} 
\,\left\{ {\dot {\vec {r}}_{m} - 2\,\vec {k} \wedge \vec {r}_{m} } 
\right\}^{2} + b_{m}^{\left( {yu} \right)} \,r_{m} ^{2}} \right]} 
\\ \quad{}
 + a^{\left( {yf} \right)}\,\left\{ {\dot {\vec {f}} - 2\,\vec {k} \wedge 
\vec {f}} \right\}^{2} + b^{\left( {yf} \right)}\,f^{2} + a^{\left( {yg} 
\right)}\,\left\{ {\dot {\vec {g}} - 2\,\vec {k} \wedge \vec {g}} 
\right\}^{2} + b^{\left( {yg} \right)}\,g^{2}
\\ \quad{}
 + a^{\left( {yh} \right)}\,\left\{ {\dot {\vec {h}} - 2\,\vec {k} \wedge 
\vec {h}} \right\}^{2} + b^{\left( {yh} \right)}\,h^{2}\left. {} \right\}\ ,
\tag{5.45h}
\end{gather*}
where
\begin{equation*}
\vec {k} \equiv \vec {f} + \eta \,\vec {g} + \theta \,\vec {h} - \vec {g} 
\wedge \vec {f} + \vec {h} \wedge \vec {y},\,\,\,\,\,\gamma \equiv \vec {g} 
\cdot \vec {v} + \vec {h} \cdot \vec {y}\ .\tag{5.45i}
\end{equation*}

These equations of motion determine the evolution of the $N + 5$ 
three-vectors $\vec {r}_{n} \left( {t} \right)$,\, $\vec {f}\left( {t} 
\right)$,\, $\vec {g}\left( {t} \right)$,\, $\vec {h}\left( {t} 
\right)$,\, $\vec {v}\left( {t} \right)$,\, $\vec {y}\left( {t} \right)$ 
and of 
the $2$ scalars $\eta \left( {t} \right)$,\, $\theta \left( {t} \right)$. 
They feature $(2\,N^{2} + 20\,N + 42)$ arbitrary ``coupling 
constants,'' hence they include many special cases, corresponding to 
appropriate choices of these coupling constants, many of which could be set 
to zero to obtain simpler systems; but the exploration of these cases is 
beyond the scope of this paper.


The diligent reader may write the more general equations of motion involving 
$N + 5$ scalar quantities, as well as $N + 5$ three-vectors, which obtain by 
using the parameterization (4.1) for all the $N + 5$ matrices that evolve 
according to (3.66), rather than for only those 2, $\underline {V} $ and 
$\underline {Y} $ , for which a parameterization (4.1), rather than (4.3), is 
mandated by the very structure of (3.23).


Finally we display the equations of motion of the scalar/vector 
\textit{solvable many-body problem }that is obtained by applying the 
parameterization (4.1) to the matrix ODEs (3.96):
\begin{gather*}
\ddot {\rho }_{nm} = a_{nm} + \sum\limits_{m_{1} = 1}^{N} {\left[ {b_{m_{1} 
m} \,\rho _{nm_{1} } + c_{m_{1} m} \,\dot {\rho }_{nm_{1} } } \right]} - 
\sum\limits_{m_{1} ,m_{2} = 1}^{N} {d_{m_{1} m_{2} } }\,
\\ \qquad{}
\left[ \dot {\rho }_{nm_{1} } \,\rho _{m_{2} m} - 
\left( {\dot {\vec {r}}}_{nm_{1} } \cdot \vec 
{r}_{m_{2} m}  \right) + 2\,\rho _{nm_{1} } \,\dot {\rho }_{m_{2} m} - 
2\,\left( {\vec {r}_{nm_{1} } }\cdot \dot {\vec {r}}_{m_{2} m}  \right)
\right]
\\ \qquad{}
 + \sum\limits_{m_{1} ,m_{2} ,m_{3} = 1}^{N} {d_{m_{1} m_{2} } \,c_{m_{3} m} 
\,\left[ {\rho _{nm_{1} } \,\rho _{m_{2} m_{3} } - \left( {\vec {r}_{nm_{1} 
} \cdot \vec {r}_{m_{2} m_{3} } } \right)} \right]}
\\ \qquad{}
- \sum\limits_{m_{1} 
,m_{2} ,m_{3} ,m_{4} = 1}^{N} {d_{m_{1} m_{2} } \,d_{m_{3} m_{4} } \,\left[ 
{} \right.} \rho _{nm_{1} } \,\rho _{m_{2} m_{3} } \,\rho _{m_{4} m}
 - \rho _{nm_{1} } \left( {\vec {r}_{m_{2} m_{3} } \cdot \vec {r}_{m_{4} m} 
} \right) 
\\ \qquad{}
- \rho _{m_{2} m_{3} } \left( {\vec {r}_{nm_{1} } \cdot \vec 
{r}_{m_{4} m} } \right)
- \rho _{m_{4} m} \,\left( {\vec {r}_{nm_{1} } \cdot 
\vec {r}_{m_{2} m_{3} } } \right) + \left( {\vec {r}_{nm_{1} } \wedge \vec 
{r}_{m_{2} m_{3} } } \right) \cdot \vec {r}_{m_{4} m} \left. {} \right]\ ,
\tag{5.46a}
\\
\ddot {\vec {r}}_{nm} = \sum\limits_{m_{1} = 1}^{N} {\left[ {b_{m_{1} m} 
\,\vec {r}_{nm_{1} } + c_{m_{1} m} \,\dot {\vec {r}}_{nm_{1} } } \right]} 
\\ \qquad{}
+\sum\limits_{m_{1} ,m_{2} = 1}^{N} {d_{m_{1} m_{2} } \,\left[ {\dot {\vec 
{r}}_{nm_{1} } \wedge \vec {r}_{m_{2} m} + 2\,\vec {r}_{nm_{1} } \wedge \dot 
{\vec {r}}_{m_{2} m} } \right]} 
\\ \qquad{}
 - \sum\limits_{m_{1} ,m_{2} ,m_{3} = 1}^{N} {d_{m_{1} m_{2}} \,c_{m_{3} m} 
\,\vec {r}_{nm_{1}} \wedge \vec {r}_{m_{2} m_{3}}}
\\ \qquad{}
- \sum\limits_{m_{1} 
,m_{2} ,m_{3} ,m_{4} = 1}^{N} {} d_{m_{3} m_{4}} d_{m_{1} m_{2}}
\\ \qquad{}
\left[ 
{} \right.\vec {r}_{nm_{1} } \,\left\{ {\rho _{m_{2} m_{3} } \,\rho _{m_{4} 
m} - \left( {\vec {r}_{m_{2} m_{3} } \cdot \vec {r}_{m_{4} m} } \right)} 
\right\}
 + \vec {r}_{m_{2} m_{3} } \,\left\{ {\rho _{nm_{1} } \,\rho _{m_{4} m} - 
\left( {\vec {r}_{nm_{1} } \cdot \vec {r}_{m_{4} m} } \right)} \right\} 
\\ \qquad{}
+\vec {r}_{m_{4} m} \,\left\{ {\rho _{nm_{1} } \,\rho _{m_{2} m_{3} } - 
\left( {\vec {r}_{nm_{1} } \cdot \vec {r}_{m_{2} m_{3} } } \right)} 
\right\}\left. {} \right] \  .\tag{5.46b}
\end{gather*}


These equations of motion describe the evolution of the $N^{2}$ scalars 
$\rho _{nm} \left( {t} \right)$ and of the $N^{2}$ three-vectors $\vec 
{r}_{nm} \left( {t} \right)$; they feature $4\,N^{2}$ arbitrary ``coupling 
constants.''

\section{Outlook}
\label{VI}


This paper may have appeared extremely long to the researcher, if any, who 
had the stamina to read through it; yet it is clear that here we merely 
\textit{introduced }our topic. Much remains to be done, in the way of more 
systematic exploration of solvable and/or integrable and/or linearizable 
many-body problems in 3-dimensional space, as well as analysis of their 
detailed behavior and identification of interesting applications. Indeed, as 
emphasized in Section \ref{I}, the purpose and scope of this paper has been to 
\textit{introduce }various techniques for the identification of such models, 
rather than to explore completely their potentialities. We are moreover 
aware of other developments [10,11] which might well provide additional 
starting points for applying the techniques described herein but which have 
not been discussed above to avoid our paper becoming excessively long. 

Let us end this section by indicating three additional developments beyond 
those we just mentioned, which are also suggested as a follow-up to our 
findings. One is the detailed study of all cases in which the many-body 
Newtonian equations of motion exhibited above, as well as others that may be 
obtained by analogous techniques, can be fitted into a Hamiltonian 
framework; their investigation is important in itself, given the special 
interest of Hamiltonian systems, and even more so inasmuch as it opens the 
way to quantization, with the expectation that many-body problems treatable 
in a classical context be also amenable to exact analysis in the quantum 
case. Secondly, a vast vista of generalizations is implied by the 
possibility to consider discretized problems (in space and/or time). And 
thirdly, there is the possibility to manufacture many-body problems which 
feature Newtonian equations of motion \textit{with velocity-independent 
forces}, and are amenable to exact treatments (solvable, integrable, 
linearizable) \textit{only for a subset of initial conditions }(typically, 
with arbitrary initial positions, but with initial velocities determined by 
the initial positions): these are the systems characterized by first-order 
equations of motion, in which the second-order (``Newtonian'') equations of 
motion are obtained by time-differentiating the first-order equations and 
then, in the second-order equations, using the first-order equations 
equations to eliminate the (once-differentiated) terms.

\section*{Appendix A. \ On the matrix ODE \  $\underline {\ddot {U}} = \left[ 
{\underline {\dot {U}} ,\,\underline {\tilde {f}} \left( {\underline {U} } 
\right)} \right]$}


In this appendix we mainly focus on ($2 \times 2$)-matrices, and in this 
context we treat (indeed, in some cases, \textit{solve}) the matrix ODE 
displayed in the title. But first we tersely discuss the theoretical 
background to the \textit{linearizability }of this matrix ODE, and indeed 
more general versions of it, without any restriction on its rank (see 
Section \ref{III.B} for an analogous, albeit more general, treatment).

Let the (time-dependent) matrices $\underline {u} \left( {t} \right)$and 
$\underline {U} \left( {t} \right)$ be related by a (time-dependent) 
similarity transformation:
\begin{equation*}
\underline {u} \left( {t} \right) = \underline {W} \left( {t} 
\right)\,\underline {U} \left( {t} \right)\,\left[ {\underline {W} \left( 
{t} \right)} \right]^{ - 1},
\quad
\underline {U} \left( {t} \right) = \left[ {\underline {W} \left( {t} 
\right)} \right]^{ - 1}\,\underline {u} \left( {t} \right)\,\underline {W} 
\left( {t} \right) \  ,\tag{A.1}
\end{equation*}
with the time-evolution of the (invertible!) matrix $\underline {W} \left( 
{t} \right)$ characterized by the matrix ODE
\begin{gather*}
\underline {\dot {W}} \left( {t} \right) = \underline {W} \left( {t} 
\right)\,\underline {\tilde {f}} \left( {\underline {U} \left( {t} \right)} 
\right) \  ,\tag{A.2a}
\\
\underline {\dot {W}} \left( {t} \right) = \underline {\tilde {f}} \left( 
{\underline {u} \left( {t} \right)} \right)\,\underline {W} \left( {t} 
\right) \  .\tag{A.2b}
\end{gather*}

The choice of the function $\underline {\tilde {f}} \left( {\underline {U} } 
\right)$ remains our privilege; but note that, as entailed by the 
consistency of (A.2a) with (A.2b) via (A.1), we assume $\underline {\tilde 
{f}} \left( {\underline {U} } \right)$ to be a scalar/matrix function (i.e., 
a function depending on scalar quantities and on the matrix $\underline {U} 
$ , but on no other matrix, so that $\underline {\tilde {f}} \left( 
{\underline {u} } \right) = \underline {W} \,\underline {\tilde {f}} \left( 
{\underline {U} } \right)\,\underline {W} ^{ - 1}$, $\underline {\tilde {f}} 
\left( {\underline {U} } \right) = \underline {W} ^{ - 1}\,\underline 
{\tilde {f}} \left( {\underline {u} } \right)\,\underline {W} $, see (A.1) 
).

The matrix evolution equation (A.2) defines the matrix $\underline {W} 
\left( {t} \right)$ up to the assignment of an initial condition,
\begin{equation*}
\underline {W} \left( {0} \right) = \underline {W} _{0} \ ,\tag{A.3}
\end{equation*}
whose choice remains our privilege, except for the requirement that the 
matrix $\underline {W} _{0} $ be invertible, see (A.1). Generally it will be 
convenient to make the very simple choice $\underline {W} _{0} = \underline 
{1} $ (see below).

It is now easily seen that (A.2a) entails
\begin{gather*}
\underline {\dot {u}} = \underline {W} \,\underline {\dot {U}} \,\underline 
{W} ^{ - 1}\,,\,\,\,\,\underline {\dot {U}} = \,\underline {W} ^{ - 
1}\,\underline {\dot {u}} \,\underline {W} \ ,\tag{A.4}
\\
\underline {\ddot {u}} = \underline {W} \,\left\{ {\underline {\ddot {U}} + 
\left[ {\underline {\tilde {f}} \left( {\underline {U} } 
\right),\,\underline {\dot {U}} } \right]} \right\}\,\underline {W} ^{ - 
1}\,,\,\,\,\,\,\underline {\ddot {U}} + \left[ {\underline {\tilde {f}} 
\left( {\underline {U} } \right),\,\underline {\dot {U}} } \right] = 
\underline {W} ^{ - 1}\,\underline {\ddot {u}} \,\underline {W} 
\ .\tag{A.5}
\end{gather*}

Here and throughout $\left[ {\underline {A} ,\underline {B} } \right]$ 
denotes the \textit{commutator }of the two matrices $\underline {A} $ and 
$\underline {B} $, $\left[ {\underline {A} ,\,\underline {B} } \right] 
\equiv \underline {A} \,\underline {B} - \underline {B} \,\underline {A} 
$.

Assume now the matrix $\underline {u} \left( {t} \right)$to evolve in time 
according to a second-order \textit{linearizable }ODE, say
\begin{equation*}
\underline {\ddot {u}} = \underline {\tilde {u}} \left( {\underline {u} 
,\underline {\dot {u}} } \right)\ .\tag{A.6}
\end{equation*}
Here the choice of the function \textit{ }$\underline {\tilde {u}} \left( 
{\underline {u} ,\underline {\dot {u}} }\right)$ remains again our 
privilege, subject to the restriction that (A.6) be \textit{linearizable 
}(or perhaps even \textit{solvable}; indeed the case on which we will mainly 
focus below corresponds to the simply choice $\underline {\tilde {u}} \left( 
{\underline {u} ,\underline {\dot {u}} } \right) = 0$); but we must again 
assume that $\underline {\tilde {u}} $ is a scalar/matrix function of its 2 
matrix arguments (namely that it depends on scalars and on the 2 matrices 
$\underline {u} $and $\underline {\dot {u}} $, but on no other matrices, so 
that $\underline {W} ^{ - 1}\,\underline {\tilde {u}} \left( {\underline {u} 
,\underline {\dot {u}} } \right)\,\underline {W} = \underline {\tilde {u}} 
\left( {\underline {U} ,\underline {\dot {U}} } \right)$, see (A.1) and 
(A.4)). It is then clear that the corresponding evolution of $\underline {U} 
\left( {t} \right)$, characterized by the matrix evolution equation (implied 
by (A.6) via (A.5), (A.1) and (A.4))
\begin{equation*}
\underline {\ddot {U}} = \underline {\tilde {u}} \left( {\underline {U} 
,\,\underline {\dot {U}} } \right) + \left[ {\underline {\dot {U}} 
,\,\underline {\tilde {f}} \left( {\underline {U} } \right)} 
\right]\ ,\tag{A.7}
\end{equation*}
is \textit{linearizable}. Indeed its solution can be obtained via the 
following steps: (i) given the initial data $\underline {U} \left( {0} 
\right),\,\,\underline {\dot {U}} \left( {0} \right)$, evaluate, via (A.1) 
and (A.4) with (A.3), the initial data $\underline {u} \left( {0} 
\right),\,\,\underline {\dot {u}} \left( {0} \right)$ (note that the 
convenient choice $\underline {W} _{0} = \underline {1} $ entails simply 
$\underline {u} \left( {0} \right) = \underline {U} \left( {0} 
\right),\,\,\,\underline {\dot {u}} \left( {0} \right) = \underline {\dot 
{U}} \left( {0} \right)$ ); (ii) evaluate $\underline {u} \left( {t} 
\right)$ from the (by assumption \textit{linearizable}) evolution equation 
(A.6); (iii) evaluate $\underline {W} \left( {t} \right)$ by integrating 
the\textit{ linear nonautonomous} matrix ODE (A.2b) (with (A.3)); (iv) 
evaluate $\underline {U} \left( {t} \right)$ from the second of the (A.1).

For instance by assuming that $\underline {u} \left( {t} \right)$ satisfies 
the \textit{solvable }matrix ODE (3.1), namely by setting in (A.6)(up to 
trivial notational changes): 
\begin{equation*}
\underline {\tilde {u}} = 2\,a\,\underline {\dot {u}} + b\,\underline {u} + 
k\,\underline {\dot {u}} \,\underline {u} ^{ - 1}\,\underline {\dot {u}} 
\  ,\tag{A8a}
\end{equation*}
(A.7) becomes
\begin{equation*}
\underline {\ddot {U}} = 2\,a\,\underline {\dot {U}} + b\,\underline {U} + 
k\,\underline {\dot {U}} \,\underline {U} ^{ - 1}\,\underline {\dot {U}} + 
\left[ {\underline {\dot {U}} ,\,\underline {\tilde {f}} \left( {\underline 
{U} } \right)} \right]\ ,\tag{A.8b}
\end{equation*}
and more specifically (to make explicit the connection with the treatment of 
Section \ref{III.B}) by moreover assuming $k = 0\,,\,\tilde {f}\left( {\underline 
{U} } \right) = c\,\underline {U} $ so that $\underline {u} \left( {t} 
\right)$ satisfies the \textit{solvable }(linear!)\textit{ }matrix ODE 
(3.37a), namely:
\begin{equation*}
\underline {\ddot {u}} = 2a\underline {\dot {u}} + b\underline {u} \  
,\tag{A.9a}
\end{equation*}
(A.7) becomes (3.36), namely:
\begin{equation*}
\underline {\ddot {U}} = 2\,a\,\underline {\dot {U}} + b\,\underline {U} + 
c\,\left[ {\underline {\dot {U}} ,\,\underline {U} } \right] \  
.\tag{A.9b}
\end{equation*}
The solution of equation (A.9a) (with $a$ and $b$ two arbitrary constants) 
reads
\begin{equation*}
\underline {u} \left( {t} \right) = {\rm exp}\left( {a\,t} \right)\,\left[ 
{\underline {u} \left( {0} \right) {\rm cosh}\left( {\Delta \,t} \right) + 
\underline {\dot {u}} \left( {0} \right)\,\Delta ^{ - 1} {\rm sinh}\left( {\Delta 
\,t} \right)} \right] \  ,\tag{A.10a}
\end{equation*}
\begin{equation*}
\Delta = \left( {a^{2} + b} \right)^{1/2}\ . \tag{A.10b}
\end{equation*}


Hence the corresponding
evolution equation for $\underline {W} $, see 
(A.2b), reads 
\begin{equation*}
\underline {\dot {W}} = {\rm exp}\left( {a\,t} \right)\,\left[ {\underline {A} 
{\rm cosh}\left( {\Delta \,t} \right) + \underline {B} \,\Delta ^{ - 1} 
{\rm sinh}\left( 
{\Delta \,t} \right)} \right]\,\,\underline {W} \  ,\tag{A.11}
\end{equation*}
with $\underline {A} = c\,\underline {u} \left( {0} \right),\,\,\,\underline 
{B} = c\,\underline {\dot {u}} \left( {0} \right)$ two constant matrices (if 
also the scalar $c$ is a constant).


In Appendix B an additional analysis is given of the linearizable matrix 
evolution equation (A.9b), and in particular it is shown that, if $a = 
\left( {3/2} \right)\,i\omega ,\,\,b = 2\,\omega ^{2}$ , or if $a = i\omega 
/2,\,\,b = 0$, with $\omega $ an arbitrary (nonvanishing) \textit{real 
}constant, (A.9b) is in fact \textit{solvable} and \textit{all }its 
solutions are \textit{completely periodic}, with period $T = 2\pi /\omega $.


The exploration of additional versions of (A.7), corresponding to other 
choices of the (scalar/matrix) function $\underline {\tilde {u}} \left( 
{\underline {U} ,\underline {\dot {U}} } \right)$ and $\underline {\tilde 
{f}} \left( {\underline {U} } \right)$, is left as an amusing task for the 
diligent reader.


Hereafter in this appendix we focus on the simple choice
\begin{equation*}
\tilde {f}\left( {\underline {u} } \right) = \frac{{1}}{{2}}h\left( {t} 
\right)\,\underline {u} \  ,\tag{A.12a}
\end{equation*}
and $\underline {\tilde {u}} \, = \,0$, entailing (see (A.6)) 
\begin{equation*}
\underline {\ddot {u}} = 0 \  ,\tag{A.12b}
\end{equation*}
hence
\begin{equation*}
\underline {u} \left( {t} \right) = \underline {u} \left( {0} \right) + 
t\,\underline {\dot {u}} \left( {0} \right) \  .\tag{A.12c}
\end{equation*}

Then (A.7) reads:
\begin{equation*}
\underline {\ddot {U}} = \frac{{1}}{{2}}h\left( {t} \right)\,\left[ 
{\underline {\dot {U}} ,\,\underline {U} } \right] \  
.\tag{A.12d}
\end{equation*}

Hereafter our attention is restricted to ($2 \times 2$) matrices. Indeed our 
main aim now is to solve the 3-vector Newtonian equation of motion 
(corresponding to (A.12d) via (4.3))
\begin{equation*}
\ddot {\vec {r}}\left( {t} \right) = \hat {h}\left( {r^{2}} \right)\,\vec 
{r}\left( {t} \right) \wedge \dot {\vec {r}}\left( {t} \right) \  
,\tag{A.13}
\end{equation*}
\noindent
where the function $\hat {h}\left( {r^{2}} \right)$ is defined below, see 
(A.15).

This evolution equation clearly entails $\dot {\vec {r}} \cdot \ddot {\vec 
{r}} = \vec {r} \cdot \ddot {\vec {r}} = 0$ , from which we easily get
\begin{gather*}
\left| {\dot {\vec {r}}\left( {t} \right)} \right| = b \  
,\tag{A.14a}
\\
r^{2}\left( {t} \right)\,\, \equiv \,\,\left| {\vec {r}\left( {t} \right)} 
\right|^{2} = \chi \left( {t} \right)\,\, = \,\,\alpha + 2\beta \,t + 
b^{2}t^{2}\ ,\tag{A.14b}
\\
\vec {r}\left( {t} \right) \cdot \dot {\vec {r}}\left( {t} \right) = \beta + 
b^{2}t\, = \,\frac{{1}}{{2}}\dot {\chi }\left( {t} 
\right)\ ,\tag{A.14c}
\end{gather*}
with the 3 constants $\alpha ,\beta $ and $b$ defined in terms of the 
initial data as follows:
\begin{equation*}
\alpha = \left| {\vec {r}\left( {0} \right)} \right|^{2},\,\,\,\beta = \vec 
{r}\left( {0} \right) \cdot \dot {\vec {r}}\left( {0} \right),\,\,\,b = 
\left| {\dot {\vec {r}}\left( {0} \right)} \right| \  
.\tag{A.14d}
\end{equation*}

(\textit{Beware}: the quantity $b$ introduced here, see (A.14a) and (A.14d), 
has no relation to the ``coupling constant'' $b$ in (A.8) or (A.9)). Taking 
into account (A.14b), it is clear that (A.13) is obtained from (A.12d) via 
(4.3), with
\begin{equation*}
h\left( {t} \right)\, = \,\hat {h}\left( {\chi \left( {t} \right)} \right)\, 
= \,\,\hat {h}\left( {r^{2}} \right) \  .\tag{A.15}
\end{equation*}
If we now time-differentiate (A.13), taking into account (A.14b) and (A.15), 
we get
\begin{equation*}
{\vec {r}} = \,\left( {\dot {h}/h} \right)\ddot {\vec {r}} + h^{2}\left[ 
{\vec {r}\left( {\vec {r} \cdot \dot {\vec {r}}} \right) - \dot {\vec 
{r}}\left| {\vec {r}} \right|^{2}} \right] \  .\tag{A.16}
\end{equation*}
We thereby see, via (A.14), that each component of the 3-vector $\vec 
{r}$ satisfies the following \textit{third-order linear nonautonomous }ODE 
(written for the dependent-variable $F\left( {t} \right)$):
\begin{equation*}
{F}\left( {t} \right) = \,\left[ {\dot {h}\left( {t} \right)/h\left( {t} 
\right)} \right]\ddot {F}\left( {t} \right) + h^{2}\left( {t} 
\right)\,\left[ {\frac{{1}}{{2}}\,\dot {\chi }\left( {t} \right)\,F\left( 
{t} \right) - \chi \left( {t} \right)\,\dot {F}\left( {t} \right)} \right] 
\  .\tag{A.17}
\end{equation*}
Let us define 3 independent solutions, $F_{0} ,F_{1} ,F_{2} $, of the above 
ODE, via the following initial conditions:
\begin{gather*}
F_{0} \left( {0} \right) = 1,\,\,\,\dot {F}_{0} \left( {0} \right) = 
0,\,\,\,\ddot {F}_{0} \left( {0} \right) = 0 \  ,\tag{A.18a}
\\
F_{1} \left( {0} \right) = 0,\,\,\,\dot {F}_{1} \left( {0} \right) = 
1,\,\,\,\ddot {F}_{1} \left( {0} \right) = 0 \  ,\tag{A.18b}
\\
F_{2} \left( {0} \right) = 0,\,\,\,\dot {F}_{2} \left( {0} \right) = 
0,\,\,\,\ddot {F}_{2} \left( {0} \right) = \hat {h}\left( {\chi \left( {0} 
\right)} \right) = \hat {h}\left( {\alpha } \right)\, = \,h\left( {0} 
\right) \  .\tag{A.18c}
\end{gather*}

It is then easily seen that the solution of (A.13) is given by the formula
\begin{equation*}
\vec {r}\left( {t} \right) = \vec {r}\left( {0} \right)\,F_{0} \left( {t} 
\right) + \dot {\vec {r}}\left( {0} \right)\,F_{1} \left( {t} \right) + \vec 
{r}\left( {0} \right) \wedge \dot {\vec {r}}\left( {0} \right)\,F_{2} \left( 
{t} \right) \  .\tag{A.19}
\end{equation*}

The task to solve (A.17) directly is daunting: nevertheless formula (A.19) 
provides an useful representation of the solution $\vec {r}\left( {t} 
\right)$ of (A.13) in terms of the functions $F_{0\,} ,\,F_{1} \,,\,F_{2} $. 
In order to find them, instead of solving directly (A.17) it is convenient 
to use the technique outlined at the beginning of this appendix (and, more 
generally, in Section \ref{II.B}) . Let us summarize the procedure: the solution 
of the 3-vector equation (A.13) is obtained from the solution of the matrix 
equation (A.12d) (using the representation (4.3)); the matrix solution of 
(A.12d) is given by (see (A.1))
\begin{equation*}
\underline {U} \left( {t} \right) = \left[ {\underline {W} \left( {t} 
\right)} \right]^{ - 1}\underline {u} \left( {t} \right)\,\,\underline {W} 
\left( {t} \right)\tag{A.20a}
\end{equation*}
with (see (A.2a), (A12a))
\begin{gather*}
\underline {\dot {W}} \left( {t} \right) = \frac{{1}}{{2}}h\left( {t} 
\right)\,\underline {u} \left( {t} \right)\,\underline {W} \left( {t} 
\right)\ ,\tag{A.20b}
\\
\underline {W} \left( {0} \right) = \underline {1} \  
,\tag{A.20c}
\end{gather*}
and (see (A12.c) and (4.3),(A.20a)):
\begin{equation*}
\underline {u} \left( {t} \right) = i\,\left[ {\vec {r}\left( {0} \right) + 
\dot {\vec {r}}\left( {0} \right)\,t} \right] \cdot \underline {\vec {\sigma 
}} \  .\tag{A.20d}
\end{equation*}
Hence we must now solve the ($2 \times 2$)-matrix equation (A.20b) with
\begin{equation*}
\underline {u} \left( {t} \right) = \,i\,\left( {{\begin{array}{*{20}c}
 {a + bt} \hfill & {c} \hfill \\
 {c} \hfill & { - \left( {a + bt} \right)} \hfill \\
\end{array} }} \right) \  ,\tag{A.21a}
\end{equation*}
which corresponds to (A.20d) via the convenient choice we make for the 
cartesian reference system, so that its \textit{z}-axis point in the 
direction of the 3-vector $\tau \, = \,\,\left( {\alpha + 2\,\beta \,t + 
b^{2}t^{2}} \right)^{1/2}\, = \,\left[ {\chi \left( {t} \right)} 
\right]^{1/2}$, and its \textit{xz}-plane contain the 3-vector $\vec 
{r}\left( {0} \right)$ (as well, of course, as the 3-vector $\dot {\vec 
{r}}\left( {0} \right)$), namely
\begin{equation*}
\vec {r}\left( {0} \right) = \left( {x\left( {0} \right)\,,\,0\,,z\left( {0} 
\right)} \right)\,\,\,,\,\,\,\dot {\vec {r}}\left( {0} \right) = \left( 
{0\,,\,0\,,b} \right) \  ,\tag{A.21b}
\end{equation*}
entailing
\begin{gather*}
z\left( {0} \right) \equiv a = \vec {r}\left( {0} \right) \cdot \dot {\vec 
{r}}\left( {0} \right)/\left| {\dot {\vec {r}}\left( {0} \right)} \right| = 
\beta /b \  ,\tag{A.21c}
\\
x\left( {0} \right) \equiv c = \left| {\vec {r}\left( {0} \right) \wedge 
\dot {\vec {r}}\left( {0} \right)} \right|/\left| {\dot {\vec {r}}\left( {0} 
\right)} \right| = \left( {\alpha - a^{2}} \right)^{1/2} \  
,\tag{A.21d}
\end{gather*}
with $\alpha ,\beta $ and $b$ defined by (A.14d). This choice, using the 
representation (4.3) and (A.19), also entails
\begin{equation*}
\underline {U} \left( {t} \right) = i\,\,\left( {{\begin{array}{*{20}c}
 {a\,F_{0} \left( {t} \right) + b\,F_{1} \left( {t} \right)} \hfill & 
{c\,F_{0} \left( {t} \right) - i\,b\,c\,F_{2} \left( {t} \right)} \hfill \\
 {c\,F_{0} \left( {t} \right) + i\,b\,c\,F_{2} \left( {t} \right)} \hfill & 
{ - a\,F_{0} \left( {t} \right) - b\,F_{1} \left( {t} \right)} \hfill \\
\end{array} }} \right) \  .\tag{A.21e}
\end{equation*}

(\textit{Beware}: the quantities $a$ and $c$ introduced here, see (A.21a), 
(A21c), (A.21d) and (A.21e) have no relation to the ``coupling constants'' 
$a$ and $c$ in (A.9b)). The standard matrix formula
\begin{equation*}
\left( {d/dt} \right) {\rm log}\left\{ {{\rm det}\left[ {\underline {W} \left( {t} 
\right)} \right]} \right\} = {\rm  Trace}\left\{ {\,\left[ {\underline {\dot {W}} 
\left( {t} \right)} \right]\,\left[ {\,\underline {W} \left( {t} \right)} 
\right]^{ - 1}} \right\}\tag{A.22}
\end{equation*}
implies, see (A.20b) and (A.21a), that the determinant of the matrix 
$\underline {W} \left( {t} \right)$ is time-inde\-pen\-dent. Hence from (A.20c) 
we get
\begin{equation*}
{\rm det}\left[ {\underline {W} \left( {t} \right)} \right] = \underline {1} \  
,\tag{A.23}
\end{equation*}
entailing that for $\underline {W} \left( {t} \right)$ and its inverse we 
can use the convenient parameterizations
\begin{equation*}
\underline {W} \left( {t} \right) = \left( {{\begin{array}{*{20}c}
 {f\left( {t} \right)} \hfill & { - \tilde {g}\left( {t} \right)} \hfill \\
 {g\left( {t} \right)} \hfill & {\tilde {f}\left( {t} \right)} \hfill \\
\end{array} }} \right)\,\,,\,\,\,\,\,\,\,\,\,\,\,\,\left[ {\underline {W} 
\left( {t} \right)} \right]^{ - 1} = \left( {{\begin{array}{*{20}c}
 {\tilde {f}\left( {t} \right)} \hfill & {\tilde {g}\left( {t} \right)} 
\hfill \\
 { - g\left( {t} \right)} \hfill & {f\left( {t} \right)} \hfill \\
\end{array} }} \right) \ ,\tag{A.24}
\end{equation*}
with
\begin{equation*}
f\left( {t} \right)\tilde {f}\left( {t} \right) + g\left( {t} \right)\tilde 
{g}\left( {t} \right) = 1 \  .\tag{A.25}
\end{equation*}

(Beware: the scalar function $\tilde{f} (t)$ introduced here should 
not be confused with the scalar/matrix function $\underline {\tilde {f}} $ 
introduced above, see for instance (A.2a)).


Moreover, from (A.24) , (A.21) and (A.20) we get
\begin{gather*}
\dot {f}\left( {t} \right) = i\left[ {h\left( {t} \right)/2} \right]\left[ 
{\left( {a + bt} \right)f\left( {t} \right) + cg\left( {t} \right)} \right] 
\ ,
\\
\dot {g}\left( {t} \right) = i\left[ {h\left( {t} \right)/2} \right]\left[ { 
- \left( {a + bt} \right)g\left( {t} \right) + cf\left( {t} \right)} \right] 
\  ,\tag{A.26a}
\\
\dot {\tilde {f}}\left( {t} \right) = - i\left[ {h\left( {t} \right)/2} 
\right]\left[ {\left( {a + bt} \right)\tilde {f}\left( {t} \right) + c\tilde 
{g}\left( {t} \right)} \right] \  ,
\\
\dot {\tilde {g}}\left( {t} \right) = - i\left[ {h\left( {t} \right)/2} 
\right]\left[ { - \left( {a + bt} \right)\tilde {g}\left( {t} \right) + 
c\tilde {f}\left( {t} \right)} \right] \  ,\tag{A.26b}
\\
f\left( {0} \right) = \tilde {f}\left( {0} \right) = 1,\,\,\,\,\,g\left( {0} 
\right) = \tilde {g}\left( {0} \right) = 0 \  .\tag{A.26c}
\end{gather*}
Hence, if $h\left( {t} \right)$ is real, as we hereafter assume, we see that
\begin{equation*}
\tilde {f}\left( {t} \right) = f^\ast \left( {t} \right),\,\,\,\,\,\tilde 
{g}\left( {t} \right) = g^\ast \left( {t} \right),\tag{A.27}
\end{equation*}
so that the remaining task is to solve (A.26a), of course with the initial 
conditions (A.26c). Once this task is achieved we get (from (A.20a), 
(A.21a), (A.21e) and (A.24) with (A.27))
\begin{gather*}
F_{0} \left( {t} \right) = \frac{{Á1}}{{2c}}\{  - 2\left( {a + bt} 
\right)\left[ f\left( {t} \right)g\left( {t} \right) + f^\ast \left( {t} 
\right)g^\ast \left( {t} \right) \right]
\\ \qquad{}
+ c\,\left[ f^{2}\left( {t} 
\right) + f^{\ast 2}\left( {t} \right) - g^{2}\left( {t} \right) - 
g^{\ast 2}\left( {t} \right) \right] \}\ ,\tag{A.28a}
\\
F_{1} \left( {t} \right) = \frac{{1}}{{b}} \{  - a F_{0} \left( {t} 
\right) + \left( {a + bt} \right) \left[ f\left( {t} \right) f^\ast \left( {t} 
\right) - g\left( {t} \right) g^\ast \left( {t} \right) \right]
\\ \qquad{}
+ c\,\left[ 
f\left( {t} \right)g^\ast \left( {t} \right) + f^\ast \left( {t} 
\right)g\left( {t} \right) \right]\} \  ,\tag{A.28b}
\\
F_{2} \left( {t} \right) = \frac{{i}}{{2cb}}\{ 2\left( {a + bt} 
\right)\left[ f\left( {t} \right)g\left( {t} \right) - f^\ast \left( {t} 
\right)g^\ast \left( {t} \right) \right]
\\ \qquad{}
- c\,\left[ f^{2}\left( {t} 
\right) - f^{\ast 2}\left( {t} \right) - g^{2}\left( {t} \right) + 
g^{\ast 2}\left( {t} \right) \right] \}\ .\tag{A.28c}
\end{gather*}

To express these formulas in terms of $f$ only, we moreover use the 
relations
\begin{equation*}
g\left( {t} \right)g^\ast \left( {t} \right) = 1 - f\left( {t} \right)f^\ast 
\left( {t} \right)\tag{A.29}
\end{equation*}
(see (A.25) and (A.27)), as well as
\begin{equation*}
g\left( {t} \right) = - \left[ {\left( {a + bt} \right)f\left( {t} \right) + 
i\,\left[ {2/h\left( {t} \right)} \right]\,\dot {f}\left( {t} \right)} 
\right]/c\tag{A.30}
\end{equation*}
(see the first of the (A.26a)), and its complex conjugate. We thus obtain 
the following expressions:
\begin{gather*}
F_{0} \left( {t} \right) = \frac{{1}}{{2c^{2}h^{2}\left( {t} 
\right)}}\left\{ {h^{2}\left( {t} \right)\left[ {c^{2} + \left( {a + bt} 
\right)^{2}} \right]\left[ {f^{2}\left( {t} \right) + f^{\ast 2}\left( {t} 
\right)} \right] + 4\left[ {\dot {f}^{2}\left( {t} \right) + \dot 
{f}^{\ast 2}\left( {t} \right)} \right]} \right\} \  ,\tag{A.31a}
\\
F_{1} \left( {t} \right) = \frac{{1}}{{b}}\left\{ { - \left( {a + bt} 
\right) - aF_{0} \left( {t} \right) + i\left[ {2/h\left( {t} \right)} 
\right]\left[ {f\left( {t} \right)\dot {f}^\ast \left( {t} \right) - f^\ast 
\left( {t} \right)\dot {f}\left( {t} \right)} \right]} \right\}
,\tag{A.31b}
\\
F_{2} \left( {t} \right) = - \frac{{i}}{{2bc^{2}h^{2}\left( {t} 
\right)}}\left\{ {h^{2}\left( {t} \right)\left[ {c^{2} + \left( {a + bt} 
\right)^{2}} \right]\left[ {f^{2}\left( {t} \right) - f^{\ast 2}\left( {t} 
\right)} \right] + 4\left[ {\dot {f}^{2}\left( {t} \right) - \dot {f} 
^{\ast 2}\left( {t} \right)} \right]} \right\}. \tag{A.31c}
\end{gather*}

These formulas provide explicit expression of the three functions $F_{j} 
\left( {t} \right)$ in terms of the initial data (see (A.14a), (A.14b), 
(A.14d), (A.21c) and (A.21d)) and of the single function $f\left( {t} 
\right)$; they provide the solution to the initial value problem for (A.13) 
via (A.19). 

There finally remains to calculate $f\left( {t} \right)$. The most 
convenient route is by inserting the expression (A.30) in the second of the 
(A.26a), obtaining thereby the following \textit{linear second-order ODE}:
\begin{equation*}
\ddot {f} = \left( {\dot {h}/h} \right)\dot {f} + \left[ { - \left( {\alpha 
+ 2\beta \,t + b^{2}t^{2}} \right)\left( {h/2} \right)^{2} + i\,b\,h/2} 
\right]f \  ,\tag{A.32}
\end{equation*}
with the initial conditions (see (A.26c), (A.30)):
\begin{equation*}
f\left( {0} \right)\, = \,1 \  ,
\quad
\dot {f}\left( {0} \right) = \,i\,a\,h\left( {0} \right)\,/\,2 \  
,\tag{A.32b}
\end{equation*}
or equivalently, via the position
\begin{equation*}
f\left( {t} \right) = \left[ {h\left( {t} \right)} \right]^{1/2}\varphi 
\left( {t} \right) \  ,\tag{A.33}
\end{equation*}
the following linear second-order ODE of (stationary) Schroedinger type:
\begin{equation*}
\ddot {\varphi } = \varphi \left[ {ibh/2 - \left( {\alpha + 2\beta t + 
b^{2}t^{2}} \right)\left( {h/2} \right)^{2} + \frac{{3}}{{4}}\left( {\dot 
{h}/h} \right)^{2} - \frac{{1}}{{2}}\ddot {h}/h} \right] \  
,\tag{A.34}
\end{equation*}
with the initial conditions:
\begin{equation*}
\varphi \left( {0} \right)\, = \,\left( {h\left( {0} \right)} \right)^{ - 
1/2},
\quad
\dot {\varphi }\left( {0} \right)\, = \,\left( {h\left( {0} \right)} 
\right)^{ - 3/2}\left[ {\,i\,a\,h\left( {0} \right) - \dot {h}\left( {0} 
\right)} \right]\,/\,2\ .\tag{A.34b}
\end{equation*}
The constants $\alpha $ ,$\beta $ and $b$ are defined by (A.14d), and we have 
used (A.21c,d). The function $h\left( {t} \right)$ is of course defined by 
(A.15) with (A.14b), with $\hat {h}\left( {r^{2}} \right)$ defined by the 
original equation (A.13) we set out to solve.


For the simplest choice
\begin{equation*}
\hat {h}\left( {r^{2}} \right)\, = \,h\left( {t} \right) = k \  
,\tag{A.35a}
\end{equation*}
 (A.32a) reads:
\begin{equation*}
\ddot {f} = \frac{{k}}{{4}}\left[ { - k\left( {\alpha + 2\beta \,t + 
b^{2}t^{2}} \right) + 2ib} \right]f \  ,\tag{A.35b}
\end{equation*}
and the corresponding initial conditions (A.32b) read
\begin{equation*}
f\left( {0} \right) = 1\ ,\,\,\,\,\dot {f}\left( {0} \right) = iak/2 \  
,\tag{A.35c}
\end{equation*}
It is amusing to note that (A.35b) is, up to trivial changes (involving 
however the introduction of complex parameters), just the stationary 
Schroedinger equation for a one-dimensional harmonic oscillator, of course 
with the time variable playing the role of (one-dimensional) space variable. 
Its solution satisfying the initial conditions (A.35c) can be easily 
expressed in terms of the parabolic cylinder function $D_{p} \left( {\tau } 
\right)$(see f.i.\ eq.\ 9.255 of [10]) and it reads:
\begin{equation*}
f\left( {t} \right)\, = \,A\,D_{p} \left( {\tau } \right) + B\,D_{p} \left( 
{ - \tau } \right) \  ,\tag{A.36a}
\end{equation*}
where:
\begin{gather*}
p = \, - \left( {1 + ikc^{2}/b} \right) \  ,\tag{A.36b}
\\
\tau \left( {t} \right) = \,\left( {ikb} \right)^{1/2}\,\left( {t - \beta 
/b^{2}} \right) \  ,\tag{A.36c}
\\
A\, = {D'_{p} (-\tau (0)) - \tilde{k} D_{p} (-\tau (0))\over D_{p} 
(\tau (0)) D'_{p} (-\tau (0)) - D_{p} (-\tau (0)) D'_{p} (\tau (0))}\ ,
\tag{A.36d}
\\
B\, ={\tilde{k}D_{p} (-\tau (0)) - D'_{p} (-\tau (0))\over D_{p} 
(\tau (0)) D'_{p} (-\tau (0)) - D_{p} (-\tau (0)) D'_{p} (\tau (0))}\ , \tag{A.36e}
\\
\tilde {k}\, = \,\frac{{a}}{{2}}\left( {ik/b} \right)^{1/2} \  
.\tag{A.36f}
\end{gather*}
Note that this case was studied in [3] where moreover it is mentioned that 
E.V. Ferapontov already uncovered by geometric methods the relation with 
parabolic cylinder functions.


The second simple choice we consider here is
\begin{equation*}
\hat {h}\left( {r^{2}} \right) = \,\frac{{k}}{{r^{2}}} \  
,\tag{A.37a}
\end{equation*}
entailing (see (A.14c), (A.15))
\begin{equation*}
h\left( {t} \right) = \,\,k/\left[ {\alpha + 2\beta \,t + b^{2}t^{2}} 
\right] = \,k/\chi \left( {t} \right) \  .\tag{A.37b}
\end{equation*}
Then setting:
\begin{gather*}
t = - \left( {\,2ibc\,z + \beta - ibc} 
 \right)\,/\,b^{2}\ ,\tag{A.38a}
\\
f\left( {t} \right) = \,f\left( {t\left( {z} \right)} \right) = 
\mathord{\buildrel{\lower3pt\hbox{$\scriptscriptstyle\frown$}}\over {f}} 
\left( {z} \right)\ ,\tag{A.38b}
\end{gather*}
it is easily seen that (A.32a) reduces to the hypergeometric 
differential equation :
\begin{equation*}
z\left( {1 - z} 
\right){\mathord{\buildrel{\lower3pt\hbox{$\scriptscriptstyle\frown$}}\over 
{f}} }''\left( {z} \right) + \left( {1 - 2z} 
\right){\mathord{\buildrel{\lower3pt\hbox{$\scriptscriptstyle\frown$}}\over 
{f}} }'\left( {z} \right) - k\left( {k - 2ib} \right)\left( {2b} \right)^{ - 
2}\mathord{\buildrel{\lower3pt\hbox{$\scriptscriptstyle\frown$}}\over {f}} 
\left( {z} \right) = 0 \  ,\tag{A.38c}
\end{equation*}
where the primes denote derivatives with respect to $z$. Thus the solution 
$f\left( {t} \right)$ of (A.32), with the initial conditions (A32b), can be 
easily obtained , via (A38a,b), in terms of the two independent solutions of 
(A38c) (see 9.151 and 9.153.2 of [10]).


The next case worthy of consideration corresponds to the choice
\begin{equation*}
h\left( {t} \right) = \,\,k\left( {\alpha + 2\beta \,t + b^{2}t^{2}} 
\right)^{ - 3/2}\, = \,k\chi ^{ - 3/2} \  ,\tag{A.39}
\end{equation*}  
This case is significant from a physical point of view, since it is easily 
seen that, with this choice, (A.13) becomes just the Newton equation of an 
electrical charge in the magnetic (radial) field of a magnetic monopole (or 
of a magnetic monopole in the electrical field of an electrical charge):
\begin{equation*}
\ddot {\vec {r}}\left( {t} \right) = \,k\,\left[ {r\left( {t} \right)} 
\right]^{ - 3}\,\vec {r}\left( {t} \right) \wedge \dot {\vec {r}}\left( {t} 
\right) \  .\tag{A.40}
\end{equation*}
Via the change of independent variable
\begin{equation*}
\tau \, = \,\,\left( {\alpha + 2\,\beta \,t + b^{2}t^{2}} \right)^{1/2}\, = 
\,\left[ {\chi \left( {t} \right)} \right]^{1/2}\tag{A.41}
\end{equation*}
(A.32a) reads:
\begin{equation*}
\tau ^{2}\left( {\tau ^{2} - c^{2}} \right)\hat {{{f}'}'}\left( {\tau } 
\right) + \tau \left( {3\tau ^{2} - 2c^{2}} \right)\hat {{f}'}\left( {\tau } 
\right) - \frac{{k}}{{4b^{2}}}\left( {2ib\tau - k} \right)\hat {f}\left( 
{\tau } \right) = 0 \  ,\tag{A.42}
\end{equation*}
where of course $\hat {f}\left( {\tau } \right)\, = \,f\left( {t} \right)$ 
and primes denote derivatives with respect to $\tau $. An investigation of 
this case is postponed to a subsequent paper.


Let us finally report the solution of (A.32), hence of (A12d), with
\begin{equation*}
h\left( {t} \right)\,\, = \,2k/\left( {1 + 2\,\omega \,t} 
 \right)\,\tag{A.43}
\end{equation*}
 (a motivation for this choice will be clear in the next Appendix B, see 
(B.7d)). Let us set
\begin{equation*}
t\, = \,\,\left( {2i\,z - bk} \right)\,/\,\left( {2\,bk\omega } 
\right).\tag{A.44a}
\end{equation*}
Then (A.34a) becomes the Whittaker equation (see eq.\ 9.220.1 of  
[10]) 
\begin{equation*}
\hat {{{\varphi }'}'}\left( {z} \right)\, + \,\left( {\frac{{1}}{{4}}\left( 
{\frac{{1}}{{z^{2}}} - 1} \right) + \frac{{\lambda }}{{z}} - \frac{{\mu 
^{2}}}{{z^{2}}}} \right)\,\hat {\varphi }\left( {z} \right)\,\, = \,\,0 
\  ,\tag{A.44b}
\end{equation*}
where the primes denote derivatives with respect to $z$ and
\begin{gather*}
\hat {\varphi }\left( {z} \right)\, = \,\varphi \left( {t\left( {z} \right)} 
\right)\, = \,\varphi \left( {t} \right),\tag{A.44c}
\\
\lambda \,\, = \left[ {2b\omega ^{2} - i\,k\,\left( {b^{2} - 2\beta \,\omega 
} \right)} \right]\,/\,\left( {4b\omega ^{4}} \right) \  
,\tag{A.44d}
\\
\mu \,^{2} = \,k^{2}\,\left( {4\beta \omega - b^{2} - 4\alpha \omega ^{2}} 
\right)\,/\,\left( {4\omega ^{4}} \right) \  .\tag{A.44d}
\end{gather*}
Thus, through (A.33), (A.43) and (A.44), the solution $f\left( {t} \right)$ 
of (A.32) can be easily expressed in terms of the two independent solutions 
of (A.44b), namely the Whittaker functions $W_{\lambda ,\mu } \left( 
{z\left( {t} \right)} \right)$ and $W_{ - \lambda ,\mu } \left( { - z\left( 
{t} \right)} \right)$ (see eq.\ 9.220.4 of [10]).

\section*{Appendix B. \ On the matrix evolution equation \ $\underline {\ddot 
{U}} = 2\,a\,\underline {\dot {U}} + b\,\underline {U} + c\,\left[ 
{\underline {\dot {U}} ,\,\underline {U} } \right]$}


In this appendix we discuss the \textit{linearizable} matrix evolution 
equation (see (3.35))
\begin{equation*}
\underline {\ddot {U}} = 2\,a\,\underline {\dot {U}} + b\,\underline {U} + 
 c\,\left[ {\underline {\dot {U}} ,\,\underline {U} } 
 \right]\ ,\tag{B.1}
\end{equation*}
 with $a,b$ and $c$ three arbitrary ``coupling constants'' 
 ($c$ could of course be rescaled away), and we identify two cases in which, 
 at least for $2 \times 2$matrices, this equation can be explicitly solved 
 in terms of known special functions.  We set
\begin{equation*}
\underline {U} \left( {t} \right) = \varphi \left( {t} 
 \right)\,\,\underline {V} \left( {\tau } \right) \  , \quad \tau \, = 
 \,\tau \left( {t} \right) \  ,\tag{B.2}
\end{equation*}
and thereby obtain
\begin{gather*}
\underline {V} ^{''} = \underline {V} ^{\prime }\left[ {2a\dot {\tau } - 
 2\dot {\tau }\left( {\dot {\varphi }/\varphi } \right) - \ddot {\tau }} 
 \right]/\dot {\tau }^{2}
\\ \qquad{}
+ \underline {V} \left[ {b + 2a\left( {\dot 
 {\varphi }/\varphi } \right) - \left( {\ddot {\varphi }/\varphi } \right)} 
 \right]/\dot {\tau }^{2} + c\left( {\varphi /\dot {\tau }} \right)\left[ 
 {\underline {V} ^{\prime },\underline {V} } \right] \  
 .\tag{B.3}
\end{gather*}
Here and throughout this appendix primes denote derivatives with respect 
to $\tau $.

A choice naturally suggested by this equation is
\begin{equation*}
\dot {\tau }\left( {t} \right) = \varphi \left( {t} \right) \  
 ,\tag{B.4a}
\end{equation*}
 which entails that $\underline {V} \left( {\tau } \right)$ 
 satisfies the matrix ODE
\begin{equation*}
\underline {V} ^{''} = \varphi ^{ - 2}\left[ {2\,a\,\varphi - 3\,\dot 
 {\varphi }} \right]\,\underline {V} ^{\prime } + \varphi ^{ - 3}\,\left[ 
 {2\,a\,\dot {\varphi } + b\,\varphi - \ddot {\varphi }} \right]\,\underline 
 {V} + c\,\left[ {\underline {V} ^{\prime },\,\underline {V} } \right] \  
 ,\tag{B.4b}
\end{equation*}
which we report here for future memory.  However, we prefer now to focus on 
two special cases of (B.1), which allow a completely explicit solution (at 
least in the case of ($2 \times 2$)-matrices).  Hence we set
\begin{gather*}
\varphi \left( {t} \right) = {\rm exp}\left( {\mu \,t} \right) \  
,\tag{B.5a}
\\
\dot {\tau }\left( {t} \right) = {\rm exp}\left( {\nu \,t} \right) 
\  ,\tag{B.5b}
\end{gather*}
so that (B.3) becomes
\begin{equation*}
\underline {V}^{'"} = p \,\,{\rm exp}\left( { - \nu \,t} \right)\,\underline {V} 
 ^{\prime } + q\, {\rm exp}\left( { - 2\,\nu \,t} \right)\,\underline {V} + 
 c\, {\rm exp}\left[ {\left( {\mu - \nu } \right)\,\,t} \right]\left[ {\underline 
 {V} ^{\prime },\,\underline {V} } \right] \  ,\tag{B.5c}
\end{equation*}
 with 
\begin{gather*}
p = 2\left( {a - \mu } \right) - \nu \ ,\tag{B.5d}
\\
q = b + 2\,a\,\mu 
 - \mu ^{2} \  .\tag{B.5e}
\end{gather*}
 The first case we consider is characterized by 
 the condition
\begin{equation*}
b = - \left( {8/9} \right)a^{2},\tag{B.6a}
\end{equation*}
 to which we associate the choices $\mu = \nu = 2a/3$, which entail, via 
 (B.5d), $p = 0$ and, via (B.5e) and (B.6a), $q = 0$ as well.  Hence in this 
 case the position
\begin{gather*}
\underline {U} \left( {t} \right) = {\rm exp}\left( {2\,a\,t/3} 
 \right)\,\underline {V} \left( {\tau } \right)\ ,\tag{B.6b}
\\
\tau \left( {t} 
 \right) = \left[ {{\rm exp}\left( {2\,a\,t/3} \right) - 1} \right]/\left( 
 {2\,a/3} \right) \  ,\tag{B.6c}
\end{gather*}
yields
\begin{equation*}
\underline {V} ^{''} = c\,\left[ {\underline {V} ^{\prime },\,\underline 
 {V} } \right] \  ,\tag{B.6d}
\end{equation*}
 which, up to a notational change and 
 a trivial rescaling of the dependent matrix variable, is just the matrix 
 evolution equation discussed in Appendix A (\textit{solvable }in the $2 
 \times 2$case, see (A.12d), (A.35a), (4.3), (A.19), (A.28) and (A.36) ).  
 Hence a convenient prescription to solve (B.1) with (B.6a) is to solve 
 instead (B.6d), and then perform the change of dependent and independent 
 variables (B.6b) with (B.6c).  Clearly, this implies that \textit{all} 
 solutions of (B.1) are completely periodic with period $T = 2\pi /\omega $ 
 if
\begin{equation*}
a = \left( {3/2} \right)\,i\omega ,\,\,\,b = 2\,\omega ^{2} \  
,\tag{B.6e}
\end{equation*}
with $\omega $ an arbitrary (nonvanishing) \textit{real } constant.  Note 
however that in this case the matrix evolution equation (B.1) is 
\textit{complex}.

The second case we consider is characterized by the restriction 
\begin{equation*}
b = 0 \  .\tag{B.7a}
\end{equation*}
 
 In this case we set $\mu = 0,\nu = 2a$, which again 
 entails, via (B.5d), $p = 0$ and, via (B.5e) and (B.6a), $q = 0$ as well.  
 Hence in this case the position
\begin{gather*}
\underline {U} \left( {t} \right) = \underline {V} \left( {\tau } 
 \right)\ ,\tag{B.7b}
\\
\tau \left( {t} \right) = \left[ {{\rm exp}\left( 
 {2\,a\,t} \right) - 1} \right]/\left( {2\,a} \right) \  
 ,\tag{B.7c}
\end{gather*}
 yields
\begin{equation*}
\underline {V}^{''} = c\,\left( {1 + 2\,a\tau } \right)^{ - 1}\left[ 
 {\underline {V} ^{\prime },\,\underline {V} } \right] \  
 ,\tag{B.7d}
\end{equation*}
which, up to a notational change, a trivial rescaling of the dependent 
matrix variable, and a redefinition of the independent ``time'' variable, 
is the \textit{solvable} matrix evolution equation discussed at the end of 
Appendix A (see (A.12d), (A.43), (4.3), (A.19), (A.28) and (A.44)).  Hence 
a convenient prescription to solve (B.1) with (B.7a) is (at least in the 
case of ($2 \times 2$)-matrices), to solve instead (B.7d), and then perform 
the change of dependent and independent variables (B.7b) with (B.7c).  
Clearly, this again implies that \textit{all} solutions of (B.1) are 
completely periodic with period $T = 2\pi /\omega $ if 
\begin{equation*}
a = i\omega \,/2,\,\,\,b = 0 \  .\tag{B.7e}
\end{equation*}
Of course in this case as well the matrix 
evolution equation (B.1) is complex.

\section*{Appendix C. \ Proof of a formula}


In this appendix we prove a formula relevant in connection with the first 
of the two techniques introduced at the end of Section \ref{III.E} to get, from 
an evolution equation involving one, or a few, functions, (coupled) 
evolution equations involving many more functions (which might be 
matrix-valued).

Let $f\left( {z} \right)$ be an analytic function of $z$ , so that, at 
least for small enough $\left| {z} \right|$ ,
\begin{equation*}
f\left( {z} \right) = \sum\limits_{m = 0}^{\infty } {f^{\left( {m} 
 \right)}\,z^{m}/m!} \tag{C.1}
\end{equation*}
 is well defined.

Let $\eta _{j} ,\,\,j = 1,...,J$ be $J$ elements of an Abelian algebra such 
that
\begin{equation*}
\eta _{j} \eta _{k} = \eta _{k} \eta _{j} = \eta _{j + k} ,\,\,\,\,\,\eta 
 _{J} = 1,\,\,\,\,\,\eta _{j \pm J} = \eta _{j} \  .\tag{C.2}
\end{equation*}
 
 The last of these equations is consistent with the assumption made throughout this 
 appendix, that all indices are always defined ${\rm mod} \left( {J} \right).$

Then clearly the formula
\begin{equation*}
f\left( {\sum\limits_{k = 1}^{J} {\eta _{k} \,z_{k} } } \right) = 
 \sum\limits_{j = 1}^{J} {\eta _{j} \,f_{j} \left( {z_{1} ,...,z_{J} } 
 \right)} \tag{C.3}
\end{equation*}
defines uniquely the $J$ functions $f_{j} \left( {z_{1} ,...,z_{J} } 
\right)$ .  The purpose of this appendix is to derive an explicit formula 
to compute these functions.  But before doing so, let us emphasize that 
(C.3) and (C.2) entail (with self-evident notation)
\begin{gather*}
f\left( {\sum\limits_{k_{1} = 1}^{J} {\eta _{k_{1} } \,z_{k_{1} } } } 
 \right)\,g\left( {\sum\limits_{k_{2} = 1}^{J} {\eta _{k_{2} } z_{k_{2} } } 
 } \right) = \sum\limits_{j = 1}^{J} {\eta _{j} \sum\limits_{k = 1}^{J} 
 {f_{j - k} \left( {z_{1} ,...,z_{J} } \right)} } \,g_{k} \left( {z_{1} 
 ,...,z_{J} } \right) \  , \tag{C.4a}
\\
f\left( {\sum\limits_{k_{1} = 1}^{J} {\eta _{k_{1} } z_{k_{1} } } } 
 \right)\,g\left( {\sum\limits_{k_{2} = 1}^{J} {\eta _{k_{2} } z_{k_{2} } } 
 } \right)\,h\left( {\sum\limits_{k_{3} = 1}^{J} {\eta _{k_{3} } z_{k_{3} } 
 } } \right) 
\\ \qquad{}
= \sum\limits_{j = 1}^{J} {\eta _{j} \sum\limits_{k = 1}^{J} 
 {\sum\limits_{{k}' = 1}^{J} {f_{j - k - {k}'} \left( {z_{1} ,...,z_{J} } 
 \right)\,g_{k} \left( {z_{1} ,...,z_{J} } \right)\,h_{{k}'} \left( {z_{1} 
 ,...,z_{J} } \right)} } } \  ,\tag{C.4b}
\end{gather*}
and so on.

To compute the functions $f_{j} \left( {z_{1} ,...,z_{J} } \right)$ , see 
(C.3), we note that the algebra (C.2) admits the following $J$ 
realizations:
\begin{equation*}
\eta _{j}^{\left( {K} \right)} = {\rm exp}\left( {2\,\pi \,i\,j\,K/J} 
 \right),\,\,\,j = 1,...,J \  ,\tag{C.5}
\end{equation*}
with $K = 1,...,J$.  These realizations are all different iff $J$ is prime.  
Hence (C.3) entails
\begin{equation*}
f\left( {\sum\limits_{k = 1}^{J} {\eta _{k}^{\left( {K} \right)} \,z_{k} 
 } } \right) = \sum\limits_{j = 1}^{J} {\,\eta _{j}^{\left( {K} \right)} 
 \,f_{j} \left( {z_{1} ,...,z_{J} } \right)} \  ,\tag{C.6a}
\end{equation*}
namely
\begin{equation*}
f\left( {\sum\limits_{k = 1}^{J} {z_{k} } {\rm exp}\left( {2\,\pi \,i\,k\,K/J} 
 \right)} \right) = \sum\limits_{j = 1}^{J} {\,f_{j} \left( {z_{1} 
 ,...,z_{J} } \right)} \,{\rm exp}\left( {2\,\pi \,i\,j\,K/J} \right) \  
 .\tag{C.6b}
\end{equation*}

Multiplication by $exp\left( { - 2\,\pi \,i\,k\,K/J} \right)$ yields
\begin{gather*}
{\rm exp}\left( { - 2\,\pi \,i\,kK/J} \right)\,f\left( {\sum\limits_{{k}' = 
 1}^{J} {\,z_{{k}'} } {\rm exp}\left( {2\,\pi \,i\,{k}'K/J} \right)} \right) 
\\ \qquad{}
=  \sum\limits_{j = 1}^{J} {\,f_{j} \left( {z_{1} ,...,z_{J} } \right)} 
 \,{\rm exp}\left[ {2\,\pi \,i\,\left( {j - k} \right)\,K/J} \right] \  
 .\tag{C.7}
\end{gather*}
We sum now over $K$ from $1$ to $J$ , and use the formula
\begin{equation*}
\sum\limits_{K = 1}^{J} {{\rm exp}\left( {2\,\pi \,i\,j\,K/J} \right) = 
 J\,\delta _{j,J} } ,\,\,\,\,\,j = 1,...,J \  ,\tag{C.8}
\end{equation*}
getting thereby
\begin{equation*}
f_{j} \left( {z_{1} ,...,z_{J} } \right) = J^{ - 1}\,\sum\limits_{K = 
 1}^{J} {\,{\rm exp}\left( { - 2\,\pi \,i\,j\,K/J} \right)} \,f\left( 
 {\sum\limits_{k = 1}^{J} {\,z_{k} } {\rm exp}\left( {2\,\pi \,i\,k\,K/J} \right)} 
 \right) \  ,\tag{C.9}
\end{equation*}
which is the formula we intended to get.

The generalization to the case when $f$ is matrix-valued is obvious, as 
well as to the case when it depends on more than one argument, say $f 
\equiv f\left( {x;y} \right)$; then
\begin{equation*}
f\left( \sum\limits_{k_{1} = 1}^{J} \eta _{k_{1}} x_{k_{1}} 
;\sum\limits_{k_{2}}^{J} \eta _{k_{2}} y_{k_{2}} \right) = 
\sum\limits_{j = 1}^{J} \,\eta _{j} \,f_{j} \left(x_{1} ,...,x_{J} 
;y_{1} ,...,y_{J}\right)\tag{C.10}
\end{equation*}
defines the $J$ functions $f_{j} \left( {x_{1} ,...,x_{J} ;y_{1} ,...,y_{J} 
} \right)$, and
\begin{gather*}
 f_{j} \left( {x_{1} ,...,x_{J} ;y_{1} ,...,y_{J} } \right) = J^{ - 
1}\,\sum\limits_{K = 1}^{J} {\,{\rm exp}\left( { - 2\,\pi i\,j\,K/J} 
\right)\,\,}
\\ \qquad{}
 f\left( {\sum\limits_{k_{1} = 1}^{J} {\,x_{k_{1} } } 
exp\left( {2\,\pi \,i\,k_{1} \,K/J} \right);\,\,\sum\limits_{k_{2} = 1}^{J} 
{\,y_{k_{2} } } {\rm exp}\left( {2\,\pi \,i\,k_{2} \,K/J} \right)} 
\right).\tag{C.11}
\end{gather*}


\section*{Appendix D. \  Algebra of $\underline{\sigma}$-matrices, 
and a useful 3-vector identity}


In this appendix we display some standard formulas for the $\underline 
{\sigma } $-matrices, and a useful 3-vector identity.
\begin{gather*}
\underline {\sigma } _{x} = \left( {{\begin{array}{*{20}c} {0} \hfill & 
 {1} \hfill \\ {1} \hfill & {0} \hfill \\ \end{array} }} 
 \right),\,\,\,\,\,\underline {\sigma } _{y} = \left( 
 {{\begin{array}{*{20}c} {0} \hfill & {i} \hfill \\ { - i} \hfill & {0} 
 \hfill \\ \end{array} }} \right),\,\,\,\,\,\underline {\sigma } _{z} = 
 \left( {{\begin{array}{*{20}c} {1} \hfill & {0} \hfill \\ {0} \hfill & { - 
 1} \hfill \\ \end{array} }} \right) \  .\tag{D.1}
\\
\left[ {\rho + i\,\left( {\vec {r} \cdot \underline {\vec {\sigma }} } 
\right)} \right]^{ - 1} = \left[ {\rho - i\,\left( {\vec {r} \cdot 
\underline {\vec {\sigma }} } \right)} \right]/\left( {\rho ^{2} + r^{2}} 
\right) \  ,\tag{D.2}
\\
\left( {\vec {r}^{\left( {1} \right)} \cdot \underline {\vec {\sigma }} } 
\right)\,\left( {\vec {r}^{\left( {2} \right)} \cdot \underline {\vec 
{\sigma }} } \right) = \left( {\vec {r}^{\left( {1} \right)} \cdot \vec 
{r}^{\left( {2} \right)}} \right) + i\,\left( {\vec {r}^{\left( {1} 
\right)} \wedge \vec {r}^{\left( {2} \right)}} \right) \cdot \underline 
{\vec {\sigma }} \  ,\tag{D.3}
\\
\left[ {\left( {\vec {r}^{\left( {1} \right)} \cdot \underline {\vec 
{\sigma }} } \right),\,\left( {\vec {r}^{\left( {2} \right)} \cdot 
\underline {\vec {\sigma }} } \right)} \right] = 2\,i\,\left( {\vec 
{r}^{\left( {1} \right)} \wedge \vec {r}^{\left( {2} \right)}} \right) 
\cdot \underline {\vec {\sigma }} \  ,\tag{D.4}
\\
 \left( {\vec {r}^{\left( {1} \right)} \cdot \underline {\vec {\sigma }} 
} \right)\,\left( {\vec {r}^{\left( {2} \right)} \cdot \underline {\vec 
{\sigma }} } \right)\,\left( {\vec {r}^{\left( {3} \right)} \cdot 
\underline {\vec {\sigma }} } \right) = i\,\left( {\vec {r}^{\left( {1} 
\right)} \wedge \vec {r}^{\left( {2} \right)}} \right) \cdot \vec 
{r}^{\left( {3} \right)}
\\ \qquad{}
 + \left[ {\vec {r}^{\left( {1} \right)}\left( {\vec {r}^{\left( {2} 
\right)} \cdot \vec {r}^{\left( {3} \right)}} \right) - \vec {r}^{\left( 
{2} \right)}\left( {\vec {r}^{\left( {1} \right)} \cdot \vec {r}^{\left( 
{3} \right)}} \right) + \vec {r}^{\left( {3} \right)}\left( {\vec 
{r}^{\left( {1} \right)} \cdot \vec {r}^{\left( {2} \right)}} \right)} 
\right] \cdot \underline {\vec {\sigma }} \  ,\tag{D.5}
\\
 \left( {\vec {r}^{\left( {1} \right)} \cdot \underline {\vec {\sigma }} 
} \right)\,\left( {\vec {r}^{\left( {2} \right)} \cdot \underline {\vec 
{\sigma }} } \right)\,\left( {\vec {r}^{\left( {3} \right)} \cdot 
\underline {\vec {\sigma }} } \right) + \left( {\vec {r}^{\left( {3} 
\right)} \cdot \underline {\vec {\sigma }} } \right)\,\left( {\vec 
{r}^{\left( {2} \right)} \cdot \underline {\vec {\sigma }} } 
\right)\,\left( {\vec {r}^{\left( {1} \right)} \cdot \underline {\vec 
{\sigma }} } \right)  
\\ \qquad{}
 = \,\,2\,\left[ {\vec {r}^{\left( {1} \right)}\left( {\vec {r}^{\left( 
{2} \right)} \cdot \vec {r}^{\left( {3} \right)}} \right) - \vec 
{r}^{\left( {2} \right)}\left( {\vec {r}^{\left( {1} \right)} \cdot \vec 
{r}^{\left( {3} \right)}} \right) + \vec {r}^{\left( {3} \right)}\left( 
{\vec {r}^{\left( {1} \right)} \cdot \vec {r}^{\left( {2} \right)}} 
\right)} \right] \cdot \underline {\vec {\sigma }} \  ,\tag{D.6}
\\
\left( {\vec {r}^{\left( {1} \right)} \cdot \underline {\vec {\sigma }} } 
\right)\,\left( {\vec {r}^{\left( {2} \right)} \cdot \underline {\vec 
{\sigma }} } \right)\,\left( {\vec {r}^{\left( {1} \right)} \cdot 
\underline {\vec {\sigma }} } \right) = \,\left[ {2\,\vec {r}^{\left( {1} 
\right)}\left( {\vec {r}^{\left( {1} \right)} \cdot \vec {r}^{\left( {2} 
\right)}} \right) - \vec {r}^{\left( {2} \right)}\left( {\vec {r}^{\left( 
{1} \right)} \cdot \vec {r}^{\left( {1} \right)}} \right)} \right] \cdot 
\underline {\vec {\sigma }} \  ,\tag{D.7}
\\
 \left( {\rho + i\,\vec {r} \cdot \underline {\vec {\sigma }} } 
\right)^{1/2} \equiv \tilde {\rho } + i\,\vec {\tilde {r}} \cdot \underline 
{\vec {\sigma }} ,\,\,\,\,\,\tilde {\rho } = \left[ {\left( {\rho + i\,r} 
\right)^{1/2} + \left( {\rho - i\,r} \right)^{1/2}} \right]/2\ ,
\\ \qquad{}
\vec {\tilde {r}} = - i\,\vec {r}\left[ {\left( {\rho + i\,r} 
\right)^{1/2} - \left( {\rho - i\,r} \right)^{1/2}} \right]/\left( {2\,r} 
\right) \  ,\tag{D.8}
\\
{\rm exp}\left( {i\,\vec {r} \cdot \underline {\vec {\sigma }} } \right) = 
{\rm cos}\left( {r} \right) + i\,\left( {\vec {r} \cdot \underline {\vec {\sigma 
}} } \right)\,r^{ - 1}\,{\rm sin}\left( {r} \right) \  ,\tag{D.9}
\\
f\left( {\vec 
{r} \cdot \underline {\vec {\sigma }} } \right) = \left\{ {f\left( {r} 
\right) + f\left( { - r} \right) + \left[ {f\left( {r} \right) - f\left( { 
- r} \right)} \right]\vec {r} \cdot \underline {\vec {\sigma }} /r} 
\right\}/2 \  .\tag{D.10}
\end{gather*}

The 3-vector formula
\begin{equation*}
a\vec {r} + \vec {r} \wedge \vec {b} + \left( {\vec {r} \cdot \vec {d}} 
\right)\,\vec {c}\, = \vec {f}\tag{D.11a}
\end{equation*}
entails
\begin{gather*}
\vec {r} = \left( {a^{2} + b^{2}} \right)^{ - 1}\left[ {a\,\vec {f} + 
\vec {b} \wedge \vec {f} + \gamma \,\vec {b} \wedge \vec {c} + a\,\gamma 
\,\vec {c} + a^{ - 1}\left( {\vec {f} \cdot \vec {b} + \gamma \,\vec {c} 
\cdot \vec {b}} \right)\,\vec {b}} \right] \  ,\tag{D.11b}
\\
\gamma = - \left[{a^{2}\vec {f} \cdot \vec {d} + a\left( {\vec {f} 
\wedge \vec {d}} \right) \cdot \vec {b} + \left({\vec {f} \cdot \vec {b}} 
\right)\left( {\vec {d} \cdot \vec {b}} \right)} \right]
\\ \qquad{}
\left[ {a\,\left( 
{a^{2} + b^{2}} \right) + a^{2}\vec {c} \cdot \vec {d} + a\left( {\vec {c} 
\wedge \vec {d}} \right) \cdot \vec {b} + \left( {\vec {c} \cdot \vec {b}} 
\right)\left( {\vec {d} \cdot \vec {b}} \right)} \right]^{ - 1}.\tag{D.11c}
\end{gather*}

The subcases $\vec {b} = 0$ , $\vec {c} = 0$ (or $\vec {d} = 0$), $a = 0$, 
while easily obtainable from the above formulas, deserve separate display.  
Case $\vec {b} = 0$:
\begin{equation*}
\vec {r}\, = \,a^{ - 1}\left[ {\vec {f} - \left( {\vec {f} \cdot \vec {d}} 
 \right)\left( {a + \vec {c} \cdot \vec {d}} \right)^{ - 1}\vec {c}} 
 \right]\ .\tag{D.11d}
\end{equation*}
 
Case $\vec {c} = 0$ (or $\vec {d} = 0$):
\begin{equation*}
\vec {r} = \left( {a^{2} + b^{2}} \right)^{ - 1}\left[ {a\,\vec {f} + 
 \vec {b} \wedge \vec {f} + a^{ - 1}\left( {\vec {f} \cdot \vec {b}} 
 \right)\,\vec {b}} \right] \  .\tag{D.11e}
\end{equation*}
 
 Case $a = 0$ (note that it requires $\vec {b} \cdot \vec {c} \ne 0$ as well as $\vec {b} \cdot \vec 
 {d} \ne 0$ ):
\begin{gather*}
 \vec {r} = b^{ - 2}\{\vec {b} \wedge \vec {f} - \left( {\vec {b} \cdot 
\vec {c}} \right)^{ - 1}\left( {\vec {b} \cdot \vec {f}} \right)\,\vec {b} 
\wedge \vec {c} + \left( {\vec {b} \cdot \vec {c}} \right)^{ - 1}\left( 
{\vec {b} \cdot \vec {d}} \right)^{ - 1}
\\ \qquad{}
\left[ {b^{2}\left( {\vec {b} \cdot \vec {f}} \right) - \left( {\vec {b} 
\cdot \vec {c}} \right)\left( {\vec {b} \wedge \vec {f}} \right) \cdot \vec 
{d} + \left( {\vec {b} \cdot \vec {f}} \right)\left( {\vec {b} \wedge \vec 
{c}} \right) \cdot \vec {d}} \right]\,\,\vec {b}\}\tag{D.11f}
\end{gather*}


\section*{Appendix E. \ Some solvable and/or linearizable matrix evolution 
equations }


In this appendix we report several \textit{solvable and/or linearizable 
}matrix evolution equations.  We also outline their derivation, although 
none of these results is really quite new (see, for instance, [11]).  Most 
of these equations are of ``nearest-neighbour'' type; as always in this 
paper we ignore here the question of the boundary conditions to be assigned 
at the extremal values of $n$ (say, for $n = 0$ and $n = N + 1$).
\textbf{ }

Let us take as starting point the following \textit{solvable }linear matrix 
equation with constant (time-independent, matrix) coefficients:
\begin{equation*}
\underline {\dot {W}} _{n} \left( {t} \right) = \underline {A} _{n} 
 \,\underline {W} _{n} \left( {t} \right) + \underline {B} _{n} \,\underline 
 {W} _{n + 1} \left( {t} \right) + \underline {C} _{n} \,\underline {W} _{n 
 - 1} \left( {t} \right)\ ,\tag{E.1}
\end{equation*}
and let us set
\begin{equation*}
\underline {V} _{n} = \underline {W} _{n + 1} \,\underline {W} _{n} 
 ^{ - 1}\tag{E.2}
\end{equation*}
(here and below we often omit, for notational simplicity, the explicit 
indication of the time dependence).  Then the matrix $\underline {V} _{n} 
\left( {t} \right)$ evolves according to the nonlinear equation
\begin{gather*}
 \underline {\dot {V}} _{n} \left( {t} \right) = \underline {A} _{n + 1} 
\,\underline {V} _{n} \left( {t} \right) - \underline {V} _{n} \left( {t} 
\right)\,\underline {A} _{n} + \underline {B} _{n + 1} \,\underline {V} _{n 
+ 1} \left( {t} \right)\,\underline {V} _{n} \left( {t} \right)
\\ \qquad{}
 - \underline {V} _{n} \left( {t} \right)\,\underline {B} _{n} \,\underline 
{V} _{n} \left( {t} \right) + \underline {C} _{n + 1} - \underline {V} _{n} 
\left( {t} \right)\,\underline {C} _{n} \,\left[ {\underline {V} _{n - 1} 
\left( {t} \right)} \right]^{ - 1} \  .\tag{E.3}
\end{gather*}

The special case of this equation with $\underline {A} _{n} = \underline 
{A} ,\,\,\,\,\underline {B} _{n} = \underline {B} $ and $\underline {C} 
_{n} = 0$ is the first nontrivial evolution equation of the so-called 
discrete Burger's hierarchy [11].

There are now various ways to derive, from this first-order 
\textit{solvable} matrix evolution equation, \textit{solvable }or 
\textit{linearizable }second-order matrix evolution equations.  We describe 
two of them.

A first trick is to separate the odd/even labeled matrices, by setting, 
say, $\underline {V} _{2m} = \underline {U} _{m} ,$\textit{ }$\underline 
{V} _{2m + 1} = \underline {\tilde {U}} _{m} $, $\underline {A} _{2m} = 
\underline {\hat {A}} _{m} $, $\underline {A} _{2m + 1} = \underline 
{\tilde {A}} _{m} $, and so on.  This yields
\begin{gather*}
\underline {\dot {U}} _{n} = \underline {\tilde {A}} _{n} \,\underline {U} 
 _{n} - \underline {U} _{n} \,\underline {\hat {A}} _{n} + \underline 
 {\tilde {B}} _{n} \underline {\tilde {U}} _{n} \,\underline {U} _{n} - 
 \underline {U} _{n} \,\underline {\hat {B}} _{n} \,\underline {U} _{n} + 
 \underline {\tilde {C}} _{n} - \underline {U} _{n} \,\underline {\hat {C}} 
 _{n} \,\left[ {\underline {\tilde {U}} _{n - 1} } \right]^{ - 
 1},\tag{E.4a}
\\
\underline {\dot {\tilde {U}}} _{n} = \underline {\hat {A}} _{n + 1} 
 \,\underline {\tilde {U}} _{n} - \underline {\tilde {U}} _{n} \,\underline 
 {\tilde {A}} _{n} + \underline {\hat {B}} _{n + 1} \,\underline {U} _{n + 
 1} \,\underline {\tilde {U}} _{n} - \underline {\tilde {U}} _{n} 
 \,\underline {\tilde {B}} _{n} \,\underline {\tilde {U}} _{n} + 
 \underline {\hat {C}} _{n + 1} - \underline {\tilde {U}} _{n} \,\underline 
 {\tilde {C}} _{n} \,\left[ {\underline {U} _{n} } \right]^{ - 1} 
 \  .\tag{E.4b}
\end{gather*}
 
 We then time-differentiate the first of these two equations, use the 
 second to eliminate $\underline {\dot {\tilde {U}}} $ , and use the first 
 (undifferentiated) to eliminate $\underline {\tilde {U}} $ after having 
 set, for simplicity's sake, $\underline {\hat {C}} _{n} = 0$ (the diligent 
 reader is welcome to work out the more general result that is obtained 
 forsaking this simplification).  We thus get:
\begin{gather*}
 \underline {\ddot {U}} _{n} = \underline {\tilde {A}} _{n} \,\underline 
{\dot {U}} _{n} - \underline {\dot {U}} _{n} \,\underline {\hat {A}} _{n} - 
\underline {\dot {U}} _{n} \,\underline {\hat {B}} _{n} \,\underline {U} 
_{n} - \underline {U} _{n} \,\underline {\hat {B}} _{n} \,\underline {\dot 
{U}} _{n} 
\\ \qquad{}
+ \underline {\tilde {B}} _{n} \,\left[ {\underline {\hat {A}} _{n + 1} + 
\underline {\hat {B}} _{n + 1} \,\underline {U} _{n + 1} } \right]\,\left[ 
{\underline {\tilde {B}} _{n} } \right]^{ - 1}\,\left[ {\underline {\dot 
{U}} _{n} - \underline {\tilde {A}} _{n} \,\underline {U} _{n} + \underline 
{U} _{n} \,\underline {\hat {A}} _{n} + \underline {U} _{n} \,\underline 
{\hat {B}} _{n} \,\underline {U} _{n} - \underline {\tilde {C}} _{n} } 
\right]
\\ \qquad{}
 - \left[ {\underline {\dot {U}} _{n} - \underline {\tilde {A}} _{n} 
\,\underline {U} _{n} + \underline {U} _{n} \,\underline {\hat {A}} _{n} + 
\underline {U} _{n} \,\underline {\hat {B}} _{n} \,\underline {U} _{n} - 
\underline {\tilde {C}} _{n} } \right]\,\left[ {\underline {\hat {A}} _{n} 
+ \underline {\hat {B}} _{n} \,\underline {U} _{n} } \right] \  
.\tag{E.5}
\end{gather*}

Let us now consider the simpler case in which the constant matrices are 
replaced by scalars, namely we set $\underline {A} _{n} = a_{n} 
\,\underline {1} $, $\underline {\tilde {A}} _{n} = \tilde {a}_{n} 
\underline {1} $ , and so on.  We thus get
\begin{gather*}
 \underline {\ddot {U}} _{n} = \left[ {a_{n} - a_{n + 1} } 
\right]\,\tilde {c}_{n} + \left[ {a_{n + 1} - a_{n} } \right]\,\left[ 
{a_{n} - \tilde {a}_{n} } \right]\,\underline {U} _{n} + \tilde {c}_{n} 
\,\left[ {b_{n} \,\underline {U} _{n} - b_{n + 1} \,\underline {U} _{n + 1} 
} \right] 
\\ \qquad{}
 + \left[ {a_{n} - \tilde {a}_{n} } \right]\,b_{n + 1} \,\underline {U} 
_{n + 1} \,\underline {U} _{n} - 2\,b_{n} \,\underline {\dot {U}} _{n} 
\,\underline {U} _{n} 
\\ \qquad{}
+ \left[ {\tilde {a}_{n} + a_{n} - 2\,a_{n} + b_{n + 
1} \,\underline {U} _{n + 1} - b_{n} \,\underline {U} _{n} } 
\right]\,
\left[ {\underline {\dot {U}} _{n} + b_{n} \,\underline {U} _{n} 
^{2}} \right] \  .\tag{E.6}
\end{gather*}

Let us also record the special case of this \textit{solvable }matrix 
evolution equation which corresponds to the position (consistent with (E.6) 
) $\underline {U} _{n} = \underline {U} ,\,\,\,\tilde {c}_{n} = - c/\alpha 
,\,\,\,b_{n} = b,\,\,\,a_{n} = \tilde {a} + \alpha \,n,\,\,\,\tilde {a}_{n} 
= \tilde {a} - a + \alpha \,n$.  This yields
\begin{equation*}
\underline {\ddot {U}} = c + a\,\alpha \,\underline {U} + \left( {\alpha 
 - a} \right)\,\underline {\dot {U}} + b\,\alpha \,\underline {U} ^{2} - 
 2\,b\,\underline {\dot {U}} \,\underline {U} \  ,\tag{E.7}
\end{equation*}
a solvable evolution equation which contains the 4 arbitrary constants 
$c,\,a,\,\alpha ,\,b$.

A second method to obtain a second-order equation from the \textit{solvable 
}first-order equation (E.3) is by setting
\begin{equation*}
\underline {V} _{n} \left( {t} \right) = \underline {\dot {M}} _{n} 
 \left( {t} \right)\,\left[ {\underline {M} _{n} \left( {t} \right)} 
 \right]^{ - 1},\,\,\,\,\,\underline {\dot {M}} _{n} \left( {t} \right) = 
 \underline {V} _{n} \left( {t} \right)\,\underline {M} _{n} \left( {t} 
 \right)\ .\tag{E.8}
\end{equation*}

One obtains thereby the following evolution equation for the matrix 
$\underline {M} _{n} \left( {t} \right)$:
\begin{gather*}
 \underline {\ddot {M}} _{n} = \underline {\dot {M}} _{n} \,\left[ 
{\underline {M} _{n} } \right]^{ - 1}\,\underline {\dot {M}} _{n} + 
\underline {A} _{n + 1} \,\underline {\dot {M}} _{n} - \underline {\dot 
{M}} _{n} \,\left[ {\underline {M} _{n} } \right]^{ - 1}\,\underline {A} 
_{n} \,\underline {M} _{n} 
\\ \qquad{}
+ \underline {B} _{n + 1} \,\underline {\dot {M}} _{n + 1} \,\left[ 
{\underline {M} _{n + 1} } \right]^{ - 1}\,\underline {\dot {M}} _{n} - 
\underline {\dot {M}} _{n} \,\left[ {\underline {M} _{n} } \right]^{ - 
1}\,\underline {B} _{n} \,\underline {\dot {M}} _{n} 
\\ \qquad{}
 + \underline {C} _{n + 1} \,\underline {M} _{n} - \underline {\dot {M}} 
_{n} \,\left[ {\underline {M} _{n} } \right]^{ - 1}\,\underline {C} _{n} 
\,\underline {M} _{n - 1} \,\left[ {\underline {\dot {M}} _{n - 1} } 
\right]^{ - 1}\,\underline {M} _{n} \  .\tag{E.9}
\end{gather*}

This should be categorized as a \textit{linearizable }system of matrix 
evolution equations, since to solve it one must solve, in addition to the 
\textit{solvable }equation (E.1) (to get $\underline {W} _{n} \left( {t} 
\right)$ and then, via (E.2), $\underline {V} _{n} \left( {t} \right)$), a 
\textit{linear nonautonomous }matrix evolution equation (to get $\underline 
{M} _{n} \left( {t} \right)$; see the second of the (E.8)).

Let us call attention to the presence, in this equation, of a 
time-differentiated matrix \textit{in the denominator }(see the last term 
in the right-hand-side).

As an obvious consequence of the way this equation, (E.9), has been 
derived, see (E.8), it is invariant under the transformation $\underline 
{M} _{n} \to \underline {M} _{n} \underline {D} _{n} $ with $\underline {D} 
_{n} $ arbitrary \textit{constant }matrices, $\underline {\dot {D}} _{n} \, 
= \,0$.

When all the constant matrices are replaced by scalars, namely if we set in 
(E.9) $\underline {A} _{n} = a_{n} \underline {1} $, $\underline {B} _{n} = 
b_{n} \underline {1} $, $\underline {C} _{n} = c_{n} \underline {1} $, we 
obtain the following \textit{linearizable }matrix evolution equation:
\begin{gather*}
\underline {\ddot {M}} _{n} = \left[ {a_{n + 1} - a_{n} - b_{n} } 
\right]\,\underline {\dot {M}} _{n} + \left\{ {\underline {\dot {M}} _{n} 
\,\left[ {\underline {M} _{n} } \right]^{ - 1} + b_{n + 1} \,\underline 
{\dot {M}} _{n + 1} \,\left[ {\underline {M} _{n + 1} } \right]^{ - 1}} 
\right\}\,\underline {\dot {M}} _{n} 
\\ \qquad{}
 + c_{n + 1} \,\underline {M} _{n} - c_{n} \,\underline {\dot {M}} _{n} 
 \,\left[ {\underline {M} _{n} } \right]^{ - 1}\,\underline {M} _{n - 1} 
 \,\left[ {\underline {\dot {M}} _{n - 1} } \right]^{ - 1}\,\underline {M} 
 _{n} \  .\tag{E.10}
\end{gather*}

Let us end by noting that the (compatible) position $\underline {M} _{n} 
(t) = \underline {M} (t)$, $a_{n} = n\,(2a + c - 1)$, $b_{n} = c - 1$, 
$c_{n} = n\,b$ yields for the 
matrix $\underline {M} \left( {t} \right)$ precisely the \textit{solvable 
}evolution equation (3.1).

\section*{Appendix F. \ A linearizable matrix system, its solvable 
versions and a class of solvable Hamiltonian many-body problems}


In this appendix we show that the system of matrix evolution equations 
\begin{gather*}
\underline {\dot {U}} _{n} = \sum\limits_{m = 1}^{N} {\left( {a_{nm} 
\,\underline {U} _{m} + b_{nm} \,\underline {V} _{m} } \right)} + \left[ 
{\underline {U} _{n} ,\,\underline {\tilde {f}} \left( {\underline {U} _{j} 
,\underline {V} _{j} ;t} \right)} \right] \  ,\tag{F.1a}
\\
\underline {\dot {V}} _{n} = \sum\limits_{m = 1}^{N} {\left( {c_{nm} 
\,\underline {U} _{m} + d_{nm} \,\underline {V} _{m} } \right)} + \left[ 
{\underline {V} _{n} ,\,\underline {\tilde {f}} \left( {\underline {U} _{j} 
,\underline {V} _{j} ;t} \right)} \right] \  ,\tag{F.1b}
\end{gather*}
is \textit{linearizable}. Here the $4\,N^{2}$ quantities $a_{nm} ,\,b_{nm} 
,\,c_{nm} ,\,d_{nm} $ are arbitrary (they could also be time-dependent 
functions), and $F$ is an arbitrary function of the $2\,N$ matrices 
$\underline {U} _{m} ,\,\underline {V} _{m} $ and of the time $t$; note 
however that the same $\underline {\tilde {f}} $ enters in (F1a) and (F.1b), 
that this quantity is independent of the index $n$, and that it is a 
scalar/matrix function of its arguments, namely it satisfies the property 
$\underline {W} \,\underline {\tilde {f}} \left( {\underline {U} _{j} 
,\underline {V} _{j} ;t} \right)\,\underline {W} ^{ - 1} = \underline 
{\tilde {f}} \left( {\underline {W} \,\underline {U} _{j} \,\underline {W} 
^{ - 1},\,\underline {W} \,\underline {V} _{j} \,\underline {W} ^{ - 1};t} 
\right)$.

To prove that (F.1) is \textit{linearizable }we proceed as in Section \ref{III.B}, 
namely we set (see (3.11) and (3.13); beware of the notational changes!)
\begin{gather*}
\underline {u} _{n} \left( {t} \right) = \underline {W} \left( {t} 
\right)\,\underline {U} _{n} \left( {t} \right)\,\left[ {\underline {W} 
\left( {t} \right)} \right]^{ - 1},
\quad
\underline {U} _{n} \left( {t} \right) = \left[ {\underline {W} \left( {t} 
\right)} \right]^{ - 1}\,\underline {u} _{n} \left( {t} \right)\,\underline 
{W} \left( {t} \right)\ ,\tag{F.2a}
\\
\underline {v} _{n} \left( {t} \right) = \underline {W} \left( {t} 
\right)\,\underline {V} _{n} \left( {t} \right)\,\left[ {\underline {W} 
\left( {t} \right)} \right]^{ - 1},
\quad
\underline {V} _{n} \left( {t} \right) = \left[ {\underline {W} \left( {t} 
\right)} \right]^{ - 1}\,\underline {v} _{n} \left( {t} \right)\,\underline 
{W} \left( {t} \right)\ ,\tag{F.2b}
\\
\underline {\dot {W}} \left( {t} \right) = \underline {W} \left( {t} 
\right)\,\underline {\tilde {f}} \left[ {\underline {U} _{j} \left( {t} 
\right),\,\underline {V} _{j} \left( {t} \right);t} \right],\,\,\,\underline 
{\dot {W}} \left( {t} \right) = \underline {\tilde {f}} \left[ {\underline 
{u} _{j} \left( {t} \right),\,\underline {v} _{j} \left( {t} \right);t} 
\right]\,\underline {W} \left( {t} \right)\ .\tag{F.3}
\end{gather*}
Time-differentiation of the first of the (F.2) yields (using the first of 
the (F.3))
\begin{gather*}
\underline {\dot {u}} _{n} = \underline {W} \,\left\{ {\underline {\dot {U}} 
_{n} - \left[ {\underline {U} _{n} ,\,\underline {\tilde {f}} \left( 
{\underline {U} _{j} ,\,\underline {V} _{j} ;t} \right)} \right]} 
\right\}\,\underline {W} ^{ - 1}\ ,\tag{F.4a}
\\
\underline {\dot {v}} _{n} = \underline {W} \,\left\{ {\underline {\dot {V}} 
_{n} - \left[ {\underline {V} _{n} ,\,\underline {\tilde {f}} \left( 
{\underline {U} _{j} ,\,\underline {V} _{j} ;t} \right)} \right]} 
\right\}\,\underline {W} ^{ - 1}\ ,\tag{F.4b}
\end{gather*}
and from these equations and (F.1) , (F2) we see that the matrices 
$\underline {u} _{n} \left( {t} \right),\,\underline {v} _{n} \left( {t} 
\right)$ satisfy the \textit{linear }evolution equations
\begin{equation*}
\underline {\dot {u}} _{n} = \sum\limits_{m = 1}^{N} {\left( {a_{nm} 
\,\underline {u} _{m} + b_{nm} \,\underline {v} _{m} } \right)} \  ,
\quad
\underline {\dot {v}} _{n} = \sum\limits_{m = 1}^{N} {\left( {c_{nm} 
\,\underline {u} _{m} + d_{nm} \,\underline {v} _{m} } \right)} 
.\tag{F.5}
\end{equation*}
The linearizability of (F.1) is thereby proven, since its solution can be 
achieved by solving firstly (F.5), then the second of the (F.3) (a linear 
equation for the matrix $\underline {W} \left( {t} \right)$), and then 
recovering $\underline {U} _{n} (t),\,\,\underline {V} _{n} (t)$ from the 
second of the (F.2a,b).

It is moreover easily seen that in some cases, see below, the system (F.1) 
is not only \textit{linearizable, }it is in fact \textit{solvable}. An 
obvious (and rather trivial) case is\textit{ }if the quantities $a_{nm} 
,\,b_{nm} ,\,c_{nm} ,\,d_{nm} $ all vanish, $a_{nm} = b_{nm} = c_{nm} = 
d_{nm} = 0$, and $\underline {\tilde {f}} $ does not depend explicitly on 
the time $t$, $\underline {\tilde {f}} \left( {\underline {U} _{j} 
,\,\underline {V} _{j} ;t} \right) \equiv \underline {\tilde {f}} \left( 
{\underline {U} _{j} ,\,\underline {V} _{j} } \right)$. In such a case (F.5) 
entails that the matrices $\underline {u} _{n} \left( {t} 
\right),\,\underline {v} _{n} \left( {t} \right)$ are in fact 
time-independent, $\underline {u} _{n} (t) = \underline {u} _{n} 
(0),\,\,\underline {v} _{n} (t) = \underline {v} _{n} (0)$, and the second 
of the (F.3) becomes explicitly solvable. Hence one concludes that the 
equations
\begin{equation*}
\underline {\dot {U}} _{n} = \left[ {\underline {U} _{n} ,\,\underline 
{\tilde {f}} \left( {\underline {U} _{j} ,\,\underline {V} _{j} } \right)} 
\right] \  ,
\quad
\underline {\dot {V}} _{n} = \left[ {\underline {V} _{n} ,\,\underline 
{\tilde {f}} \left( {\underline {U} _{j} ,\,\underline {V} _{j} } \right)} 
\right] \  ,\tag{F.6a}
\end{equation*}
are explicitly \textit{solvable}:
\begin{gather*}
\underline {U} _{n} \left( {t} \right) = {\rm exp}\left[ { - t\,\underline {\tilde 
{f}} \left( {\underline {U} _{j} \left( {0} \right),\,\underline {V} _{j} 
\left( {0} \right)} \right)} \right]\,\underline {U} _{n} \left( {0} 
\right)\,exp\,\left[ {t\,\underline {\tilde {f}} \left( {\underline {U} _{j} 
\left( {0} \right),\,\underline {V} _{j} \left( {0} \right)} \right)} 
\right] \  ,\tag{F.6b}
\\
\underline {V} _{n} \left( {t} \right) = {\rm exp}\left[ { - t\,\underline {\tilde 
{f}} \left( {\underline {U} _{j} \left( {0} \right),\,\underline {V} _{j} 
\left( {0} \right)} \right)} \right]\,\underline {V} _{n} \left( {0} 
\right)\,exp\left[ {t\,\underline {\tilde {f}} \left( {\underline {U} _{j} 
\left( {0} \right),\,\underline {V} _{j} \left( {0} \right)} \right)} 
\right] \ .\tag{F.6c}
\end{gather*}
Another, perhaps less trivial, case in which the equations (F.1) are in fact 
also \textit{solvable }obtains if 
\begin{equation*}
\underline {\tilde {f}} = \sum\limits_{j = 1}^{N} {\lambda _{j} \,\left[ 
{\underline {U} _{j} ,\,\underline {V} _{j} } \right]} \  
,\tag{F.7a}
\end{equation*}
and the quantities $a_{nm} ,\,b_{nm} ,\,c_{nm} ,\,d_{nm} ,\,\lambda _{n} $ 
are all time-independent and satisfy the constraints
\begin{equation*}
\lambda _{n} \,a_{nm} + \lambda _{m} \,d_{mn} = 0,\,\,\,\,\lambda _{n} 
\,c_{nm} - \lambda _{m} \,c_{mn} = 0,\,\,\,\lambda _{n} \,b_{nm} - \lambda 
_{m} \,b_{mn} = 0 \  .\tag{F.7b}
\end{equation*}

Indeed it is easily seen that these conditions are sufficient to guarantee, 
via (F.1), that $\underline {\tilde {f}} $ is time-independent, $\underline 
{\dot {\tilde {f}}} = 0$, so that there holds again the explicit solution 
(F.6b,c), of course with $\underline {\tilde {f}} $ given by (F.7a). An 
interesting case (see below) is that with $\lambda _{n} = \lambda $, so that 
$d_{nm} = - a_{mn} $ and (F.1) read
\begin{gather*}
\underline{\dot {U}} _{n} = \sum\limits_{m = 1}^{N} \left( a_{nm} 
\,\underline {U}_{m} + b_{nm} \,\underline{V} _{m} + \lambda \,\left[ 
\underline{U}_{n} ,\,\,\left[ \underline{U}_{m} ,\,\underline{V} _{m} 
 \right] \right] \right)\ ,
\\
\underline{\dot {V}} _{n} = 
\sum\limits_{m = 1}^{N} \left( c_{nm} \,\underline {U} _{m} - a_{mn} 
\,\underline{V}_{m}\right. + \lambda \,[ \underline{V} _{n} ,\,[ 
\underline{U}_{m} ,\,\underline{V} _{m}]] )
\ ,\tag{F.7c}
\end{gather*}
with
\begin{equation*}
b_{nm} = b_{mn} ,\,\,\,\,\,c_{nm} = c_{mn} \  .\tag{F.7d}
\end{equation*}
 
 \textit{Linearizable }and \textit{solvable} evolution equations in ordinary 
(3-dimensional) space can be obtained from the above matrix equations, by 
using (compatible) parameterizations in terms of 3-vectors, see 
Section \ref{IV}. 
We only display the equations that are obtained from the \textit{solvable 
}equation (F.7c) with (F.7d) via the parameterization (4.3), namely by 
setting $\underline {U} _{n} = i\,\vec {q}_{n} \cdot \vec {\sigma }$, 
$\underline {V} _{n} = i\,\vec {p}_{n} \cdot \vec {\sigma }$. They read:
\begin{gather*}
\dot {\vec {q}}_{n} = \sum\limits_{m = 1}^{N} {\left( {a_{nm} \,\vec {q}_{m} 
+ b_{nm} \,\vec {p}_{m} + 4\,\lambda \,\left[ {\vec {q}_{n} \wedge \left[ 
{\vec {q}_{m} \wedge \vec {p}_{m} } \right]} \right]} \right)}\ ,
\\
\dot {\vec {p}}_{n} =\sum\limits_{m = 1}^{N} {\left( {c_{nm} 
\,\vec {q}_{m} - a_{mn} \,\vec {p}_{m} + 4\,\lambda \,\left[ {\vec {p}_{n} 
\wedge \left[ {\vec {q}_{m} \wedge \vec {p}_{m} } \right]} \right]} \right)} 
\ .\tag{F.8a}
\end{gather*}
Remarkably, these \textit{solvable }equations of motion, with (F.7d), are 
precisely the equations of motion entailed by the Hamiltonian
\begin{gather*}
H\left( {\vec {q}_{j}},\vec{p}_{j}  \right) = \sum\limits_{n,m = 1}^{N} 
\{ a_{nm} \,\vec{q}_{n} \cdot \vec{p}_{m} + b_{nm} \,\vec{p}_{n} 
\cdot \vec{p}_{m} /2 - c_{nm} \,\vec{q}_{n} \cdot \vec {q}_{m} /2
\\ \qquad{}
+ 
4\,\lambda \,( \vec{q}_{n} \wedge \vec{p}_{n} ) \cdot ( 
\vec{q}_{m} \wedge \vec{p}_{m} ) \} \ .\tag{F.8b}
\end{gather*}
Note that the $\left( {2\,N^{2} + N + 1} \right)$ constants that appear in 
this Hamiltonian, and in the equations of motion (F.8a), are arbitrary. But 
the solvability of this Hamiltonian system has a rather trivial origin; 
indeed, the evolution equations (F.8a) are hardly nonlinear, since they 
entail, as it can be easily verified by direct computation (and it is of 
course implied by our treatment), that the quantity $\sum\limits_{m = 1}^{N} 
{\left( {\vec {q}_{m} \wedge \vec {p}_{m} } \right)} $ is a constant of the 
motion.

Nevertheless, even for $N = 1$, when the Hamiltonian reads simply
\begin{equation*}
H\left( {\vec {q},\vec {p}} \right) = a\vec {q} \cdot \vec {p} + bp^{2}/2 - 
cq^{2}/2 + 2\lambda \left( {\vec {q} \wedge \vec {p}} 
\right)^{2}\ ,\tag{F.9a}
\end{equation*}
and the corresponding equations read simply
\begin{equation*}
\dot {\vec {q}} = a\,\vec {q} + b\,\vec {p} + 4\,\lambda \,\left[ {\vec {q} 
\wedge \left[ {\vec {q} \wedge \vec {p}} \right]} \right],\,\,\,\,\,\dot 
{\vec {p}} = c\,\vec {q} - a\,\vec {p} + 4\,\lambda \,\left[ {\vec {p} 
\wedge \left[ {\vec {q} \wedge \vec {p}} \right]} \right]\ ,\tag{F.9b}
\end{equation*}
this is a rather interesting system, whose detailed analysis, however, 
exceeds the scope of the present paper.

We end this appendix by reporting the many-body generalization of the last 
two equations, (F.9b), that are obtained by using the first of the two 
\textit{multiplication }techniques described at the end of Section \ref{III.E} (see 
also Appendix C). Hence we set
\begin{equation*}
\vec {p} = \sum\limits_{n = 1}^{N} {\eta _{n} \,\vec {p}_{n} } 
,\,\,\,\,\,\vec {q} = \sum\limits_{n = 1}^{N} {\eta _{n} \,\vec {q}_{n} } 
\ ,\tag{F.10}
\end{equation*}
with analogous formulas for the ``coupling constants'' $a,\,b,\,c,\,\lambda 
$ (see (F.9)), as well as
\begin{equation*}
H(\vec {p},\vec {q}) = \sum\limits_{n = 1}^{N} {\eta _{n} \,H_{n} (\vec 
{p}_{1} ,...,\vec {p}_{N} ;\vec {q}_{1} ,...,\vec {q}_{N} )} \  
,\tag{F.11a}
\end{equation*}
with (see (F.9a))
\begin{gather*}
H_{n} (\vec{p}_{1} ,...,\vec{p}_{N} ;\vec{q}_{1} ,...,\vec{q}_{N} ) 
= \sum\limits_{n_{1} ,n_{2} = 1}^{N} \left[ a_{n - n_{1} - n_{2} 
} \,( \vec{p}_{n_{1}} \cdot \vec{q}_{n_{2}}) \right.
\\ \qquad{}
+ b_{n - n_{1} - n_{2}} \,( \vec{p}_{n_{1} } \cdot \vec{p}_{n_{2} } 
)/2 + c_{n - n_{1} - n_{2} } \,( \vec{q}_{n_{1} } \cdot \vec 
{q}_{n_{2} } )/2] 
\\ \qquad{}
+ 2\sum\limits_{n_{1} ,n_{2} ,n_{3} ,n_{4} = 1}^{N} {\lambda _{n - n_{1} - 
n_{2} - n_{3} - n_{4} } \,\left( {\vec {p}_{n_{1} } \wedge \vec {q}_{n_{2} } 
} \right) \cdot \left( {\vec {p}_{n_{3} } \wedge \vec {q}_{n_{4} } } 
\right)} \  .\tag{F.11b}
\end{gather*}

Here and below all the indices $n,n_{1} ,...,n_{4} $ are defined 
${\rm mod}\left( 
{N} \right)$.

It is then easily seen that the $N$ Hamiltonians $H_{n} $ are in involution, 
and that the Hamiltonian $H_{N} $ yields the \textit{solvable N-body 
equations of motion }entailed via (F.10) from (F.9b), namely
\begin{gather*}
\dot {\vec {q}}_{n} = \sum\limits_{n_{1} = 1}^{N} {\left( {a_{n - n_{1} } 
\,\vec {q}_{n_{1} } + b_{n - n_{1} } \,\vec {p}_{n_{1} } } \right)} + 
4\sum\limits_{n_{1} ,n_{2} ,n_{3} = 1}^{N} {\left[ {\lambda _{n - n_{1} - 
n_{2} - n_{3} } \,\left( {\vec {p}_{n_{1} } \wedge \vec {q}_{n_{2} } } 
\right) \wedge \vec {q}_{n_{3} } } \right]} \ ,\tag{F.12a}
\\
\dot {\vec {p}}_{n} = \sum\limits_{n_{1} = 1}^{N} {\left( {c_{n - n_{1} } 
\,\vec {q}_{n_{1} } - a_{n - n_{1} } \,\vec {p}_{n_{1} } } \right)} + 
4\sum\limits_{n_{1} ,n_{2} ,n_{3} = 1}^{N} {\left[ {\lambda _{n - n_{1} - 
n_{2} - n_{3} } \,\left( {\vec {p}_{n_{1} } \wedge \vec {q}_{n_{2} } } 
\right) \wedge \vec {p}_{n_{3} } } \right]} \ .\tag{F.12b}
\end{gather*}


But of course, as emphasized in Section \ref{III.E}, these coupled equations of 
motion can be decoupled by linear transformations of type (3.92).


\section*{Appendix G. \ Example: first-order and second-order linearizable matrix ODEs, 
and a linearizable one-body problem}


In this appendix we illustrate, via a simple example, the possibility, in 
the context of the technique of Section \ref{III.B}, to restrict firstly attention 
to \textit{first-order }matrix ODEs and to obtain subsequently 
\textit{second-order }ODEs by appropriate additional steps. We only consider 
an illustrative, very simple, example; this allows, at very little cost in 
terms of repetitiveness, a completely self-contained presentation, which can 
be understood without any previous knowledge of the treatment of Section 
\ref{III.B}; 
but for the diligent reader interested in the connection we note that 
the case considered in this appendix corresponds, up to a trivial notational 
change ($\underline {u} _{1} \to \underline {u} ,\,\underline {u} _{2} \to 
\underline {v} $), to the treatment of Section \ref{III.B} with $N = 
2,\,\underline {\tilde {U}} _{1} = \underline {\dot {u}} _{1} - \left( 
{\alpha \,\underline {u} _{1} + \beta \,\underline {u} _{2} } 
\right),\,\underline {\tilde {U}} _{2} = \underline {\dot {u}} _{2} - \left( 
{\gamma \,\underline {u} _{1} + \delta \,\underline {u} _{2} } 
\right),\,\underline {\tilde {F}} = \underline {f} - \left( {a\,\underline 
{u} _{1} + b\,\underline {u} _{2} + c\,\left[ {\underline {u} _{1} 
,\,\underline {u} _{2} } \right]} \right)$ and $\underline {g} = \underline 
{h} = \underline {v} = \underline {y} = 0$. At the end we display the 
(highly nonlinear) \textit{linearizable one-body problem }which corresponds 
to the second-order linearizable matrix ODE via the parameterization (4.3). 

Let us set
\begin{gather*}
\underline {u} = \underline {W} \,\underline {U} \,\underline {W} ^{ - 
1},\,\,\,\,\,\underline {U} = \underline {W} ^{ - 1}\,\underline {u} 
\,\underline {W} \  ,\tag{G.1a}
\\
\underline {v} = \underline {W} \,\underline {V} \,\underline {W} ^{ - 
1},\,\,\,\,\,\underline {V} = \underline {W} ^{ - 1}\,\underline {v} 
\,\underline {W} \  ,\tag{G.1b}
\\
\underline {\dot {W}} = \underline {W} \,\left\{ {a\,\underline {U} + 
b\,\underline {V} + c\,\left[ {\underline {U} ,\,\underline {V} } \right]} 
\right\} = \left\{ {a\,\underline {u} + b\,\underline {v} + c\,\left[ 
{\underline {u} ,\,\underline {v} } \right]} \right\}\,\underline {W} \  
,\tag{G.2}
\\
\underline {\dot {u}} = \alpha \,\underline {u} + \beta \,\underline {v} 
,\,\,\,\,\,\underline {\dot {v}} = \gamma \,\underline {u} + \delta 
\,\underline {v} \  ,\tag{G.3}
\end{gather*}
with $a,b,c,\alpha ,\beta ,\gamma ,\delta $ arbitrary constants. 

This entails for the matrices $\underline {U} $ and $\underline {V} $ the 
first-order nonlinear ODEs
\begin{gather*}
\underline {\dot {U}} = \alpha \,\underline {U} + \beta \,\underline {V} + 
b\,\left[ {\underline {U} ,\,\underline {V} } \right] + c\,\left[ 
{\underline {U} ,\,\left[ {\underline {U} ,\,\underline {V} } \right]} 
\right] \  ,\tag{G.4a}
\\
\underline {\dot {V}} = \gamma \,\underline {U} + \delta \,\underline {V} + 
a\,\left[ {\underline {V} ,\,\underline {U} } \right] + c\,\left[ 
{\underline {V} ,\,\left[ {\underline {V} ,\,\underline {U} } \right]} 
\right] \  .\tag{G.4b}
\end{gather*}

These nonlinear matrix ODEs are of course \textit{linearizable}, since their 
solution can be achieved via the following steps: (i) set (for simplicity) 
$\underline {W} \left( {0} \right) = \underline {1} $ as initial condition 
to complement (G.2); (ii) note that this entails $\underline {u} \left( {0} 
\right) = \underline {U} \left( {0} \right)$, $\underline {v} \left( {0} 
\right) = \underline {V} \left( {0} \right)$ (see (G.1a,b)); (iii) evaluate 
$\underline {u} \left( {t} \right)$ and $\underline {v} \left( {t} \right)$ 
from the (explicitly solvable) evolution equation (G.3), taking into account 
the appropriate initial conditions, see (ii); (iv) evaluate $\underline {W} 
\left( {t} \right)$ by solving the second of the (G.2), with initial 
condition $\underline {W} \left( {0} \right) = \underline {1} $, see (i) 
(note that this is a \textit{linear nonautonomous }matrix ODE, entailing the 
solutions of $M$analogous systems of $M$ linear first-order coupled 
nonautonomous ODEs -- assuming we are dealing with ($M \times M$)-matrices); 
(v) finally evaluate $\underline {U} \left( {t} \right)$ and $\underline {V} 
\left( {t} \right)$ from the second of the (G.1a,b).

Let us now derive, from the 2 \textit{first-order }ODEs (G.4) satisfied by 
the 2 matrices $\underline {U} \left( {t} \right)$ and $\underline {V} 
\left( {t} \right)$, a single \textit{second-order} ODE for one of these two 
matrices, say for $\underline {U} \left( {t} \right)$. This is easily 
obtained by time-differentiating (G.4a), thereby obtaining (using (G.4b)) 
the \textit{second-order linearizable ODE} 
\begin{gather*}
\underline {\ddot {U}} = \alpha \,\underline {\dot {U}} + \beta \,\gamma 
\,\underline {U} + \beta \,\delta \,\underline {V} + \left( {b\delta - 
a\beta } \right)\,\left[ {\underline {U} ,\,\underline {V} } \right] + 
b\,\left[ {\underline {\dot {U}} ,\,\underline {V} } \right] + \left( 
{c\delta - a\beta } \right)\,\left[ {\underline {U} ,\,\left[ {\underline 
{U} ,\,\underline {V} } \right]} \right]
\\ \qquad{}
 + c\,\left\{ {\left[ {\underline {U} ,\,\left[ {\underline {\dot {U}} 
,\,\underline {V} } \right]} \right] + \left[ {\underline {\dot {U}} 
,\,\left[ {\underline {U} ,\,\underline {V} } \right]} \right] - \beta 
\,\left[ {\left[ {\underline {U} ,\,\underline {V} } \right],\,\underline 
{V} } \right]} \right\} - c\,\{ a\,\left[ \underline {U} ,\,\left[ 
\underline{U} ,\,\left[ \underline {U} ,\,\underline {V}  \right]
\right] \right]
\\ \qquad{}
+ b\,\left[ \underline {U} ,\,\left[ \left[ \underline 
{U} ,\,\underline {V}  \right],\,\underline {V}  \right] \right]
\} - c^{2}\,\left[ \underline {U} ,\,\left[ \underline {U} ,\,\left[ \left[ 
\underline {U} ,\,\underline {V}  \right],\,\underline {V}  \right]
\right] \right] \  ,\tag{G.5a}
\end{gather*}
where the matrix $\underline {V} $ should be expressed in terms of 
$\underline {U} $ and $\underline {\dot {U}} $ by solving for $\underline 
{V} $ the (nondifferential) linear matrix equation (G.4a), namely
\begin{equation*}
\beta \,\underline {V} - b\,\left[ {\underline {V} ,\,\underline {U} } 
\right] + c\,\left[ {\left[ {\underline {V} ,\,\underline {U} } 
\right],\,\underline {U} } \right] = \underline {\dot {U}} - \alpha 
\,\underline {U} \  .\tag{G.5b}
\end{equation*}

We end this appendix by displaying the \textit{linearizable one-body problem 
}that corresponds, via the parameterization (4.3) (and by using (D.11)), to 
the linearizable matrix ODE (G.5):
\begin{gather*}
\ddot {\vec {r}} = \alpha \,\dot {\vec {r}} + \beta \,\gamma \,\vec {r} + 
\beta \,\delta \,\vec {\tilde {r}} + 2\,\left( {\beta a - b\delta } 
\right)\,\,\vec {r} \wedge \vec {\tilde {r}} + 2b\,\vec {\tilde {r}} \wedge 
\dot {\vec {r}} 
\\ \qquad{}
+ 4\left( {c\,\delta - a\,\beta } \right)\,\vec {r} \wedge 
\left( {\vec {r} \wedge \vec {\tilde {r}}} \right) + 4c\,\left\{ {\,\vec {r} 
\wedge \left( {\dot {\vec {r}} \wedge \vec {\tilde {r}}} \right) + \,\dot 
{\vec {r}} \wedge \left( {\vec {r} \wedge \vec {\tilde {r}}} \right)} 
\right\}
\\ \qquad{}
 + 4c\,\{\beta \,\,\left(\tilde{r} \wedge \vec {r} \right) \wedge 
\vec {r} + 2a\,c\,\left[ \vec{r} \wedge \left(\vec {r} \wedge \vec 
{\tilde {r}} \right) \right] \wedge \vec {r} 
\\ \qquad{}
+ 2b\,\,\vec {r} \wedge 
\left[{\vec{\tilde {r}}} \wedge \left(\vec {r} \wedge \vec {\tilde {r}} 
\right) \right] + 16c^{2}\,\vec {r} \wedge \left[\vec {r} \wedge \left( 
{\vec {\tilde{r}}} \wedge \left( \vec {\tilde {r}} \wedge \vec {r} 
\right) \right) \right]\}\ ,\tag{G.6a}
\end{gather*}
where
\begin{gather*}
\vec {\tilde {r}} = \left[ {\left( {\beta - 4c\,r^{2}} \right)^{2} + 
4b^{2}r^{2}} \right]^{ - 1}\{ {2\,b\,\,\vec {r} \wedge \dot {\vec {r}} 
+ } \left( {\beta - 4c\,r^{2}} \right)\,\,\dot {\vec {r}}
\\ \qquad{}
+ \left[ 
{4b^{2}\left( {\beta - 4c\,r^{2}} \right)^{ - 1}\left( {\vec {r} \cdot \dot 
{\vec {r}}} \right. + \left( {4c\,\chi - \alpha } \right)\,\left. {r^{2}} 
\right)} \right.
\\ \qquad{}
 + \left( {\beta - 4c\,r^{2}} \right)\left. {\left( {4c\,\chi - \alpha } 
\right)} \right]\,\,\, {\vec {r}}\}\tag{G.6b}
\end{gather*}
and
\begin{gather*}
\chi = \left[ {\left( {\beta - 4c\,r^{2}} \right)^{3} + 8b\,c\,r^{2}\left( 
{\beta - 4c\,r^{2}} \right)^{2} + 4b^{2}\beta \,r^{2}} \right]^{ - 
1}
\\ \qquad{}
\left[ 
{\left( {\beta - 4c\,r^{2}} \right)^{2} + 4b^{2}r^{2}} \right]\,\,\left( 
{\alpha \,r^{2} - \vec {r} \cdot \dot {\vec {r}}} \right).\tag{G.6c}
\end{gather*}


\section*{Appendix H. \ Solvable matrix ODEs, and a solvable many-body problem}


In this appendix we manufacture a set of \textit{solvable }matrix evolution 
equations, and we obtain the corresponding \textit{solvable }equations of 
motion for 3-vectors by applying to the matrices the parameterization (4.3).

We start from the equations
\begin{gather*}
\underline {\dot {U}} _{n} = \sum\limits_{m = 1}^{N} {\left\{ {a_{nm} 
\,\underline {V} _{m} + b_{nm} \,\underline {U} _{m} + c_{nm} \,\left[ 
{\underline {U} _{m} ,\,\underline {V} _{m} } \right]} \right\}} \  
,\tag{H.1a}
\\
\underline {\dot {V}} _{n} = \alpha _{n} \,\underline {\dot {U}} _{n} \  
.\tag{H.1b}
\end{gather*}
Here, for simplicity, we assume the $N(3\,N + 1)$ quantities $a_{nm} 
,\,b_{nm} ,\,c_{nm} ,\,\alpha _{n} $ to be \textit{time-independent 
}``coupling constants,'' although the more general case in which they are 
given functions of the time $t$ could be easily treated as well.

Firstly, let us obtain the \textit{second-order }evolution equations 
satisfied by the $N$ matrices $\underline {U} _{n} \equiv \underline {U} 
_{n} \left( {t} \right)$; these equations are \textit{solvable}, because we 
show next that the equations (H.1) can be solved by performing only 
algebraic operations.

Time-differentiation of (H.1a) yields, using (H.1b),
\begin{equation*}
\underline {\ddot {U}} _{n} = \sum\limits_{m = 1}^{N} {\left\{ {\lambda 
_{nm} \,\underline {\dot {U}} _{m} + c_{nm} \,\left[ {\underline {\dot {U}} 
_{m} ,\,\underline {V} _{m} } \right] + c_{nm} \,\alpha _{m} \,\left[ 
{\underline {U} _{m} ,\,\underline {\dot {U}} _{m} } \right]} \right\}} 
\  ,\tag{H.2a}
\end{equation*}
where we have introduced the convenient notation
\begin{equation*}
\lambda _{nm} = a_{nm} \,\alpha _{m} + b_{nm} \  .\tag{H.2b}
\end{equation*}
In (H.2a), the matrices $\underline {V} _{n} $ are supposed to be expressed 
in terms of $\underline {U} _{m} $ and $\underline {\dot {U}} _{m} $ by 
solving the (algebraic) equations (H.1a), not the (differential) equations 
(H.1b). This can always be done algebraically (up to obvious restrictions, 
see below), but hereafter we restrict attention to the simpler case 
characterized by the restriction
\begin{equation*}
a_{nm} = \tilde {d}_{nm} \,a_{m} ,\,\,\,\,\,\,c_{nm} = \tilde {d}_{nm} 
\,c_{m} \  ,\tag{H.3a}
\end{equation*}
which expresses the $2\,N^{2}$ constants $a_{nm} ,\,c_{nm} $ in terms of the 
$N\,\left( {N + 2} \right)$ (arbitrary) constants $\tilde {d}_{nm} ,\,a_{n} 
,\,c_{n} $. We moreover assume that the ($N \times N$)-matrix $\underline 
{\tilde {D}} $, with matrix elements $\tilde {d}_{nm} $, is 
\textit{invertible}, and we term $d_{nm} $ the matrix elements of the 
inverse matrix $\underline {\tilde {D}} ^{ - 1} \equiv \underline {D} $:
\begin{equation*}
\tilde {d}_{nm} \equiv \left( {\underline {\tilde {D}} } \right)_{nm} 
,\,\,\,\,\,\,d_{nm} \equiv \left( {\underline {\tilde {D}} ^{ - 1}} 
\right)_{nm} \equiv \left( {\underline {D} } \right)_{nm} \  
.\tag{H.3b}
\end{equation*}
Then clearly from (H.1a) we get
\begin{equation*}
a_{n} \,\underline {V} _{n} + c_{n} \,\left[ {\underline {U} _{n} 
,\,\underline {V} _{n} } \right] = \sum\limits_{m = 1}^{N} {\left\{ {d_{nm} 
\,\left[ {\underline {\dot {U}} _{m} - \sum\limits_{{m}' = 1}^{N} {b_{m{m}'} 
\,\underline {U} _{{m}'} } } \right]} \right\}} \  ,\tag{H.4}
\end{equation*}
a linear (nondifferential!) equation for the matrices $\underline {V} _{n} 
$which can be conveniently solved in explicit form [6] (see below).

Let us now show how to solve (H.2), or rather, equivalently, (H.1). From 
(H.1b) we get
\begin{equation*}
\underline {V} _{n} \left( {t} \right) = \alpha _{n} \,\underline {U} _{n} 
\left( {t} \right) + \underline {C} _{n} \  ,\tag{H.5a}
\end{equation*}
with the \textit{constant }matrices $\underline {C} _{n} $ given in terms of 
the initial data:
\begin{equation*}
\underline {C} _{n} = \underline {V} _{n} \left( {0} \right) - \alpha _{n} 
\,\underline {U} _{n} \left( {0} \right) \  .\tag{H.5b}
\end{equation*}

Insertion of (H.5a) in (H.1a) yields the following set of \textit{linear 
ODEs with constant coefficients }for the matrices $\underline {U} _{n} 
\left( {t} \right)$, which can of course be explicitly solved by purely 
algebraic operations:
\begin{equation*}
\underline {\dot {U}} _{n} = \sum\limits_{m = 1}^{N} {\left\{ {a_{nm} 
\,\left[ {\alpha _{m} \,\underline {U} _{m} + \underline {C} _{m} } \right] 
+ b_{nm} \,\underline {U} _{m} + c_{nm} \,\left[ {\underline {U} _{m} 
,\,\underline {C} _{m} } \right]} \right\}} \  .\tag{H.6}
\end{equation*}

Next, let us exhibit the \textit{solvable equations of motions }which are 
obtained by applying the parameterization (4.3) to the \textit{solvable 
}matrix evolution equations (H.2a) with (H.4) (we leave as an exercise for 
the diligent reader the derivation of the more general equations that obtain 
via the parameterization (4.1) rather than (4.3)). First of all we note that 
(H.4) can now be rewritten as follows: 
\begin{equation*}
a_{n} \,\vec {v}_{n} + \gamma _{n} \,\vec {v}_{n} \wedge \vec {r}_{n} = \vec 
{w}_{n} \  ,\tag{H.7a}
\end{equation*}
where we set, for notational convenience,
\begin{equation*}
\gamma _{n} = 2\,c_{n} \  ,\tag{H.7b}
\end{equation*}
as well as
\begin{equation*}
\vec {w}_{n} = \sum\limits_{m = 1}^{N} {\left\{ {d_{nm} \,\left[ {\dot {\vec 
{r}}_{m} - \sum\limits_{{m}' = 1}^{N} {b_{m{m}'} \,\vec {r}_{{m}'} } } 
\right]} \right\}} \  ,\tag{H.7c}
\end{equation*}
and of course $\underline {U} _{n} = i\,\vec {r}_{n} \cdot \underline {\vec 
{\sigma }} ,\,\,\underline {V} _{n} = i\,\vec {v}_{n} \cdot \underline {\vec 
{\sigma }} $. Hence, using (D.11),
\begin{equation*}
\vec {v}_{n} = \left[ {\gamma _{n} ^{2}\,\vec {r}_{n} \,\left( {\vec {r}_{n} 
\cdot \vec {w}_{n} } \right) + a_{n} ^{2}\,\vec {w}_{n} + a_{n} \,\gamma 
_{n} \,\vec {r}_{n} \wedge \vec {w}_{n} } \right]/\left[ {a_{n} \,\left( 
{a_{n} + \gamma _{n} ^{2}\,\,r_{n} ^{2}} \right)} 
\right]\ .\tag{H.7d}
\end{equation*}
On the other hand, via the parameterization (4.3), the \textit{solvable 
}matrix evolution equations (H.2a) read
\begin{equation*}
\ddot {\vec {r}}_{n} = \sum\limits_{m = 1}^{N} {\left\{ {\lambda _{nm} 
\,\dot {\vec {r}}_{m} - \tilde {d}_{nm} \,\gamma _{m} \,\left[ {\dot {\vec 
{r}}_{m} \wedge \vec {v}_{m} + \alpha _{m} \,\vec {r}_{m} \wedge \dot {\vec 
{r}}_{m} } \right]} \right\}} \  ,\tag{H.8}
\end{equation*}
where we used (H.3a) and (H.7b). Here the $N$ three-vectors $\vec {v}_{n} 
$should be expressed via (H.7d,c). 

Let us write in more explicit form these equations of motion in the special 
case in which the matrix $\underline {D} $ is diagonal, indeed when we set 
(without any additional loss of generality) 
\begin{equation*}
\underline {D} = \underline {\tilde {D}} = \underline {1} ,\,\,\,\,\,d_{nm} 
= \tilde {d}_{nm} = \delta _{nm} \  .\tag{H.9}
\end{equation*}
We thus get the following \textit{solvable many-body problem}, which only 
features two-body (velocity-dependent) forces, and contains the $N\,\left( 
{N + 3} \right)$ coupling constants $a_{n}$, $\alpha _{n}$, $\gamma _{n} 
$, $b_{nm}$, which are arbitrary except for the restrictions $a_{n} \ne 0$ 
(see the following equations of motion):
\begin{gather*}
 \ddot{\vec {r}}_{n} = a_{n} \,\dot {\vec{r}}_{n} + \alpha_{n} \,\gamma 
_{n} \,\vec{r}_{n} \wedge \dot {\vec {r}}_{n} + \{ - \gamma 
_{n} \,\dot {\vec {r}}_{n} \wedge \left[ \gamma_{n}^{2}\,\left( \vec 
{r}_{n} \cdot \dot {\vec {r}}_{n}  \right)\,\vec{r}_{n} + a_{n} \,\gamma 
_{n} \,\vec{r}_{n} \wedge \dot{\vec {r}}_{n} \right]
\\ \qquad{}
- \sum\limits_{m = 1}^{N} \{ b_{nm} \,\{ \alpha _{m} \,\dot 
{\vec {r}}_{m} - \gamma_{n} \,\dot{\vec {r}}_{n} \wedge \left[ \gamma 
_{n}^{2}\,\left( \vec{r}_{n} \cdot \vec{r}_{m} \right)\,\vec{r}_{n} + 
a_{n}^{2}\,\,\vec{r}_{m} + a_{n} \,\gamma_{n} \,\vec{r}_{n} \wedge \vec {r}_{m}  
\right] \}\}\}/
\\ \qquad{}
\left[ a_{n} 
\,\left( a_{n}^{2} + \gamma_{n}^{2}\,\,r_{n}^{2} \right) \right] \ .
\tag{H.10}
\end{gather*}


\subsection*{Acknowledgements}

This research was initiated in collaboration with our colleague and 
friend Orlando Ragnisco. Unfortunately the pressure of other duties prevented him 
from taking an active part in the detailed development of the results 
reported herein, and he therefore preferred not to sign this paper, although 
he did participate in the elaboration of the main ideas contained in it. It 
is a pleasure for us to acknowledge his contribution, with the hope that he 
will be able to take a more active role in the developments of these 
results, to be reported in subsequent papers of this series.

The results reported above were partly obtained while one of us (FC) was 
visiting the University Paris VI. He would like to thank his host there, 
professor Jean-Pierre Francoise, not only for the gracious hospitality, but 
as well for several inspiring discussions, which have been instrumental in 
obtaining some of these results.

Finally, we would like to acknowledge that a number of useful ideas were 
triggered by fruitful discussions with several colleagues, including in 
particular M. Ablowitz, E. V. Ferapontov, M. Kruskal, S. V. Manakov, A. V. 
Mikhailov, A. B. Shabat and, last but not least, V. V. Sokolov, to all of 
whom we would like to express our thanks, while emphasizing that we take 
exclusive responsibility for all shortcomings of this paper.


\label{lastpage}

\end{document}